\documentclass[11pt, a4paper, oneside]{Thesis} % Paper size, default font size and one-sided paper

\graphicspath{{./Pictures/}} % Specifies the directory where pictures are stored

\usepackage[square, numbers, comma, sort&compress]{natbib} % Use the natbib reference package - read up on this to edit the reference style; if you want text (e.g. Smith et al., 2012) for the in-text references (instead of numbers), remove 'numbers' 
\hypersetup{urlcolor=blue, colorlinks=true} % Colors hyperlinks in blue - change to black if annoying
\title{\ttitle} % Defines the thesis title - don't touch this

\usepackage{epigraph}
\makeatletter
\renewcommand{\@chapapp}{}% Not necessary...
\newenvironment{chapquote}[2][2em]
  {\setlength{\@tempdima}{#1}%
   \def\chapquote@author{#2}%
   \parshape 1 \@tempdima \dimexpr\textwidth-2\@tempdima\relax%
   \itshape}
  {\par\normalfont\hfill--\ \chapquote@author\hspace*{\@tempdima}\par\bigskip}
\makeatother

\newcommand{\cd}{{\cal D}}
\renewcommand{\[}{\begin{equation}}
\renewcommand{\]}{\end{equation}}
\newcommand{\fnl}{f_{\rm NL}}

\renewcommand{\[}{\begin{equation}}
\renewcommand{\]}{\end{equation}}
\def\beq{\begin{equation}}
\def\eeq{\end{equation}}
\newcommand{\be}{\begin{eqnarray}}
\newcommand{\ee}{\end{eqnarray}}

\renewcommand{\texttt}{{}}

\def\bs{\begin{subequations}}
\def\es{\end{subequations}}

\def\Fc{\mathcal{F}}

\def\Tc{\mathcal{T}}

\newcommand{\tia}[1]{}

\newcommand{\bea}{\begin{eqnarray}}
\newcommand{\eea}{\end{eqnarray}}
\newcommand{\beas}{\begin{eqnarray*}}
\newcommand{\eeas}{\end{eqnarray*}}
\newcommand{\bal}{\begin{aligned}}
\newcommand{\eal}{\end{aligned}}

\def\({\left(}
\def\){\right)}
\newcommand{\pd}{\partial}

\newcommand{\const}{\mathrm{const}}

\begin{document}

\frontmatter % Use roman page numbering style (i, ii, iii, iv...) for the pre-content pages

\setstretch{1.3} % Line spacing of 1.3

% Define the page headers using the FancyHdr package and set up for one-sided printing
\fancyhead{} % Clears all page headers and footers
\rhead{\thepage} % Sets the right side header to show the page number
\lhead{} % Clears the left side page header

\pagestyle{fancy} % Finally, use the "fancy" page style to implement the FancyHdr headers

\newcommand{\HRule}{\rule{\linewidth}{0.5mm}} % New command to make the lines in the title page

% PDF meta-data
\hypersetup{pdftitle={\ttitle}}
\hypersetup{pdfsubject=\subjectname}
\hypersetup{pdfauthor=\authornames}
\hypersetup{pdfkeywords=\keywordnames}

%----------------------------------------------------------------------------------------
%	TITLE PAGE
%----------------------------------------------------------------------------------------
%%%%%%%%%%%% Pagina rosto-UBI style 
%Folha de rosto
\newcommand{\rostoubi}{\fontsize{14pt}{14pt}\selectfont}				%Texto a dizer UBI 14pt, normal
\newcommand{\rostotitulo}{\fontsize{18pt}{18pt}\selectfont}				%Titulo da tese: 18pt, (+negrito)
\newcommand{\rostosubtit}{\fontsize{16pt}{16pt}\selectfont}				%Titulo da tese: 16pt, (+negrito)
\newcommand{\rostonomes}{\fontsize{14pt}{14pt}\selectfont}				%Nome autor, curso: 14pt, (+negrito)
\newcommand{\rostooutros}{\fontsize{12pt}{12pt}\selectfont}				%Local e data: 12pt, (+negrito)
\newcommand{\rostofac}{\fontsize{12.5pt}{12.5pt}\selectfont}

%%%%%%%%%%%%%% For text color
\newcommand{\tcb}{\textcolor{blue}}
\newcommand{\tcr}{\textcolor{red}}
\newcommand{\tcm}{\textcolor{magenta}}

\begin{titlepage}
\begin{center}

\begin{flushleft}
\includegraphics[height=2.22cm]{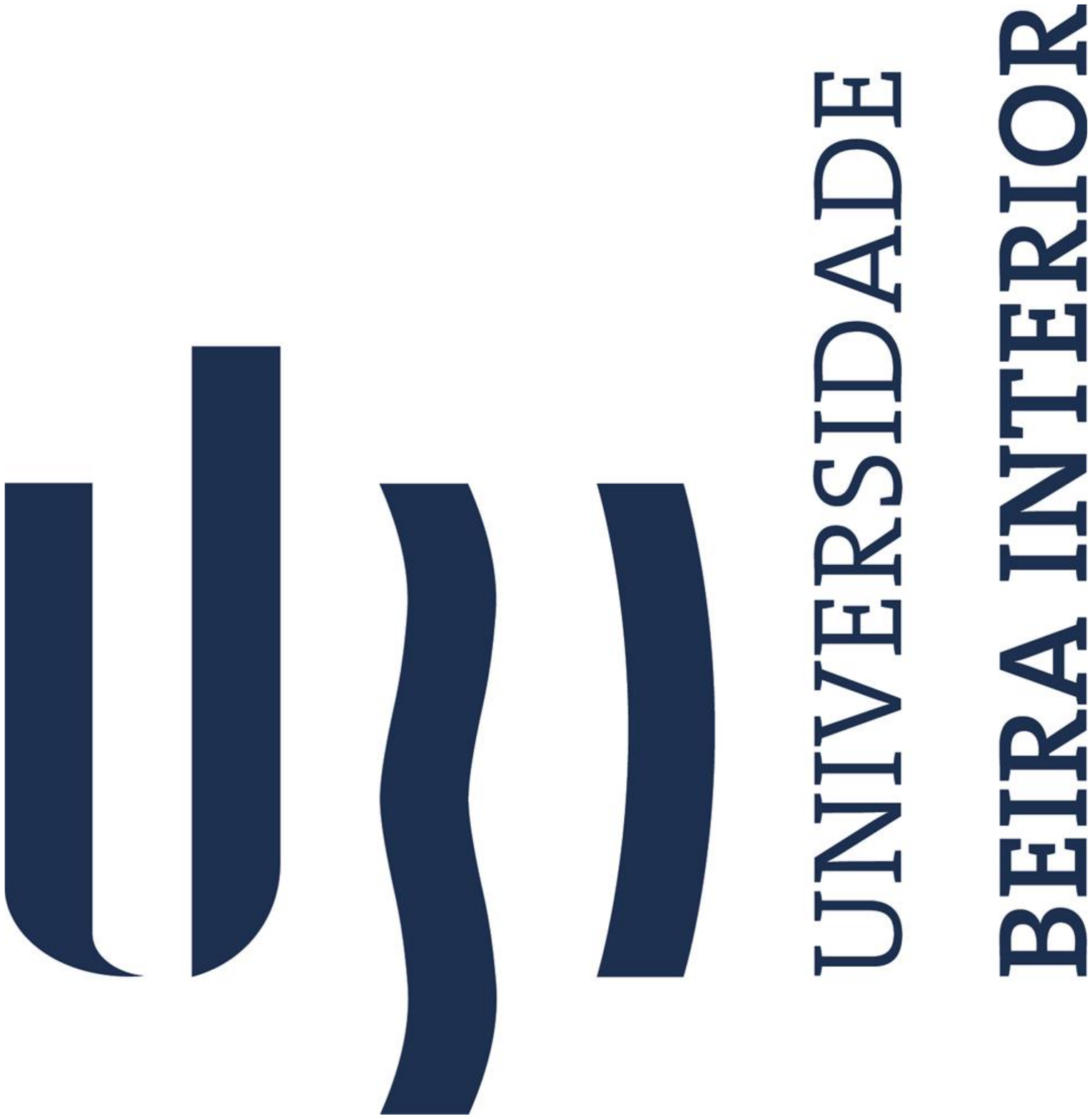}\\
\rostoubi UNIVERSIDADE DA BEIRA INTERIOR\\
\rostofac Ci\^encias\\
\end{flushleft}

\vspace{7.6cm}

\rostotitulo \textbf{Inflaton candidates:} \\
\rostosubtit \textbf{from string theory to particle physics}\\

\vspace{1.8cm}

\rostonomes \textbf{Sravan Kumar Korumilli}\\

\vspace{1.4cm}

\rostooutros Tese para obten\c{c}\~ao do Grau de Doutor em\\
\rostonomes \textbf{F\'{i}sica}\\
\rostooutros ($3^{\text{o}}$ ciclo de estudos)\\

\vspace{3.3cm}

\rostooutros Orientador: Prof. Doutor Paulo Vargas Moniz\\
%Co-orientador: Prof. Doutor Nome\\

\vspace{1.4cm}

\rostooutros \textbf{Covilh\~a, Maio 2017}

\end{center}
\end{titlepage}

\clearpage % Start a new page

%----------------------------------------------------------------------------------------
%	GRANT PAGE
%----------------------------------------------------------------------------------------

\pagestyle{empty} % No headers or footers for the following pages

\null\vfill % Add some space to move the quote down the page a bit

The research work on this dissertation was supported by the PhD scholarship from the International Doctorate Network in Particle Physics, Astrophysics and Cosmology (IDPASC) funded by 
Portuguese agency Funda\c{c}\~ao para a Ci\^encia e Tecnologia (FCT) through
the fellowship SFRH/BD/51980/2012.

\vfill\vfill\vfill\vfill\vfill\vfill\null 
%\begin{tabular}[t]\input{../../../../../../../Desktop}

%\begin{minipage}[c]{1\textwidth}%
\begin{center}
\includegraphics[width=1.9in]{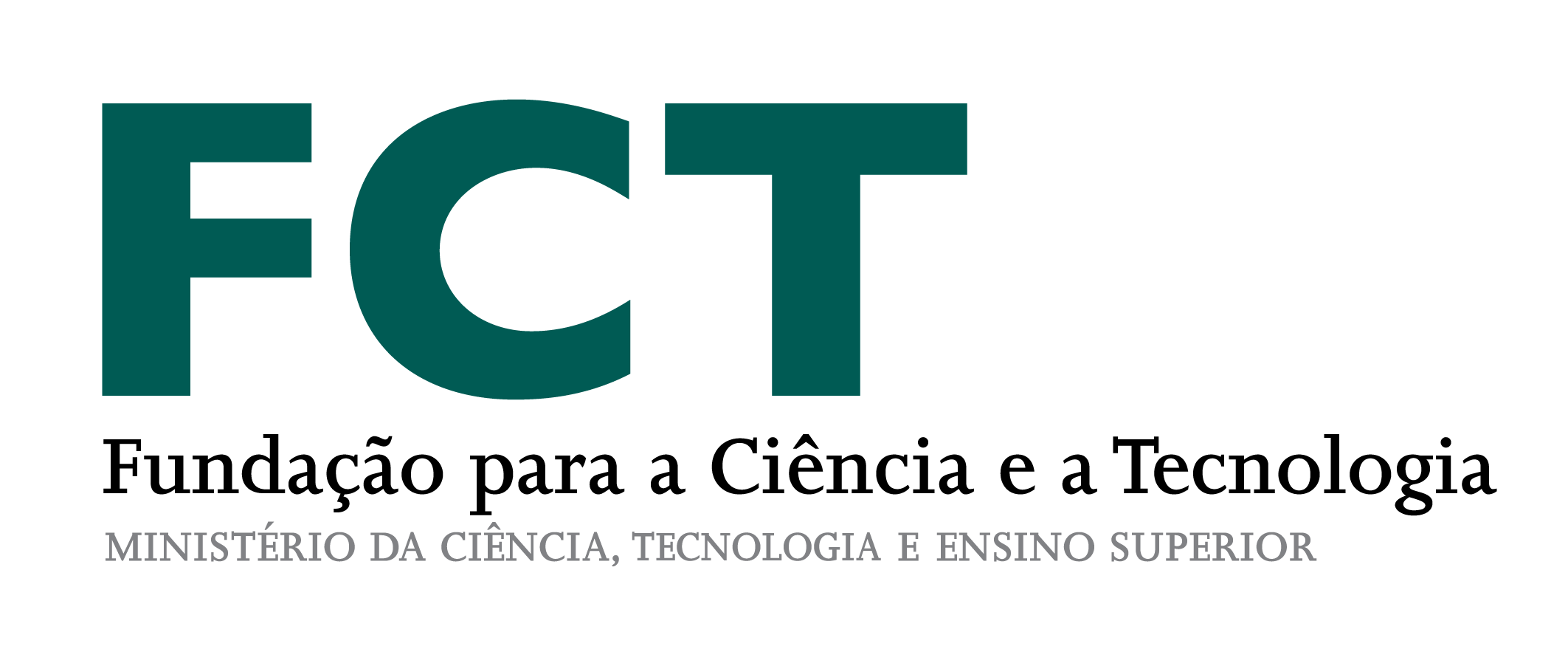} \quad{} \includegraphics[width=1.9in]{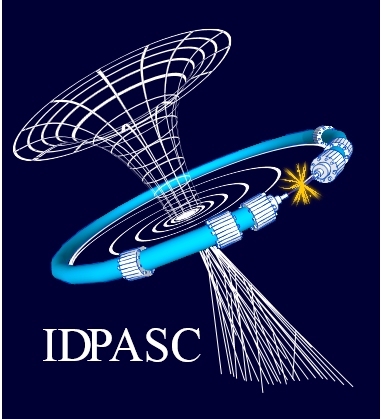} \quad{} \includegraphics[width=1.9in]{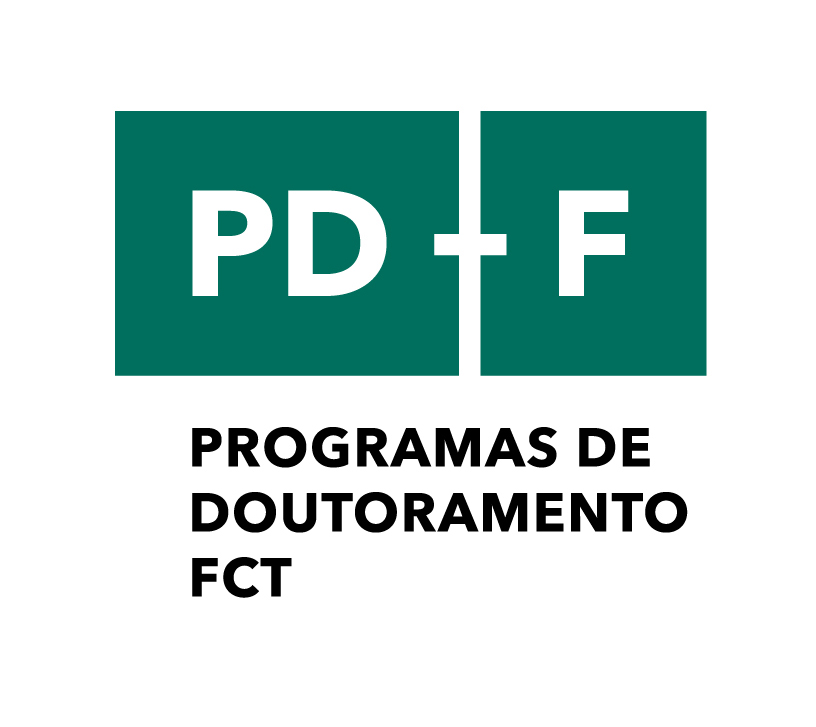}  \quad{} \includegraphics[width=1.9in]{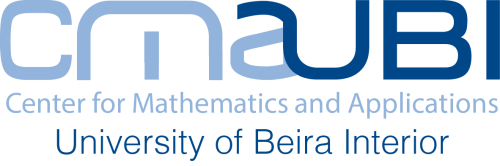}  %\quad{} %\includegraphics[width=1.2in]{Figures/LogoFondBlanc.pdf}
\par
\end{center}%

 % Start a new page

%% new page 
\newpage 
$\quad$
\vspace{5.5cm}
\vspace{0.5cm}
\begin{quote}
	\centering\it{I don't want to believe. I want to know}\\ \hspace{4cm} - Carl Sagan
\end{quote}
\begin{figure}[h!]
	\centering\includegraphics[width=3in]{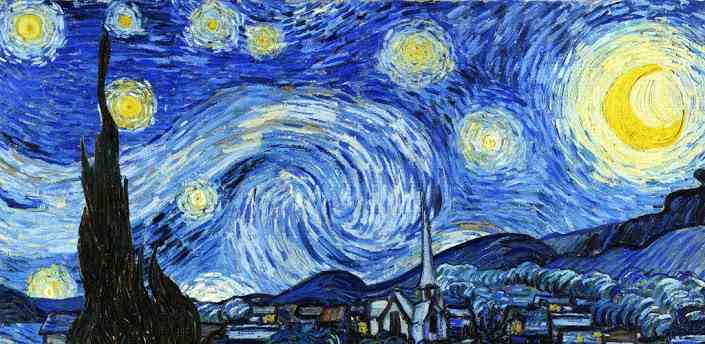}\caption*{\tiny{The Starry Night by Vincent Van Gogh, 1889.}}
	\label{vangoph}
\end{figure}

\clearpage
\pagestyle{empty}
%----------------------------------------------------------------------------------------
%	ABSTRACT PAGE
%----------------------------------------------------------------------------------------
%\pagestyle{empty}

%\addtotoc{Abstract} % Add the "Abstract" page entry to the %Contents

%% English abstract 
\section*{Abstract}

Cosmic inflation is the cornerstone of modern cosmology. In particular, following the {\it Planck} mission reports presented in 2015 regarding cosmic microwave background (CMB), there is an increasing interest in searching for inflaton candidates within fundamental theories and to ultimately test them with future CMB data. This thesis presents inflationary models 
using a methodology that can be described as venturing top-down or bottom-up along energy scales. In the top-down motivation, we study inflationary scenarios in string theory and supergravity (SUGRA), namely with (multiple) 3-forms, Dirac-Born-Infeld Galileon model, a string field theory setup and $\mathcal{N}=1$ SUGRA $\alpha-$attractor models. In the bottom-up motivation, we construct a grand unified theory based inflationary model with an additional conformal symmetry and study not only inflation but also provide predictions related to particle physics. Our research work includes various classes of inflation driven by scalar fields under a canonical, non-canonical and induced gravity frameworks. All these models are consistent with Planck data, supported by key primordial cosmological parameters such as the scalar spectral index $n_{s}$, the tensor to scalar ratio $r$, together with the primordial non-Gaussianities. Future probes aiming to detect primordial gravitational waves and CMB non-Gaussianities can further help to distinguish between them.

\paragraph*{Keywords:}

%\vspace{10mm} {\fontsize{18}{16} \textbf{Keywords}}
%\vspace{5mm}

Inflation, String theory, Supergravity, Grand unified theories, Primordial gravitational waves, Non-gaussianities.

\clearpage

%----------------------------------------------------------------------------------------
%	LIST OF PUBLICATION
%----------------------------------------------------------------------------------------

\pagestyle{empty} % No headers or footers for the following pages

\section*{List of publications}

\noindent This dissertation is based on the following publications:
\begin{itemize} 
\item[\tiny{$\blacksquare$}] K.~Sravan~Kumar, J.~Marto, N.~J. Nunes, and P.~V. Moniz, ``{Inflation in a two
  3-form fields scenario},''
  \href{http://dx.doi.org/10.1088/1475-7516/2014/06/064}{{\em JCAP} {\bfseries
  1406} (2014) 064},
\href{http://arxiv.org/abs/1404.0211}{{\ttfamily arXiv:1404.0211 [gr-qc]}}.
\item[\tiny{$\blacksquare$}] K.~Sravan~Kumar, J.~C. Bueno~S\'anchez, C.~Escamilla-Rivera, J.~Marto, and
  P.~Vargas~Moniz, ``{DBI Galileon inflation in the light of Planck 2015},''
  \href{http://dx.doi.org/10.1088/1475-7516/2016/02/063}{{\em JCAP} {\bfseries
  1602} no.~02, (2016) 063},
\href{http://arxiv.org/abs/1504.01348}{{\ttfamily arXiv:1504.01348
  [astro-ph.CO]}}.
\item[\tiny{$\blacksquare$}]  K.~Sravan~Kumar, J.~Marto, P.~Vargas~Moniz, and S.~Das, ``{Non-slow-roll dynamics
  in $\alpha-$attractors},''
  \href{http://dx.doi.org/10.1088/1475-7516/2016/04/005}{{\em JCAP} {\bfseries
  1604} no.~04, (2016) 005},
\href{http://arxiv.org/abs/1506.05366}{{\ttfamily arXiv:1506.05366 [gr-qc]}}.
\item[\tiny{$\blacksquare$}]  K.~Sravan~Kumar, J.~Marto, P.~Vargas~Moniz, and S.~Das, 
\href{http://dx.doi.org/10.1142/9789813226609_0571}{``{Gravitational waves in
		$\alpha-$attractors},''} in {\em {Proceedings, 14th Marcel Grossmann Meeting
		on Recent Developments in Theoretical and Experimental General Relativity,
		Astrophysics, and Relativistic Field Theories (MG14) (In 4 Volumes): Rome,
		Italy, July 12-18, 2015}}, vol.~4, pp.~4262--4267.
\newblock 2017.
\newblock \href{http://arxiv.org/abs/1512.08490}{{\ttfamily arXiv:1512.08490
  [gr-qc]}}.
\item[\tiny{$\blacksquare$}]  A.~S. Koshelev, K.~Sravan~Kumar, and P.~Vargas~Moniz, ``{Inflation from string
  field theory},''
\href{https://journals.aps.org/prd/abstract/10.1103/PhysRevD.96.103503}{Phys.Rev. D96 (2017) no.10, 103503},\href{http://arxiv.org/abs/1604.01440}{{\ttfamily arXiv:1604.01440 [hep-th]}}.
\item[\tiny{$\blacksquare$}]  K.~Sravan~Kumar, D.~J. Mulryne, N.~J. Nunes, J.~Marto, and P.~Vargas~Moniz,
  ``{Non-Gaussianity in multiple three-form field inflation},''
\href{http://journals.aps.org/prd/abstract/10.1103/PhysRevD.94.103504}{Phys. Rev. D 94, 103504 (2016)},\href{http://arxiv.org/abs/1606.07114}{arXiv:1606.07114 [astro-ph.CO]}.
  
\item[\tiny{$\blacksquare$}]  K.~Sravan~Kumar and P.~Vargas~Moniz, ``{Conformal GUT inflation, proton life time and non-thermal leptogenesis},'' \href{http://arxiv.org/abs/1806.09032}{arXiv:arXiv:1806.09032 [astro-ph.CO]}
%\item[\tiny{$\blacksquare$}]  To appear
\end{itemize}

\noindent Other publications during this doctoral studies:
\begin{itemize} 
\item[\tiny{$\blacksquare$}] A.~Burgazli, A.~Zhuk, J.~Morais, M.~Bouhmadi-L\'opez and K.~Sravan~Kumar, ``{Coupled scalar fields
  in the late Universe: The mechanical approach and the late cosmic
  acceleration},'' \href{http://dx.doi.org/10.1088/1475-7516/2016/09/045}{{\em
  JCAP} {\bfseries 1609} no.~09, (2016) 045},
\href{http://arxiv.org/abs/1512.03819}{{\ttfamily arXiv:1512.03819 [gr-qc]}}.
\item[\tiny{$\blacksquare$}] M.~Bouhmadi-L\'opez, K.~S. Kumar, J.~Marto, J.~Morais, and A.~Zhuk,
  ``{$K$-essence model from the mechanical approach point of view: coupled
  scalar field and the late cosmic acceleration},''
  \href{http://dx.doi.org/10.1088/1475-7516/2016/07/050}{{\em JCAP} {\bfseries
  1607} no.~07, (2016) 050},
\href{http://arxiv.org/abs/1605.03212}{{\ttfamily arXiv:1605.03212 [gr-qc]}}.
\item[\tiny{$\blacksquare$}] J.~Morais, M.~Bouhmadi-L\'opez, K.~Sravan~Kumar, J.~Marto, and
  Y.~Tavakoli, ``{Interacting 3-form dark energy models: distinguishing
  interactions and avoiding the Little Sibling of the Big Rip},''
  \href{http://dx.doi.org/10.1016/j.dark.2016.11.002}{{\em Phys. Dark Univ.}
  {\bfseries 15} (2017) 7--30},
\href{http://arxiv.org/abs/1608.01679}{{\ttfamily arXiv:1608.01679 [gr-qc]}}.
\item[\tiny{$\blacksquare$}]  Shreya Banerjee, Suratna Das, K.~Sravan~Kumar and T.P. Singh, ``{On signatures of spontaneous collapse dynamics modified single field inflation},'', \href{https://journals.aps.org/prd/abstract/10.1103/PhysRevD.95.103518}{Phys. Rev. D 95, 103518 (2017)}
\href{https://arxiv.org/abs/1612.09131}{{\ttfamily arXiv:1612.09131 [astro-ph.CO]}}.
\item[\tiny{$\blacksquare$}]  Alexey S.~Koshelev, K.~Sravan Kumar, Alexei A. Starobinsky ``{$R^2$ inflation to probe non-perturbative quantum gravity},'' \href{https://link.springer.com/article/10.1007%2FJHEP03%282018%29071}{JHEP 1803 (2018) 071}
\href{https://arxiv.org/abs/1711.08864}{{\ttfamily arXiv:1711.08864 [hep-th]}}.
\end{itemize}

%----------------------------------------------------------------------------------------
%	RESUMO
%----------------------------------------------------------------------------------------
\clearpage
\pagestyle{empty} % No headers or footers for the following pages

\section*{Resumo}

A infla\c{c}\~ao constitui um paradigma essencial na cosmologia moderna. Em particular, e de acordo com os 
comunicados da missão {\it Planck} em 2015, acerca da medi\c{c}\~ao da radia\c{c}\~ao c\'osmica de fundo,
h\'a um interesse crescente na procura de candidatos a inflat\~ao extra\'idos de teorias fundamentais e em testar estas propostas.
Esta tese apresenta modelos inflacion\'arios que podem ser classificados numa abordagem descendente ou ascendente nas escalas de energia.
Na abordagem descendente, apresentamos estudos de cen\'arios inflacion\'arios ligados \`a teoria de cordas e \`a supergravidade (SUGRA),
seja com campos (m\'ultiplos) 3-formas, com o modelo Dirac-Born-Infeld Galileon, no contexto de uma teoria de campos para cordas ou ainda
no modelo $\alpha-$atrator SUGRA $\mathcal{N}=1$. Na abordagem ascendente, propomos a constru\c{c}\~ao de um modelo inflacion\'ario baseado numa teoria de grande unifica\c{c}\~ao, complementada com uma simetria conforme, em que estudamos, n\~ao s\'o a infla\c{c}\~ao, mas tamb\'em implica\c{c}\~oes no campo da f\'isica de part\'iculas. O nosso trabalho de investiga\c{c}\~ao inclui diferentes classes de infla\c{c}\~ao governadas por campos escalares can\'onicos, n\~ao can\'onicos ou ainda em contexto de gravidade induzida. A totalidade destes modelos \'e consistente com os dados obtidos na miss\~ao {\it Planck} e suportados por par\^ametros cosmol\'ogicos cruciais como o \'indice espectral escalar $n_{s}$, a raz\~ao tensor para escalar $r$ ou ainda a n\~ao-Gaussianidade primordial. O estudo abordado nesta tese refor\c{c}a a espectativa que futuras miss\~oes observacionais, cujo objectivo seja detetar ondas gravitacionais primordiais e a n\~ao-Gaussianidade da radia\c{c}\~ao c\'osmica de fundo, possam ajudar a melhor distinguir os modelos inflacion\'arios
considerados.
 
%{\small 

%% Write abstract in Portuguese 

\paragraph*{Palavras-chave:}

Infla\c{c}\~ao, Teoria de cordas, Supergravidade, Teoria de grande unifica\c{c}\~ao, Ondas gravitacionais primordiais, N\~ao-Gaussianidades.

%% Repeat those that were in english previously
%}

\clearpage

%----------------------------------------------------------------------------------------
%	ACKNOWLEDGEMENTS
%----------------------------------------------------------------------------------------

\setstretch{1.1} % Reset the line-spacing to 1.3 for body text (if it has changed)

\acknowledgements{\addtocontents{toc}{\vspace{1em}} % Add a gap in the Contents, for aesthetics
First, my sincere thanks to the Jury members of this thesis Dr. J. P. Mimoso, Dr. J. G. Rosa, Dr. N. J. Nunes, Dr. C. A. R. Herdeiro, Dr. J. P.  Marto, Dr. P. A. P. Parada and Dr. P. R. L. V. Moniz.

My sincere and deepest gratitude to my PhD supervisor Prof.~Paulo Vargas Moniz for the guidance, patience and invaluable suggestions. I am grateful for all the efforts he has done introducing me to the exciting area of theoretical cosmology. 

I am greatly thankful and indebted to Dr.~Jo\~ao Marto for his very generous and abundant help in all aspects during my PhD. I always enjoyed very fruitful scientific discussions with him and also heavily benefitted in learning numerical techniques such as $Mathematica$ from him. Thanks you for invaluable help in writing this dissertation. 

I am specially thankful to Dr.~Alexey S. Koshelev, Dr.~Mariam Bouhmadi L\'opez, Dr.~Suratna Das, Dr.~Yaser Tavakoli and Dr.~Paulo Parada for reading and providing valuable feedbacks during the writing of this thesis. I am very thankful to Dr.~Yaser Tavakoli for helping me with \LaTeX ~program during the writing of this thesis. 

During my PhD I greatly benefited from invaluable discussions with many scientists. In particular, I am grateful to Dr.~Nelson J. Nunes for very fruitful collaboration. I learned a lot from him during my doctoral research. My sincere gratitude to Dr.~Mariam Bouhmadi L\'opez for her excellent support, collaboration and for introducing me to the research on late time cosmology. I am very thankful to  Prof.~Alexander Zhuk for the opportunity to work with him during his visit to UBI. I am indebted to Dr.~Alexey S. Koshelev with whom I have been having very delighted scientific discussions and wonderful collaboration. I am thankful to him for introducing me to the exciting area of non-local gravity. I am very grateful to Prof.~Alexei A. Starobinsky for the opportunity to work with him during PhD. I am grateful to Prof.~Qaisar shafi for illuminating me on particle physics and cosmology. I am thankful to Prof.~Anupam  Mazumdar and Prof. Leonardo Modesto for the very useful discussions and on going exciting collaborations. I thank Dr.~C. Pallis and Dr.~John Ward for useful comments on the drafts of my work. 

I thank Dr.~David J. Mulryne,  Dr.~Suratna Das, Dr.~Celia Escamilla-Rivera, Dr.~Juan C. Bueno S\'anchez, Prof.~T.P. Singh, Shreya Banerjee, Jo\~ao Morais and A.~Burgazli for very interesting dicussions and collaborations during these years. I acknowledge Dr.~S. M. M. Rasouli for his help during my PhD. I also thank Dr.~C\'esar Silva for useful discussions. 

I acknowledge the hospitality of Institute of astrophysics (IA) at University of Lisbon, Department of Theoretical Physics of University of Basque country (UPV/EHU) and Indian Institute of Technology (IIT) Kanpur, where part of my research during PhD has been carried out. 

I greatly acknowledge Center of Mathematics and Applications (CMA-UBI) for the support and encouragement to attend conferences and present my research. I am also very delighted to express my gratitude to the people in department of physics of UBI, especially grateful to Dr.~Jos\'e Amoreira for his moral support and unforgettable help in many aspects. I also acknowledge the support of Dr. Fernando Ferreira, Dr. Jorge Maia and Dr. Jose Velhinho in different times. Last but not the least I am very much thankful to the secretaries of physics, mathematics, faculty of sciences and Vice-reitoria, Ms. Dulce H. Santos, Ms. Filipa Raposo and Ms. Cristina Gil and also Ms. Paula Fernandes, respectively for their help in administrative tasks. 

The seed for desire of pursuing PhD dates back to the time of my masters at University of Hyderabad (UoH). I thank all my wonderful teachers and high energy physics (HEP) group at UoH. I am highly grateful to Prof.~E. Harikumar for inspiring on HEP and being always supportive with valuable suggestions. I am so grateful to Prof.~M. Sivakumar from whom I learned a lot and got trained in doing in theoretical physics. I greatly acknowledge my teacher Prof.~A. K. Kapoor for imparting his great knowledge and insights in quantum and mathematical physics. I am also very thankful to Prof.~Bindu A. Bambah, Prof.~Rukmani Mohanta, Prof.~Ashok Chaturvedi, Prof.~P.K. Suresh and Dr. Soma Sanyal for motivating me towards HEP. My great thanks to Prof.~Debajyoti Choudhury, Dr.~Sukanta Dutta and Dr.~Mamta Dahia for keeping me excited on HEP during my summer project period at University of Delhi.  

A very heartfelt thanks to my friends A. Narsireddy, Dr.~Shyam Sumanta Das, R. Srikanth, Dr.~K.N. Deepthi, Imanol Albarran, Francisco Cabral and Jo\~ao Morais who have been very supportive and encouraging in many occasions including tough times during PhD. We shared very good memories during my stay in Portugal. Very special thanks to G. Trivikram, A. Murali for their encouraging friendship. Thanks to Dr.~K.N. Deepthi for valuable suggestions in the difficult times. I am thankful to all my friends in UoH, also to my school and college friends for their encouragement.  

It would have been impossible for me to pursue a career in physics without the support of my teachers in school and college. I especially acknowledge my beloved teacher Mr.~P.~Ramamurthy for the immense help, motivation and inspiration in the early years.

It would have been impossible to pursue my PhD here in Portugal being so far away from home without the substantial support and blessings of my family. In particular, my heartfelt gratitude to my cousins G. Subbarao, G. S. Ranganath, V. V. S. R. N. Raju and N. Kameswara Rao who were in numerous occasions helped me throughout my life. I owe a lot to them for so much believing in me. I am whole heartedly thankful to my aunts G. Nagamani, V. Subbalakshmi and to my uncle Dr. S. Sriramachandra Murthy and his family for their unconditional love and affection on me. I am especially grateful to my uncle V. Badarinadh who has been motivating, encouraging and supportive in crucial parts of my life and career. 
I miss my uncle (late) S. Nooka Raju who would have been very happy now with my thesis.

I thank each and every person who were directly or indirectly helped and influenced to pursue my studies. 

Last but not the least my beloved parents, my mother Padmavathi Devi, my father Radhakrishna whose love, affection, patience and support always been enormous strength for me to flourish in life and career. 

I dedicate my work to my grandparents Korumilli Krishnaveni and Korumilli Sanyasiraju who unfortunately passed away recently. 
}

\clearpage % Start a new page

%----------------------------------------------------------------------------------------
%	LIST OF CONTENTS/FIGURES/TABLES PAGES
%----------------------------------------------------------------------------------------

\pagestyle{fancy} % The page style headers have been "empty" all this time, now use the "fancy" headers as defined before to bring them back

\lhead{\emph{Contents}} % Set the left side page header to "Contents"
\tableofcontents % Write out the Table of Contents

\lhead{\emph{List of Figures}} % Set the left side page header to "List of Figures"
\listoffigures % Write out the List of Figures

\lhead{\emph{List of Tables}} % Set the left side page header to "List of Tables"
\listoftables % Write out the List of Tables

%----------------------------------------------------------------------------------------
%	ABBREVIATIONS
%----------------------------------------------------------------------------------------

\clearpage % Start a new page

\setstretch{1.5} % Set the line spacing to 1.5, this makes the following tables easier to read

\lhead{\emph{Abbreviations}} % Set the left side page header to "Abbreviations"
\listofsymbols{ll} % Include a list of Abbreviations (a table of two columns)
{
\textbf{$\Lambda$CDM} & $\Lambda$ cold dark matter \\
\textbf{AdS/dS} & (Anti-) de Sitter \\
%\textbf{BSM} & Beyond SM \\
\textbf{CAM}  & Cosmological attractor models \\
\textbf{CMB} & Cosmic Microwave Background \\
\textbf{CW} & Coleman-Weinberg \\
\textbf{DBIG} & Dirac-Born-Infeld Galileon \\
\textbf{EFT} & Effective field theory  \\
\textbf{GL}  & Goncharov-Linde \\
\textbf{GS} & Gong and Sasaki \\
\textbf{GUTs} & Grand unified theories \\
\textbf{FLRW} & Friedmann-Lema\^itre-Robertson-Walker \\
\textbf{LSS} & Large Scale Structure  \\
\textbf{RHNs} & Right handed neutrinos \\
\textbf{SBCS} & Spontaneous breaking of conformal symmetry \\
\textbf{SM} & Standard model \\
\textbf{SFT} & String field theory \\
\textbf{SUSY} & Supersymmetry \\
\textbf{SUGRA} & Supergravity \\
\textbf{SV} & Shafi-Vilenkin \\
\textbf{TC} &  Tachyon condensation \\
%\textbf{BAO} & baryon acoustic oscillations \\
%\textbf{WMAP} & Wilkinson Microwave Anisotropic Probe \\
}

%----------------------------------------------------------------------------------------
%	PHYSICAL CONSTANTS/OTHER DEFINITIONS
%----------------------------------------------------------------------------------------

%\clearpage % Start a new page

%\lhead{\emph{Physical Constants}} % Set the left side page header %to "Physical Constants"
%
%\listofconstants{lrcl} % Include a list of Physical Constants (a %four column table)
%{
%Speed of Light & $c$ & $=$ & $2.997\ 924\ 58\times10^{8}\ %\mbox{ms}^{-\mbox{s}}$ (exact)\\
% Constant Name & Symbol & = & Constant Value (with units) \\
%}

%----------------------------------------------------------------------------------------
%	SYMBOLS
%----------------------------------------------------------------------------------------

\clearpage % Start a new page

%\lhead{\emph{Symbols}} % Set the left side page header to %"Symbols"

%\listofnomenclature{lll} % Include a list of Symbols (a three %column table)
%{
%$a$ & distance & m \\
%$P$ & power & W (Js$^{-1}$) \\
%% Symbol & Name & Unit \\
%
%& & \\ % Gap to separate the Roman symbols from the Greek
%
%$\omega$ & angular frequency & rads$^{-1}$ \\
%% Symbol & Name & Unit \\
%}

%----------------------------------------------------------------------------------------
%	DEDICATION
%----------------------------------------------------------------------------------------

\setstretch{1.3} % Return the line spacing back to 1.3

\pagestyle{empty} % Page style needs to be empty for this page

\dedicatory{\textbf{This thesis is dedicated to my parents}}%\\ $\quad$ \\ {Mr. \& Mrs. Radhakrishna \& Padmavathi}} %Mr and Mrs. K.V. Radhakrishna and Padmavathi Devi} % Dedication text

\addtocontents{toc}{\vspace{2em}} % Add a gap in the Contents, for aesthetics

%----------------------------------------------------------------------------------------
%	THESIS CONTENT - CHAPTERS
%----------------------------------------------------------------------------------------
\clearpage

\mainmatter % Begin numeric (1,2,3...) page numbering

\pagestyle{fancy} % Return the page headers back to the "fancy" style

% Include the chapters of the thesis as separate files from the Chapters folder
% Uncomment the lines as you write the chapters
\chapter{Introduction}
%\chapter{Semiclassical  collapse arising from LQG}
\label{Intro}

%\epigraph{one two three infinity}{G. Gamow}
\begin{chapquote}{George Gamow, \textit{The creation of the Universe}}
It took less than an hour to make the atoms, a few hundred million years to make the stars and planets, but five billion years to make man
\end{chapquote}

\lhead{\bf Chapter 1. \emph{Introduction}} %

The classical Big Bang cosmology scenario proposed by G. Gamow in 1946 \cite{Gamow}, was supported by the first detection of Cosmic Microwave Background (CMB) reported by A. A. Penzias and R. W. Wilson in 1965 \cite{Penzias1965}. However, such setting suffered from serious difficulties that became known as horizon and flatness problems \cite{Dicke}. Moreover, the development of Grand Unified Theories (GUTs) in the late 70's \cite{Georgei-Gl} predicting the unification of strong, electromagnetic and weak interactions at the energy scales $\sim10^{16}$ GeV, revealed the possible over production of magnetic monopoles, in the early Universe, which was known as monopole problem \cite{Hooft}. These problems could be solved by means of an accelerated (near de Sitter) expansion of the Universe, as proposed by A. A. Starobinsky and A. H. Guth \cite{Starobinsky:1980te,Guth:1980zm}, which is designated as {\it the theory of cosmological inflation}. This theory was subsequently improved by the proposals of,  A. D. Linde \cite{Linde:1982uu,Linde:1983gd}, A. Albrecht and P. J. Steinhardt \cite{Albrecht:1982wi}, which were known as {\it chaotic inflationary scenario} and {\it new inflationary scenario}, respectively. Afterwards, V. F. Mukhanov, G. V. Chibisov and S. W. Hawking \cite{Mukhanov:1981xt,Hawking:1982cz} provided an explanation for the Large Scale Structure (LSS) formation seeded by primordial quantum fluctuations, which made the theory of inflation observationally attractive. In order to have an adequate particle production at the end of inflation, 
a reheating process \cite{Kofman:1994rk,Kofman:1997yn} is expected, constituting an intermediate stage in the evolution of the Universe, subsequently leading into {\it a radiation dominated era and then a matter dominated era.} Currently, the {\it inflationary paradigm} has been widely accepted and stands as an essential mechanism abridging the epoch of quantum gravity, the theory of all fundamental interactions and elementary particles (i.e, physics near the Planck energy scale), to our present day understanding of particle physics. In summary, the theory of inflation combines features imported from particle physics, astrophysics and cosmology to border and connect to a theory of everything. 

Once the exponential expansion begins, the Universe rapidly becomes homogeneous, isotropic and spatially flat, which can be described by a Friedmann-Lema\^itre-Robertson-Walker (FLRW) metric.
During the inflationary regime, the Universe  scale factor $a(t)$ (which is a function of cosmic time $t$) increases exponentially, leaving the Hubble parameter $H=\frac{1}{a}\frac{da}{dt}$ almost constant \cite{Baumann:2009ds}. This implies the comoving Hubble radius $(aH)^{-1}$ to decrease during this period i.e., $\frac{d}{dt}\left(\frac{1}{aH}\right)<0$ for the time $t_{*}<t_{e}$, where $t_{*}$, $t_{e}$ mark the beginning and the end of inflation, respectively. To solve the horizon and flatness problems it is essential that the scale factor during inflation should increase at least $N=50-60$ number of $e$-foldings where $N=\ln \left(\frac{a\left(t_{e}\right)}{a\left(t_{*}\right)}\right)$ \cite{Baumann:2009ds}. 

The required inflationary dynamics can be retrieved either by modifying General Relativity or by the addition of hypothetical matter fields which means modifying either left hand or right hand side of the Einstein equations, given by

\begin{equation}
R_{\mu\nu}-\frac{1}{2}g_{\mu\nu}R= \frac{1}{m_{\rm P}^{2}} T_{\mu\nu}\,, \label{GReq}
\end{equation}
where we fix the units $\hbar=1,\,c=1,\,$ $m_{\rm P}^{2}=\frac{1}{8\pi G}$ with the value of reduced Planck mass $m_{\rm P}=2.43\times10^{18}$ GeV. Here $R_{\mu\nu}$ is the Ricci tensor, $R$ is the Ricci scalar, $T_{\mu\nu}$ is the energy-momentum tensor and $g_{\mu\nu}$ is the spacetime metric tensor\footnote{Throughout the thesis, we set the metric signature $(-,+,+,+)$, small
Greek letters are the fully covariant indexes.}. The scalar field responsible for inflation is usually named as inflaton, when it is a hypothetical matter field or scalaron when it emerges from modified gravity. In this thesis, we are mainly interested in finding inflaton candidates. 

An adequate period of exponential expansion ending in a reheating epoch can be met when the so called slow-roll parameters $\epsilon,\,\eta$ satisfy the following conditions during inflation \cite{Mukhanov:2013tua}

\begin{equation}
\epsilon=-\frac{\dot{H}}{H^{2}}\ll1\,,\quad\quad \eta=\frac{\dot{\epsilon}}{H\epsilon}\ll1\,,
\label{SLRconditions}
\end{equation}
where over dot indicates the differentiation with respect to $t$.  

The temperature fluctuations in the CMB are caused by the primordial quantum fluctuations of the scalar degrees of freedom during inflation \cite{Baumann:2009ds}. In other words, the source of inflationary expansion and LSS of the Universe can be traced back to the dynamics and nature of one or more scalars. Inflationary background fluctuations (primordial modes) are created quantum mechanically at subhorizon scales $k\gg aH$, where $k$ is the comoving wavenumber. The CMB temperature anisotropy and LSS can be explained by the evolution of those fluctuations on superhorizon scales $k\ll aH$.

The quantum fluctuations during inflation can depicted by the curvature perturbation $\zeta$ in the comoving gauge\footnote{The details of other gauge choices can read from \cite{Baumann:2009ds}. Throughout this thesis we use the notation for curvature perturbation either $\zeta$ or $\mathcal{R}$ \cite{Baumann:2009ds}, bearing the fact that the curvature perturbation defined on uniform density hypersurfaces $\zeta$ and the comoving curvature perturbation $\mathcal{R}$ are nearly equal in the slow-roll inflation (see \cite{Baumann:2009ds} for details).}.  In the case of single field inflation, $\zeta$ gets conserved on superhorizon scales. Therefore, the fluctuations are adiabatic and the power spectrum measured at the time of horizon exit $k\sim aH$, is related to the temperature anisotropies in the CMB \cite{Sasaki:1995aw,Lyth:2009zz}. Whereas in the case of multifield inflation, $\zeta$ evolves on the superhorizon scales as it is additionally sourced by isocurvature modes. In this case $\zeta$ is computed either by using transfer functions or the so called $\delta N$ formalism \cite{Wands:2007bd,Sasaki:2012ss,Tanaka:2010km,Byrnes:2010em,Elliston:2011dr}.

The key observables of inflationary scenarios are related to the two-point and higher order correlation functions of curvature perturbation (see Appendix.~\ref{curvedef} for a brief review). The two-point correlation function of $\zeta$ defines the scalar power spectrum which predicts the Gaussian distribution of density fluctuations. Inflationary expansion obeying conditions (\ref{SLRconditions}) predicts that the scalar power spectrum would departure from exact scale invariance. This is quantified by a parameter named scalar spectral index, or scalar tilt, $n_{s}$ that should differ from unity. 

The other prediction of inflationary theory is the primordial gravitational wave power spectra, that can be defined in a similar way to the scalar power spectrum as a two point correlation function of tensor modes. The ratio of tensor to scalar power spectrum $r$ and the tensor tilt $n_{t}$, defined in a similar way as $n_{s}$, are crucial to test any model of inflation against observations. In the context of the recent results from {\it Planck} satellite in 2015 \cite{Ade:2015lrj,Ade:2015ava} and
the joint analysis of BICEP2/Keck Array and {\it Planck} (BKP) \cite{Ade:2015tva},
the single field inflationary paradigm, emerges as adequate to generate the observed adiabatic,
nearly scale invariant and the highly Gaussian density fluctuations imprinted as the CMB temperature anisotropies.
Moreover, the data is very much consistent with $\Lambda$CDM model\footnote{$\Lambda$ stands for the cosmological constant and the CDM means the Cold Dark Matter.} of the current Universe and the results also indicate that we live in a spatially flat Universe \cite{Ade:2015xua}.

The CMB observations from {\it Planck} 2015 \cite{Ade:2015lrj}, constrains the scalar spectral index and the tensor to scalar ratio as 

\begin{equation}
n_{s}=0.968\pm0.006\,,\qquad r<0.09\,,
\end{equation}
with respect to {\it Planck} TT+lowP+WP at 95\% confidence level\footnote{Here TT+lowP+WP indicates the combined results of {\it Planck}'s angular power spectrum of temperature fluctuations with low-$l$ polarization (of CMB radiation) likelihood analysis and polarization data from Wilkinson microwave anisotropy Probe (WP) \cite{Hinshaw:2012aka}.} (CL) which rules out scale invariance at more than $5\sigma$ \cite{Ade:2015lrj}. Furthermore, the data suggests a small running of the spectral index $dn_{s}/d\ln{k}=-0.003\pm0.007$, which is consistent with the prediction from single field models of inflation
\cite{Lyth:2009zz}. So far, there is no significant detection of
primordial tensor modes (from the value of $r$), which is crucial to fully confirm the inflationary paradigm. A concrete measurement of $r$ relates to an observation of the so called B-mode polarization amplitude, which can only be caused by primordial tensor modes in the CMB radiation. 
Moreover, the latest results
suggest so far no evidence for a blue tilt of the gravitational wave power
spectra i.e., $n_{t}>0$ from a very preliminary statistical analysis \cite{Ade:2015lrj,Ade:2015tva}. The proposed post-{\it Planck} satellites CMBPol, COrE, Prism, LiteBIRD
and many other ground based experiments such as Keck/BICEP3 \cite{Creminelli:2015oda,Errard:2015cxa,Huang:2015gca,Finelli:2016cyd}
are expected to reach enough sensitivity to detect B-modes and
establish if $r\sim\mathcal{O}\left(10^{-3}\right)$. 

There is a considerable
variety of different models of inflation that can be motivated theoretically, but the degeneracy of the
predictions from various models of inflation is an ongoing problem
for cosmologists \cite{Martin:2013tda,Martin:2015dha}. One way to probe further the nature of the inflaton field
is to study the statistics of the perturbations it produces beyond
the two-point correlation function \cite{Maldacena:2002vr,Chen:2006nt,Komatsu:2009kd}, starting
with the three-point function. The latter is parametrized in Fourier space
by the bipsectrum (defined in Appendix.~\ref{curvedef}),
a function of the amplitude of three wave vectors that sum to zero
as a consequence of momentum conservation. The bispectrum of \char`\``{}local shape\char`\"{},
is a function of three wave numbers that peaks in the squeezed
limit where two wave numbers are much larger than the third. The
bispectrum of \char`\``{}equilateral shape\char`\"{}
tends to zero in the squeezed limit, but peaks when all three wave numbers
are similar in size. A third shape is often considered that peaks on folded
triangles, where two wave numbers are approximately half of the third. 
Introducing three parameters $\fnl^{{\rm loc}}$, $\fnl^{{\rm equi}}$
and $\fnl^{{\rm ortho}}$, which parametrize the overall amplitude
of a local, equilateral and orthonormal shapes for the bispectrum,
{\it Planck} 2015 data \cite{Ade:2015ava} informs us that 
\begin{equation}
f_{\rm NL}^{\rm loc}=0.8\pm 5.0 \,,\quad f_{\rm NL}^{\rm equi}=-4\pm 43 \,,
\quad f_{\rm NL}^{\rm ortho}=-26\pm 21,%
\end{equation}
at $68\%$ CL.

Any confirmation of non-Gaussianities in the CMB would be a significant information about the nature of the inflaton field \cite{Komatsu:2009kd}. For example, establishing through unequivocal observations and data analysis a local non-Gaussianity would rule out all the single field models of inflation \cite{Creminelli:2004yq}. 

\section{Standard scalar field inflation and Planck data}

\label{stdslr}

The simplest standard mechanism to set up inflationary expansion is conveyed by a (canonical) scalar field minimally coupled to Einstein gravity dictated by the following action 

\begin{equation}
S=\int d^4x\,\sqrt{-g}\left[\frac{m_{\rm P}^2}{2}R-\frac{1}{2}\partial_{\mu}\varphi\partial^{\mu}\varphi-V(\varphi)\right]\,,
\end{equation}
where $g$ is the determinant of the metric $g_{\mu\nu}$.

To sustain inflationary expansion long enough, the general ingredient has been that the potential $V(\varphi)$ needs to dominate over the kinetic term $-\frac{1}{2}\partial_{\mu}\varphi\partial^{\mu}\varphi$, for which the inflaton is required to be almost constant during inflation. This is achieved by the slow-roll approximation, which can be expressed in terms of potential slow-roll parameters\footnote{These are related to the general slow-roll parameters in (\ref{SLRconditions}) as $\epsilon\approx \epsilon_{V},\,\eta\approx4\epsilon_{V}-2\eta_{V}$ \cite{Baumann:2009ds}.} as
\begin{equation}
\epsilon_{V}\equiv\frac{m_{\rm P}^2}{2}\frac{V^\prime(\varphi)}{V(\varphi)}\ll1\,,\quad \eta_{V}=m_{\rm P}^2\frac{V^{\prime\prime}(\varphi)}{V(\varphi)}\ll1\,,
\end{equation}
where `a prime' denotes differentiation with respect to the argument $\varphi$.
The scalar spectral index $n_{s}$ and the tensor to scalar ratio $r$ read as \cite{Baumann:2009ds}
\begin{equation}
n_{s}=1-6\epsilon^{\ast}_{V}+2\eta^{\ast}_{V}\,,\qquad r=16\epsilon^{\ast}_{V}\,,\label{ns-r-standard}
\end{equation}
where "$\ast$" denotes the quantities evaluated at the horizon exit. The energy scale of inflation can be estimated as $M_{\rm inf}\equiv V_{*}^{(1/4)}\simeq M_{\rm GUT}r^{(1/4)}$ and the range of values that the field can take during inflation can be determined by the Lyth bound \cite{Lyth:1996im,Garcia-Bellido:2014wfa,Linde:2016hbb}. In Appendix.~\ref{slacs}, we summarize the observational tests of standard single field inflation. 

\begin{figure}[h!]
\centering\includegraphics[width=5in]{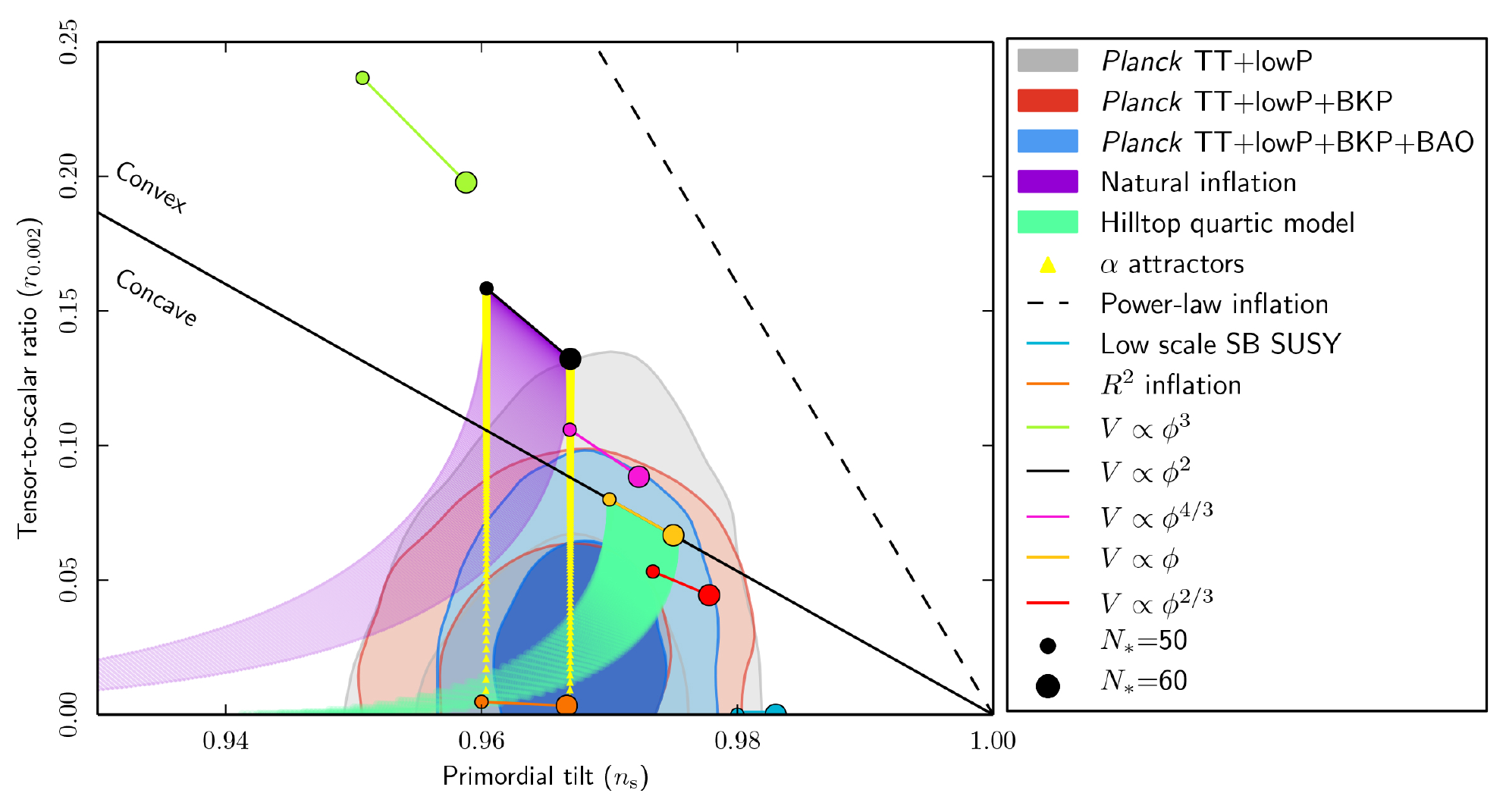}\caption{Marginalized joint 68\,\% and 95\,\%~CL regions for $n_{s}$ and $r$ at the pivot scale $k_{*}=0.002 {\rm Mpc}^{-1}$ from Planck in combination with other data sets, compared to the theoretical predictions of selected inflationary models.
}\label{nsrPl15}
\end{figure}

The constrains from  {\it Planck} and BICEP2/Keck array data \cite{Ade:2015lrj} rule out several potentials for a standard scalar field (see Fig.~\ref{nsrPl15}), nevertheless the flat ones of the following form 
\begin{equation}
V\sim\left(1-e^{-\sqrt{2/3B}\varphi}\right)^{2n}\,,\label{flatpot}
\end{equation}
became successful candidates for the description of inflation and
appeared in various scenarios \cite{Martin:2013tda,Linde:2014nna,Martin:2015dha}.
The parameter $B$, in the above potential, can lead to
any value of $r<0.09$ with a fixed value for $n_s$, namely
\begin{equation}
n_s=1-\frac 2N\, ,\quad\quad r=\frac{12B}{N^{2}}\,.\label{attractorpred}
\end{equation}
We note that the potentials of the form in (\ref{flatpot}) cannot be easily justified field theoretically in the standard scalar description. 

In the case\footnote{The potentials with $B\neq 1$  requires more complicated realization of inflation in a fundamental theory which we will discuss later in Sec. \ref{TBBTA}. 
} of $B=1$ the potential is the same as the one in the Einstein frame description of Starobinsky's $R+R^2$ inflation \cite{Starobinsky:1980te,Mukhanov:1990me} and also in the Higgs inflation with a non-minimal coupling \cite{Bezrukov:2007ep}. Although, these two models occupy a privileged position in the $n_{s}-r$ plane of {\it Planck} 2015, it is still not possible to distinguish these two models observationally. The difference between these models is greatly expected to be found at the reheating phase \cite{Bezrukov:2011gp}, whose observational reach is uncertain in the near future \cite{Martin:2016oyk}.

%\begin{itemize}
%\item The energy scale of inflation is given by $M_{inf}\equiv V_{*}^{(1/4)}\simeq M_{\text GUT}r^{(1/4)}$ which is known as the Lyth bound \cite{Lyth:1996im,Garcia-Bellido:2014wfa,Linde:2016hbb}.
%\end{itemize}

\section{Beyond standard scalar inflation?}

\begin{figure}[h!]
\centering\includegraphics[width=2.5in]{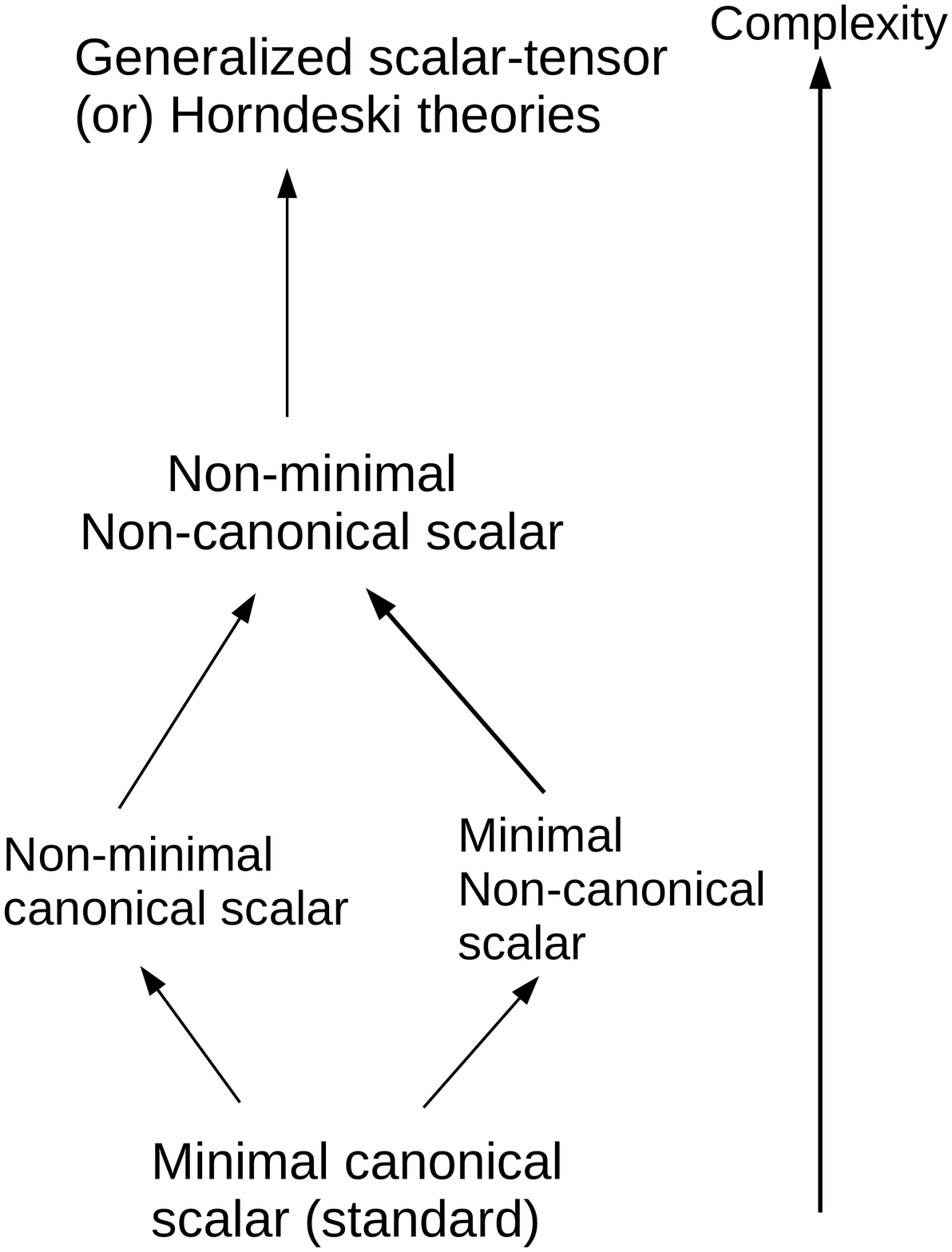}\caption{\label{tree-complex} In this tree diagram we present the ways towards more elaborated inflationary model building.}
\label{compscalar}
\end{figure}

The standard scalar field action can be extended (cf. Fig.~\ref{compscalar}) either with a non-minimal coupling to gravity (e.g., Higgs inflation \cite{Bezrukov:2007ep}) or with a non-canonical kinetic term\footnote{In general, we can also add additional matter fields to play crucial role along with the inflaton e.g., in the case of Warm inflation, radiation plays crucial role \cite{Bartrum:2013fia}}. Furthermore, a general scalar-tensor theory was written and is known as Horndeski theory \cite{Horndeski:1974wa}, which was shown to be equivalent to generalized Galileon model (G-inflation) \cite{Kobayashi:2011nu}. The details about the G-inflationary action and calculations of perturbation spectra are presented in Appendix.~\ref{G-infapp}. 

The standard single field models predicts very much a Gaussian landscape, where any small non-Gaussianities are suppressed by the slow-roll parameters \cite{Maldacena:2002vr}, whereas the non-canonical and multified models predict detectable levels of non-Gaussianities \cite{Babich:2004gb,Chen:2010xka,Byrnes:2014pja,Vennin:2015vfa}. In Ref.~\cite{DeFelice:2013ar} shapes of non-Gaussianities in the general scalar-tensor theories were worked out, however the specific predictions are model dependent. 

%\section{Top-down vs bottom-up approach}
\section{Top-down vs bottom-up motivations}

\label{TBBTA}

\begin{figure}[h!]
\centering\includegraphics[width=2.5in]{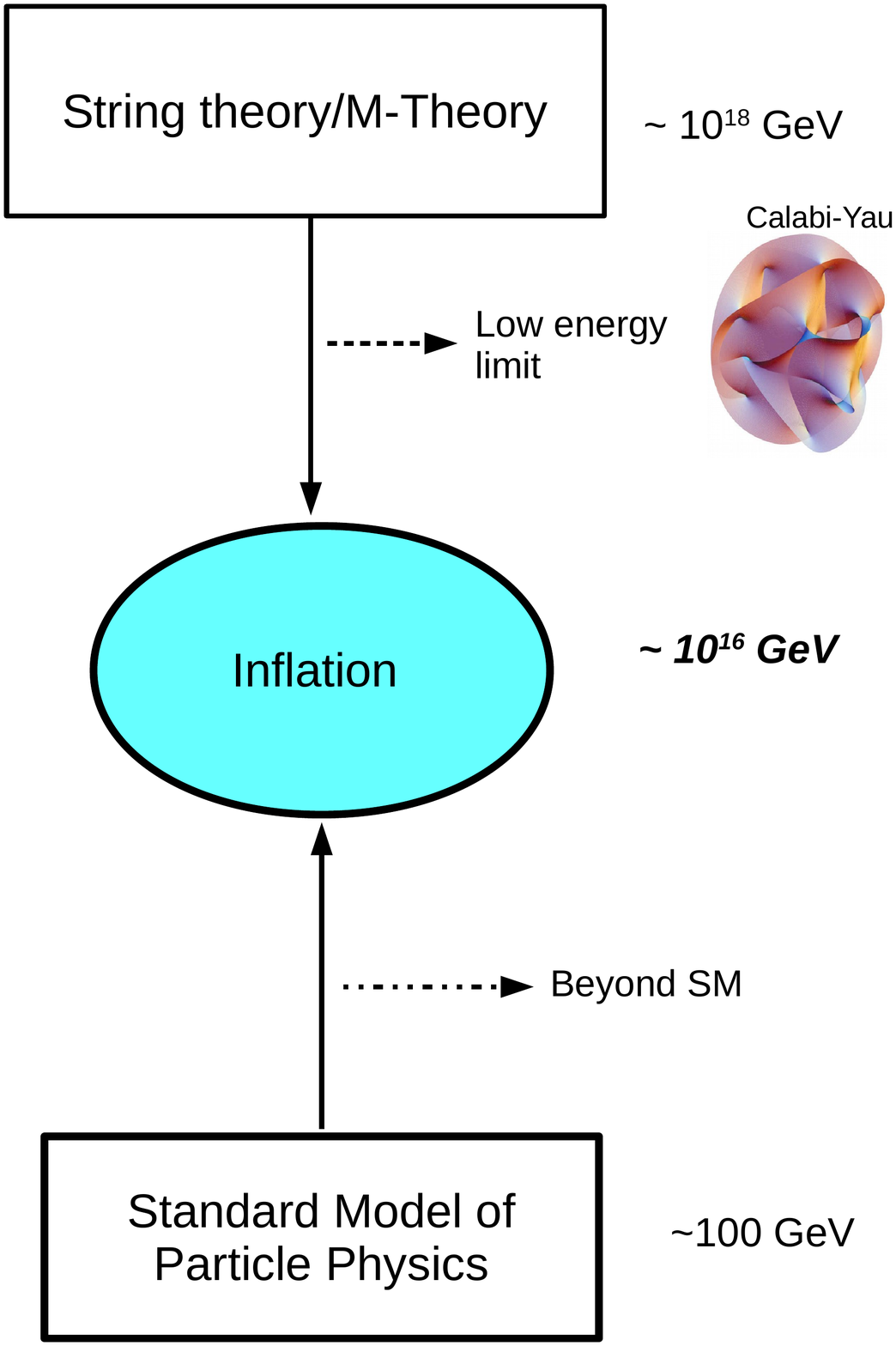}\caption{\label{tree-complex} In the top-down motivation we build models in the low EFTs of string theory/M-theory which can be realized via compactifications on Calabi-Yau manifolds. In the bottom-up motivation we build models based on the physics beyond the SM of particle physics e.g., in GUTs and MSSM.}
\label{TDBU}
\end{figure}

Inflationary models can be phenomenologically realized in top-down or bottom-up (cf.~Fig. \ref{TDBU}) motivations as described below.
 
\subsection{Top-down: Inflation in string theory/supergravity}

\label{StrInGRA}
 
According to the present observations, the Hubble parameter during inflation can be as large as $10^{13-14}$
GeV, suggesting the scale of inflation to be of the order of $M_{\rm inf}\gtrsim10^{15}$
GeV. These energy scales are acceptable in theories of gravity promising ultraviolet (UV) completion, such as string theory/M-theory and supergravity (SUGRA), hence argued to play a crucial role in inflation \cite{Cicoli:2010yj}. Therefore, during the last years there have been many attempts to
understand the inflationary picture from the low energy effective field theories (EFTs) motivated from such fundamental approaches \cite{Linde:2014nna,Silverstein:2015mll,Burgess:2013sla,Baumann:2014nda}.
The interest of studying such inflationary scenarios is that it gives the best framework to get some observational indication of these fundamental theories. There has been a plethora of inflationary models in the literature, based on several modifications of the matter or gravity sector inspired from string theory/SUGRA \cite{Martin:2013tda}. Given our ignorance on the relation between a UV complete theory and its low energy effective limit, there is a plenty of room to construct models \cite{Burgess:2013sla,Martin:2015dha} and aim to falsify them against current and future CMB observations. 

\begin{figure}[h!]
\centering\includegraphics[width=5in]{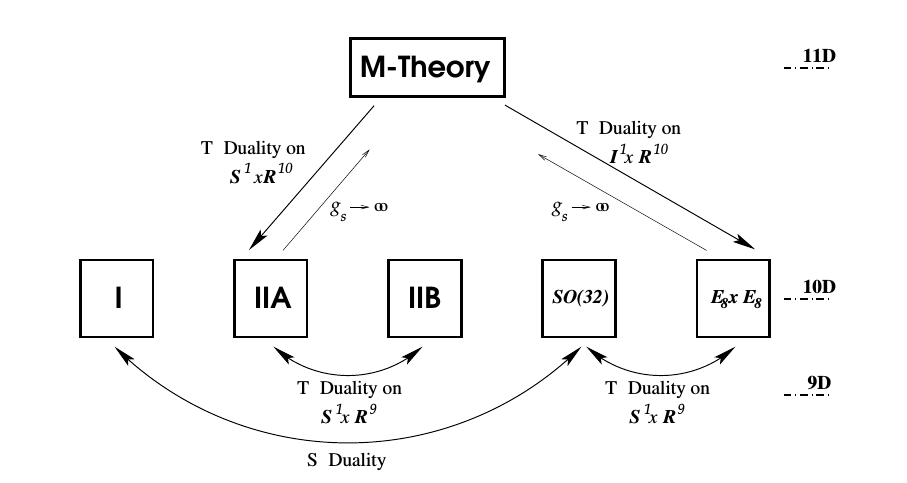}\caption{\label{Mtheory} The various duality transformations that relate the superstring theories in nine
and ten dimensions. T-Duality inverts the radius $\mathbf{R}$ of the circle $\textbf{S}^{1}$ or the length of the
finite interval $\textbf{I}^{1}$, along which a single direction of the spacetime is compactified, i.e.
$\mathbf{R}\to l_{\text P}^2/\mathbf{R}$. S-duality inverts the (dimensionless) string coupling constant $g_{s}$, $g_{s}\to 1/g_{s}$,
and is the analog of electric-magnetic duality (or strong-weak coupling duality) in four-
dimensional gauge theories. M-Theory originates as the strong coupling limit of either
the Type IIA or $E_{8}\times E_{8}$ heterotic string theories.}
\end{figure}

M-theory, believed to explain all fundamental interactions including gravity, that describes the physics near Planck energy scale, is defined in a 11 Dimensional (11D) spacetime and claims to unify all five versions of superstring theories\footnote{Which are related by T-, S- dualities \cite{Schwarz:1982jn,Becker:2007zj,Erdmenger:2009zz}}  \cite{Becker:2007zj,Erdmenger:2009zz}, as presented in Fig.~\ref{Mtheory} (taken from Ref.~\cite{Szabo:2011sd}). The idea of building inflationary models within string theory allows to test possible 4D low energy EFTs of these five superstring theories. In principle, to obtain a low energy limit of any version of superstring theory into 4D, we need to compactify six extra dimensions on small internal manifold such as Calabi-Yau\footnote{Moreover, these compactifications have to be well stabilized to accommodate sufficient conditions for inflation to happen in the resultant EFT. For example, this was successfully prescribed in type IIB string theory through KKLT and KKLMMT scenarios \cite{Kachru:2003sx,Kachru:2003aw,Linde:2005dd}. } and we thus are generically left with many possibilities to construct 4D EFTs \cite{Susskind:2003kw,Westphal:2014ana}. Studying inflation in these theories is therefore most pertinent 
%in order to get some observational viability for our theoretical understandings
 \cite{Erdmenger:2009zz,Bachas:2012zz,Ibanez:2012zz}.

Broadly, inflationary scenarios in string theory can be divided into two categories:
\begin{enumerate}
\item Open string inflation (e.g., Brane/Anti-brane inflation)\,;
\item Closed string inflation (e.g., Moduli inflation)\,. 
\end{enumerate}
A detailed review and recent observational status (with respect to {\it Planck} 2015 data) of several of these inflationary scenarios, driven by closed and open string fields, can be found in \cite{StGC-cqg,Martin:2013tda,Baumann:2014nda,Burgess:2013sla}. Inflation in string theory contains several types of scalar field terms (e.g., Dirac-Born-Infeld (DBI) inflation where the scalar field is non-canonical), with fundamentally motivated choices of potentials, which can be tested by inflationary observables. With more precise CMB data, in the future we may aim to establish the role of string theory in inflationary dynamics, and help to fulfill our understanding of a fundamental theory \cite{Baumann:2014cja}. 

Supergravity (SUGRA) is a gauge theory that is an extension of General relativity where we impose a local (gauged) supersymmetry (SUSY) and most SUGRA settings constitute a low energy limit of superstring theory. 
There are several versions of SUGRA, characterized by the number of massless gravitinos $\mathcal{N}=1,..,8$. In particular, $D=4$, $\mathcal{N}=1$ SUGRA could be an intermediate step between superstring theory and the supersymmetric standard model of particle physics that we hope to observe at low energies \cite{Cerdeno:1998hs,Gates:1983nr,VargasMoniz:2010zz,Moniz:2010fea}. Therefore, it is realistic to construct EFT of inflation in $D=4$, $\mathcal{N}=1$ SUGRA from high scale SUSY breaking\footnote{If SUSY is not found in the current collider experiments, then high scale SUSY breaking is a natural expectation in a UV complete theory; inflation can thus be a testing ground for high scale SUSY breaking \cite{Baumann:2011nk}.}. Several inflationary models in the past have been constructed in $\mathcal{N}=1$ SUGRA and they stand out to be an interesting possibility in regard of {\it Planck} data \cite{Yamaguchi:2011kg,Linde:2014nna}. Moreover, inflation in SUGRA has the interesting feature of predicting particle DM candidates (e.g., massive gravitino\footnote{Gravitino is the supersymmetric partner of graviton with spin$=\frac{3}{2}$ which can gain mass due to SUSY breaking.}) \cite{Arnowitt:2000sb}. Inflation in SUGRA is usually described by the K\"ahler potential as well as superpotentials, which depend on the chiral superfields \cite{Cerdeno:1998hs,Bailin:1994qt}. A brief discussion of SUSY breaking mechanisms for different SUGRA inflationary scenarios can be found in \cite{Abe:2014opa}. 

As mentioned previously, the observational data provided a special
stimulus to study inflation with flat
potentials of the form (\ref{flatpot}), which
became successful candidates \cite{Martin:2013tda,Linde:2014nna,Martin:2015dha}.
Such potentials are so far shown to occur in the low energy
effective models of string theory/SUGRA and modified gravity \cite{Ellis:2013nxa,Ellis:2015xna,Kallosh:2013xya,Kallosh:2013yoa,Nastase:2015pua,Diamandis:2015xra,Ozkan:2015iva}. A generic structure of K\"ahler potentials in SUGRA suitable for inflation
and a possible connection to the open/closed string theory were studied
in \cite{Roest:2013aoa}. 

\subsection{Bottom-up: Inflation and particle physics}

Inflation has convincingly abridge cosmology with our present knowledge of particle physics, through the process of reheating: the scalar mode that drives expansion settles to a (true) vacuum, leading to particle production through a mechanism that depends on how the inflaton oscillates when it reaches to the minimum of the potential \cite{Kofman:1994rk, Kofman:1997yn}. This has strongly motivated the construction of inflationary models within the standard model (SM) of particle physics and beyond. Therefore, in a bottom-up motivation, several particle physics models were proposed including Higgs inflation, within grand unified theories (GUT) \cite{Bezrukov:2007ep,Shafi:1983bd,Linde:2005ht,Mazumdar:2010sa} and Minimally Supersymmetric extensions of SM (MSSM) \cite{Martin:2013tda}. The interesting feature of a bottom-up motivation is that these models can be tested outside the scope of CMB e.g., at collider experiments. 

\begin{figure}[h!]
\centering\includegraphics[width=3.5in]{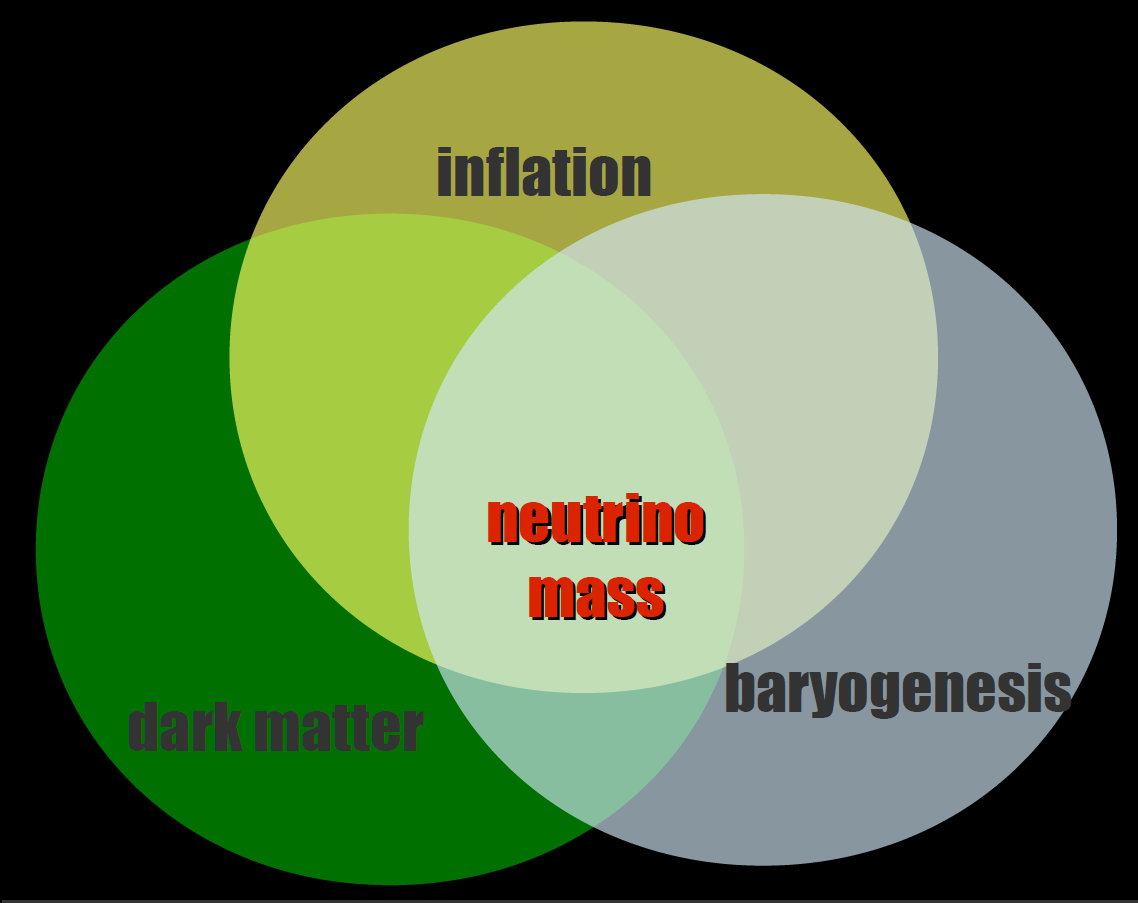}\caption{\label{venn-BSM} Inflation in particle physics motivated models such as GUTs and MSSM are particularly interesting, when considering neutrino masses, DM and baryogenesis. Neutrinos are worthy elements beyond SM particle physics.}
\end{figure}

In the particle physics context, SM Higgs inflation \cite{Bezrukov:2007ep} is particularly interesting due to the fact that Higgs
was the only scalar so far found at LHC \cite{Aad:2012tfa}. Nevertheless, for Higgs to be a candidate for inflaton, it requires a large non-minimal coupling\footnote{It was known that a scalar field with large non-minimal
coupling gives rise to a $R^{2}$ term considering 1-loop quantum corrections. Consequently, renormalization group (RG) analysis shows that Higgs inflation is less preferable compared to Starobinsky model \cite{Salvio:2015kka,Calmet:2016fsr}.}. On the other hand, SM is known to be incomplete due to the mass hierarchy problems e.g., the Higgs mass being very low (125 GeV) compared to GUT scale, plus nearly but not quite negligible neutrino masses ($\sim 0.1\text{eV}$). Furthermore, observed matter anti-matter asymmetry and dark matter find no explanation within SM. In this regard, inflationary models beyond SM physics i.e., GUTs and MSSM, are quite natural to explore \cite{Lazarides:1991wu,Allahverdi:2010xz} (see Fig.~\ref{venn-BSM} which is taken from \cite{Valle:2015mma}).
The main advantage of studying inflation in SM extension theories is that in these constructions it is more natural to accommodate the reheating process after inflation and moreover, we can expand the observational tests beyond CMB, something that is more difficult to achieve when we consider models in the string theory/SUGRA. However, on the other hand there is a hope that the GUTs and MSSM can be UV completed in heterotic superstring theories \cite{Witten:1985xc,Greene:1986qva,Abe:2015mua}.

\section{Overview of the thesis}

In this thesis, we study the following inflationary scenarios based on string/SUGRA and GUTs, which we show to be compatible with constraints from Planck data \cite{Ade:2015lrj}. 

\begin{itemize}
\item Chapter \ref{In-3-forms}: Multiple 3-form field inflation and non-Gaussianity

$p$-form fields\footnote{A covariant tensor of rank $p$, which is anti-symmetric under exchange
of any pair of indices is called $p$-form.} are part of type IIA string theory \cite{Becker:2007zj,Erdmenger:2009zz}
where they generically appear as the gauge fields of SUSY multiplets.
$p-$form fields are massless in the standard Calabi-Yau compactifications
while in the low energy effective theories, $p-$form fields can gain
mass via the St\"uckelberg mechanism \cite{Louis:2002ny,Grimm:2004uq}. In the broader class of $p-$form inflation \cite{Germani:2009iq,Kobayashi:2009hj}, 3-forms are realized to be viable alternative to scalar field inflation \cite{Koivisto:2009ew,Koivisto:2009fb}. Moreover, in $D=4$, $3-$form fields are relevant and they have been studied especially in $\mathcal{N}=1$ SUSY theories with quadratic potentials\footnote{The 3-form potential can be generated by SUSY breaking which also breaks the gauge symmetry.} \cite{Groh:2012tf,Gubser:2000vg,Frey:2002qc}. 
%For $D=4$, the degrees of freedom (dof) of different $p-$forms are $0-$form which is a scalar field (one dof), $1-$form which is massive vector field (3-dof), $2-$form which is a massless tensor field (2-dof) and $3-$form which again has only one degree of freedom like a scalar field. Moreover, in $D=4$, $3-$form fields are relevant and they have been studied especially in $\mathcal{N}=1$ SUSY theories with quadratic potentials \cite{Groh:2012tf,Gubser:2000vg,Frey:2002qc}. 
In this chapter, we generalize the single 3-form inflation with multiple 3-form fields and find suitable (phenomenological) choice of potentials
compatible with observations. We also compute the corresponding generation of non-Gaussianities in this model. 

\item Chapter \ref{DBIGc}: DBI Galileon inflation

D-branes are fundamental objects in string theory to which open strings are attached, satisfying Dirichlet boundary condition \cite{Becker:2007zj}. Inflationary scenarios involving D-branes are associated with the motion of the branes in internal dimensions. These models are promising ones in string cosmology \cite{Erdmenger:2009zz}. In particular, the Dirac-Born-Infeld (DBI) inflation has gained substantial attention in recent years \cite{Silverstein:2003hf,Alishahiha:2004eh,Chen:2005fe,Shandera:2006ax,Kobayashi:2007hm,Baumann:2006cd,Becker:2007ui,
Deser:1998rj,Deffayet:2009mn} via low energy effective versions of type IIB string theory and $\mathcal{N}=1$ SUGRA \cite{Westphal:2014ana,Baumann:2014nda,Cecotti:2014ipa,Dalianis:2015fpa}. In this chapter we study a well motivated extension of this model, known as DBI Galileon inflation, and show that it enables a wider compatibility with {\it Planck} data. 

\item Chapter \ref{SFTin}: Effective models of inflation from an SFT inspired framework

Assuming stringy energy scales are relevant at inflation, the field theory of interacting strings i.e., 
string field theory (SFT) would perhaps be crucial to be accounted \cite{Siegel:1988yz,Arefeva:2001ps}. There were early attempts of considering
inflation in SFT studied with $p-$adic strings \cite{Barnaby:2006hi,Biswas:2012bp}. In this chapter, admitting non-locality being the distinct feature of SFT which is associated with how the string fields interact (see Appendix. \ref{AppSFT} for details), we introduce a framework motivated from open-closed string field
theory coupling; the open string tachyon condensation ends up in an inflationary (in general a constant curvature) 
background with a stabilized dilaton field. We demonstrate that this configuration leads to interesting effective and viable models of inflation. 

\item Chapter \ref{CAM}: Non-slow-roll dynamics in $\alpha-$attractors

The so-called $\alpha-$attractor models are very successful with Planck data, predicting any value of $r<0.09$ with $n_{s}=0.968$ for $N=60$. The predictions of this model are strongly connected to the mathematical features of the inflaton's kinetic term \cite{Galante:2014ifa}. These models were first proposed in $\mathcal{N}=1$ SUGRA in the context of superconformal symmetries \cite{Kallosh:2013xya}. In this chapter, we study the model in non-slow-roll (or) Hamilton-Jacobi formalism \cite{Muslimov:1990be,Gong:2015ypa}, which is different from standard slow-roll approximation discussed in Sec. \ref{stdslr}. 

\item Chapter \ref{CGUTin}: Conformal GUT inflation

Coleman-Weinberg (CW) inflation proposed by Q. Shafi and A. Vilenkin \cite{Shafi:1983bd,Linde:2005ht} was the first model of inflation proposed in the context of GUTs such as $\text{SU}(5)$ and $\text{SO}(10)$, where inflation is the result of GUT symmetry breaking. In this chapter, we generalize this model with conformal symmetry whose spontaneous symmetry breaking, in addition to the GUT symmetry, flattens the CW potential. As a result we obtain $n_{s}\sim0.96-0.967$ and $r\sim0.003-0.005$ for $50-60$ number of $e$-foldings. We compute the predictions for proton life time and get values above the current experimental bound \cite{Rehman:2008qs}. We implement type I seesaw mechanism by coupling the inflaton field to the right handed neutrinos. We further study the reheating and baryogenesis in this model through non-thermal leptogenesis. 
\end{itemize}

%Type IIA string theory contains stable Dp-branes with p even ,  while type IIB string theory contains stable Dp-branes with p odd. 
% Chapter 2

\chapter{Multiple 3-form field inflation and non-Gaussianity} 
\label{In-3-forms} 
\begin{chapquote}{Edward Witten}
One of the basic things about a string theory is that it can vibrate in different shapes or forms, which gives music its beauty
\end{chapquote}
\setstretch{1.3}
\lhead{\bf Chapter 2. \emph{Multiple 3-form field inflation and non-Gaussianity}} % This is for the header on each page - perhaps a shortened title

Considered as a suitable alternative to the conventional
scalar field, single 3-form inflation has been introduced
and studied in 
Ref.~\cite{Koivisto:2009ew,Koivisto:2009fb,DeFelice:2012jt,Mulryne:2012ax}.
In \cite{DeFelice:2012jt} a suitable choice of the potential for
the $3-$form has been proposed in order to avoid ghosts and Laplacian
instabilities; the authors have shown that potentials showing a quadratic
dominance, in the small field limit, would introduce sufficient oscillations
for reheating \cite{DeFelice:2012jt} and would be free of ghost instabilities. In \cite{Mulryne:2012ax}, it was shown that single 3-form field is dual to a non-canonical scalar field whose kinetic term can be determined from the form of 3-form potential. Therefore, similar to the non-canonical scalar field 3-form field perturbations propagate with a sound speed $ 0<c_{s}\lesssim1 $ which produces effects into inflationary observables. In \cite{Mulryne:2012ax}, single 3-form inflation was shown to be consistent with $ n_{s}=0.97 $ for power law and exponential potentials and the corresponding generation of large non-Gaussianities were studied for small values of sound speed.

In this chapter, we extend the single 3-form framework to $\mathbb{N}$ 3-forms and explore their subsequent inflationary dynamics. We particularly focus on two 3-forms scenario for which we compute the power spectra relevant for the observations at CMB. More concretely, we obtain the inflationary observables for suitable choice of potentials and aim to falsify the two 3-forms inflationary scenario. In this regard,  this chapter is divided into two main sections. The first section is dedicated to  study of the different type of inflationary scenarios driven by two 3-forms. We study the evolution of curvature perturbation on superhorizon scales $ \left(c_{s}k\ll aH\right) $ effected by the dynamics of isocurvature perturbations. For this we compute the transfer functions that measure the sourcing of isocurvature modes to the curvature modes on superhorizon scales. We obtain the observables such as scalar spectral tilt and its running, tensor to scalar ratio. 

In the second section, we compute the non-Gaussianities generated by two 3-forms dynamics. We compute the bispectrum using the fact that 3-form fields are dual to a non-canonical scalar fields. We compute reduced bispectrum $ \fnl $ on superhorizon scales using our prescription of $ \delta N $ formalism applied to the 3-forms. We predict the values of $\fnl$ parameter in different limits of 3-momenta for the same choice of potentials studied in the first section. Finally, in Sec.~\ref{sum3-form} we confirm, in particular, that two 3-forms inflationary scenario is compatible with current observational constraints. 

In this chapter we follow the units $m_{\rm P}=1$. 

\section{Inflation with multiple 3-forms and primordial power spectrum}

\label{partIchap3-f}

This section is organized as follows. In Sec.~\ref{N-form} we identify
basic features of $\mathbb{N}$ 3-forms slow-roll solutions, which
can be classified into two types. We also discuss how the inflaton
mass can be brought to lower energy scales, for large values of $\mathbb{N}$.
In Sec.~\ref{two 3-form} we examine the possible inflationary
solutions, when two 3-forms are present. There are two classes; solutions not able to generate isocurvature perturbations (type I);
and solutions with inducing isocurvature effects (type II). We show that, using a dynamical system analysis\footnote{Details presented in Appendix \ref{AI3-forms}.}, the type I solutions does not bring any new interesting features than single 3-form inflation \cite{DeFelice:2012jt,Koivisto:2009fb}. Type II case, however,
characterizes a new behaviour, through curved trajectories in field
space. Moreover, type II inflation is clearly dominated by the gravity
mediated coupling term which appears in the equations of motion. We
present and discuss type II solutions for several classes of potentials,
which are free from ghost instabilities \cite{DeFelice:2012jt} and
show evidence of a consistent oscillatory behavior at the end of
the two 3-forms driven inflation period. %We provide analytical
%arguments to explain the oscillatory behavior present in this choice
%of potentials. 
In addition, we calculate the speed of sound, $c_{s}^{2}$,
of adiabatic perturbations for two 3-forms and show it has significant
variations during inflation for type II solutions. Therefore, our major
objective in this chapter is to understand and explore the cosmological consequences
of type II solutions. %More concretely, we will show that type II brings
%about interesting effects. %In particular, we use a dynamical system analysis and conclude that type I case is very similar to single 3-form inflation \cite{DeFelice:2012jt,Koivisto:2009fb}. 
In Sec.~\ref{powerspectra}
we discussed adiabatic and entropy perturbations for two 3-form fields,
using a dualized action \cite{Mulryne:2012ax,Langlois:2008mn}. We
distinguish, type I and type II solutions with respect to isocurvature
perturbations and calculate the power spectrum expression \cite{Lyth:2009zz}.
In Sec.~\ref{sec:Planck} we present how our inflationary setting
can fit the tensor to scalar ratio, spectral index and its running provided
by the {\it Planck} data \cite{Planck:2013jfk}. 
%In Sec.~\ref{sum3-form}summarize and discuss the model herein investigated and propose subsequent lines to explore. 

In this chapter, we follow the units $m_{\rm P}=1$. 

\subsection{$\mathbb{N}$ 3-form fields model}

\label{N-form}

In this section, we generalize the background equations
associated to a single 3-form field, which has been studied in \cite{Koivisto:2009ew,Koivisto:2009fb,DeFelice:2012jt},
to $\mathbb{N}$ 3-form fields. 
We take a flat FLRW cosmology, described with the metric 
\begin{equation}
ds^{2}=-dt{}^{2}+a^{2}(t)d\boldsymbol{x}^{2}\,,\label{FRW-metric1-1}
\end{equation}

The general action for Einstein gravity and\textit{\emph{ $\mathbb{N}$}}
3-form fields is written as 
\begin{equation}
S=-\int d^{4}x\sqrt{-g}\left[\frac{1}{2}R-\sum_{I=1}^{\mathbb{N}}\left(\frac{1}{48}F_{I}^{2}+V(A_{I}^{2})\right)\right]\,,\label{N3F-action-1}
\end{equation}
where $A_{\beta\gamma\delta}^{(I)}$ is the $I$th 3-form field and
we have squared the quantities by contracting all the indices. The
strength tensor of the 3-form is given by\footnote{Throughout this chapter, the Latin index $I$ will be used to refer
the number of the quantity (or the 3-form field) or the $I$th quantity/field.
The other Latin indices, which take the values $i,j=1,2,3$, will
indicate the three dimensional quantities; whereas the Greek indices
will be used to denote four-dimensional quantities and they stand
for $\mu,\nu=0,1,2,3$.} 
\begin{equation}
F_{\alpha\beta\gamma\delta}^{(I)}\equiv4\nabla_{[\alpha}A_{\beta\gamma\delta]}^{(I)}\,,\label{N3f-Maxw-1}
\end{equation}
where anti-symmetrization is denoted by square brackets. As we have
assumed a homogeneous and isotropic universe, the 3-form fields depend
only on time and hence only the space like components will be dynamical,
thus their non-zero components (for FLRW background) are given by 
\begin{equation}
A_{ijk}^{(I)}=a^{3}(t)\epsilon_{ijk}\chi_{I}(t)\quad\Rightarrow A_{I}^{2}=6\chi_{I}^{2}\,,\label{NNZ-comp-1-1}
\end{equation}
where $\chi_{I}(t)$ is a comoving field associated to the $I$th
3-form field and $\epsilon_{ijk}$ is the standard three dimensional
Levi-Civita symbol. Also note that by introducing the more convenient
field $\chi_{I}(t)$, which is related to the corresponding 3-form
field by the above relation, we have, subsequently, the following
system of equations of motion for $\mathbb{N}$ 3-form fields 
\begin{eqnarray}
\ddot{\chi}_{I}+3H\dot{\chi}_{I}+3\dot{H}\chi_{I}+V_{,\chi_{I}} & =0\,,\label{NDiff-syst-1-1}
\end{eqnarray}
where $V_{,\chi_{I}}\equiv\frac{dV}{d\chi_{I}}$.
For each value of $I$ , each of the (\ref{NDiff-syst-1-1})
are not independent: it is straightforward to see that a peculiar
coupling is present through the Hubble parameter derivative, $\dot{H}$.
This fact will play a crucial role, establishing different classes
of inflationary behavior when more than one 3-from field is employed.
In this setting, the gravitational sector equations are given by 
\begin{equation}
\begin{aligned}
H^{2} & =\frac{1}{3}\left\{ \frac{1}{2}\sum_{I=1}^{\mathbb{N}}
\left[\left(\dot{\chi}_{I}
+3H\chi_{I}\right){}^{2}+
2V(\chi_{I})\right]\right\}\,,\\
\dot{H} & =-\frac{1}{2}\left[\sum_{I=1}^{\mathbb{N}}\chi_{I}V_{,\chi_{I}}\right]\,.
\end{aligned}
\label{NFriedm-1-1}
\end{equation}
%\begin{equation}
%\dot{H}=-\frac{1}{2}\left[\sum_{I=1}^{\mathbb{N}}\chi_{I}V_{,\chi_{I}}\right]\,.\label{NHdot-1-1}
%\end{equation}
Therefore, the mentioned (gravity mediated) coupling between the several
$\mathbb{N}$ 3-form fields will act through the gravitational sector
of the equations of motion. The total energy density and pressure
of the $\mathbb{N}$ 3-form fields read 
\begin{equation}
\begin{aligned}
\rho_{\mathbb{N}} & =\frac{1}{2}\sum_{I=1}^{\mathbb{N}}\left[\left(\dot{\chi}_{I}+3H\chi_{I}\right){}^{2}+2V(\chi_{I})\right]\,,\\
p_{\mathbb{N}} & =-\frac{1}{2}
\sum_{I=1}^{\mathbb{N}}\left[\left(\dot{\chi}_{I}+3H\chi_{I}\right){}^{2}+2V(\chi_{I})-2\chi_{I}V_{,\chi_{I}}\right]\,.
\label{Ndens-1-1}
\end{aligned}
\end{equation}
%\begin{equation}
%p_{\mathbb{N}}=-\frac{1}{2}
%\sum_{I=1}^{\mathbb{N}}\left[\left(\dot{\chi}_{I}+3H\chi_{I}\right){}^{2}+2V(\chi_{I})-2\chi_{I}V_{,\chi_{I}}\right]\,.\label{Npress-1-1}
%\end{equation}
We rewrite (\ref{NDiff-syst-1-1}) as 
\begin{eqnarray}
\ddot{\chi}_{I}+3H\dot{\chi}_{I}+V_{,\chi_{I}}^{\textrm{eff}} & =0\,,\label{NDiff-syst-1-1-2}
\end{eqnarray}
where 
\begin{equation}
V_{,\chi_{I}}^{\textrm{eff}}\equiv3\dot{H}\chi_{I}+V_{,\chi_{I}}=V_{,\chi_{I}}\left[1-\frac{3}{2}\chi_{I}^{2}\right]-\frac{3}{2}\chi_{I}\left[\overset{\mathbb{N}}{\underset{\underset{I\neq J}{J=1}}{\sum}}\chi_{I}V_{,\chi_{I}}\right]\,.\label{NGen-X-Pot-1}
\end{equation}
In order to describe the dynamics of the 3-form fields, we express
the equations of motion in terms of the variable
\begin{equation}
w_{I} \equiv \frac{\chi'_{I}+3\chi_{I}}{\sqrt{6}}\,,\label{NQuad-omega-1}
\end{equation}
where $\chi'_{I}\equiv d\chi_{I}/dN$ in which the number of e-folds of
inflationary expansion is $N=\ln\,a(t)$. Thus, we get 
\begin{eqnarray}
H^{2}\chi_{I}''+\left(3H^{2}+\dot{H}\right)\chi_{I}'+V_{,\chi_{I}}^{\textrm{eff}} & = & 0\,.\label{NQuad-X-diff-1}
\end{eqnarray}
The Friedmann constraint is written as 
\begin{equation}
H^{2}=\frac{1}{3}\frac{V(\chi_{I})}{\left(1-w^{2}\right)}\,,\label{NQuad-Fried-1}
\end{equation}
where 
\[
w^{2}\equiv\sum_{I=1}^{\mathbb{N}}w_{I}^{2}\,.
\]
Employing the dimensionless variables (\ref{NQuad-omega-1}), the
equations of motion (\ref{NQuad-X-diff-1}) can be rewritten in the
autonomous form as 
\begin{equation}
\begin{aligned}
\chi'_{I} &=3\left(\sqrt{\frac{2}{3}}w_{I}
-\chi_{I}\right)\,\\
w'_{I} & =  \frac{3}{2}\frac{V_{,\chi_{I}}}{V}\left(1-w^{2}\right)\left(\chi_{I}w_{I}-
\sqrt{\frac{2}{3}}\right)+\frac{3}{2}\left(1-w^{2}\right)\frac{1}{V}w_{I}
\overset{\mathbb{N}}{\underset{\underset{I\neq J}{J=1}}{\sum}}\chi_{J}V_{,\chi_{J}}\,,
\label{Nauton-n-n}
\end{aligned}
\end{equation}
%\begin{eqnarray}
%w'_{I} & = & \frac{3}{2}\frac{V_{,\chi_{I}}}{V}\left(1-w^{2}\right)\left(\chi_{I}w_{I}-
%\sqrt{\frac{2}{3}}\right)+\frac{3}{2}\left(1-w^{2}\right)\frac{1}{V}w_{I}
%\overset{\mathbb{N}}{\underset{\underset{I\neq J}{J=1}}{\sum}}\chi_{J}V_{,\chi_{J}}\,,\label{Nwprime-n}
%\end{eqnarray}

In the whole chapter, we study the sum separable potentials of the form 
\begin{equation}
 V=\underset{I}{\sum}\,V_{I}(\chi_{I})\,. 
\end{equation}

\subsubsection{Dual action for $\mathbb{N}$ 3-forms}

\label{dual-Sec}

In general, any $p$-form in $D$ dimensions has a dual of $(D-p)-$form
\cite{Germani:2009iq,Mulryne:2012ax}. In our case 3-form field $\left(A\right)$
and its field tensor four-form $(F)$ are dual to a vector and a scalar
field respectively which can be expressed as \cite{Mulryne:2012ax}
\begin{equation}
A_{\mu\nu\rho}=\epsilon_{\alpha\mu\nu\rho}B^{\alpha}\,,\quad F_{\mu\nu\rho\sigma}=-\epsilon_{\mu\nu\rho\sigma}\phi\,,
\label{dual definitions}
\end{equation}
where $\epsilon_{\mu\nu\rho\sigma}$ is an antisymmetric tensor.

The corresponding action for the scalar field dual representation
of the $\mathbb{N}$ 3-forms is \cite{Kumar:2014oka,Mulryne:2012ax}
\begin{equation}
S=-\int d^{4}x\sqrt{-g}\left[\frac{1}{2}R+P\left(X,\phi_{I}\right)\right]\,,
\end{equation}
where 
\begin{equation}
P\left(X,\phi_{I}\right)=\sum_{I=1}^{\mathbb{N}}\left(\chi_{I}V_{I,\chi_{I}}-V\left(\chi_{I}\right)-\frac{\phi_{I}^{2}}{2}\right)\,,
\label{dual action}
\end{equation}
with $X=-\frac{1}{2}G^{IJ}\left(\phi\right)\partial_{\mu}\phi_{I}\partial^{\mu}\phi_{J}$. 

The dual fields are related to the 3-forms through the following relation \cite{Kumar:2014oka,Mulryne:2012ax} 

\begin{equation}
X_{I}=-\frac{1}{2}\partial_{\mu}\phi_I\partial^{\mu}\phi_I=\frac{1}{2}V_{,\chi_{I}}^{2}\,.
\label{relation}
\end{equation}

For a background unperturbed FLRW cosmology, we can use the dualities
defined in (\ref{relation}) to write the following relation
between a 3-form field and its dual scalar field 
\begin{equation}
\phi_{I}=\dot{\chi}_{I}+3H\chi_{I}\,.\label{3-formdual}
\end{equation}

Using the above relations, in the Lagrangian (\ref{dual action}) we can identify the kinetic
term to be
\begin{equation}
K(X_{I})=\overset{\mathbb{N}}{\underset{I=1}{\sum}}\left(\chi_{I}V_{,\chi_{I}}
-V(\chi_{I})\right)\,.\label{kinetic term-1}
\end{equation}
Since this kinetic term is only a function of $\chi_{I}$ and not of
$\phi_{I}$, this means that the field metric is $G_{IJ}(\phi_{J})=1$. Therefore, we have $X=\sum X_{I}$. The 3-from fields present
on the right-hand side of (\ref{dual action}) should be viewed
as functions of the kinetic terms $X_{I}$ though the inverse of the
relation (\ref{relation}). 

Following the above relations we compute here the following quantities which we use later in our study
\begin{equation}
P_{,X}\equiv\sum_{I}P_{,\chi_{I}}=\sum_{I}P_{,\chi_{I}}\left(\frac{\partial\chi_{I}}{\partial X_{I}}\right)=\sum_{I}\frac{\chi_{I}}{V_{\chi_{I}}}\,.\label{PX}
\end{equation}
And similarly 
\begin{eqnarray}
P_{,X_{I}X_{I}} & = & \frac{1}{V_{,\chi_{I}\chi_{I}}V_{,\chi_{I}}^{2}}-\frac{\chi_{I}}{V_{,\chi_{I}}^{3}}\,.\label{PX23}\\
P_{,X_{I}X_{I}X_{I}} & = & -\frac{V_{,\chi_{I}\chi_{I}\chi_{I}}}{V_{,\chi_{I}\chi_{I}}^{3}V_{,\chi_{I}}^{2}}+\frac{3\chi_{I}}{V_{,\chi_{I}}^{5}}-\frac{3}{V_{,\chi_{I}}^{4}V_{,\chi_{I}\chi_{I}}}\,.\label{PX33}\\
P_{,I} & = & -\phi_{I}=-\sqrt{6}Hw_{I}\,.\label{PI}
\end{eqnarray}

Considering the {large amount of non-canonical scalar fields studies} in cosmology,
it might be tempting to think that given a 3-form theory the best
way to proceed would be to simply pass to the dual scalar field theory
and work solely with scalar field quantities. {However, starting} from a set
of massive 3-form fields {makes the task of analytically
writing the dual scalar field theory very difficult,} except for very particular potentials \cite{Mulryne:2012ax}.
{This can be seen by noting but the technical difficulty found when one tries to invert (\ref{relation}).}
Yet, in a similar manner to that advocated in Ref.~\cite{Mulryne:2012ax}
for the single field case, we will see that we can still make use
of the dual theory indirectly.

\subsubsection{Initial conditions and slow-roll inflation}

\label{sub:Initial cond.}

Analogous to the scalar field \cite{Liddle:1998jc} as well as single
3-form \cite{Koivisto:2009ew,DeFelice:2012jt} inflationary models,
the so-called slow-roll parameters are taken as $\epsilon\equiv-\dot{H}/H^{2}=-d\ln H/dN$
and $\eta\equiv\epsilon'/\epsilon-2\epsilon$, which, for our model,
are given by\footnote{Equivalently solely in terms of $\chi_{I}$ and $w_{I}$, $\eta=\frac{\sum_{I=1}^{\mathbb{N}}3\left(\sqrt{\frac{2}{3}}w_{I}-\chi_{I}\right)
\left(V_{,\chi_{I}}+V_{,\chi_{I}\chi_{I}}\chi_{I}\right)}{\sum_{I=1}^{\mathbb{N}}V_{,\chi_{I}}\chi_{I}}$}

\begin{equation}
\epsilon=\frac{3}{2}\frac{\sum_{I=1}^{\mathbb{N}}\chi_{I}V_{,\chi_{I}}}{V}
\left(1-w^{2}\right)\,,\label{N-3F-epsilon-1}
\end{equation}

\begin{equation}
\eta=\frac{\sum_{I=1}^{\mathbb{N}}\chi'_{I}\left(V_{,\chi_{I}}+\chi_{I}V_{,\chi_{I}\chi_{I}}\right)}
{\sum_{I=1}^{\mathbb{N}}\chi_{I}V_{,\chi_{I}}}\,.\label{N-3F-eta-1}
\end{equation}
We can see from (\ref{N-3F-epsilon-1}) and (\ref{N-3F-eta-1}) that,
for $\mathbb{N}$ 3-form fields, one manner to establish a sufficient
condition for inflation (with the slow-roll parameters $\epsilon\ll1$
and $\eta\ll1$) is by means of 
\begin{equation}
\begin{cases}
1-\sum_{I=1}^{\mathbb{N}}w_{I}^{2} & \approx0\,,\\
\chi'_{I} & \approx0\,.
\end{cases}\label{Ident-sl-cond}
\end{equation}
It is important, however, to also consider another (albeit less obvious)
possibility, which is to have instead 
\begin{equation}
\begin{cases}
1-\sum_{I=1}^{\mathbb{N}}w_{I}^{2}\approx0\,,\\
\sum_{I=1}^{\mathbb{N}}\chi'_{I}\left(V_{,\chi_{I}}+\chi_{I}V_{,\chi_{I}\chi_{I}}\right) & \approx0\,.
\end{cases}\label{NonIdent-sl-cond}
\end{equation}
The condition expressed in (\ref{NonIdent-sl-cond}) means that the
inclusion of more than one 3-form field allows the emergence of an
inflationary scenario without even requiring that $\chi_{I}'\approx0$.
Therefore, we can expect to have different behaviors, in contrast
to the ones usually found in models with just one 3-form. The different
$\mathbb{N}$ 3-form fields will evolve in an intricate correlated
way in order to satisfy (\ref{NonIdent-sl-cond}). This possibility
will deserve a more detailed analysis in the next sections. We should
note that all the derived equations in this section reduce to the
single one 3-form case when $\mathbb{N}=1$, as expected.

\subsubsection{Inflaton mass}

\label{low3formmass}Returning to the condition (\ref{Ident-sl-cond}),
we have

\[
\sum_{I=1}^{\mathbb{N}}\chi_{I}^{2}\thickapprox\,\frac{2}{3}\,.
\]
Note that the Friedmann constraint (\ref{NQuad-Fried-1}) does not
hold precisely at $\overset{\mathbb{N}}{\underset{I=0}{\sum}}\chi_{I}^{2}=\,2/3$,
and $\chi'_{I}=0$. If we assume a symmetric situation, where all $w_{I}$
are equal during inflation, i.e, if all fields come to the same value
during inflation, then $\chi_{I}\left(N\right)$ will take a constant
value 
\begin{equation}
\chi_{p}=\,\sqrt{\frac{2}{3\mathbb{N}}}\,.
\label{N3formsymplateau}
\end{equation}
In this symmetric situation, all the 3-form fields will behave identically
during inflation. If $\mathbb{N}$ is very large, the plateau of $\chi_{I}\left(\textrm{N}\right)$
converges towards zero ($\chi_{p}=\sqrt{\frac{2}{3\mathbb{N}}}\rightarrow0$
as $\mathbb{N\rightarrow\infty}$, for the symmetric case where all
$w_{n}$ are equal). The initial conditions for the single 3-form
inflation case were discussed in \cite{DeFelice:2012jt}. The reduction
of the plateau energy scale for $\mathbb{N}$ 3-forms can have a nontrivial
consequence, which is to bring the inflaton mass well below Planck
mass. This is illustrated by the following analysis. 
%Expressing the scale of the $\mathbb{N}$ 3-forms plateau (\ref{N3formsymplateau}) in terms\footnote{An equation similar to (\ref{inflaton mass}) was also proposed to study the case of $\mathbb{N}$-flation (multiple axion inflation) model in \cite{Dimopoulos:2005ac}. } of the reduced Planck mass $\left(m_{\rm P}=1/\sqrt{8\pi G}\right)$,
%\begin{equation}
%\chi_{p}\thickapprox\sqrt{\frac{2}{3\mathbb{N}}}m_{\rm P},\label{inflaton mass}
%\end{equation}
let us assume that all the 3-form fields behave in the same way, reaching
a constant value $\chi_{p}$ during inflation and starting to oscillate
by the end of inflation. Subsequently, we rewrite the Friedmann constraint
(\ref{NQuad-Fried-1}) for this case as,

\begin{equation}
H^{2}=\frac{1}{3}\frac{V}{\left(1-w^{2}\right)}\,.\label{N3form mass}
\end{equation}

Taking $V_{I}=V_{0I}\,f_{I}\left(\chi_{I}\right)$
, and where $f_{I}\left(\chi_{I}\right)$ are dimensionless functions.
Comparing (\ref{N3form mass}) with the Friedmann constraint
of a single 3-form field case, we get

\begin{equation}
V=\tilde{V}_1\,,\label{Mass label}
\end{equation}
where $\tilde{V}_{1}=\tilde{V}_{01}\,\tilde{f}_{1}(\chi_{1})$ is the potential for the single
3-form field case. If we choose $V_{01}=V_{02}=\cdots=V_{0n}=V_{0\mathbb{N}}$,
which means that the energy scales of the potentials are the same,
and also assume that $\chi_{I}=\chi_{p}$ for all $I$ in (\ref{Mass label}),
we get 
\begin{equation}
\frac{V_{0\mathbb{N}}}{\tilde{V}_{01}}=\frac{\tilde{f}_{1}\left(\chi_{1}\right)}
{\mathbb{N}f_{\mathbb{N}}\left(\chi_{\mathbb{N}}\right)}\,.\label{energy lable}
\end{equation}
Let us consider the power law potential $f=\chi^{l}$ , for a 3-form.
If we substitute the corresponding value of the plateau for $\mathbb{N}$
3-form $\left(\chi_{p}=\sqrt{\frac{2}{3\mathbb{N}}}\right)$ and of the
single 3-form case $\left(\chi_{1p}=\sqrt{\frac{2}{3}}\right)$ in 
(\ref{energy lable}), then we can have the following ratio of energy
scales for the potentials, of $\mathbb{N}$ 3-forms and single 3-form

\begin{equation}
\frac{V_{0\mathbb{N}}}{\tilde{V}_{01}}=\mathbb{\mathbb{N}}^{-1
	+\frac{l}{2}}\,.\label{energy scale}
\end{equation}
We can translate this argument in terms of the inflaton mass, which
is defined to be the square root of the second derivative of the potential.
Therefore, the ratio between the inflaton masses corresponding to
the $\mathbb{N}$ 3-forms potential $\left(m_{\mathbb{N}}\right)$,
and the single 3-form $\left(\tilde{m}_{1}\right)$, for a power law potential
$\left(\chi_{I}^{l}\right)$, is given by

\begin{equation}
\frac{m_{\mathbb{N}}}{\tilde{m}_{1}}\equiv\sqrt{\frac{V_{0\mathbb{N}}\,\chi_{p}^{l-2}}{\tilde{V}_{01}\,\chi_{1p}^{l-2}}}=\mathbb{N}^{-\frac{l}{2}}\,.
\label{m-scale}
\end{equation}

%In the case of quadratic potentials $\left(l=2\right)$, there is no reduction in the inflaton mass, similar to the single field case.
%The 3-form dual action (\ref{dual action}), can be shown to be equivalent
%to a canonical scalar field inflation. This dual action picture enhance
%the fact that the 3-form inflation can encompass both features of
%canonical and non canonical models. 
Therefore, it is possible to bring down the mass of the inflaton to lower energy
scales by increasing the number of 3-form fields.

%%%%%%%%%%%%%%%%%%%%%%%%%%%%%%%%%%%%%%%

\subsection{Two 3-form fields model}

\label{two 3-form}In this subsection, we would like to concentrate
on the case where only two 3-form fields are present.
%\footnote{A generalization from two to $\mathbb{N}$ 3-form fields is left for a future work, although, whenever possible, we will point out the main aspects that can straightforwardly be generalized.}. 
Accordingly, we will rewrite some of the equations as follows. Thus,
the non-zero components of (\ref{NNZ-comp-1-1}) are 
\begin{equation}
A_{ijk}^{\left(1\right)}= a^{3}(t)\epsilon_{ijk}\chi_{1}(t),\hspace{1cm}A_{ijk}^{\left(2\right)}=a^{3}(t)\epsilon_{ijk}\chi_{2}(t)\,,\\
\label{2NZ-comp-1}
\end{equation}
which implies $A_{1}^{2}= 6\chi_{1}^{2}, A_{2}^{2}=6\chi_{2}^{2}\,.$
Also, we rewrite equations of motion (\ref{NQuad-X-diff-1}) in terms
of our dimensionless variables as 
\begin{eqnarray}
H^{2}\chi''_{1}+\left(3H^{2}+\dot{H}\right)\chi'_{1}+V_{,\chi_{1}}^{\textrm{eff}} & = & 0\,,\label{2Quad-X-diff}\\
H^{2}\chi''_{2}+\left(3H^{2}+\dot{H}\right)\chi'_{2}+V_{,\chi_{2}}^{\textrm{eff}} & = & 0\,,\label{2Quad-Y-diff}
\end{eqnarray}
where the Friedmann and acceleration equations are given by 
\begin{equation}
\begin{aligned}H^{2} & =\frac{1}{3}\frac{V_{1}(\chi_{1})+V_{2}(\chi_{2})}{\left(1-w_{1}^{2}-w_{2}^{2}\right)}\,,
\end{aligned}
\label{2Quad-Fried}
\end{equation}
\begin{equation}
\dot{H}=-\frac{1}{2}\left(\chi_{1}V_{,\chi_{1}}+\chi_{2}V_{,\chi_{2}}\right)\,.\label{2Hdot-1}
\end{equation}

In order to further discuss suitable initial conditions, the slow
roll conditions $\epsilon,\left|\eta\right|\ll1$ suggests the equation
of a circle (of unit radius), as 
\begin{equation}
w_{1}^{2}+w_{2}^{2}\approx1\,,\label{eps -constr-1}
\end{equation}
which we rewrite in terms of trivial parametric relations as 
\begin{equation}
\begin{aligned}w_{1}\approx\:\cos\theta\,,\\
w_{2}\approx\:\sin\theta\,.
\end{aligned}
\label{a-value}
\end{equation}
Subsequently, from (\ref{Nauton-n-n}), we can establish the initial
conditions for the field derivatives 
\begin{eqnarray}
\begin{cases}
\chi'_{1}\approx3\left(\sqrt{\frac{2}{3}}\:\cos\theta-\chi_{1}\right)\,,\\
\chi'_{2}\approx3\left(\sqrt{\frac{2}{3}}\;\sin\theta-\chi_{2}\right)\,.
\end{cases}\label{x1x2plateau}
\end{eqnarray}
Since (\ref{a-value}) can be satisfied by assigning many different
continuous values of the new parameter $\theta$, we, therefore, anticipate
to investigate diverse solutions. More precisely, a particular choice
of this parameter will affect the way (\ref{N-3F-eta-1}), (i.e,
the value of $\eta$) will depend on the two 3-form fields. Before
proceeding, let us mention that for two 3-forms inflation, we choose
herein initial conditions for the fields above or below the value
given by (\ref{N3formsymplateau}), which are expected to influence
the number of $e$-foldings. In particular, we will investigate the asymmetric
situation, when each $w_{I}$ is different\footnote{For example, if we take $w_{1}=1$ and $w_{I>1}=0$, we then find
a scenario similar to single 3-form field driving the inflation
and where all the other fields approach zero.}, which will provide a new behavior with respect to inflation.

%%%%%%%%%%%%%%%%%%%%%%%%

\subsubsection{Type I inflation ($\chi_{I}^{\prime}\approx0$)}

\label{sub: Type I}As we have established in Sec.~\ref{sub:Initial cond.},
the slow-roll conditions enable us to find two types of inflationary
solutions, according to relation (\ref{Ident-sl-cond})-(\ref{NonIdent-sl-cond}).
In the following, we investigate them in more
detail. In type I solution, the 3-form
fields which are responsible for driving the inflationary period,
will be displaying $\chi'_{I}\approx0$. The following is a stability
analysis for this type, presented in a dynamical system context. Whenever
necessary, we will complement this study by a numerical discussion.

Let us remind the autonomous system of equations for the field $\chi_{1}$,
\begin{align}
\chi'_{1}= & 3\left(\sqrt{\frac{2}{3}}w_{1}-\chi_{1}\right)\,,\label{Dyn-x1}\\
w'_{1}= & \frac{3}{2}\left(1-\left(w_{1}^{2}+w_{2}^{2}\right)\right)\left(\lambda_{1}\left(\chi_{1}w_{1}-\sqrt{\frac{2}{3}}\right)+\lambda_{2}\chi_{2}w_{1}\right)\,,\label{Dyn-w1}
\end{align}
and also for $\chi_{2}$, 
\begin{align}
\chi'_{2}= & 3\left(\sqrt{\frac{2}{3}}w_{2}-\chi_{2}\right)\,,\label{Dyn-x2}\\
w'_{2}= & \frac{3}{2}\left(1-\left(w_{1}^{2}+w_{2}^{2}\right)\right)
\left(\lambda_{2}\left(\chi_{2}w_{2}-\sqrt{\frac{2}{3}}\right)+\lambda_{1}\chi_{1}w_{2}\right)\,,\label{Dyn-w2}
\end{align}
where $\lambda_{I}=V_{,\chi_{I}}/V$. Notice that, (\ref{Dyn-x1})-(\ref{Dyn-w1})
are coupled with (\ref{Dyn-x2})-(\ref{Dyn-w2}). With the variables
$(\chi_{I},w_{n})$, let $f_{1}:=d\chi_{1}/dN$, $f_{2}:=dw_{1}/dN$, $f_{3}:=d\chi_{2}/dN$
and $f_{4}:=dw_{2}/dN$. The critical points are located at the field
space coordinates $\left(x_{c}\right)$ and are obtained by setting
the condition $(f_{1},\,f_{2},\,f_{3},\,f_{4})|_{x_{{\rm c}}}=0$.

To determine the stability of the critical points, we need to perform
linear perturbations around each of them by using $x(t)=x_{{\rm c}}+\delta x(t)$;
this results in the equations of motion $\delta x'={\cal M}\delta x$,
where ${\cal M}$ is the Jacobi matrix of each critical point whose
components are ${\cal M}_{ij}=(\partial f_{i}/\partial x_{j})|_{x_{{\rm c}}}$.
A critical point is called stable (unstable) whenever the eigenvalues
$\zeta_{i}$ of ${\cal M}$ are such that $\mathsf{Re}(\zeta_{i})<0$
($\mathsf{Re}(\zeta_{i})>0$) \cite{Khalil}. If $\mathsf{Re}(\zeta_{i})=0$,
then other methods should be employed to further assess the stability
of the critical point. Among different approaches, we have the center
manifold theorem \cite{Carr,Guckenheimer,Khalil,Boehmer:2011tp} or,
alternatively, we can consider a perturbative expansion to nonlinear
order as in Refs.~\cite{Koivisto:2009ew,DeFelice:2012jt}. In this
work we will follow the last mentioned method, whenever necessary.

The autonomous dynamical (\ref{Dyn-x1})-(\ref{Dyn-w2}) fixed
points are given by 
\begin{equation}
\begin{aligned}\chi_{1c}= & \sqrt{\frac{2}{3}}w_{1},\hspace{1cm}w_{1c}=\sqrt{\frac{2}{3}}
\frac{\lambda_{1}}{\lambda_{1}\chi_{1}+\lambda_{2}\chi_{2}}\,,\\
\chi_{2c}= & \sqrt{\frac{2}{3}}w_{2},\hspace{1cm}w_{2c}=\sqrt{\frac{2}{3}}\frac{\lambda_{2}}{\lambda_{1}\chi_{1}+\lambda_{2}\chi_{2}}\,.
\end{aligned}
\label{fixed-P1}
\end{equation}
If $\lambda_{1}\neq0$ and $\lambda_{2}\neq0$ (otherwise, $V_{1,\chi_{1}}=0$
and $V_{2,\chi_{2}}=0$ ), (\ref{fixed-P1}) can be rewritten as
\begin{equation}
\begin{aligned}
\chi_{1c}= & \sqrt{\frac{2}{3}}w_{1},\hspace{1cm}w_{1c}=\frac{\lambda_{1}}{\sqrt{\lambda_{1}^{2}+\lambda_{2}^{2}}}\,,\\
\chi_{2c}= & \sqrt{\frac{2}{3}}w_{2},\hspace{1cm}w_{2c}=\frac{\lambda_{2}}{\sqrt{\lambda_{1}^{2}+\lambda_{2}^{2}}}\,.
\end{aligned}
\label{fixed-P2}
\end{equation}

Generically, with a inflationary stage as a target, the fixed points
coordinates must satisfy $w_{1c}^{2}+w_{2c}^{2}\simeq1$. This last
condition is required to satisfy the slow-roll condition (\ref{Ident-sl-cond}).
Therefore, we can define as well 
\begin{equation}
\begin{aligned}w_{1c}=\:\cos\theta\,,\\
w_{2c}=\:\sin\theta\,.
\end{aligned}
\label{Type I omega}
\end{equation}
Note that, when we consider the field coordinates in (\ref{fixed-P1}),
also $\chi_{1c}$ and $\chi_{2c}$ are constrained by $\chi_{1c}^{2}+\chi_{2c}^{2}=2/3$.
Consequently, the dynamical system (\ref{Dyn-x1})-(\ref{Dyn-w2})
has fixed points with only two independent degrees of freedom, which
can be chosen to be the pair $\left(\chi_{1c},\,w_{1c}\right)$. Therefore,
the critical points or inflationary attractors are found by solving
the following expression $w_{1c}^{2}+w_{2c}^{2}=1$, which upon substitution
gives, 
\begin{equation}
\left(\sqrt{\frac{2}{3}}\frac{\lambda_{1}}{\lambda_{1}\chi_{1c}+\lambda_{2}\chi_{2c}}\right)^{2}+
\left(\sqrt{\frac{2}{3}}\frac{\lambda_{2}}{\lambda_{1}\chi_{1c}+\lambda_{2}\chi_{2c}}\right)^{2}=1\,.\label{theta label}
\end{equation}
It is clear from (\ref{theta label}) that the location of
critical points depends on the choice of 3-form potentials. For example
let us take $V_{1}=\chi_{1}^{n}$ and $V_{2}=\chi_{2}^{m}$. It follows
that\textcolor{blue}{{} } 
\begin{equation}
\left(\frac{n\left(\sqrt{\frac{2}{3}}\right)^{n}\left(\cos\theta\right)^{n-1}}{n\left(\sqrt{\frac{2}{3}}\cos\theta\right)^{n}+m\left(\sqrt{\frac{2}{3}}\sin\theta\right)^{m}}\right)^{2}+\left(\frac{m\left(\sqrt{\frac{2}{3}}\right)^{m}\left(\sin\theta\right)^{m-1}}{n\left(\sqrt{\frac{2}{3}}\cos\theta\right)^{n}+m\left(\sqrt{\frac{2}{3}}\sin\theta\right)^{m}}\right)^{2}=1\,.\label{cond-V1-V2}
\end{equation}
From (\ref{cond-V1-V2}), $\theta=0$ and $\theta=\pi/2$ can
be called as trivial fixed points independent of the choice of $n,m$.
Satisfying the condition (\ref{cond-V1-V2}), for our particular choice
of potentials, allows us to also identify non trivial fixed points
in the range $0<\theta<\pi/2$. To easily identify these, we can extract
a simple constraint from (\ref{fixed-P2}), given by

\begin{equation}
\chi_{1c}/\chi_{2c}=\lambda_{1}/\lambda_{2}\,.\label{simpelconditiontype 1}
\end{equation}
Condition (\ref{simpelconditiontype 1}) is fully consistent with
(\ref{cond-V1-V2}), except for the trivial fixed points $\theta=0$
and $\theta=\pi/2$. Let us apply the example where $V_{1}=\chi_{1}^{n}$
and $V_{2}=\chi_{2}^{m}$, and substituting in (\ref{simpelconditiontype 1}),
We have 
\begin{equation}
\frac{n\left(\sqrt{\frac{2}{3}}\cos\theta\right)^{n-2}}{m\left(\sqrt{\frac{2}{3}}\sin\theta\right)^{m-2}}=1\,.\label{n,m.critical points}
\end{equation}
We can read from (\ref{n,m.critical points}) that for identical
quadratic potentials, i.e., for $n=m=2$, (\ref{simpelconditiontype 1})
is satisfied for all values of $0<\theta<\pi/2$. Identical quadratic
potentials is the only case where we can have an infinite number of
non trivial fixed points. For any other choice of potentials , i.e.,
for $n\neq m$, there will only be a finite number of non trivial
fixed points.

In Fig.~\ref{fig1} we illustrate the evolution of the fields $\chi_{1}$
and $\chi_{2}$ for quadratic potentials with $\theta=\pi/2$ and $\theta=\pi/4$.
The asymmetry in choosing $\theta\neq\pi/4$ manifests through one
of the 3-form fields having a plateau slightly higher than the other.

\begin{figure}[h!]
\centering\includegraphics[height=1.8in]{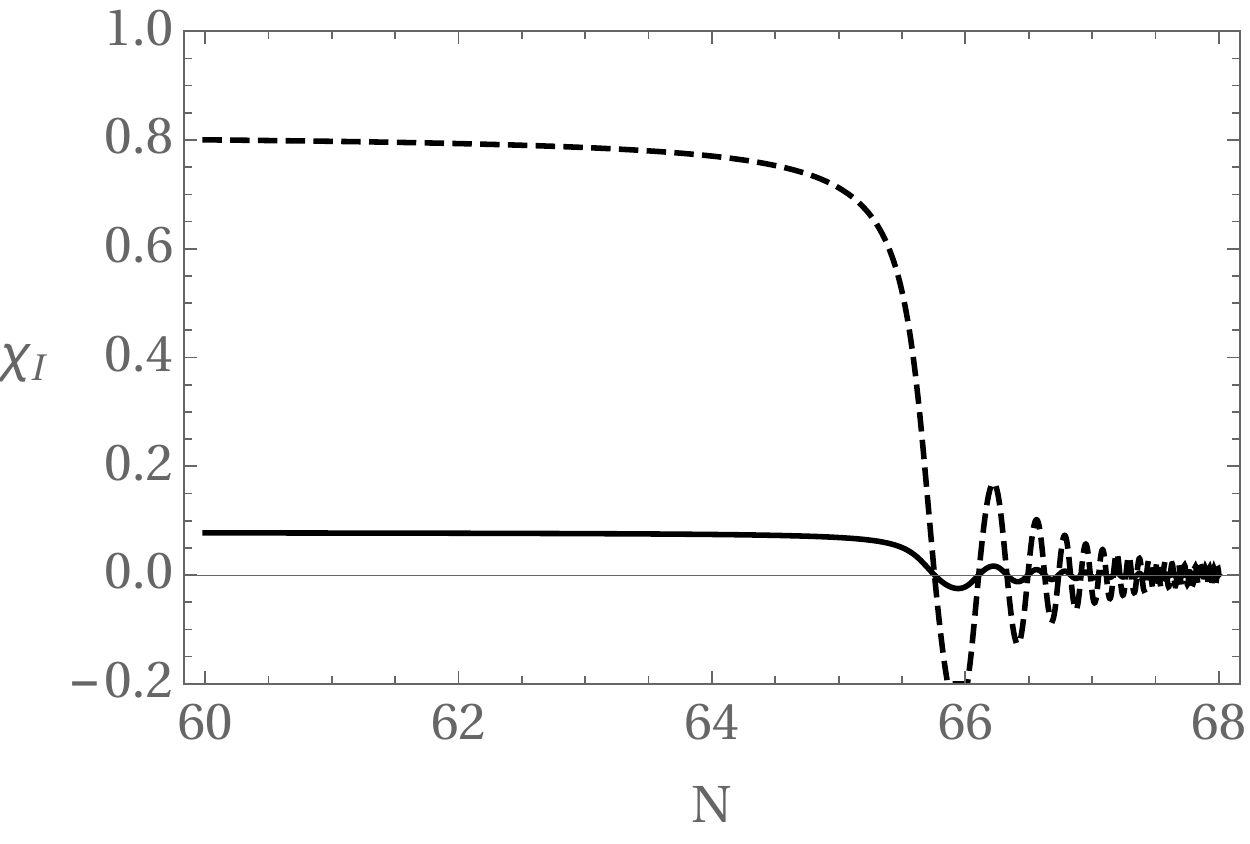}\quad{}\includegraphics[height=1.8in]{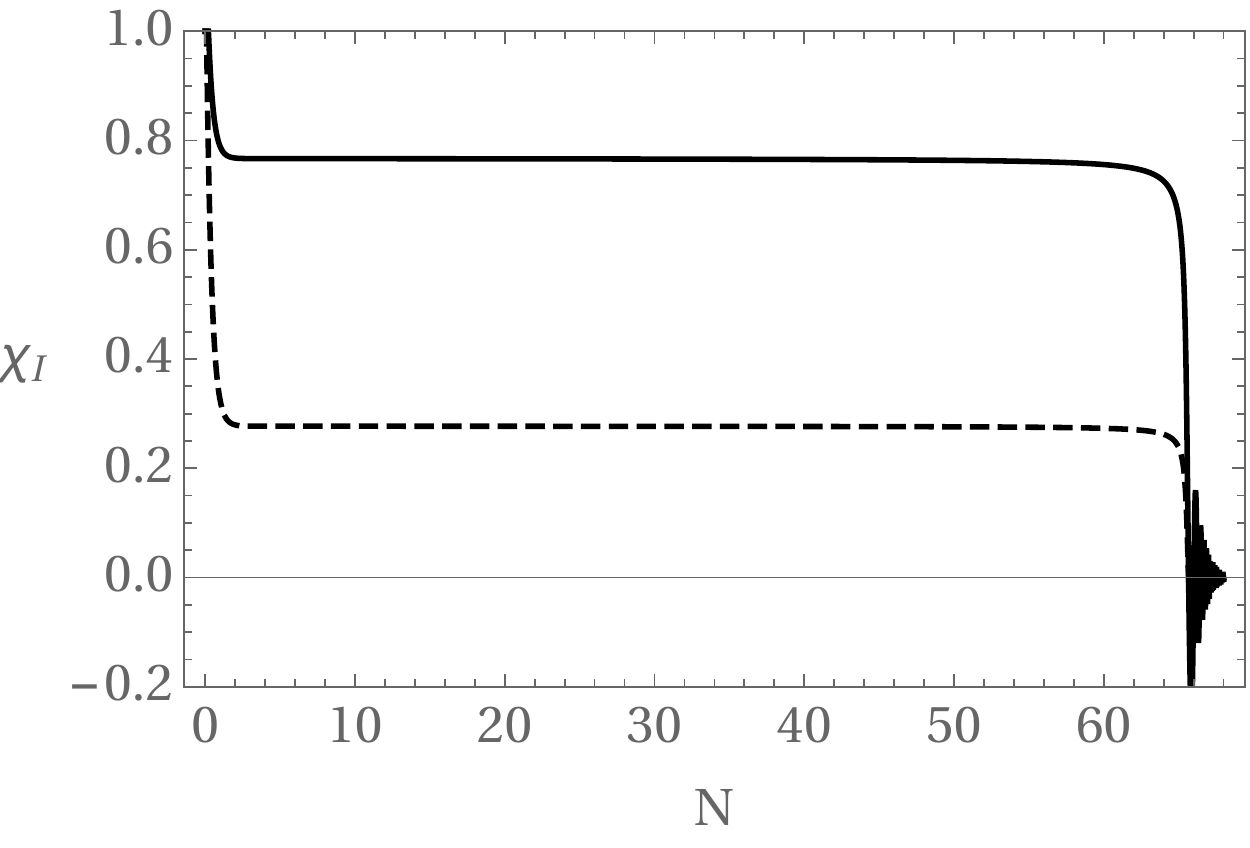}
\caption{Left panel is the graphical representation of the numerical solutions
of (\ref{2Quad-X-diff}) and (\ref{2Quad-Y-diff}) for $\chi_{1}\left(N\right)$
(full line) and $\chi_{2}\left(N\right)$ (dashed line) with $\theta\approx\dfrac{\pi}{2}$
for the potentials $V_{1}=\chi_{1}^{2}$ and $V_{2}=\chi_{2}^{2}$\textbf{.}
In the right panel, we depict the graphical representation of the
numerical solutions of (\ref{2Quad-X-diff}) and (\ref{2Quad-Y-diff})
for $\chi_{1}\left(N\right)$ (full line) and $\chi_{2}\left(N\right)$
(dashed line) with $\theta=\dfrac{\pi}{9}$. We have taken the initial
conditions as $\chi_{1}(0)=2.1\times\sqrt{\frac{1}{3}}$ and $\chi_{2}(0)=2.1\times\sqrt{\frac{1}{3}}$.}

\label{fig1} 
\end{figure}

Type I solutions, as far as the stability analysis, are very similar
to the scenario where just single 3-form field is present. The
novelty here is that we can have solutions as shown in Fig.~\ref{fig1}.
Therein, we have a case where we consider that the 3-form fields $\chi_{1}$
and $\chi_{2}$ are under the influence of the same kind of quadratic
potential, i.e, $V_{n}=\chi_{I}^{2}$.

We discuss the stability of these fixed points and their stability
in the Appendix \ref{AI3-forms} for simple potentials. Other combinations of potentials
can be tested for stability along the same lines presented there.
We summarize the results in Table \ref{potential-stability}.
\begin{center}
\begin{table}[h!]
\centering%
\begin{tabular}{c|c|c|c|c}
\hline 
{\footnotesize{}{{}~~$V\left(\chi_{1}\right)$~~}}  & {\footnotesize{}{{}~~$V\left(\chi_{2}\right)$~~}}  & {\footnotesize{}{{}existence}}  & {\footnotesize{}{{}stability}}  & {\footnotesize{}{{}Oscillatory regime}}\tabularnewline
\hline 
\hline 
{\footnotesize{}{{}$\chi_{1}^{2}$}}  & {\footnotesize{}{{}$\chi_{2}^{2}$}}  & {\footnotesize{}{{}$0<\theta<\pi/2$}}  & {\footnotesize{}{{}unstable saddle}}  & {\footnotesize{}{{}yes}}\tabularnewline
{\footnotesize{}{{}$\chi_{1}^{4}+\chi_{1}^{2}$}}  & {\footnotesize{}{{}$\chi_{2}^{4}+\chi_{2}^{2}$}}  & {\footnotesize{}{{}$\theta=\left\{ 0,\,\pi/4,\,\pi/2\right\} $}}  & {\footnotesize{}{{}unstable}}  & {\footnotesize{}{{}yes}}\tabularnewline
{\footnotesize{}{{}$\chi_{1}^{3}+\chi_{1}^{2}$}}  & {\footnotesize{}{{}$\chi_{2}^{3}+\chi_{2}^{2}$}}  & {\footnotesize{}{{}$\theta=\left\{ 0,\,\pi/4,\,\pi/2\right\} $}}  & {\footnotesize{}{{}unstable}}  & {\footnotesize{}{{}yes}}\tabularnewline
{\footnotesize{}{{}$\exp\left(\chi_{1}^{2}\right)-1$}}  & {\footnotesize{}{{}$\exp\left(\chi_{2}^{2}\right)-1$}}  & {\footnotesize{}{{}$\theta=\left\{ 0,\,\pi/4,\,\pi/2\right\} $}}  & {\footnotesize{}{{}unstable}}  & {\footnotesize{}{{}yes}}\tabularnewline
{\footnotesize{}{{}$\chi_{1}^{2}$}}  & {\footnotesize{}{{}$\chi_{2}^{4}+\chi_{2}^{2}$}}  & {\footnotesize{}{{} $\theta=\left\{ 0,\,\pi/2\right\} $}}  & {\footnotesize{}{{}unstable}}  & {\footnotesize{}{{}yes}}\tabularnewline
{\footnotesize{}{{}$\exp\left(-\chi_{1}^{2}\right)$}}  & {\footnotesize{}{{}$\exp\left(-\chi_{2}^{2}\right)$}}  & {\footnotesize{}{{}$\theta=\left\{ 0,\,\pi/4,\,\pi/2\right\} $}}  & {\footnotesize{}{{}unstable}}  & {\footnotesize{}{{}no}}\tabularnewline
{\footnotesize{}{{}$\chi_{1}^{2}$}}  & {\footnotesize{}{{}$\chi_{2}^{4}$}}  & {\footnotesize{}{{}$\theta=\left\{ 0,\,\pi/3,\,\pi/2\right\} $}}  & {\footnotesize{}{{}unstable}}  & {\footnotesize{}{{}yes}}\tabularnewline
{\footnotesize{}{{}$\chi_{1}^{n}\quad\left(n>2\right)$}}  & {\footnotesize{}{{}$\chi_{2}^{n}\quad\left(n>2\right)$}}  & {\footnotesize{}{{}$\theta=\left\{ 0,\,\pi/4,\,\pi/2\right\} $}}  & {\footnotesize{}{{}unstable}}  & {\footnotesize{}{{}no}}\tabularnewline
\hline 
\hline 
\multicolumn{1}{c}{} & \multicolumn{1}{c}{} & \multicolumn{1}{c}{} & \multicolumn{1}{c|}{} & \tabularnewline
\end{tabular}\caption{Summary of some type I solutions critical points and their properties.}

\label{potential-stability} 
\end{table}
\par\end{center}

\subsubsection{Type II inflation $\left(\chi_{I}^{\prime}\protect\not\approx0\right)$ }

\label{sub: Type II} 
\begin{figure}[h!]
\centering\includegraphics[height=1.8in]{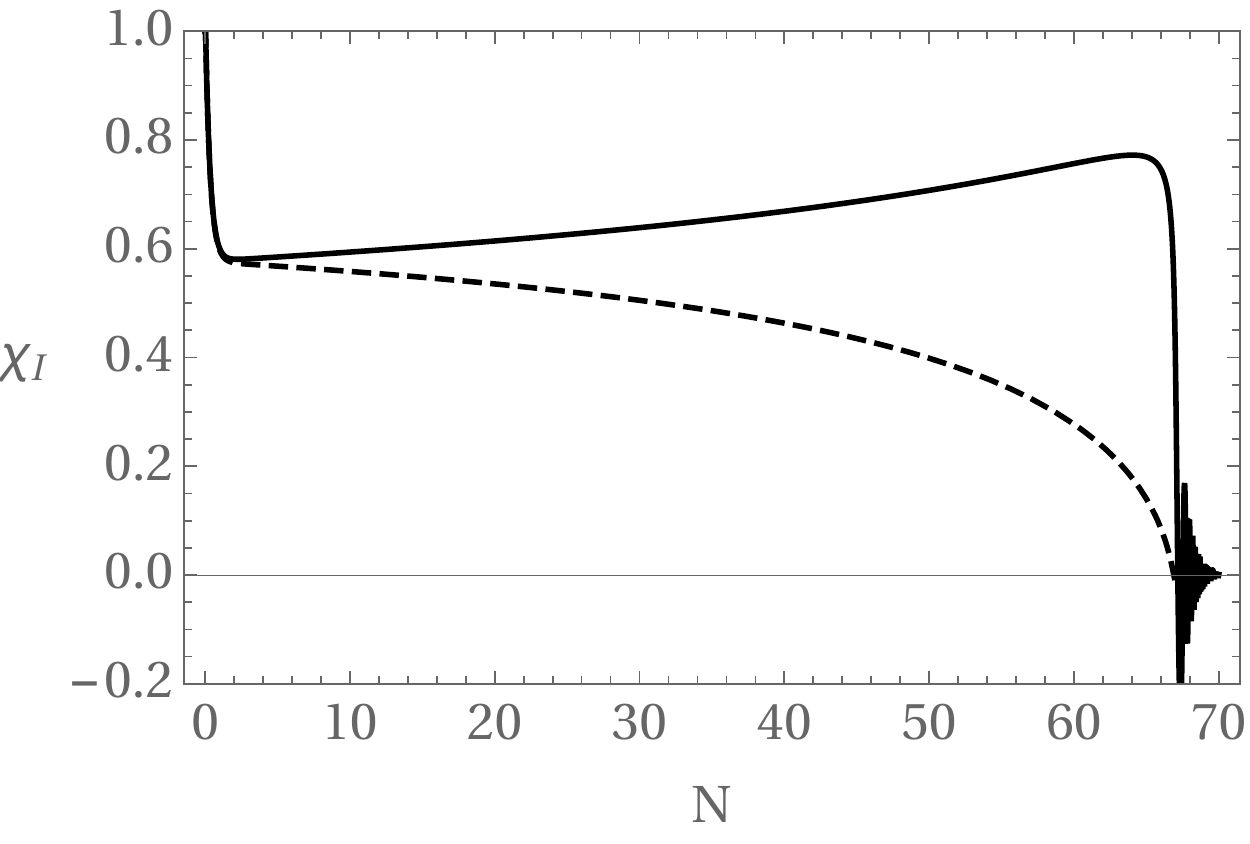}\quad{}\includegraphics[height=1.8in]{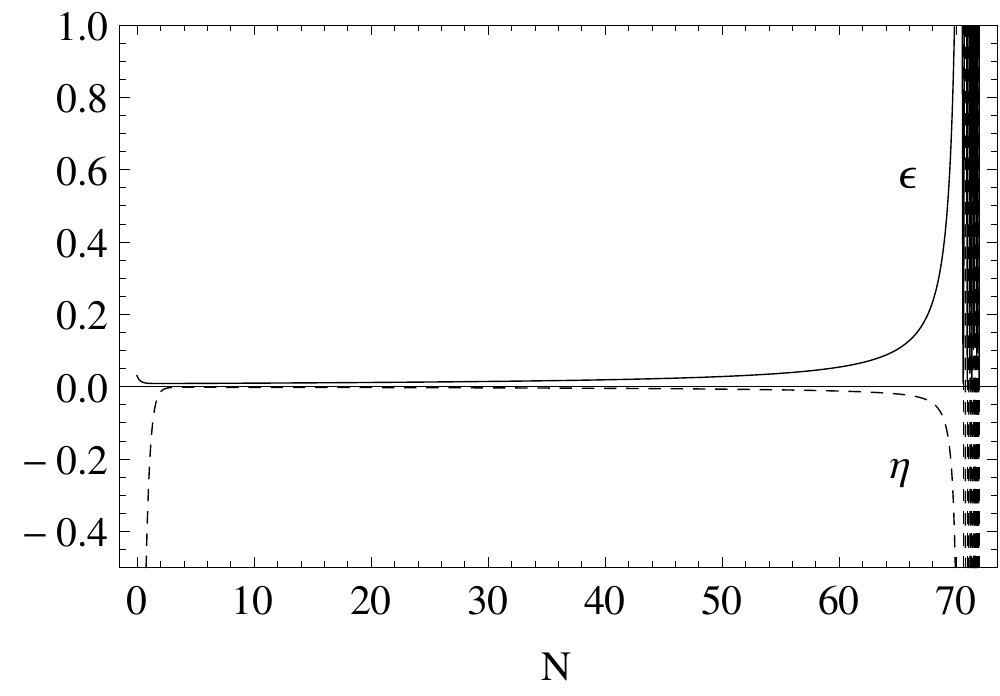}
\caption{In the left panel we have the graphical representation of the numerical
solutions of (\ref{2Quad-X-diff}) and (\ref{2Quad-Y-diff}) for $\chi_{1}\left(N\right)$
(full line) and $\chi_{2}\left(N\right)$ (dashed line) with $\theta=\dfrac{\pi}{4}$
for the potentials $V_{1}=\chi_{1}^{2}$ and $V_{2}=2\chi_{2}^{2}$. We
have taken the initial conditions as $\chi_{1}(0)=1.8\times\sqrt{\frac{1}{3}}$
and $\chi_{2}(0)=2.0\times\sqrt{\frac{1}{3}}$. In the right panel, and
for the same initial conditions, we have the graphical representation
of the numerical solutions for $\epsilon\left(N\right)$ (full line)
and $\eta\left(N\right)$ (dashed line).}

\label{fig7} 
\end{figure}

\begin{figure}[h!]
\centering\includegraphics[height=2.5in]{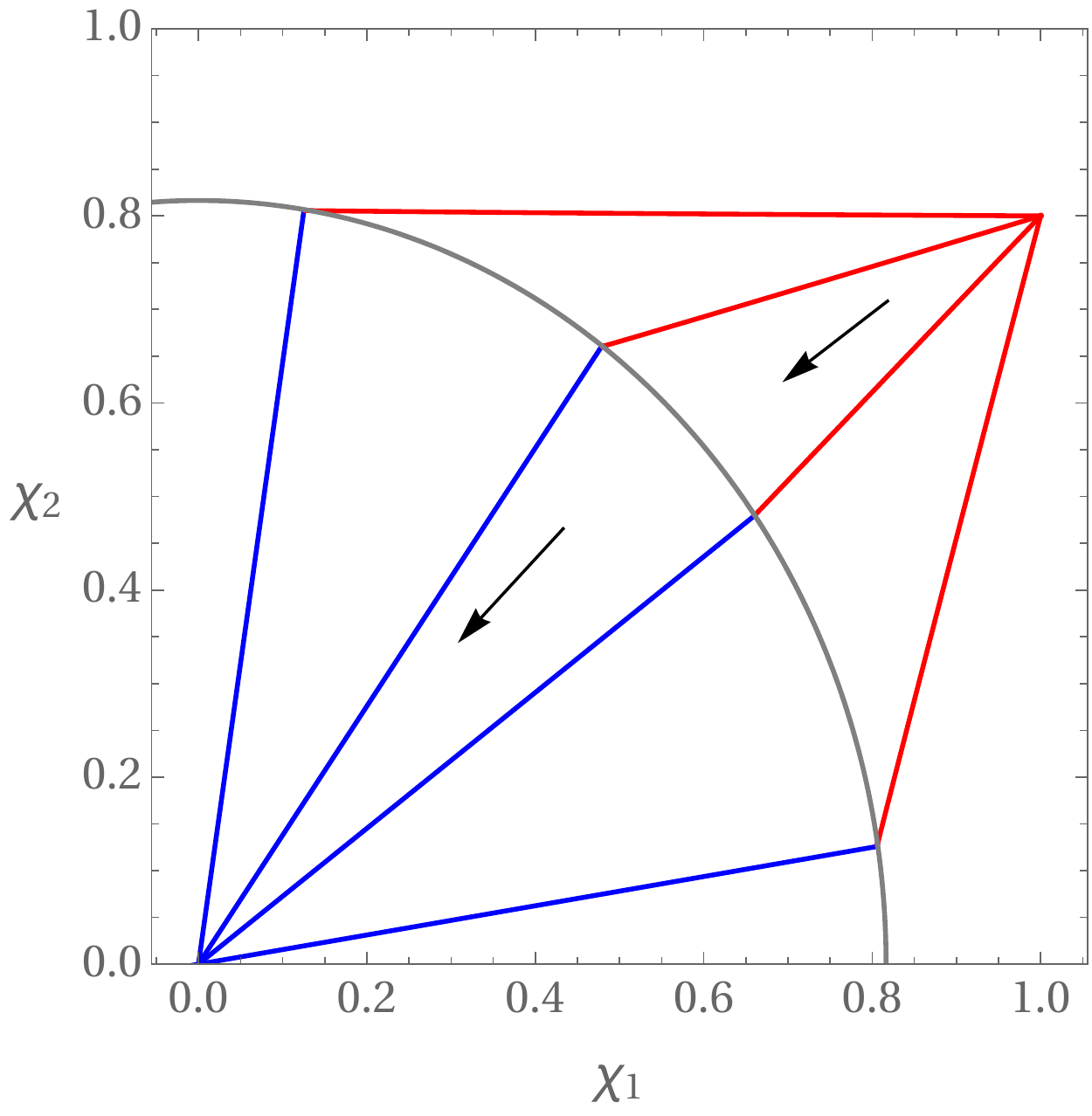}\quad{}\includegraphics[height=2.5in]{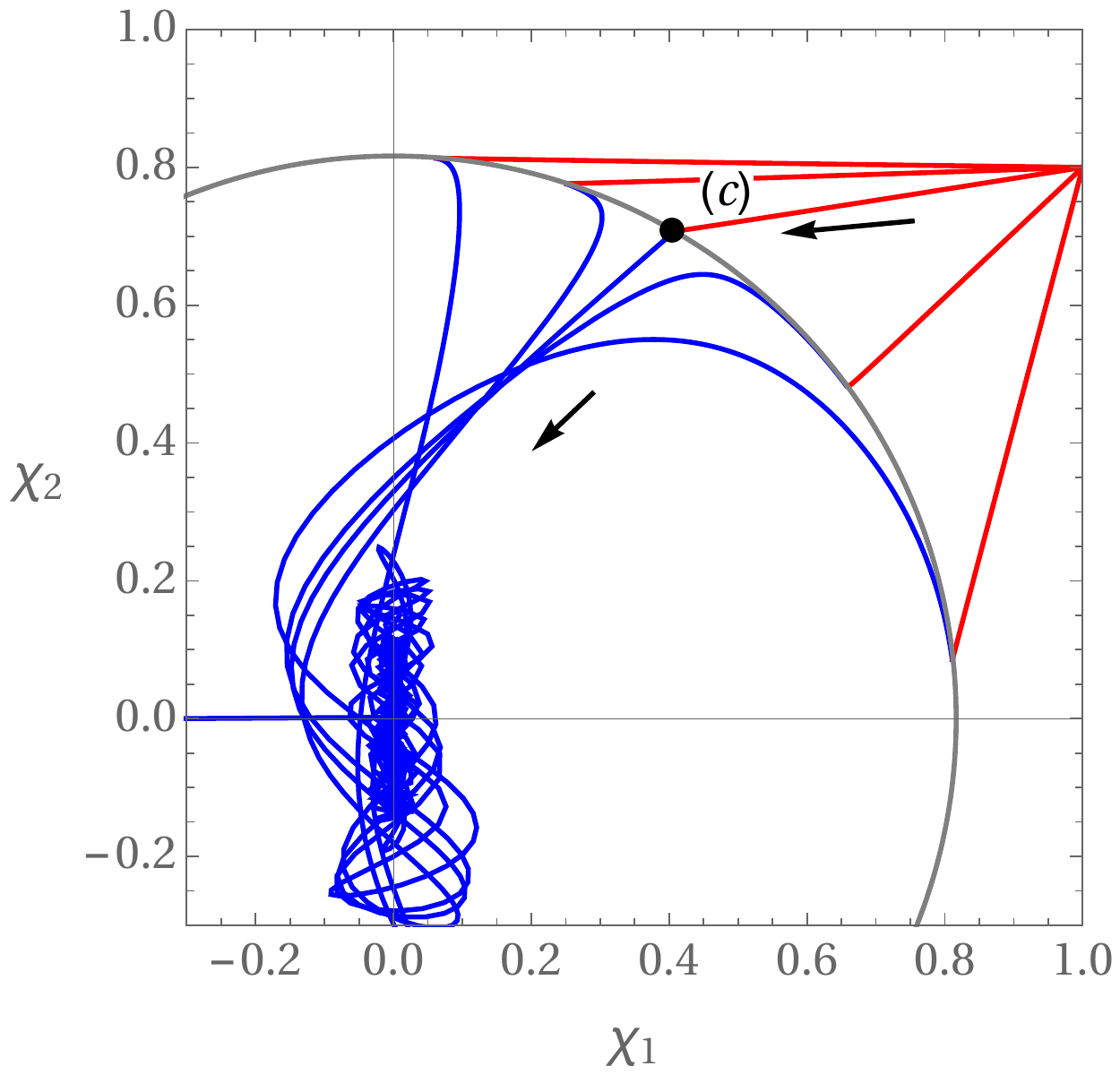}
\caption{This figure represents a set of trajectories evolving in the $\left(\chi_{1},\,\chi_{2}\right)$
space. These trajectories are numerical solutions of (\ref{2Quad-X-diff})
and (\ref{2Quad-Y-diff}) and correspond to a situation where we choose
$V_{1}=\chi_{1}^{2}$ and $V_{2}=\chi_{2}^{2}$ (left panel), as an illustrative
example only showing type I solution. All the fixed points are part
of the arc of radius $\sqrt{2/3}$ in the $\left(\chi_{1},\chi_{2}\right)$
plane. In the right panel, we have an example, where we have taken
$V_{1}=\chi_{1}^{2}$ and $V_{2}=\chi_{2}^{4}$, showing type II solutions,
except for the trajectory going close to a fixed point with $\theta=\pi/3$
(point $C$). In addition, in the right panel, we have an illustration
of two 3-form fields damped oscillations by the end of inflation.
The arrows, in the plots, indicate the direction of time in the trajectories. }

\label{fig7-sq-sq} 
\end{figure}

Let us now present the other class of inflationary
solution, which was mentioned in the Introduction. This type is associated
to the manner asymmetry is present. Let us be more specific. One way
to attain this solution consists of choosing an initial value of $\theta$
away from the fixed points previously discussed. This corresponds
to the curved trajectories in the right panel of Fig.~\ref{fig7-sq-sq}.
Another manner is by choosing different scales of the potentials i.e., $V_{01}\neq V_{02}$. In any
case, the inflationary behavior (type II) is similarly affected concerning
either way of introducing asymmetry. We should note here that there
is no analog for a type II solution within single 3-form driven inflation.
To understand this new type of inflationary scenario, let us take
$V_{1}=\chi_{1}^{2}$ and $V_{2}=2\chi_{2}^{2}$ (just different slopes),
whose numerical solutions are plotted in Fig.~\ref{fig7}.

In Fig.~\ref{fig7}, the two fields continuously evolve, and at the
same time assist each other in order to sustain a slow-roll regime.
As we can see from the left panel of Fig.~\ref{fig7}, one field continues
to slowly decrease (dashed line) and the other (full line) starts
to increase until it enters in an oscillatory regime. However, in
the right panel of Fig.~\ref{fig7}, we see that the slow-roll parameters
evolve (before oscillating) near to zero during the period of inflation.
Moreover, from (\ref{N-3F-eta-1}), the behavior of the two fields
are such that even with $\chi_{I}^{\prime}\not\approx0$, the slow-roll
conditions are consistent with inflation. The fact is that the slow
roll parameter $\eta\rightarrow0$ is now due to the constraint (\ref{NonIdent-sl-cond}).
As previously mentioned, a rather unusual cooperation between the
two 3-form fields, emphasized by the mentioned coupling (gravity mediated,
through $\dot{H}$) provides a different inflationary dynamics.

This new type of solution presents a period of inflation with an interesting
new feature. More precisely, when one 3-form field decreases, say
$\chi_{1}$, then the other field, $\chi_{2}$, is constrained to increase.
However, the increase of the second 3-form field is limited by the
fact that, as the first one inevitably approaches zero, then (\ref{Dyn-w2})
becomes 
\begin{equation}
w'_{2}\sim\frac{3}{2}\left(1-w_{2}^{2}\right)\lambda_{2}\left(\chi_{2}w_{2}-\sqrt{\frac{2}{3}}\right)\,,\label{Dyn-w2-app1}
\end{equation}
with the coupling term $\lambda_{1}\chi_{1}w_{2}$ being negligibly small.
We see that (\ref{Dyn-w2-app1}) will become zero when $w_{2}$
(which is increasing, as is $\chi_{2}$) will approach $1$. At this
stage, and inspecting (\ref{Dyn-x2}), it is clear that $\chi_{2}$
will stop increasing and start to decrease, making $\chi_{2}'<0$. This
situation is depicted in the left panel Fig.~\ref{fig7}, where the
decreasing field is reaching zero at the same period where the other
stops to increase and also converges to zero. The two 3-form fields
behave strongly correlated and assisting each other through the inflationary
period. Therefore, this more complex and correlated evolution of the
fields can provide a different observational signature when compared
to other multifield inflationary models.

The different nature of type I and type II solutions is be represented
in Fig.~\ref{fig7-sq-sq}. Therein, we have a parametric plot\footnote{Please note that Fig.~\ref{fig7-sq-sq} is $\mathit{not}$ a phase
space representation.} of $\chi_{1}(N)$ and $\chi_{2}(N)$ in the field space, where the fixed
points (cf. in particular the analysis in \ref{Quadrid} and \ref{Quadr-quartic})
are located at a pair of coordinates $(\chi_{1c},\,\chi_{2c})$, of course
associated to a situation where $(\chi_{1}',\,\chi_{2}')=0$. 

The two fields rapidly evolve towards this pair of coordinates, (cf.
the behavior illustrated in Figs.~\ref{fig1} and \ref{fig7}) settling
there for the inflationary period. Afterwards, and because these fixed
points are not stable, the two fields will eventually diverge from
it. More precisely, in the left panel of Fig.~\ref{fig7-sq-sq} we
have the particular case where the two 3-form fields are under the
influence of identical quadratic potentials. In this case, only type
I solutions are present and the inflationary epochs, occur near the
depicted circle. Those fixed points in this figure are all located
in the arc of radius $\sqrt{2/3}$ in the $\left(\chi_{1},\chi_{2}\right)$
plane. The right panel, of the same figure, constitutes an example
where only one fixed point is present (using (\ref{cond-V1-V2}))
between $\theta=0$ and $\theta=\pi/2$. This fixed point, located
at $(C)$ in the right panel, corresponds to a type I solution when
$\theta=\pi/3$, for a case where the potentials are $V\left(\chi_{1}\right)=\chi_{1}^{2}$
and $V\left(\chi_{2}\right)=\chi_{2}^{4}$. All the other depicted trajectories
are type II solutions, where the $\dot{H}$-term coupling mediation
plays a crucial role(cf. Fig \ref{fig7}). The peculiar oscillatory
regime, present the right panel of \ref{fig7-sq-sq}, is also characteristic
of the coupling term in the effective potential (\ref{NGen-X-Pot-1}).
We shall discuss the oscillatory behavior in the following.

\paragraph{Oscillatory regime after inflation}
\textcolor{blue}{\label{oscillation} }

The main purpose of here
is to present an analytical description of the oscillatory behavior,
emerging by the end of inflation for the choice of potentials presented
in Table~\ref{potential-stability}. This analysis can also be useful
for subsequent studies on reheating and particle production, as modeled
by the two 3-forms scenario which we postpone for a future work. The
interesting aspect that happens with two 3-forms is due to the presence
of the $\dot{H}$ coupling term in the effective potential (\ref{NDiff-syst-1-1-2}),
which becomes particularly dominant and produces a nontrivial interaction
between the 3-form fields in the type II case. At this point, we must
note that this property is more general, in the sense that the conclusion
drawn for two fields can be easily extended when more 3-form fields
are included. The choice of potential plays an important role regarding
the presence of a consistent oscillatory behavior, which successfully
avoid ghost instabilities by the end of inflation. This is illustrated
for single 3-form inflation in the Ref.~\cite{DeFelice:2012jt,DeFelice:2012wy}.
Based on the studies of single 3-form inflation, we chose potentials
containing quadratic behavior. Moreover, we must emphasize that the
oscillatory regime for two 3-forms case is different from single 3-form
inflation, due to the presence of the coupling term in the equations
of motion. An exception is the case of identical quadratic potentials,
i.e., taking $V_{I}=\chi_{I}^{2}$, where we can reasonably ignore
the effect of coupling. This is the special case where two 3-form
fields oscillate almost independently.

To illustrate this, let us first consider that the two fields are
subjected to quadratic potentials $V_{I}=\frac{1}{2}m_{I}^{2}\chi_{I}^{2}$.
For simplicity we work with the equations of motion in $t$ time (\ref{NDiff-syst-1-1}).
The equation of motion (\ref{NDiff-syst-1-1}) for the 3-form field
$\chi_{I}$ can be approximated in the small field limit $\left(\chi_{I}\rightarrow0\right)$
by neglecting the effect of coupling term in the effective potential
(\ref{NGen-X-Pot-1}) as, % 
\begin{equation}
\ddot{\chi}_{I}+3H\dot{\chi}_{I}+m_{I}^{2}\chi_{I}\approx0\,.\label{quad-oscil-1}
\end{equation}
From the Friedmann constraint (\ref{NFriedm-1-1}) we have that during
inflation $H$ slowly decreases, since $\dot{H}<0$. When inflation
ends, $m_{I}^{2}\thicksim H^{2}$, and subsequently the 3-form fields
begin to coherently oscillate at scales $m_{I}^{2}\gg H^{2}$. The
evolution of $\chi_{I}$ at the oscillatory phase can be studied by
changing the variable $\chi_{I}=a^{-3/2}\bar{\chi}_{I}$, so that
(\ref{quad-oscil-1}) becomes 
\begin{equation}
\ddot{\bar{\chi}}_{I}+\left(m_{I}^{2}-\frac{9}{4}H^{2}-
\frac{3}{2}\dot{H}\right)\bar{\chi}_{I}\approx0\,.\label{dampxnosci}
\end{equation}
Using the approximations $m_{I}^{2}\gg H^{2}$ and $m_{I}^{2}\gg\dot{H}$,
the solution to (\ref{dampxnosci}) can be written as 
\begin{equation}
\bar{\chi}_{I}=C\sin\left(m_{I}t\right)\,.\label{xnbarsol}
\end{equation}
where $C$ is a the maximum amplitude of the oscillations. Thus the
solution for $\chi_{I}$ can be written as 
\begin{equation}
\chi_{I}=Ca^{-3/2}\sin\left(m_{I}t\right)\,.\label{dampxnsolution}
\end{equation}

An interesting aspect arises in the small field limit when one of
the two 3-form fields potentials is not quadratic. Let us suppose
the situation described in \ref{Quadr-quartic}, with one field subjected
to a quartic potential, $V_{2}=\lambda\chi_{2}^{4}$. This discussion
is related to the oscillatory phase we see in the right panel of Fig.
\ref{fig7-sq-sq}, regarding the type II case. This combination of
potentials has the peculiar feature to induce an oscillatory regime,
more precisely, that for a single 3-form field it would be absent
under the quartic potential due to the presence of a ghost term \cite{DeFelice:2012jt}.
In the limit $\chi_{1},\,\chi_{2}\rightarrow0$, towards the oscillatory
phase, the field $\chi_{1}$ will be approximately described by 
(\ref{dampxnsolution}). Therefore the 3-form field $\chi_{1}$ undergoes
a damped oscillatory regime due to the dominance of quadratic behavior.
However, the second field $\chi_{2}$, also undergoes an oscillatory
regime, not caused by the quartic potential but due to the coupling
term, $V_{2,\chi_{2}}^{\textrm{eff}}$, dominance in (\ref{2Quad-Y-diff}).
The equation of motion (\ref{2Quad-X-diff}) for the 3-form field
becomes (in the small field limit, $\chi_{1},\,\chi_{2}\rightarrow0$,
near the oscillatory phase), 
\begin{equation}
\ddot{\chi}_{2}+3H\dot{\chi}_{2}+\left(4\lambda\chi_{2}^{3}-\dfrac{3}{2}m_{1}^{2}\chi_{1}^{2}\chi_{2}\right)\approx0\,.\label{quad-oscil-2}
\end{equation}
The nonlinear differential equation (\ref{quad-oscil-2}) is explicitly
affected by the oscillatory behavior of $\chi_{1}$, which could cause
something similar to a parametric resonance effect in particle production
\cite{DeFelice:2012wy}. The effective potential also carries a cubic
term, which turns the equation difficult to solve. However, we can
conjecture that for two 3-forms inflation, at least one of the potentials
must contain a quadratic behavior, which forces all the other fields
to undergo a consistent oscillatory phase due to the influence of
the coupling term. In the case of the single 3-form inflation, there
is no oscillatory behavior for quartic potential, a fact that the
authors in \cite{DeFelice:2012jt} explain by means of ghost instabilities.
Therefore, we present a new choice of potential
i.e., $V_{1}=\chi_{1}^{2}$ and $V_{2}=\chi_{2}^{4}$\,, which can avoid ghost
instabilities due to the presence of consistent oscillatory phase.
A similar oscillatory regime is present when assisted inflation with
two scalar fields is studied by means of an explicit quartic coupling
in the action \cite{Braden:2010wd}.

\paragraph{Varying speed of sound for two 3-form fields}
\textcolor{blue}{\label{sec: sound speed} }

In the following we examine how the type II solutions establish pressure perturbations with varying speed of sound.

Adiabatic perturbations are defined by

\[
\frac{\delta P}{\dot{P}}=\frac{\delta\rho}{\dot{\rho}}\,,
\]
where $P$ and $\rho$ are the pressure and energy density of the
system. Pressure perturbations can in general be expanded as a sum
of an adiabatic and a non adiabatic perturbations ($\delta P_{nad}$),
which is given by \cite{Huston:2011fr}

\[
\delta P=\delta P_{{\rm nad}}+c_{s}^{2}\delta\rho\,,
\]
where $c_{s}^{2}$= $\dot{P}/\dot{\rho}$ is the adiabatic sound speed
for scalar perturbations in a thermodynamic system\footnote{The distinction
between adiabatic sound speed and effective sound speed is given for scalar
field models in Ref.~\cite{Christopherson:2008ry,Piattella:2013wpa}.}. When an
adiabatic system is composed with multiple scalar fields $\phi_{n}$,
we have that

\begin{equation}
\frac{\delta\phi_{i}}{\dot{\phi_{i}}}=\frac{\delta\phi_{j}}{\dot{\phi_{j}}}\,.
\label{multadia}
\end{equation}
The condition (\ref{multadia}) is consequently valid for any two
scalar field systems. The above condition can also be applicable for
a system of $\mathbb{N}$ 3-forms because its action can (at least
formally) always be dualized and reduced to an action with $\mathbb{N}$
non canonical scalar fields \cite{Mulryne:2012ax}.

The general expression for the adiabatic sound speed for $\mathbb{N}$
3-form fields is defined as

\begin{equation}
c_{s}^{2}=\frac{\dot{P}_{\mathbb{N}}}{\dot{\rho}_{\mathbb{N}}}\,.\label{speeddef}
\end{equation}
If we take (\ref{Ndens-1-1}) within the slow
roll approximation $\chi''_{I}\ll V_{I}(\chi_{I}),$ we get, generally

\begin{equation}
c_{s}^{2}=\frac{\sum_{n=1}^{\mathbb{N}}\chi'_{I}\,\chi_{I}\,V_{,\chi_{I}\chi_{I}}}
{\sum_{I=1}^{\mathbb{N}}\chi'_{I}\,V_{,\chi_{I}}}\,,\label{speedN3form}
\end{equation}
which, in the two 3-forms case, allows the speed of sound to be explicitly
written as

\begin{equation}
c_{s}^{2}=\frac{\chi'_{1}\,\chi_{1}\,V_{,\chi_{1}\chi_{1}}+\chi'_{2}\,\chi_{2}\,V_{,\chi_{2}\chi_{2}}}{\chi'_{1}\,V_{,\chi_{1}}+\chi'_{2}\,V_{,\chi_{2}}}\,.
\label{cs23f}
\end{equation}
Unlike the single 3-form sound speed, in a two 3-forms setting the sound
speed will depend on $\chi'_{I}$. For type I inflation, for which
we have $\left(\chi'_{I}\approx0\right)$, the speed of sound (\ref{speedN3form})
becomes constant during inflation. For the type II solution, where
we have $\chi'_{I}\not\approx0$, the speed of sound, $c_{s}^{2}$,
can vary during the inflationary period. This varying speed can subsequently
exhibit a peculiar imprint in the primordial power spectrum, scale
invariance and bi-spectrum extracted from the CMB data. We are going
to explore, in the next two subsections, observational consequences,
due to a varying speed of sound, upon important quantities like the
tensor-scalar ratio, spectral index and running spectral index, by
examining particular type II solutions for suitable choice of potentials.

\subsection{Isocurvature perturbations and primordial spectra}

\label{powerspectra} One important feature of multiple field models
is the generation of isocurvature perturbations. In this subsection we
examine the effect of these perturbations in the context of two 3-form
fields scenario. More concretely, we will distinguish, type I and
type II solutions, with respect to the evolution of isocurvature perturbations.

As depicted, in the right panel of Fig.~\ref{fig7-sq-sq} type I solutions
are characterized by a straight line, whereas type II solutions follow
a curved trajectory in field space. In scalar multifield models, a
local rotation in the field space is carried to define the adiabatic
and entropy modes (or fields \cite{Gordon:2000hv}). In order to express
these adiabatic and entropy fields from two 3-form fields, we use the relation between 3-form field dual scalar field presented in Sec.~\ref{dual-Sec}.
%the first step consists in defining a dual scalar field Lagrangian for the two
%3-forms (see Ref. \cite{Mulryne:2012ax}). The second step consists
%in applying the general framework of adiabatic and entropy perturbations
%to our model. We start by presenting the $\mathbb{N}$ 3-forms Lagrangian
%in terms of a multiple non-canonical scalar field Lagrangian %$P\left(X,\phi_{I}\right)$,
%where $X=-\frac{1}{2}G^{IJ}\partial^{\mu}\phi_{I}\partial_{\mu}\phi_{J}$
%\footnote{Note that in the Ref.\cite{Mulryne:2012ax} the definition $X=-\partial^{\mu}\phi_{I}\partial_{\mu}\phi_{I}$
%differs by a $2$ factor from our notation. Accordingly, we have considered
%this difference throughout this section. }
%, and $\phi_{I}$ is the dual scalar field which is 
The motivation to work with the dual action is related to the fact that the general framework
of adiabatic and entropy perturbations for the non-canonical multifield
model has already been consistently established. In the following
we will briefly review and adopt to our case the results described
previously in \cite{Langlois:2008mn,Arroja:2008yy,Kaiser:2012ak,Mulryne:2012ax,Langlois:2009ej}.

Restricting ourselves now to a two 3-form scenario, and according
to \cite{Gordon:2000hv}, we can define the adiabatic and entropy
fields through a rotation in the two 3-form dual field space
\begin{equation}
\dot{\sigma}=\sqrt{2X_{1}}\,\cos\Theta+\sqrt{2X_{2}}\,\sin\Theta\,,\label{adiab-field1}
\end{equation}
\begin{equation}
\dot{s}=-\sqrt{2X_{1}}\,\sin\Theta+\sqrt{2X_{2}}\,\cos\Theta\,,\label{entrop-field1}
\end{equation}
where $\tan\Theta=\sqrt{X_{2}}/\sqrt{X_{1}}$, $X_{1}=\frac{1}{2}V_{1,\chi_{1}}^{2}$
and $X_{2}=\frac{1}{2}V_{2,\chi_{2}}^{2}$. Subsequently, the adiabatic
and entropy perturbations are 
\begin{equation}
Q_{\sigma}=\delta\phi_{1}\,\cos\Theta+\delta\phi_{2}\,\sin\Theta\,,\label{adiab-perturb1}
\end{equation}
\begin{equation}
Q_{s}=-\delta\phi_{1}\,\sin\Theta+\delta\phi_{2}\,\cos\Theta\,,\label{entrop-perturb1}
\end{equation}
respectively, along and orthogonal to the background classical trajectory
in dual field space.

Let us assume that the linearly perturbed metric in terms of Bardeen potentials $\Phi,\,\Psi$ which is given by \cite{Baumann:2009ds}
%\[
%ds^{2}=-(1+2\varphi)dt^{2}+2\partial_{i}\beta dx^{i}dt+a^{2}(t)\left(1-2\psi\right)dx^{2}\,.
%\]
\[
ds^{2}=-(1+2\Phi)dt^{2}+a^{2}(t)\left(1-2\Psi\right)d{\bf x}^{2}\,.
\]
We choose a flat gauge, where the dynamics of linear perturbations
are completely expressed in terms of the scalar field perturbations
$\left(\phi^{I}\rightarrow\phi_{0}^{I}+Q^{I}\right)$. Moreover, these
are defined as gauge invariant combinations given by $Q^{I}=\delta\phi^{I}+\left(\phi^{I}/H\right)\Psi$.
The comoving curvature perturbation is given by 
\begin{equation}
\mathcal{R}\equiv\Psi-\frac{H}{p+\rho}\delta q\,,
\end{equation}
where $\partial_{i}\delta q_{i}=\delta T_{i}^{0}$ and $\mathcal{R}$
purely characterizes the adiabatic part of the perturbations. The
variation of $\mathcal{R}$, in the flat gauge, is given by \cite{Langlois:2008mn}
\begin{equation}
\dot{\mathcal{R}}=\frac{H}{\dot{H}}\frac{c_{s}^{2}k^{2}}{a^{2}}\Psi+\frac{H}{\dot{\sigma}}\,\Xi\,Q_{s}\quad\textrm{with}\quad\Xi=
\frac{1}{\dot{\sigma}P_{,X}}\left(\left(1+c_{s}^{2}\right)P_{,s}-c_{s}^{2}\dot{\sigma}^{2}P_{,Xs}\right)\,,\label{rdot at all}
\end{equation}
where $\Psi$ is the Bardeen potential and 
\begin{equation}
P_{,s}=P_{,X}\dot{\sigma}\dot{\Theta}\,,\hspace{1cm}\left(\begin{array}{c}
P_{,X\sigma}\\
P_{,Xs}
\end{array}\right)=\left(\begin{array}{cc}
\cos\Theta & \sin\Theta\\
-\sin\Theta & \cos\Theta
\end{array}\right)\left(\begin{array}{c}
P_{,\chi_{1}}\\
P_{,\chi_{2}}
\end{array}\right)\,.\label{Pxs3form}
\end{equation}
For a two 3-form dual Lagrangian, extracted from (\ref{dual action}),
we can express the above quantities as functions of the 3-form fields,
i.e., 
\begin{equation}
P_{,X}\equiv P_{,X_{1}}+P_{,X_{2}}=\frac{\chi_{1}}{V_{1,\chi_{1}}}+\frac{\chi_{2}}{V_{2,\chi_{2}}}\,.\label{px3form}
\end{equation}
Using (\ref{px3form}) and (\ref{Pxs3form}) we can simplify
$\Xi$, to obtain, 
\begin{equation}
\Xi=H\left(\left(1+c_{s}^{2}\right)\,\frac{d\Theta}{dN}-c_{s}^{2}\,\frac{\dot{\sigma}}{H}\,\frac{P_{,Xs}}{P_{,X}}\right)\,.\label{cascadeNtime}
\end{equation}
The function $\Xi$ is a measure of the coupling between the entropy
and adiabatic modes.

\subsubsection{Type I inflation}

\label{type1 entropy.} In type I inflationary scenarios, where $\dot{\Theta}=0$
(as $\tan\Theta=\lambda_{2}/\lambda_{1}=\chi_{2}/\chi_{1}=$ constant in
the fixed point, cf. (\ref{fixed-P2}) and see Fig.~\ref{fig7-sq-sq}),
the classical trajectory is a straight line. This fact makes the first
term of $\Xi$, in (\ref{cascadeNtime}), to vanish.

On the other hand, the ratio $P_{,Xs}/P_{,X}$ can be expressed as
\begin{equation}
\frac{P_{,Xs}}{P_{,X}}=\frac{-\chi_{1}\,\sin\Theta+\chi_{2}\,\cos\Theta}{\chi_{1}V_{1,\chi_{1}}+\chi_{2}V_{2,\chi_{2}}}\,.\label{cascadezerofortype1}
\end{equation}
Expression (\ref{cascadezerofortype1}) vanishes for all type I solutions
since $\chi_{2}=\chi_{1}\left(\lambda_{2}/\lambda_{1}\right)=\chi_{1}\tan\Theta$.
In other words, there are no entropy perturbations sourcing the curvature
perturbations. We then recover the known relation for a single field
inflation 
\begin{equation}
\dot{\mathcal{R}}=\frac{H}{\dot{H}}\frac{c_{s}^{2}k^{2}}{a^{2}}\Psi\,
\end{equation}
and we can state that the curvature perturbation is conserved on the
large scales. We can, therefore, compute the power spectrum of curvature
perturbations in terms of quantities values at horizon exit.

\subsubsection{Type II inflation}

For type II inflation, the aforementioned effects, namely of entropy
perturbations, can be present due to the curved trajectory (cf. the
right panel of Fig.~\ref{fig7-sq-sq}) in field space $(\dot{\Theta}\neq0)$.
Due to this the curvature power spectrum could be sourced by entropy
perturbations on large scales.

In order to study quantum fluctuations of the system we must consider
the following canonically normalized fields defined by, 
\[
v_{\sigma}=\frac{a\sqrt{P_{,X}}}{c_{s}}Q_{\sigma},\hspace{1cm}v_{s}=a\sqrt{P_{,X}}Q_{s}\,,
\]
we can express the second order action for the adiabatic and entropy
modes as 
\begin{equation}
S_{(2)}=\frac{1}{2}\int d\tau d^{3}k\left[v_{\sigma}^{\prime^{2}}+v_{s}^{\prime^{2}}-2\xi v_{\sigma}^{\prime}v_{s}-k^{2}c_{s}^{2}v_{\sigma}^{2}-k^{2}v_{s}^{2}+\Omega_{\sigma\sigma}v_{s}^{2}+\Omega_{ss}v_{\sigma}^{2}+2\Omega_{s\sigma}v_{\sigma}v_{s}\right]\,,\label{2ndorder action}
\end{equation}
with 
\[
\xi=\frac{a}{c_{s}}\Xi,\quad\Omega_{\sigma\sigma}=\frac{z^{\prime\prime}}{z}\quad\textrm{and}\quad\Omega_{ss}=\frac{\alpha^{\prime\prime}}{\alpha}-a^{2}\mu_{s}^{2}\,,
\]
where $z$ and $\alpha$ are background dependent functions defined
by 
\[
z=\frac{a\dot{\sigma}\sqrt{P_{,X}}}{c_{s}H},\;\alpha=a\,\sqrt{P_{,X}}\,.
\]
The equations of motion derived from the action (\ref{2ndorder action})
are given by 
\begin{equation}
v_{\sigma}^{\prime\prime}-\xi v_{s}^{\prime}+\left(c_{s}^{2}k^{2}-\frac{z^{\prime\prime}}{z}\right)v_{\sigma}-\frac{\left(z\xi\right)^{\prime}}{z}v_{s}=0\,,\label{sigma eqn}
\end{equation}
\begin{equation}
v_{s}^{\prime\prime}+\xi v_{\sigma}^{\prime}+\left(k^{2}-\frac{\alpha^{\prime\prime}}{\alpha}+a^{2}\mu_{s}^{2}\right)v_{s}-\frac{z^{\prime}}{z}\xi v_{\sigma}=0\,,\label{s eqn}
\end{equation}
where $\mu_{s}^{2}$ is the effective mass for the entropy field given
by \cite{Langlois:2008mn} 
\begin{equation}
\mu_{s}^{2}=-\frac{P_{,ss}}{P_{,X}}-\frac{1}{2c_{s}^{2}\left(X_{1}+X_{2}\right)}\frac{P_{,s}^{2}}{P_{,X}^{2}}+2\frac{P_{,Xs}P_{,s}}{P_{,X}^{2}}
\end{equation}
and 
\begin{equation}
\left(\begin{array}{cc}
P_{,\sigma\sigma} & P_{,\sigma s}\\
P_{,s\sigma} & P_{,ss}
\end{array}\right)=\left(\begin{array}{cc}
\cos\Theta & \sin\Theta\\
-\sin\Theta & \cos\Theta
\end{array}\right)\left(\begin{array}{cc}
P_{,X_{1}X_{1}} & P_{,X_{1}X_{2}}\\
P_{,X_{2}X_{1}} & P_{,X_{2}X_{2}}
\end{array}\right)\left(\begin{array}{cc}
\cos\Theta & -\sin\Theta\\
\sin\Theta & \cos\Theta
\end{array}\right)\,.
\end{equation}
The coupling between adiabatic and entropy modes is governed by the
parameter $\xi$. In the cases where this parameter can be assumed
to be small (see \cite{Langlois:2008mn,Arroja:2008yy}) at the typical
scale of sound horizon exit\footnote{In contrast to the inflationary models where a sharp turn in field
space occurs during inflation \cite{Lalak:2007vi,Peterson:2010np,Konieczka:2014zja}.} the adiabatic and entropy modes decouple and analytical solutions
for (\ref{sigma eqn})-(\ref{s eqn}) can easily be found. In
the decoupled case the adiabatic and entropy modes evolve according
to the following equations, 
\begin{equation}
v_{\sigma}^{\prime\prime}-\left(c_{s}^{2}k^{2}-\frac{z^{\prime\prime}}{z}\right)v_{\sigma}=0\,,\label{free adiabatic}
\end{equation}
\begin{equation}
v_{s}^{\prime\prime}+\left(k^{2}-\frac{\alpha^{\prime\prime}}{\alpha}+a^{2}\mu_{s}^{2}\right)v_{s}=0\,.\label{freeentropy}
\end{equation}

In the slow-roll limit, for a speed of sound that slowly varies while
the scales of interest cross out the sound horizon, we can assume
$z^{''}/z^{'}=1/\tau^{2}$. Using this, we get as a general approximate
solutions for the adiabatic and entropy modes with Bunch-Davies vacuum
initial conditions, 
\begin{equation}
v_{\sigma k}\simeq\frac{1}{\sqrt{2kc_{s}}}\exp\left(-ikc_{s}\tau\right)\left(1-\frac{i}{kc_{s}\tau}\right)\,,\label{adiabatic solution}
\end{equation}
\begin{equation}
v_{sk}\simeq\frac{1}{\sqrt{2k}}\exp\left(-ik\tau\right)\left(1-\frac{i}{k\tau}\right)\,,\label{entropy solution}
\end{equation}
where we assume $\frac{\mu_{s}^{2}}{H^{2}}\ll1$ is valid for our
case. This means entropy modes get amplified with respect to the adiabatic
modes at the sound horizon crossing 
\begin{equation}
Q_{\sigma_{\ast}}\simeq\frac{Q_{s_{\ast}}}{c_{s_{\ast}}}\,.\label{entropy amp}
\end{equation}
The curvature and isocurvature perturbations are respectively, 
\begin{equation}
\mathcal{R}=\frac{H}{\dot{\sigma}}Q_{\sigma},\hspace{1cm}\mathcal{S}=c_{s}\frac{H}{\dot{\sigma}}Q_{s}\,.\label{adiapentp}
\end{equation}
The power spectrum of the curvature perturbation, evaluated at the
sound horizon crossing $\left(c_{s}k=aH\right)$, is given by 
\begin{equation}
\mathcal{P}_{\mathcal{R}_{\ast}}=\frac{k^{3}}{2\pi^{2}}\frac{\mid v_{\sigma k}\mid^{2}}{z^{2}}\simeq\frac{H^{4}}{8\pi^{2}XP_{,X}}=\frac{H^{2}}{8\pi^{2}\epsilon c_{s}}\Bigg|_{\ast}\,,\label{curvature power spectrum}
\end{equation}
which recovers with the single field power spectrum result at horizon
crossing \cite{Mulryne:2012ax}. However, in contrast to the single
field inflation, the function $\xi$ is not negligible and typically
varies with time. This means that there will be a transfer between
entropic and adiabatic modes on large scales but the converse is not
true. From (\ref{rdot at all}) and (\ref{adiapentp}), the evolution
of the curvature and entropy modes in the long wavelength limit can
be approximated as \cite{Langlois:2008mn} 
\[
\dot{\mathcal{R}}\approx\alpha HS,\hspace{1cm}\dot{\mathcal{S}}\approx\beta HS,
\]
where the coefficients $\alpha$ and $\beta$ are taken to be, 
\begin{equation}
\alpha=\frac{\Xi}{c_{s}H}\,,\label{alpha}
\end{equation}
\begin{equation}
\beta\simeq\frac{s}{2}-\frac{\eta}{2}-\frac{1}{3H^{2}}\left(\mu_{s}^{2}+\frac{\Xi^{2}}{c_{s}^{2}}\right)\,,\label{beta}
\end{equation}
endowed with the definition of an additional slow-roll parameter $s=\frac{\dot{c}_{s}}{Hc_{s}}$.
The evolution of curvature and isocurvature perturbations after horizon
crossing can be evaluated using transfer functions defined by 
\[
\left(\begin{array}{c}
\mathcal{R}\\
\mathcal{S}
\end{array}\right)=\left(\begin{array}{cc}
1 & \mathcal{T}_{\mathcal{R}\mathcal{S}}\\
0 & \mathcal{T}_{\mathcal{S}\mathcal{S}}
\end{array}\right)\left(\begin{array}{c}
\mathcal{R}\\
\mathcal{S}
\end{array}\right)_{\ast},
\]
where 
\begin{equation}
\mathcal{T}_{\mathcal{R}\mathcal{S}}\left(t_{\ast},t\right)=\int_{t_{\ast}}^{t}dt^{\prime}\alpha\left(t^{\prime}\right)H\left(t^{\prime}\right)T_{SS}\left(t_{\ast}\right)\,,\label{trs}
\end{equation}
and 
\begin{equation}
\mathcal{T}_{\mathcal{S}\mathcal{S}}\left(t_{\ast},t\right)=\exp\left\{ \int_{t_{\ast}}^{t}dt^{\prime}\beta\left(t^{\prime}\right)H\left(t^{\prime}\right)dt^{\prime}\right\}\,,\label{tss}
\end{equation}
In addition, the curvature perturbation power spectrum, the entropy
perturbation and the correlation between the two can be formally related
as 
\begin{equation}
\mathcal{P_{R}=}\left(1+\mathcal{T}_{\mathcal{R}\mathcal{S}}^{2}\right)\mathcal{P}_{\ast},\hspace{1cm}\mathcal{P_{S}=}
\mathcal{T}_{\mathcal{S}\mathcal{S}}^{2}\mathcal{P}_{\ast}\,,\label{full powerspectrum}
\end{equation}
\begin{equation}
\mathcal{C_{RS}}\equiv\langle\mathcal{RS}\rangle=\mathcal{T}_{\mathcal{R}S}\mathcal{T_{SS}}\mathcal{P}_{\ast}\,.\label{coupling power}
\end{equation}
In contrast to the power spectrum for the scalar perturbations, the
tensor power spectrum amplitude is the same as for a single field,
\[
\mathcal{P}_{t}=\frac{2}{\pi^{2}}\frac{H^{2}}{M_{PI}^{2}}\Bigg|_{\ast}\,.
\]
The tensor to scalar ratio defined in multifield inflation is given by

\begin{equation}
r\equiv\frac{\mathcal{P}_{t}}{\mathcal{P_{R}}}=16\epsilon c_{s}\Bigg|_{\ast}\cos^{2}\Delta\,,\label{modified tensor scalar}
\end{equation}
where $\Delta$ is the transfer angle given by 
\[
\cos\Delta=\frac{1}{\sqrt{1+\mathcal{T}_{\mathcal{R}\mathcal{S}}^{2}}}\,.
\]
Similarly, the spectral index also gets a correction, provided by
the transfer functions, 
\begin{equation}
n_{s}\equiv\frac{d\,\ln\mathcal{P_{R}}}{d\,\ln k}=n_{s}(t_{\ast})+\frac{1}{H_{\ast}}\left(\frac{\partial T_{RS}}{\partial t_{\ast}}\right)\sin\left(2\Delta\right)\,,\label{modifiedns}
\end{equation}
where 
\[
n_{s_{\ast}}=1-2\epsilon_{\ast}-\eta_{\ast}-s_{\ast}\,.
\]

The spectral index and the tensor to scalar ratio are the key observables
which not only depend on the slow-roll at horizon crossing, but also
depend on the transfer angle $\Delta$. This enables a clear distinction
between multifields and single field inflationary scenarios \footnote{However tensor to scalar ratio is more constrained by consistency relations
in case of inflation with more than two fields \cite{Wands:2002bn,Bartolo:2001rt}.} \cite{Wands:2002bn}. The transfer functions defined in (\ref{trs})
and (\ref{tss}) are allowed to evolve after the Hubble exit, even
after inflation, during the reheating and radiation dominated era
\cite{Wands:2002bn,Vernizzi:2006ve}. However the evolution of isocurvature
perturbations, during reheating and radiation dominated era, would
depend on the particular final stage of the inflationary scenario.
Consider for example, a two field scenario, if one field enters a
regime of oscillations while the second field is still inflating the
Universe. In such cases the curvature perturbation can be sourced
by entropy modes even after inflation \cite{Bartolo:2001rt}. This
kind of scenarios are known as `curvaton' or `spectator' field behavior
\cite{Elliston:2013afa,Choi:2008et} and also found in double quadratic
inflation \cite{Vernizzi:2006ve}. In the case of two 3-forms inflation,
we will assume that entropy perturbations do not grow further after
inflation. Therefore we only evaluate transfer functions from horizon
exit until the end of inflation and predict the values of $n_{s}$
and $r$ \cite{Peterson:2010np}. We can see from (\ref{modified tensor scalar})
and (\ref{modifiedns}) that if $\mathcal{T}_{\mathcal{RS}}=0$ then
our predictions match the single field result. From the Sec.~\ref{type1 entropy.}
and \ref{alpha} it is evident that $\mathcal{T}_{\mathcal{RS}}=0$
for type I inflation. Therefore to make observational contrast with
single 3-form we mainly focus on testing type II inflationary scenario
in the following subsection.

\subsection{Two 3-form fields inflation and Power spectra}

\label{sec:Planck} Based on the discussion made on the curvature
perturbation power spectrum in Sec.~\ref{powerspectra}, the main
objective is to test our two 3-forms model and predicting values of
inflationary parameters. We choose suitable potentials and initial
conditions, in order to obtain a reasonable fit with present available
experimental bounds \cite{Planck:2013jfk}. The majority of inflationary
models with a non canonical kinetic term contain a common feature
that the adiabatic fluctuations propagate with a sound speed $c_{s}^{2}<1.$
The recent {\it Planck} data restricts this speed of sound to be in the
interval $0.02\lesssim c_{s}^{2}<1$. Multiple field inflation models
allow the possibility of having a varying speed of sound, i.e, like
for the type II solution in our model (cf. Sec.~\ref{sec: sound speed}).
The speed of sound variation will therefore have implications on the
running spectral index and the scale invariance. These peculiar effects,
being a consequence of the varying speed of sound, have been studied
in a DBI context and also in modified gravity models with an effective
inflaton \cite{Khoury:2008wj,Park:2012rh,Cai:2010wt}.

We have examined all the potentials in Table \ref{potential-stability}.
We found that $\chi_{I}^{2}+b_{i}\chi_{I}^{4}$ is consistent with
observational bounds\footnote{We confront our results with $\chi_{I}^{2}+b_{I}\chi_{I}^{4}$ potential,
and one can find make similar predictions with $\chi_{I}^{2}+b_{I}\chi_{I}^{3}$
potential. We have not consider to explore quadratic potential as it
is equivalent to inflation with canonical scalar fields (in dual picture). }. It is quite difficult to constrain the speed of sound $\left(0.02\lesssim c_{s}^{2}<1\right)$
during inflation. We found that only type II solutions which are slightly
deviated from type I are suitable to maintain consistent speed of
sound during inflation. To predict values of inflationary parameters,
first we need to compute the transfer functions defined in Sec.~\ref{powerspectra} and evaluate their value at the end of inflation.

\begin{figure}[h!]
\centering\includegraphics[height=2.5in]{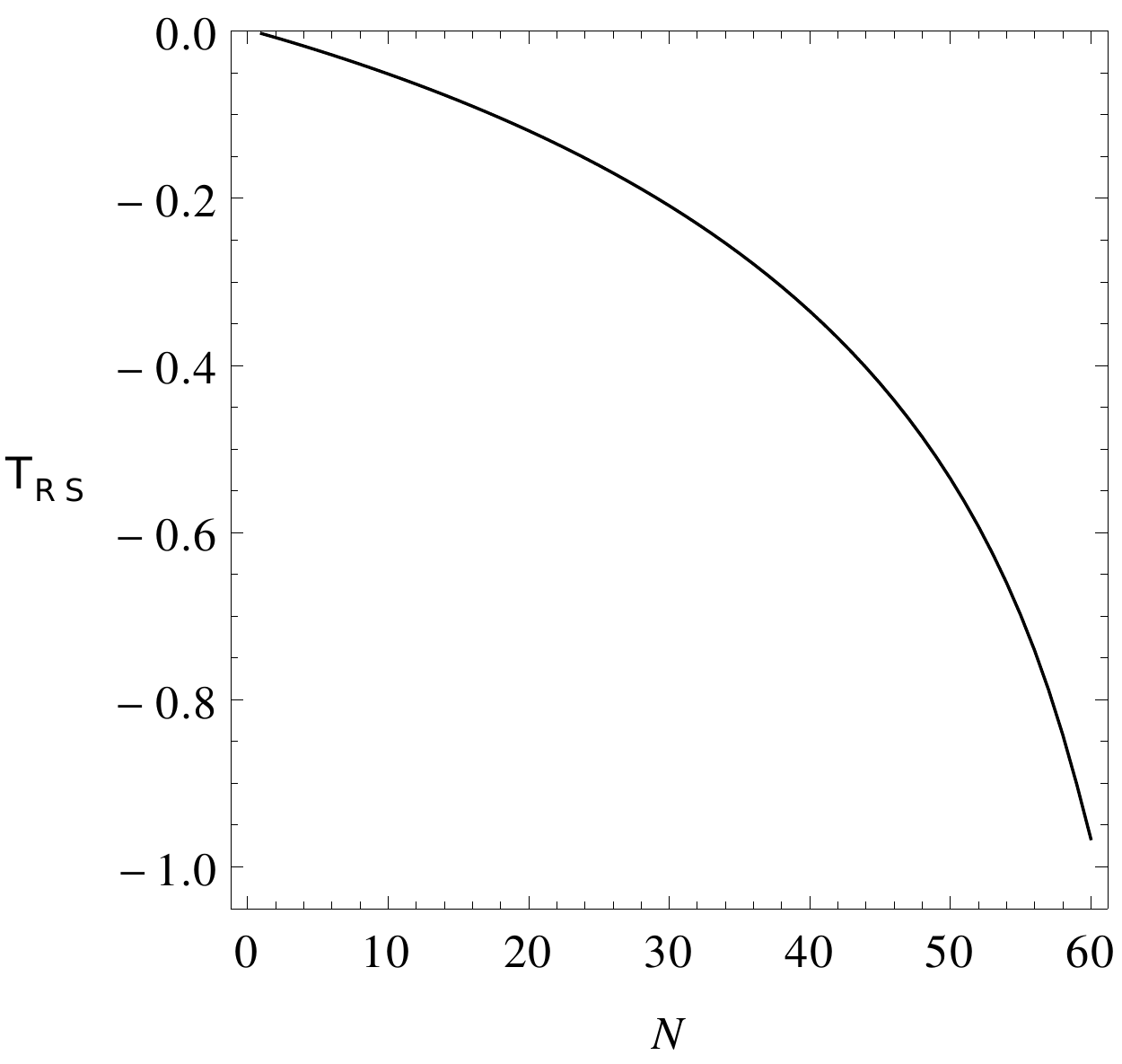}\quad{}\includegraphics[height=2.5in]{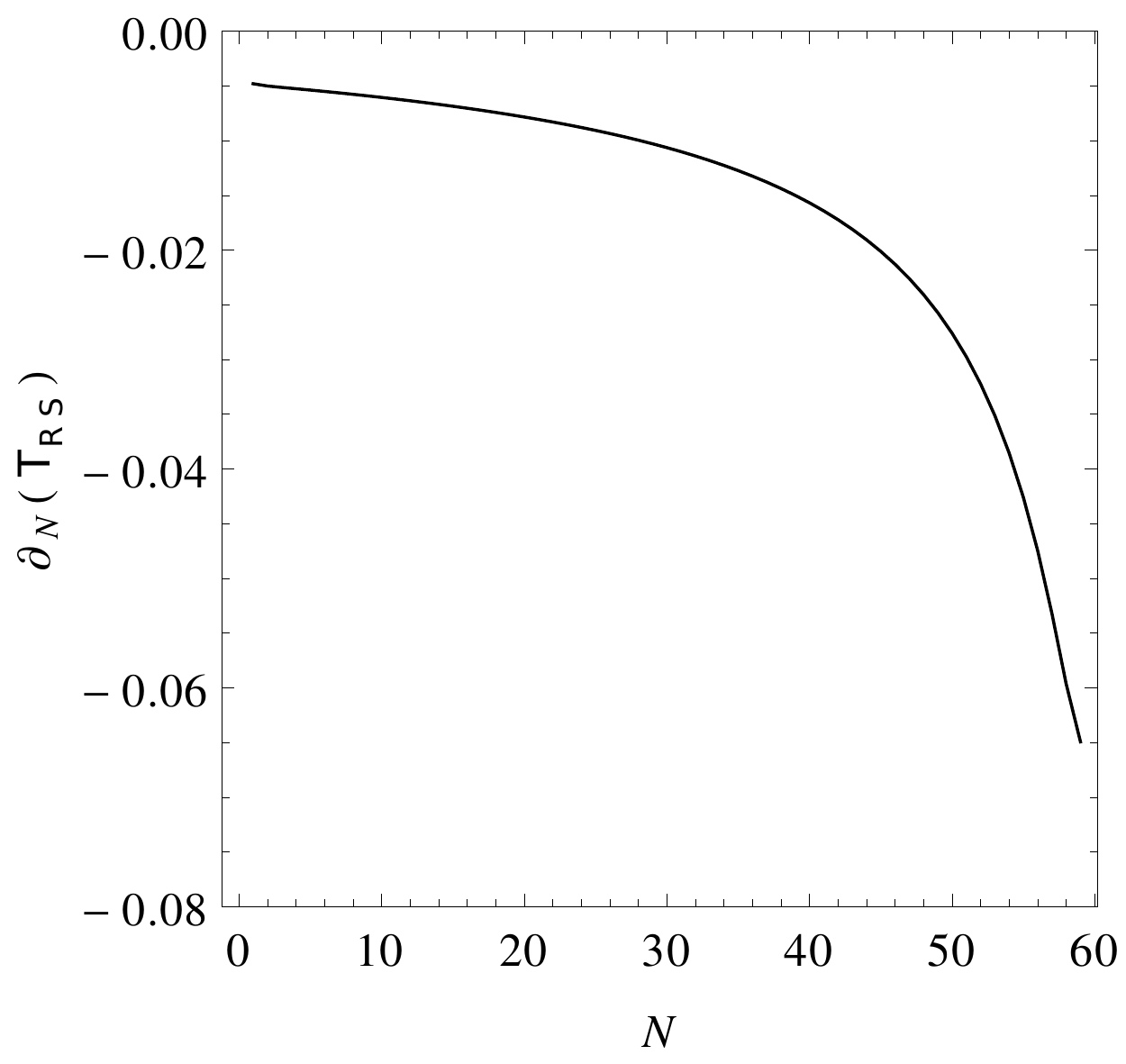}
\caption{Graphical representation of $\mathcal{T_{RS}}$ (left panel) and $\frac{d\mathcal{T_{RS}}}{dN}$
(right panel) until the end of inflation (defined for $\epsilon=1$).
We have taken $V_{1}=V_{01}(\chi_{1}^{2}+b\chi_{1}^{4})$ and $V_{2}=V_{02}(\chi_{2}^{2}+b\chi_{2}^{4})$
where $V_{01}=1,\:V_{02}=0.93,\:b=-0.35$ and with initial conditions
$\theta=\pi/4$.}
\label{tranferfunctions} 
\end{figure}

We can read from (\ref{modifiedns}) that the spectral index depends
on the derivative of $\mathcal{T_{RS}}$ at horizon crossing. From
the right panel of Fig.~\ref{tranferfunctions} it is clear that the
derivative of $\mathcal{T_{RS}}$, between $N=0$ and $N=60$, is
very small and we can, therefore, neglect it. Hence, our prediction
of spectral index only depends on the values of the slow-roll parameters
at horizon exit. 

The running of the spectral index to the lowest order in slow-roll
is now given by, regime, 
\begin{equation}
\begin{aligned}\frac{dn_{s}}{d\ln k} & = &\,\left(1+\epsilon+\frac{c_{s}^{\prime}}{c_{s}}\right)\Bigg|_{\ast}\left(n'_{s_{\ast}}+\frac{\partial\mathcal{T_{RS}}}{\partial N_{\ast}}\frac{\partial}{\partial N}\left(\frac{2\mathcal{T_{RS}}}{1+\mathcal{T}_{\mathcal{R}\mathcal{S}}^{2}}\right)
+\frac{\partial^{2}\mathcal{T_{RS}}}{\partial N_{\ast}^{2}}\sin2\Delta\right)\,.\end{aligned}
\label{Running sp}
\end{equation}
For the choice of potential in Fig.~\ref{tranferfunctions} we can
neglect the transfer function corrections to the running spectral
index (\ref{Running sp}). Therefore for this case the additional
slow-roll parameter $s=\frac{c_{s}^{\prime}}{c_{s}}$ is of relevance,
which enables us to observationally distinguish between two 3-forms
and single 3-form inflation\footnote{In the single 3-form case \cite{DeFelice:2012jt,DeFelice:2012wy}
and also in the type I solution of two 3-forms case, this additional
slow-roll parameter satisfies, $s\equiv\frac{\dot{c_{s}}}{c_{s}H}=0.$}, with respect to the running of spectral index. Expression (\ref{Running sp})
is expanded up to the first order in the slow-roll parameters. The
second order corrections are crucial if there is an abrupt path turn
in field space during horizon exit. These types of scenarios are considered
in detail in studies related with hybrid inflation and double quadratic
inflation \cite{Avgoustidis:2011em}. We can neglect these corrections
for two 3-form inflation, since the type II solutions herein considered
do not exhibit abrupt turns in field space under slow-roll conditions.

To predict tensor to scalar ratio (\ref{modified tensor scalar}) for
two 3-forms it is required to know the value of $\mathcal{T_{RS}}$
at the end of inflation. From the left panel of Fig.~\ref{tranferfunctions},
$\mathcal{T_{RS}}$ is $\mathcal{O}(1)$ at the end of inflation.
Therefore it can reduce the value of tensor to scalar ratio in contrast
to the single 3-form case.

Evidently two 3-forms inflation can be observationally distinguished
from single 3-form inflation, due to the possibility of a varying
speed of sound (cf. Sec.~\ref{sec: sound speed}) and transfer
function corrections by the end of inflation. Our method of observational
analysis are quite similar to the studies in \cite{Peterson:2010np,Kaiser:2012ak}.
In the following we confront our results against {\it Planck}+WP+BAO data
which provides $\frac{dn_{s}}{d\ln k}=-0.013\pm0.009$ for the running
of spectral index, and $\frac{d^{2}n_{s}}{d\ln k^{2}}=0.017\pm0.009$
for the running of running spectral index, both at 95\% CL, which
rules out exact scale-invariance at more than 5$\sigma$ level. Our
analysis show that for type II solution, a better fit can be achieved
given the current observational bounds (ruling out exact scale-invariance).

\begin{figure}[h!]
\centering\includegraphics[height=2.6in]{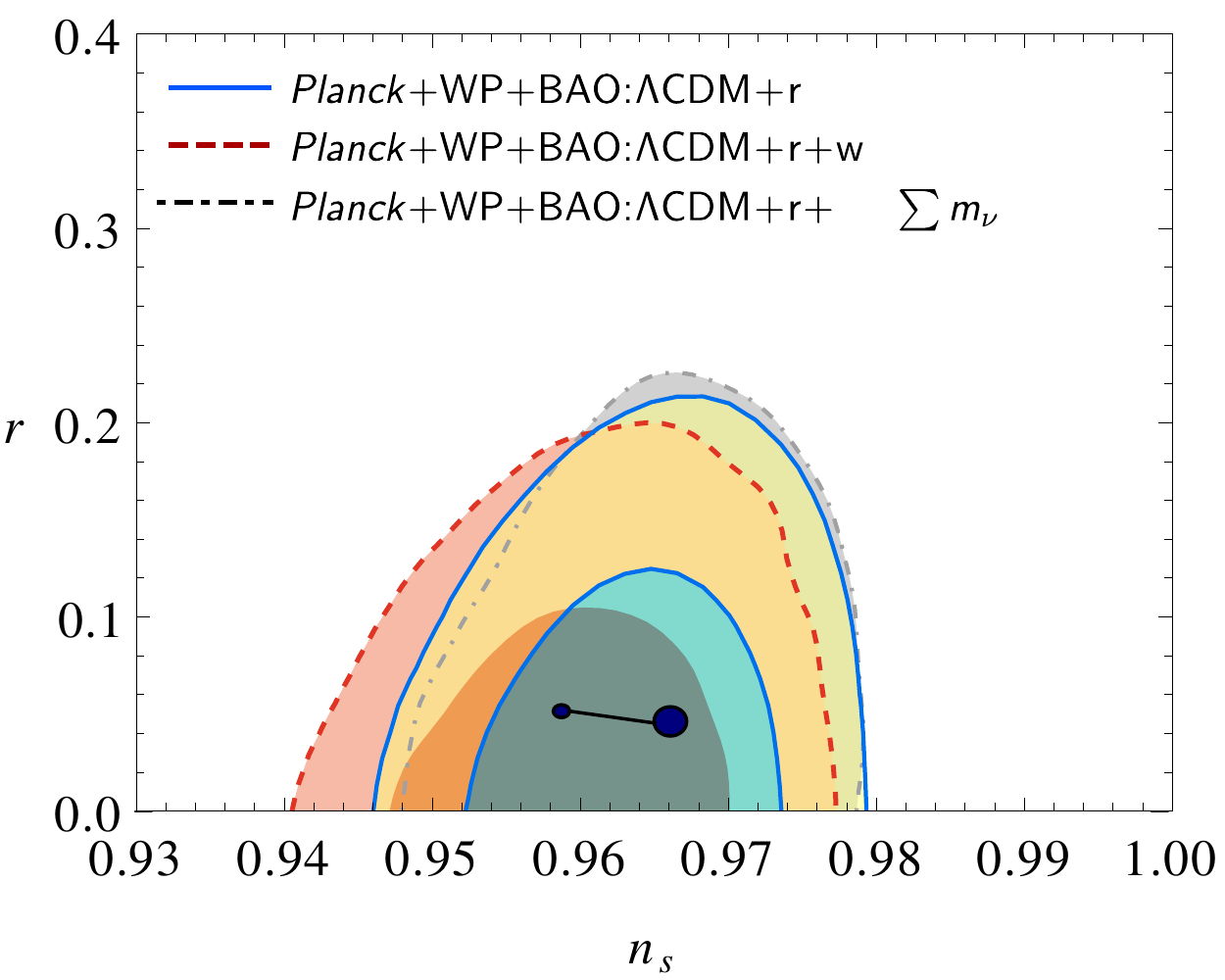}\caption{Graphical representation of the spectral index versus the tensor to
scalar ratio, in the background of {\it Planck}+WP+BAO data (left panel),
for $N=60$ number of $e$-folds before the end of inflation (large
dot) and $N=50$ (small dot). We have taken $V_{1}=V_{01}(\chi_{1}^{2}+b\chi_{1}^{4})$
and $V_{2}=V_{20}(\chi_{2}^{2}+b\chi_{2}^{4})$ where $V_{01}=1,\:V_{20}=0.93,\:b=-0.35$
for two 3-form. }

\label{fig8-1} 
\end{figure}

\begin{figure}[h!]
\centering\includegraphics[height=1.9in]{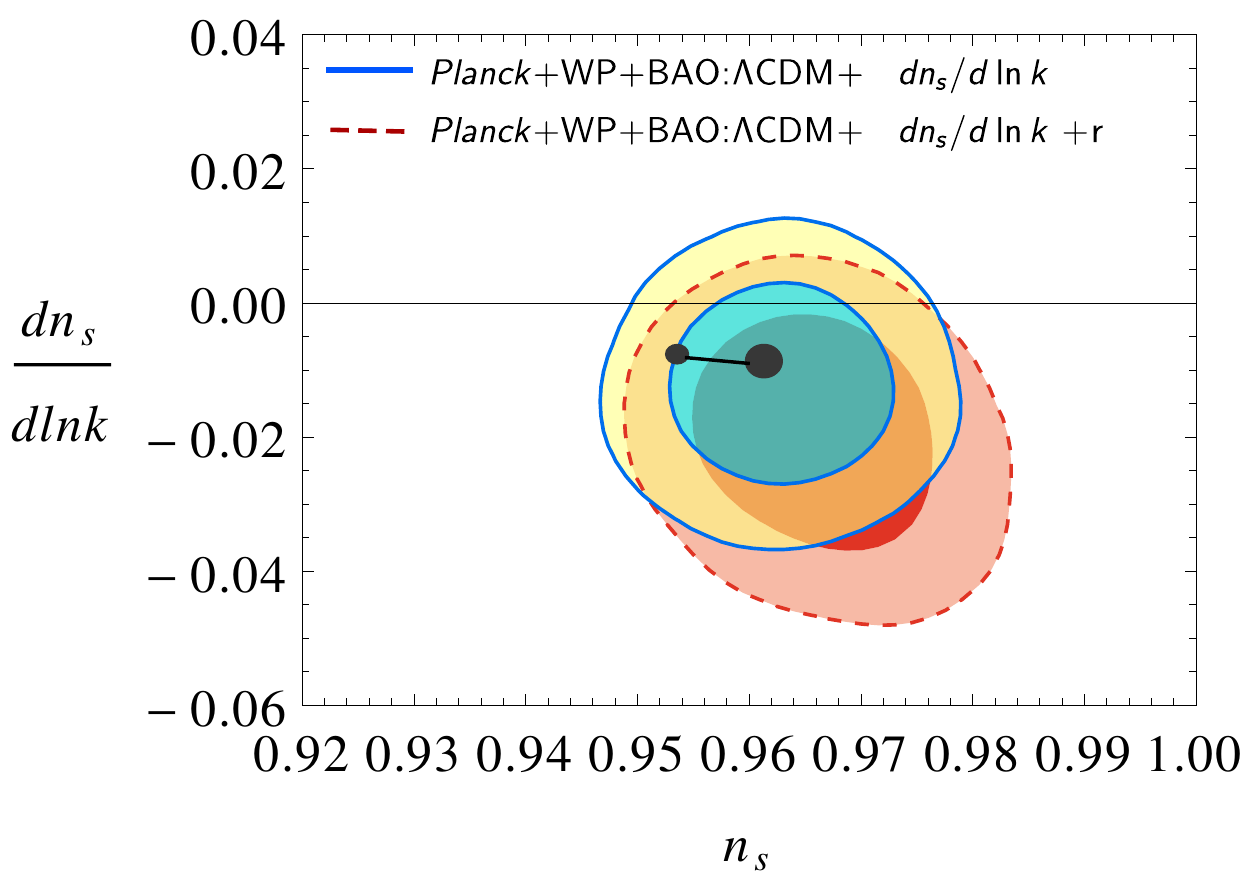}\quad{}\includegraphics[height=1.9in]{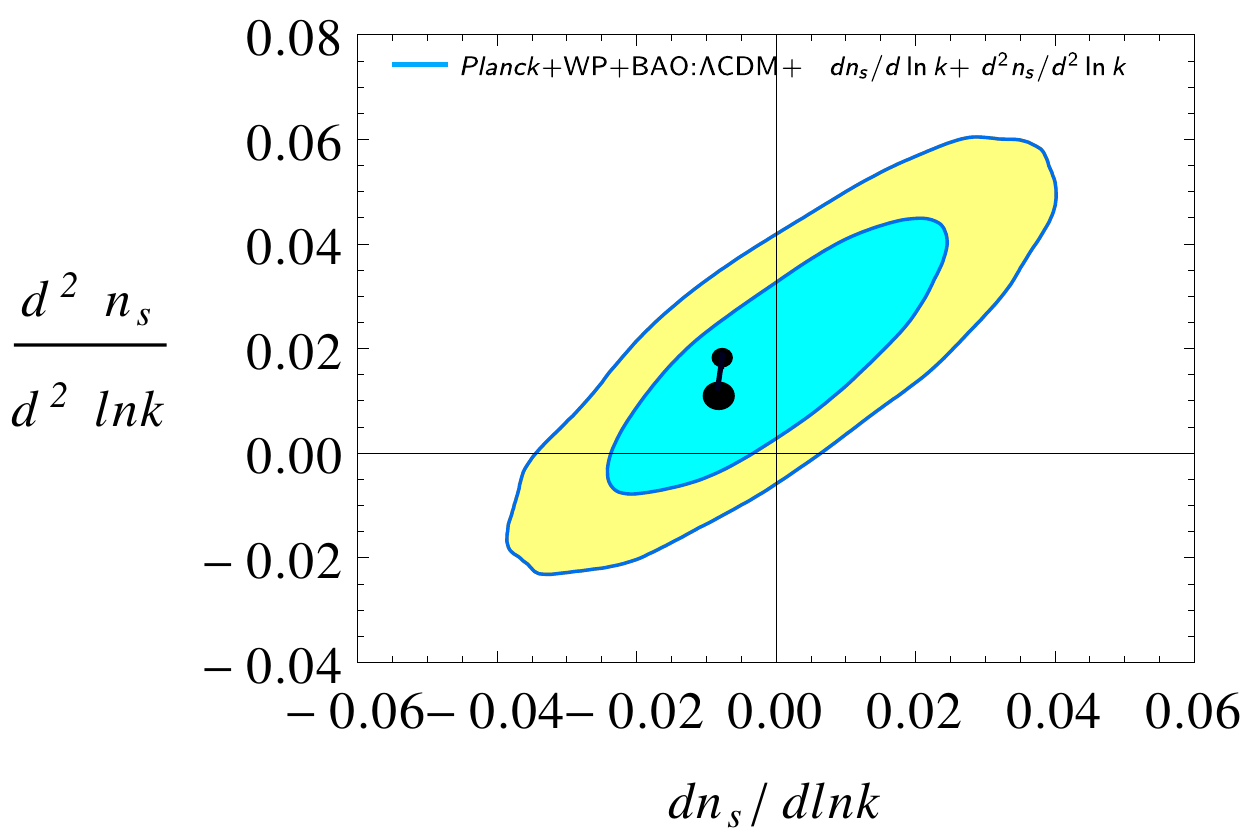}
\caption{Graphical representation of the running of the spectral index versus
the spectral index (left panel), and running of the running of the
spectral index versus the running of the spectral index (right panel)
in the background of {\it Planck}+WP+BAO data for $N=60$ number of
$e$-folds before the end of inflation (large dot) and $N=50$
(small dot). We have taken $V_{1}=V_{01}(\chi_{1}^{2}+b\chi_{1}^{4})$ and
$V_{2}=V_{20}(\chi_{2}^{2}+b\chi_{2}^{4})$ where $V_{01}=1,\:V_{20}=0.93,\:b=-0.35$
for two 3-form. This figure was also obtained by taking the initial condition $\theta=\pi/4$.}

\label{fig8-2} 
\end{figure}

In Figs.~\ref{fig8-1} and \ref{fig8-2}, obtained through suitable
data manipulating programs \cite{Lewis:2002ah,Lewis:2013hha}, we
have examined various types of potentials for a reasonable fit to
the observational constraints from {\it Planck} data. We found that potentials
such as $V_{I}=V_{0I}\left(\chi_{I}^{2}+b_{I}\chi_{I}^{4}\right)$
allow favorable contrast of two 3-forms inflation scenario against
recent observational data. The parameter $b_{i}$, in the mentioned
potential, is adequately chosen, so that the speed of sound gets bounded
by $0.02\lesssim c_{s}^{2}<1$, in order to comply with the {\it Planck}
constraint. We found that type II inflation, obtained through a small
asymmetry in the slopes of the potentials (making $V_{01}\neq V_{02}$),
is needed to fit the parameters within the bounds of the observational
data, especially for the running and running of running spectral indexes.
There are two relevant aspects that should be mentioned regarding
this comparison; one is related to the property of type II solution
for computing the running of the spectral index. This is a consequence
of the varying speed of sound, which is natural for this solution.
The other aspect is the requirement of the asymmetry between the potentials.
This leads to a mild generation of isocurvature perturbations towards
the end of inflation, which can accommodate tensor to scalar ratio values
within the present bounds of {\it Planck}. We note that solutions with large
curved trajectory in field space can lead to values for inflationary
parameters beyond the observational bounds. The presence of curvature,
in the field space trajectories, implies a peculiar imprint in the
primordial bispectrum during multiple field inflation which we will study in the next section. 

\section{Non-Gaussianities with multiple 3-forms}

In the previous section, we have computed the powerspectrum of curvature perturbations and its evolution on superhorizon scales using transfer fuctions. In this section, we compute the Bispectrum and the reduded bispectrum $\fnl$ on superhorizon scales using $\delta N$ formalism \cite{Lyth:2005fi} which is more convenient method for computing non-Gaussianities with multifields over using transfer functions \cite{Kaiser:2012ak}. However, both of these methods are equivalent and the final results are independent of the formalism we use.  

This section is organized as follows. In Sec.~\ref{DeltaNformalism}
we discuss the bispectrum and describe a procedure to adapt the $\delta N$
formalism \cite{Lyth:2005fi} to multiple 3-forms to calculate
it. We explain a numerical method for calculating derivatives of the
unperturbed number of $e$-foldings with respect to the unperturbed
3-form field values at sound horizon crossing, and show how these
derivatives can be related to those of a dual scalar field description.
In turn these can be used in combination with existing results to
compute the bispectrum. We stress that although our method utilizes
the dual scalar field description, it is not possible in general to
simply pass to that description and work solely with a scalar field
model. In Sec.~\ref{NG23forms} we consider the two 3-form inflation with the same
potentials of the previous section that provides a power-spectrum compatible
with {\it Planck} constraints and compute the bispectrum in that model.
We quantify and compare the momentum dependent contribution and momentum
independent contributions of the reduced bispectrum and plot the shape
of the bispectrum. 
%Finally, we conclude in Sec.~\ref{discussion} highlighting the important results of this chapter and discuss the status of two 3-form model with respect to the Planck data. 

\subsection{Non-Gaussianity and the $\delta N$ formalism }

\label{DeltaNformalism}

\subsubsection{The $\delta N$ formalism }

\label{delta N formalism}

%%%%%%%%%%%%%%%%%%%%%%%%%%%%%%%%%%%%%%%%%%%%%%%%%%%%%%%%%%%%%%%%%%%%%%%%%%%%%%%

%%%%%%%%%%%%%%%%%%%%%%%%%%%%%%%%%%%%%%%%%%%%%%%%%%%%%%%%%%%%%%%%%%%%%%%%%%%%%%%

The $\delta N$ formalism is based on the separate universe assumption
\cite{Lyth:1984gv,Starobinsky:1986fxa,Sasaki:1995aw,Wands2000,Lyth2004,Lyth:2005du}
and provides a powerful tool to evaluate the superhorizon evolution
of the curvature perturbation. In the case of multiple 3-forms,
however, the direct implementation of the $\delta N$ formalism would
be cumbersome. Using the formal relation between 3-forms and their
scalar field duals in Sec.~\ref{dual-Sec}, however,
one can indirectly implement the $\delta N$ formalism while still
employing only 3-form quantities that are easy to calculate.

The $\delta N$ formalism allows the evolution of the curvature perturbation
to be calculated, on scales larger than the horizon scale where one
can neglect spatial gradients, using only the evolution of unperturbed
\char`\"{}separate universes\char`\"{}. The central result is that
the difference in the number of $e$-folds that occurs from different
positions on an initial flat slice of spacetime to a final uniform
density slice, when compared with some fiducial value, is related
to the curvature perturbation. Writing the number of $e$-foldings as
a function of the initial and final time on the relevant hypersurfaces,
\begin{equation}
N\left(t,\,t_{i},\,x\right)=\int_{t_{i}}^{t}dt^{\prime}H\left(t^{\prime},\,x\right)\,,
\end{equation}
the primordial curvature perturbation can be expressed as 
\begin{equation}
\zeta\left(t,x\right)=N\left(t,\,t_{i},\,x\right)-N_{0}\left(t,\,t_{i}\right)\,,
\end{equation}
where $N_{0}\left(t,\,t_{i}\right)=\int_{t_{i}}^{t}dt^{\prime}H_{0}\left(t^{\prime}\right)$.
Taking $t_{i}=t_{\ast}$, the time corresponding to the modes exiting
the horizon $\left(kc_{s}=aH\right)$, the curvature perturbation
on superhorizon scales can be written in terms of partial derivatives
of $N$ with respect to the unperturbed scalar field values at horizon
exit, while holding the initial and final hypersurface constant. More
precisely 
\begin{equation}
\zeta\left(t,\,x\right)=\sum_{I}N_{,I}\left(t\right)\delta\phi_{\ast}^{I}(x)+\frac{1}{2}\sum_{IJ}N_{,IJ}\left(t\right)\delta\phi_{\ast}^{I}\left(x\right)\delta\phi_{\ast}^{J}\left(x\right)+\cdots\,,
\end{equation}
where $N_{,I}=\frac{\partial N}{\partial\phi_{I}^{\ast}}$. In momentum
space we have 
\begin{equation}
\zeta(k)=N_{,I}\delta\phi_{\ast}^{I}(k)+\frac{1}{2}N_{,IJ}\left[\delta\phi_{\ast}^{I}\star\delta\phi_{\ast}^{J}\right](k)+\cdots\,,\label{pertexpansion}
\end{equation}
where $\star$ indicates a convolution.

%\subsubsection{The bispectrum}

%In Fourier space the two- and three-point functions are defined, respectively, by 
%\begin{eqnarray}
%\langle\zeta\left(\mathbf{k_{1}}\right)\zeta\left(\mathbf{k_{2}}\right)\rangle & = & \left(2\pi\right)^{3}\delta^{3}\left(\mathbf{k_{1}}+\mathbf{k_{2}}\right){P}_{\zeta}\left(k_{1}\right)\,,\\
%\langle\zeta\left(\mathbf{k_{1}}\right)\zeta\left(\mathbf{k_{2}}\right)\zeta\left(\mathbf{k_{3}}\right)\rangle & = & \left(2\pi\right)^{3}\delta^{3}\left(\mathbf{k_{1}}+\mathbf{k_{2}}+\mathbf{k_{3}}\right)\mathcal{B}_{\zeta}\left(k_{1},k_{2},k_{3}\right)\,,
%\end{eqnarray}
%where $P_{\zeta}(k)$ is the power spectrum, and $B_{\zeta}\left(k_{1},k_{2},k_{3}\right)$
%the bispectrum. Often the bispectrum is normalized to form the reduced
%bispectrum $\fnl\left(k_{1},k_{2},k_{3}\right)$ 
%\begin{equation}
%\begin{aligned}\end{aligned}
%B_{\zeta}\left(k_{1},k_{2},k_{3}\right)=\frac{6}{5}f_{{\rm NL}}(k_{1},k_{2},k_{3})\biggl[P_{\zeta}\left(k_{1}\right)P_{\zeta}\left(k_{2}\right)+P_{\zeta}\left(k_{2}\right)P_{\zeta}\left(k_{3}\right)+P_{\zeta}\left(k_{3}\right)P_{\zeta}\left(k_{1}\right)\biggr]\,,%{aligned}
%\label{bispectrum}
%\end{equation}

\subsubsection{Calculating the bispectrum with $\delta N$}

The power spectrum and bispectrum of field fluctuations at horizon
crossing follow from the two- and three-point correlations of these
perturbations as 
\begin{align}
\langle\delta\phi_{\ast}^{I}(\mathbf{k_{1}})\delta\phi_{\ast}^{J}(\mathbf{k_{2}})\rangle  = & (2\pi)^{3}G^{IJ}\frac{2\pi^{2}}{k^{3}}{\cal P}^{\ast}\delta\left(\mathbf{k_{1}}+\mathbf{k_{2}}\right)\\
\langle\delta\phi_{\ast}^{I}(\mathbf{k_{1}})\delta\phi_{\ast}^{J}(\mathbf{k_{2}})\delta\phi_{\ast}^{K}(\mathbf{k_{3}})\rangle  = & (2\pi)^{3}\frac{4\pi^{4}}{\Pi_{i}k_{i}^{3}}{\cal P^{\ast}}^{2}A^{IJK}(k_{1},k_{2},k_{3})\delta\left(\mathbf{k_{1}}+\mathbf{k_{2}}+\mathbf{k_{2}}\right)\,,
\end{align}
where ${\cal P}=Pk^{3}/(2\pi^{2})$. Employing the $\delta N$ expansion
one finds that 
\begin{equation}
P_{\zeta}(k)=N_{I}N_{I}P^{\ast}
\end{equation}
and 
\begin{equation}
f_{{\rm NL}}=f_{{\rm NL}}^{(3)}+f_{{\rm NL}}^{(4)}+\cdots\,,\label{fnl}
\end{equation}
where 
\begin{equation}
\begin{aligned}f_{{\rm NL}}^{(3)} & =\frac{5}{6}\frac{N_{,I}N_{,J}N_{,K}A^{IJK}}{\left(G^{IJ}N_{,I}N_{,J}\right)^{2}\sum_{i}k_{i}^{3}},\\
f_{{\rm NL}}^{(4)} & =\frac{5}{6}\frac{G^{IK}G^{JL}N_{,I}N_{,J}N_{,KL}}{\left(G^{IJ}N_{,I}N_{,J}\right)^{2}}\,.
\end{aligned}
\label{fnl34}
\end{equation}
Here $f_{{\rm NL}}^{(3)}$ is momentum dependent, whereas $f_{{\rm NL}}^{(4)}$
is momentum independent (which is the definition of local $\fnl$)\footnote{Technically these results are valid only when there is not a large
hierarchy between the three wave numbers of the bispectrum and they
can all be assumed to cross the horizon at roughly the same time.
This provides a good approximation even for large hierarchies as long
as there is not a significant evolution between the horizon crossing
times of the three modes (see Refs.~\cite{Kenton:2015lxa,Kenton:2016abp}
for a full discussion)}. %\footnote{Technically these results are only valid when there is not a large hierarchy between the three wavenumbers of the bispectrum and they can all be assumed to cross the horizon at roughly the same time. This provides a good approximation even for large hierarchies as long as there is not significant evolution  between the horizon crossing times of the three modes (see Ref.~\cite{Kenton:2015lxa} for a full discussion)}. 
In general, the dominant contribution, $f_{{\rm NL}}^{(3)}$ or $f_{{\rm NL}}^{(4)}$,
is model dependent. For example, in the case of multiple canonical
scalar fields inflation, $f_{{\rm NL}}^{(4)}$ can become significant
. In contrast, for non-canonical models, $f_{{\rm NL}}^{(3)}$ can
become large.

For general multi-field non-canonical models in slow-roll (which is
the situation relevant to our models), utilising the In-In formalism
to calculate the statistics of the scalar field perturbations on flat
hypersurfaces at horizon crossing it was found that 
\begin{eqnarray}
P_{\ast}=\frac{H^{2}}{2k^{3}P_{,X}},\label{Pstar}
\end{eqnarray}
and that \cite{Gao:2008dt} 
\begin{equation}
A_{IJK}=\frac{1}{4}\sqrt{\frac{P_{,X}}{2}}\tilde{A}_{IJK},\label{aamplitude}
\end{equation}
with 
\begin{equation}
\begin{aligned}\tilde{A}^{IJK}= & G^{IJ}\epsilon^{K}\frac{u}{\epsilon}\left[\frac{4k_{1}^{2}k_{2}^{2}k_{3}^{2}}{K^{3}}-2\left(\mathbf{k_{1}}.\mathbf{k_{2}}\right)k_{3}^{2}\left(\frac{1}{K}+\frac{k_{1}+k_{2}}{K^{2}}+\frac{2k_{1}k_{2}}{K^{3}}\right)\right]\\
 & -G^{IJ}\epsilon^{K}\left[6\frac{k_{1}^{2}k_{2}^{2}}{K}+2\frac{k_{1}^{2}k_{2}^{2}\left(k_{3}+2k_{2}\right)}{K^{2}}+k_{3}k_{2}^{2}-k_{3}^{3}\right]\\
 & +G^{IJ}\left[\left(3\frac{u}{\epsilon}+4u+4\right)\tilde{\epsilon}^{K}+\tilde{\epsilon}_{,X}^{K}\frac{12H^{2}}{P_{,X}}\right]\times\\
 & \left[-\frac{k_{1}^{2}k_{2}^{2}}{K}-\frac{k_{1}^{2}k_{2}^{2}k_{3}}{K^{2}}+\left(\mathbf{k_{1}}.\mathbf{k_{2}}\right)\left(-K+\frac{\underset{i>j}{\sum}k_{i}k_{j}}{K}+\frac{k_{1}k_{2}k_{3}}{K^{2}}\right)\right]\\
 & +\frac{\epsilon^{IJ}}{\epsilon}\epsilon^{K}\left(\frac{2\lambda}{H^{2}\epsilon^{2}}-\frac{u}{\epsilon}\right)\frac{4k_{1}^{2}k_{2}^{2}k_{3}^{2}}{K^{3}}+{\rm perms}.\,,
\end{aligned}
\label{amplitude}
\end{equation}
where $K=k_{1}+k_{2}+k_{3}$, and the Hubble parameter $H$, the sound
speed squared $\left(c_{s}^{2}\right)$, and slow-roll parameters
$\left(\epsilon,\epsilon^{I},...,{\rm etc.}\right)$ are evaluated
at sound horizon exit $c_{s}k=aH$. Expressions for $c_{s}^{2}$,
$u$ and $\lambda$ are given in Ref.~\cite{Gao:2008dt} for non-canonical
models\footnote{{We have corrected typos in the first and third lines of (\ref{amplitude})
that were present in Ref.~\cite{Gao:2008dt}.}}. In this work, we express all of these parameters in terms of 3-form
quantities using (\ref{dual action}) and (\ref{3-formdual}).
First $u$ is defined as 
\begin{equation}
u\equiv\frac{1}{\bar{c}_{s}^{2}}-1\,,\label{u}
\end{equation}
where the effective speed of sound\footnote{We note that during the slow-roll regime effective sound speed is nearly the same as adiabatic sound speed \cite{Piattella:2013wpa}. Therefore, using the slow-roll approximation, from \ref{NQuad-X-diff-1} and \ref{speedN3form} we can deduce $\bar{c}_{s}^2\approx c_{s}^2$.} is given by 
\begin{equation}
%\begin{aligned}\end{aligned}
\bar{c}_{s}^{2}=\frac{P_{,X}}{2XP_{,XX}+P_{,X}}=\frac{\underset{I}{\sum}\frac{\chi_{I}}{V_{,\chi_{I}}}}{\underset{I}{\sum}V_{,\chi_{I}\chi_{I}}^{-1}}.%{aligned}
\label{cs2}
\end{equation}
We also define $\lambda$, such that 
\begin{equation}
%\begin{aligned}\end{aligned}
\lambda=X^{2}P_{,XX}+\frac{2}{3}X^{3}P_{,XXX}=-\sum_{I}\frac{V_{,\chi_{I}}^{3}V_{\chi_{I}\chi_{I}\chi_{I}}}{12V_{,\chi_{I}\chi_{I}}^{3}}\,.%{aligned}
\,\label{lambda}
\end{equation}
The various slow-roll quantities are defined by 
\begin{equation}
\epsilon\equiv-\frac{\dot{H}}{H^{2}}=\frac{3}{2}\frac{\underset{I}{\sum}\chi_{I}V_{,\chi_{I}}}{V}\left(1-\underset{I}{\sum}w_{I}^{2}\right)\,,\label{eps}
\end{equation}
\begin{equation}
\epsilon^{IJ}=\frac{P_{,X}\dot{\phi}^{I}\dot{\phi}^{J}}{2H^{2}}=\frac{P_{,X}\sqrt{X_{I}X_{J}}}{2H^{2}}=\epsilon^{I}\epsilon^{J}\,,\label{epsij}
\end{equation}
where 
\begin{equation}
\epsilon^{I}=\sqrt{\frac{X_{I}P_{,X}}{2H^{2}}}=\sqrt{\frac{3V_{,\chi_{I}}^{2}}{4V}\left(\underset{I}{\sum}\frac{\chi_{I}}{V_{,\chi_{I}}}\right)\left(1-\underset{I}{\sum}w_{I}^{2}\right)}\,,\label{epsi}
\end{equation}
\begin{equation}
\tilde{\epsilon}{}_{I}=-\frac{P_{,I}}{3\sqrt{2P_{,X}}H^{2}}=\frac{\sqrt{6}w_{I}}{3\sqrt{2\underset{I}{\sum}
\frac{\chi_{I}}{V_{,\chi_{I}}}}H}\,.\label{epstilde}
\end{equation}
Using the Friedmann equation in (\ref{NFriedm-1-1}) we obtain
\begin{equation}
\begin{alignedat}{1}\tilde{\epsilon}_{,X}^{I} & =-\frac{P_{,XI}}{3\sqrt{2P_{,X}}H^{2}}+P_{,I}\left[\frac{2XP_{,XX}+P_{,X}}{9\sqrt{2P_{,X}}H^{4}}+\frac{P_{,XX}}{6\sqrt{2}P_{,X}^{3/2}H^{2}}\right]\,,\\
 & =-\sqrt{6}Hw_{I}\left[\frac{\underset{I}{\sum}V_{,\chi_{I}\chi_{I}}^{-1}}{\sqrt{2\underset{I}{\sum}\frac{\chi_{I}}{V_{,\chi_{I}}}}V}+\frac{\underset{I}{\sum}\left(V_{,\chi_{I}\chi_{I}}^{-1}V_{,\chi_{I}}^{-2}-\chi_{I}V_{,\chi_{I}}^{-3}\right)}{3\sqrt{2}\left(\underset{I}{\sum}\frac{\chi_{I}}{V_{,\chi_{I}}}\right)^{3/2}V}\right]\left(1-\underset{I}{\sum}w_{I}^{2}\right)\,.
\end{alignedat}
\label{depst}
\end{equation}
Note that the dual scalar field action in (\ref{dual action})
satisfies $P_{,XI}=0.$

In the squeezed limit i.e., $k_{2}\rightarrow0$, it can be seen from
(\ref{amplitude}) that $\fnl^{(3)}$ reduces to the order of
slow-roll parameters. Therefore $f_{{\rm NL}}^{(4)}$ is expected
to be dominant in this limit if non-Gaussianity is significant.

\subsubsection{The $\delta N$ for two 3-forms}

\label{N derivatives}

The crucial step, when it comes to computing $f_{{\rm NL}}$, is the
calculation of the derivatives of $N$ with respect to the fields
at the sound horizon crossing. In general $N_{,I}$ and $N_{,IJ}$
evolve on superhorizon scales and except in a few cases (see e.g., Ref.~\cite{Vernizzi:2006ve})
the analytical computation of these quantities is not tractable. For
this reason we do our computations numerically using a method that
is explained in Sec.~\ref{NG23forms}.

First of all we must rewrite the derivatives in terms of 3-forms.
Here we do this explicitly for two 3-forms. The same procedure
can be extended trivially to $\mathbb{N}$ 3-form fields. We can
infer the following relations from (\ref{3-formdual}) and (\ref{NQuad-Fried-1})
relating two 3-forms to the two non-canonical scalar fields 
\begin{equation}
\phi_{1}=\sqrt{6}Hw_{1}\equiv\phi_{1}\left(\chi_{1},\chi_{2},w_{1},w_{2}\right)\,,\label{phi1}
\end{equation}
\begin{equation}
\phi_{2}=\sqrt{6}Hw_{2}\equiv\phi_{2}\left(\chi_{1},\chi_{2},w_{1},w_{2}\right)\,,\label{phi1-1}
\end{equation}
It is highly nontrivial to invert the relations in (\ref{phi1})
and (\ref{phi1-1}). % i.e., it is hard to express $\chi_{I}=\chi_{I}\left(\phi\right)$.However,duringinflation
While the fields are slowly rolling, one can verify that the approximation
$w_{I}\approx\sqrt{\frac{3}{2}}\chi_{I}$ is accurately satisfied. As a consequence, we express the
$N$ derivatives $N_{,I}$ and $N_{,IJ}$ in terms of the two 3-forms
$\chi_{1},\,\chi_{2}$ as 
\begin{equation}
\frac{\partial N}{\partial\phi_{1}^{\ast}}=\frac{\partial N}{\partial\chi_{1}^{\ast}}\frac{\partial\chi_{1}^{\ast}}{\partial\phi_{1}^{\ast}}+\frac{\partial N}{\partial\chi_{2}^{\ast}}\frac{\partial\chi_{2}^{\ast}}{\partial\phi_{1}^{\ast}}\,,\label{firstderivatives}
\end{equation}
%\begin{equation}%\frac{\partial N}{\partial\phi_{2}^{\ast}}=\frac{\partial N}{\partial\chi_{1}^{\ast}}\frac{\partial\chi_{1}^{\ast}}{\partial\phi_{2}^{\ast}}+\frac{\partial N}{\partial\chi_{2}^{\ast}}\frac{\partial\chi_{2}^{\ast}}{\partial\phi_{2}^{\ast}}\,.\label{firstderivatives-1}%\end{equation}
\begin{equation}
\begin{aligned}\frac{\partial^{2}N}{\partial\phi_{1}^{\ast}\partial\phi_{2}^{\ast}}= & \frac{\partial N}{\partial\chi_{1}^{\ast}}\frac{\partial^{2}\chi_{1}^{\ast}}{\partial\phi_{1}^{\ast}\partial\phi_{2}^{\ast}}+\frac{\partial N}{\partial\chi_{2}^{\ast}}\frac{\partial^{2}\chi_{2}^{\ast}}{\partial\phi_{1}^{\ast}\partial\phi_{2}^{\ast}}+\frac{\partial^{2}N}{\partial\chi_{1}^{\ast2}}\frac{\partial\chi_{1}^{\ast}}{\partial\phi_{1}^{\ast}}\frac{\partial\chi_{1}^{\ast}}{\partial\phi_{2}^{\ast}}\\
 & +\frac{\partial^{2}N}{\partial\chi_{2}^{\ast2}}\frac{\partial\chi_{2}^{\ast}}{\partial\phi_{1}^{\ast}}\frac{\partial\chi_{2}^{\ast}}{\partial\phi_{2}^{\ast}}+\frac{\partial^{2}N}{\partial\chi_{1}^{\ast}\partial\chi_{2}^{\ast}}\frac{\partial\chi_{1}^{\ast}}{\partial\phi_{1}^{\ast}}\frac{\partial\chi_{2}^{\ast}}{\partial\phi_{2}^{\ast}}+\frac{\partial^{2}N}{\partial\chi_{1}^{\ast}\partial\chi_{2}^{\ast}}\frac{\partial\chi_{1}^{\ast}}{\partial\phi_{2}^{\ast}}\frac{\partial\chi_{2}^{\ast}}{\partial\phi_{1}^{\ast}}\,,
\end{aligned}
\label{d2np12-1}
\end{equation}
\begin{equation}
\frac{\partial^{2}N}{\partial\phi_{1}^{\ast2}}=\frac{\partial N}{\partial\chi_{1}^{\ast}}\frac{\partial^{2}\chi_{1}^{\ast}}{\partial\phi_{1}^{\ast2}}+\frac{\partial N}{\partial\chi_{2}^{\ast}}\frac{\partial^{2}\chi_{2}^{\ast}}{\partial\phi_{1}^{\ast2}}+\frac{\partial^{2}N}{\partial\chi_{1}^{\ast2}}\left(\frac{\partial\chi_{1}^{\ast}}{\partial\phi_{1}^{\ast}}\right)^{2}+\frac{\partial^{2}N}{\partial\chi_{2}^{\ast2}}\left(\frac{\partial\chi_{2}^{\ast}}{\partial\phi_{1}^{\ast}}\right)^{2}+2\frac{\partial^{2}N}{\partial\chi_{1}^{\ast}\partial\chi_{2}^{\ast}}\frac{\partial\chi_{1}^{\ast}}{\partial\phi_{1}^{\ast}}\frac{\partial\chi_{2}^{\ast}}{\partial\phi_{1}^{\ast}}\,.\label{d2np12}
\end{equation}
%\begin{equation}%\frac{\partial^{2}N}{\partial\phi_{2}^{\ast2}}=\frac{\partial N}{\partial\chi_{1}^{\ast}}\frac{\partial^{2}\chi_{1}^{\ast}}{\partial\phi_{2}^{\ast2}}+\frac{\partial N}{\partial\chi_{2}^{\ast}}\frac{\partial^{2}\chi_{2}^{\ast}}{\partial\phi_{2}^{\ast2}}+\frac{\partial^{2}N}{\partial\chi_{1}^{\ast2}}\left(\frac{\partial\chi_{1}^{\ast}}{\partial\phi_{2}^{\ast}}\right)^{2}+\frac{\partial^{2}N}{\partial\chi_{2}^{\ast2}}\left(\frac{\partial\chi_{2}^{\ast}}{\partial\phi_{1}^{\ast}}\right)^{2}+2\frac{\partial^{2}N}{\partial\chi_{1}^{\ast}\partial\chi_{2}^{\ast}}\frac{\partial\chi_{1}^{\ast}}{\partial\phi_{2}^{\ast}}\frac{\partial\chi_{2}^{\ast}}{\partial\phi_{2}^{\ast}}\,.\label{d2np22}%\end{equation}andequivalentlyfor
derivatives of $\phi_{2}$. These equations define the relations among
the $N$ derivatives ($N_{,I}$ and $N_{,IJ}$) with respect to scalar
field $\phi_{I}^{\ast}$ to the $N$ derivatives with respect to 3-form
fields at horizon crossing $\frac{\partial N}{\partial\chi_{1}^{\ast}}\,,\,\frac{\partial N}{\partial\chi_{2}^{\ast}}\,,\,\frac{\partial^{2}N}{\partial\chi_{1}^{\ast}\partial\chi_{2}^{\ast}}\,,\,\frac{\partial^{2}N}{\partial\chi_{1}^{\ast2}}\,,\,\frac{\partial^{2}N}{\partial\chi_{2}^{\ast2}}$.
In other words, we have indirectly transported the $\delta N$ formalism
from scalar fields to 3-form fields. However, we still need to
calculate the derivatives of the 3-form fields with respect to
the dual scalar fields. For this purpose we differentiate the relations
(\ref{phi1}) and (\ref{phi1-1}) keeping in mind that $\phi_{1}$
and $\phi_{2}$ are independent fields. Then we have that 
\begin{equation}
\frac{d\phi_{1}}{d\phi_{1}}=\frac{1}{\sqrt{6}w_{1}}\frac{\partial H}{\partial\phi_{1}}+\frac{1}{\sqrt{6}H}\frac{\partial w_{1}}{\partial\phi_{1}}=1\,.\label{dph11}
\end{equation}
\begin{equation}
\frac{d\phi_{1}}{d\phi_{2}}=\frac{1}{\sqrt{6}w_{1}}\frac{\partial H}{\partial\phi_{2}}+\frac{1}{\sqrt{6}H}\frac{\partial w_{1}}{\partial\phi_{2}}=0\,.\label{dph11-1}
\end{equation}
\begin{equation}
\frac{d\phi_{2}}{d\phi_{1}}=\frac{1}{\sqrt{6}w_{2}}\frac{\partial H}{\partial\phi_{2}}+\frac{1}{\sqrt{6}H}\frac{\partial w_{2}}{\partial\phi_{2}}=1\,.\label{dph11-1-1}
\end{equation}
\begin{equation}
\frac{d\phi_{2}}{d\phi_{2}}=\frac{1}{\sqrt{6}w_{2}}\frac{\partial H}{\partial\phi_{1}}+\frac{1}{\sqrt{6}H}\frac{\partial w_{2}}{\partial\phi_{1}}=0\,.\label{dph11-1-1-1}
\end{equation}
Solving (\ref{dph11})-(\ref{dph11-1-1-1}) for a potential
of the form $V=V\left(\chi_{1}\right)+V\left(\chi_{2}\right)$, we
obtain 
\begin{equation}
\begin{aligned}\frac{\partial\chi_{1}}{\partial\phi_{1}}= & \frac{\chi_{2}V_{,\chi_{2}}+H^{2}\left(6-9\chi_{1}^{2}\right)}{3H\left(6H^{2}+\chi_{1}V_{,\chi_{1}}+\chi_{2}V_{,\chi_{2}}\right)}\\
\frac{\partial\chi_{1}}{\partial\phi_{2}}= & -\frac{\chi_{1}\left(V_{,\chi_{2}}+9H^{2}\chi_{2}\right)}{3H\left(6H^{2}+\chi_{1}V_{,\chi_{1}}+\chi_{2}V_{,\chi_{2}}\right)}
\end{aligned}
\label{Qf}
\end{equation}
\begin{equation}
\begin{aligned}\frac{\partial^{2}\chi_{1}}{\partial\phi_{1}^{2}}= & \frac{-1}{9H^{2}\left(6H^{2}+\chi_{1}V_{,\chi_{1}}+\chi_{2}V_{,\chi_{2}}\right)^{3}}
\Bigg\lbrace\chi_{1}V_{,\chi_{1}}^{2}\left[\chi_{2}\left(\chi_{2}
V_{,\chi_{2}\chi_{2}}+2V_{,\chi_{2}}\right)
+H^{2}\left(9\chi_{1}^{2}-6\right)\right]\\
 & -2V_{,\chi_{1}}\left[-3H^{2}\chi_{2}\left(3V_{,\chi_{2}\chi_{2}}\chi_{1}^{2}
 \chi_{2}+6V_{,\chi_{2}}\chi_{1}^{2}+4V_{,\chi_{2}}\right)-V_{,\chi_{2}}^{2}
 \chi_{2}^{2}+18H^{4}\left(3\chi_{1}^{2}-2\right)\right]\\
 & +\chi_{1}V_{,\chi_{1}\chi_{1}}\left(\chi_{2}V_{,\chi_{2}}+H^{2}\left(6-9\chi_{1}^{2}\right)\right)^{2}\\
 & -9\chi_{1}H^{2}\left(-3H^{2}\chi_{2}\left(3V_{,\chi_{2}\chi_{2}}\chi_{1}^{2}
 \chi_{2}+12V_{,\chi_{2}}\right)-3V_{,\chi_{2}}^{2}\chi_{2}^{2}+54H^{4}
 \left(3\chi_{1}^{2}-2\right)\right)\Bigg\rbrace\,.
\end{aligned}
\label{Q1F1F1}
\end{equation}
%\begin{equation}%\begin{aligned}\frac{\partial^{2}\chi_{2}}{\partial\phi_{2}^{2}}= & \frac{-1}{9H^{2}\left(6H^{2}+\chi_{1}V_{,\chi_{1}}+\chi_{2}V_{,\chi_{2}}\right)^{3}}\lbrace\chi_{2}V_{,\chi_{2}}^{2}\left[\chi_{1}\left(\chi_{1}V_{,\chi_{1}\chi_{1}}+2V_{,\chi_{1}}\right)+H^{2}\left(9\chi_{2}^{2}-6\right)\right]\\ & -2V_{,\chi_{2}}\left[-3H^{2}\chi_{1}\left(3V_{,\chi_{1}\chi_{1}}\chi_{2}^{2}\chi_{1}+6V_{,\chi_{1}}\chi_{2}^{2}+4V_{,\chi_{1}}\right)-V_{,\chi_{1}}^{2}\chi_{1}^{2}+18H^{4}\left(3\chi_{2}^{2}-2\right)\right]\\ & +\chi_{2}V_{,\chi_{2}\chi_{2}}\left(\chi_{1}V_{,\chi_{1}}+H^{2}\left(6-9\chi_{2}^{2}\right)\right)^{2}\\ & -9\chi_{2}H^{2}\left(-3H^{2}\chi_{1}\left(3V_{,\chi_{1}\chi_{1}}\chi_{2}^{2}\chi_{1}+12V_{,\chi_{1}}\right)-3V_{,\chi_{1}}^{2}\chi_{1}^{2}+54H^{4}\left(3\chi_{2}^{2}-2\right)\right)\rbrace\,.%\end{aligned}%\label{q2f2f2}%\end{equation}
\begin{equation}
\begin{aligned}
\frac{\partial^{2}\chi_{1}}{\partial\phi_{2}^{2}}= & \frac{-1}{9H^{2}\left(6H^{2}+\chi_{1}V_{,\chi_{1}}+\chi_{2}V_{,\chi_{2}}\right)^{3}}
\Bigg\lbrace\chi_{1}\left[18V_{,\chi_{2}}H^{2}\chi_{2}\left(V_{,\chi_{1}\chi_{1}}\chi_{1}^{2}-18H^{2}\right)-2V_{,\chi_{2}}^{3}\chi_{2}\right]\\
 & +\chi_{1}V_{,\chi_{2}}^{2}\left[\chi_{1}\left(\chi_{1}V_{,\chi_{1}\chi_{1}}-2V_{,\chi_{1}}\right)-3H^{2}\left(3\chi_{2}^{2}+10\right)\right]\\
 &+\chi_{1}V_{,\chi_{2}\chi_{2}}\left[V_{,\chi_{1}}\chi_{1}
 +H^{2}\left(6-9\chi_{2}^{2}\right)\right]^{2}\\
 & +9\chi_{1}H^{2}\left[3H^{2}\chi_{1}\left(3V_{,\chi_{1}\chi_{1}}
 \chi_{1}\chi_{2}^{2}+4V_{,\chi_{1}}\right)+V_{,\chi_{1}}^{2}\chi_{1}^{2}-18H^{4}
 \left(9\chi_{2}^{2}-2\right)\right]\Bigg\rbrace\,.
\end{aligned}
\label{q1f2f2}
\end{equation}
%\begin{equation}%\begin{aligned}\frac{\partial^{2}\chi_{2}}{\partial\phi_{1}^{2}}= & \frac{-1}{9H^{2}\left(6H^{2}+\chi_{1}V_{,\chi_{1}}+\chi_{2}V_{,\chi_{2}}\right)^{3}}\lbrace\chi_{2}\left[18V_{,\chi_{1}}H^{2}\chi_{1}\left(V_{,\chi_{2}\chi_{2}}\chi_{2}^{2}-18H^{2}\right)-2V_{,\chi_{1}}^{3}\chi_{1}\right]\\ & +\chi_{2}V_{,\chi_{1}}^{2}\left[\chi_{2}\left(\chi_{2}V_{,\chi_{2}\chi_{2}}-2V_{,\chi_{2}}\right)-3H^{2}\left(3\chi_{1}^{2}+10\right)\right]+\chi_{2}V_{,\chi_{1}\chi_{1}}\left[V_{,\chi_{2}}\chi_{2}+H^{2}\left(6-9\chi_{1}^{2}\right)\right]^{2}\\ & +9\chi_{2}H^{2}\left[3H^{2}\chi_{2}\left(3V_{,\chi_{2}\chi_{2}}\chi_{2}\chi_{1}^{2}+4V_{,\chi_{2}}\right)+V_{,\chi_{2}}^{2}\chi_{2}^{2}-18H^{4}\left(9\chi_{1}^{2}-2\right)\right]\rbrace\,.%\end{aligned}%\label{q2f1f1}%\end{equation}
\begin{equation}
\begin{aligned}\frac{\partial^{2}\chi_{1}}{\partial\phi_{1}\partial\phi_{2}}= & \frac{1}{9H^{2}\left(6H^{2}+\chi_{1}V_{,\chi_{1}}+\chi_{2}V_{,\chi_{2}}\right)^{3}}
\Bigg\lbrace\\
& -V_{,\chi_{2}}^{3}\chi_{2}^{2}+V_{,\chi_{2}}^{2}\chi_{2}
\left[V_{,\chi_{1}\chi_{1}}\chi_{1}^{2}+3H^{2}\left(-4+3\chi_{1}^{2}-3\chi_{2}\right)
\right]\\
 & +V_{,\chi_{2}}\left[3H^{2}\chi_{1}\left(V_{,\chi_{1}\chi_{1}}\chi_{1}\left(-3\chi_{1}^{2}+3\chi_{2}^{2}+2\right)+3V_{,\chi_{1}}\left(\chi_{1}^{2}-\chi_{2}^{2}+2\right)\right)+V_{,\chi_{1}}^{2}\chi_{1}^{2}\right]\\
 & +36V_{,\chi_{2}}H^{4}\left(6\chi_{1}^{2}-3\chi_{2}^{2}-1\right)+
 \text{\ensuremath{\chi_{2}}}V_{,\chi_{2}\chi_{2}}V_{,\chi_{1}}^{2}\chi_{1}^{2}\\
 &+3\chi_{2}V_{,\chi_{2}\chi_{2}}V_{,\chi_{1}}H^{2}\chi_{1}\left(3\chi_{1}^{2}-3\chi_{2}^{2} +2\right)\\
 &+ 162\chi_{2}H^{6}\left(9\chi_{1}^{2}-2\right)\\
 &+ 27\chi_{2}H^{4}\chi_{1}\left(\chi_{1}
 \left(-3V_{,\chi_{1}\chi_{1}}\chi_{1}^{2}+2V_{,\chi_{1}\chi_{1}}-3V_{,\chi_{2}
 \chi_{2}}\chi_{2}^{2}+2V_{,\chi_{2}\chi_{2}}\right)+4\chi_{2}V_{,\chi_{1}}\right)
 \Bigg\rbrace\,.
\end{aligned}
\label{q1f1f2}
\end{equation}
The remaining derivatives can be obtained from these by interchanging
$1\leftrightarrow2$. Following (\ref{firstderivatives})-(\ref{d2np12})
%\ref{d2np22}) 
the quantities obtained in (\ref{Qf})-(\ref{q1f1f2}) %\ref{q2f1f2}) 
are to be evaluated at $kc_{s}=aH$. However, the derivatives of $N$
with respect to the 3-form fields evolve on superhorizon scales.

In the squeezed limit i.e., $k_{2}\rightarrow0$, it can be seen from
(\ref{amplitude}) that $\fnl^{(3)}$ reduces to the order of
slow-roll parameters. Therefore $f_{{\rm NL}}^{(4)}$ is expected
to be dominant in this limit if non-Gaussianity is significant.

\subsection{Non-Gaussianities in two 3-form inflation}

\label{NG23forms}

In this subsection, we aim further update the observational status of two
3-form inflation by means of calculating
the reduced bispectrum $f_{{\rm NL}}$. We consider, the type II solutions with the potentials, $V\left(\chi_{1},\,\chi_{2}\right)=V_{01}f\left(\chi_{1}\right)+V_{20}f\left(\chi_{2}\right)$,
where $f\left(\chi_{I}\right)=\chi_{I}^{2}+b\chi_{I}^{2n}$, which were studied in  Sec.~\ref{partIchap3-f} and were shown to be consistent with {\it Planck} data, predicting the scalar spectral index $n_{s}\sim0.967$ and the tensor to scalar ratio $r\sim0.0422$.

The observational prediction of non-Gaussianity for multifield inflation
is deeply associated with the evolution of isocurvature perturbations.
In the single field inflation the statistics of the curvature pertrubation
evaluated at horizon exit can be confronted with the observation.
This is because the curvature perturbation is conserved on superhorizon
scales if the system is adiabatic \cite{Lyth2004,Rigopoulos:2003ak,Christopherson:2008ry}.
Whereas for multifield models, the statistics evolve on superhorizon
scales and non-Gaussianity can be generated as a consequence of the
presence of isocurvature perturbations. This can happen in two regimes,
namely, (i) during inflation \cite{Byrnes:2008wi,Peterson:2010np,Elliston:2011dr,Elliston:2012wm} and
(ii) after inflation such as in the curvaton model \cite{Mollerach:1989hu,Linde:1996gt,Enqvist:2001zp,Lyth:2001nq,Moroi:2001ct,Enqvist:2005pg,Linde:2005yw,Malik:2006pm,Sasaki:2006kq,Meyers:2013gua,Elliston:2014zea,Byrnes:2014xua}.
In general the statistics continue to evolve until all isocurvature
perturbations decay, the so-called adiabatic limit \cite{Elliston:2011dr}.
We evaluate $f_{{\rm NL}}$ at the end of inflation, this is
a good approximation as long as reheating proceeds quickly, and curvaton
type effects do not occur. 
\begin{figure}[h!]
\centering\includegraphics[width=4in]{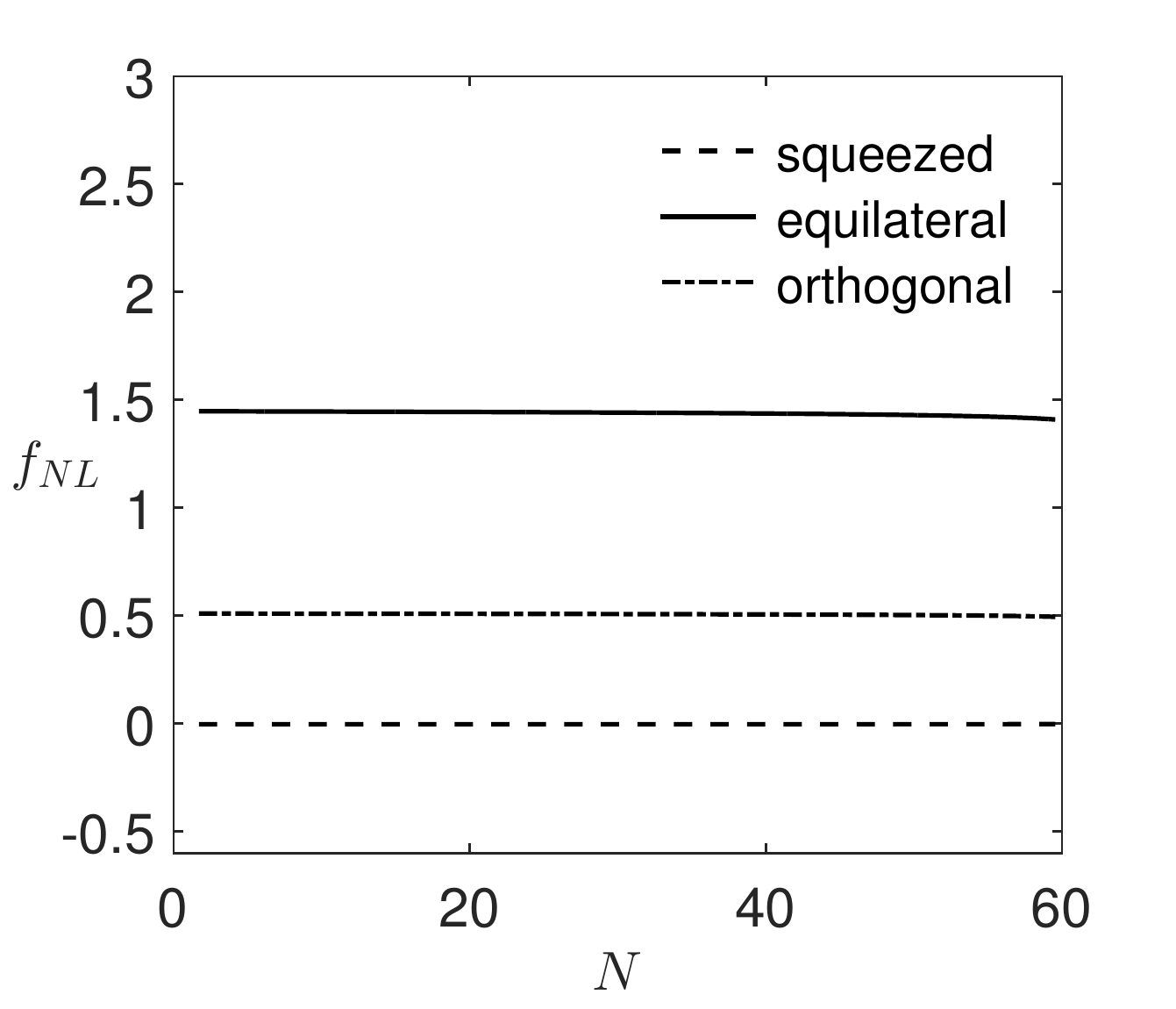}\caption{\label{fNLshape} In this plot we depict $f_{{\rm NL}}$ against $N$
for squeezed $\left(k_{2}\ll k_{1}=k_{3}\right)$ equilateral $\left(k_{1}=k_{2}=k_{3}\right)$
and orthogonal $\left(k_{1}=2k_{2}=2k_{3}\right)$ configurations.
We have considered the potentials $V_{1}=V_{01}\left(\chi_{1}^{2}+b_{1}\chi_{1}^{4}\right)$
and $V_{2}=V_{20}\left(\chi_{2}^{2}+b_{2}\chi_{2}^{4}\right)$ with
$V_{01}=1,\:V_{20}=0.93,\:b_{1,2}=-0.35$ and taken the initial conditions
$\chi_{1}\left(0\right)\approx0.5763,\,\chi_{2}\left(0\right)\approx0.5766,\,\chi_{1}^{\prime}\left(0\right)=-0.000224,\,\chi_{2}^{\prime}\left(0\right)=0.00014$.}
\end{figure}

To calculate $f_{{\rm NL}}$ given in (\ref{fnl}), we need to
compute the $N$ derivatives with respect to the initial conditions
of 3-form fields defined in (\ref{firstderivatives})-(\ref{d2np12}).
To compute these numerically, we define the following discrete derivatives
that can in principle, be extended to any number of fields, 
\begin{equation}
\begin{aligned}N_{,\chi_{1}^{\ast}} & =\frac{N\left(\chi_{1}^{\ast}+\Delta\chi_{1}\,,\,\chi_{2}^{\ast}\right)-N\left(\chi_{1}^{\ast}-\Delta\chi_{1}\,,\,\chi_{2}^{\ast}\right)}{2\Delta\chi_{1}},\\
N_{,\chi_{1}^{\ast}\chi_{1}^{\ast}} & =\frac{N\left(\chi_{1}^{\ast}+\Delta\chi_{1}\,,\,\chi_{2}^{\ast}\right)-2N\left(\chi_{1}^{\ast}\right)+N\left(\chi_{1}^{\ast}+\Delta\chi_{1}\,,\,\chi_{2}^{\ast}\right)}{\Delta\chi_{1}^{2}},\\
N_{,\chi_{1}^{\ast}\chi_{2}^{\ast}} & =\left[N\left(\chi_{1}^{\ast}+\Delta\chi_{1}\,,\,\chi_{2}^{\ast}+\Delta\chi_{2}\right)-N\left(\chi_{1}^{\ast}+\Delta\chi_{1}\,,\,\chi_{2}^{\ast}-\Delta\chi_{2}\right)-\right.\\
 & ~~~\left.N\left(\chi_{1}^{\ast}-\Delta\chi_{1}\,,\,\chi_{2}^{\ast}+\Delta\chi_{2}\right)+N\left(\chi_{1}^{\ast}-\Delta\chi_{1}\,,\,\chi_{2}^{\ast}-\Delta\chi_{2}\right)\right](4\Delta\chi_{1}^{2})^{-1},%N_{,\chi_{I}\chi_{J}}
\end{aligned}
\label{discretederivatives}
\end{equation}
and similarly we can obtain the remaining derivatives by interchanging
$1\leftrightarrow2$. In the above expression, $N\left(\chi_{1},\chi_{2}\right)$
is the number of $e$-foldings that occur starting at initial conditions
$\{\chi_{1}^{\ast},\chi_{2}^{\ast}\}$ and ending at a given final energy
density. This final energy density is defined by the condition that
$N\left(\chi_{1},\chi_{2}\right)=60.35$ at the point $\epsilon=1$.
That is the central point in the finite difference represents a trajectory
that undergoes $60$ $e$-folds of inflation, from the initial field
value until inflation ends, and the density at that time is used as
the final density for all the other points in the difference scheme.
These other points therefore represent slightly different amounts
of inflation, and we note that their associated trajectories do
not end exactly at the point $\epsilon=1$. In our numerical results
we take $\Delta\chi_{I}\sim10^{-5}$. Using the $N$ derivatives calculated
from (\ref{discretederivatives}) and evaluating the amplitude given
by (\ref{aamplitude}), we compute $f_{{\rm NL}}$ in (\ref{fnl}).
We obtain the momentum independent contribution $\fnl^{(4)}$ in
(\ref{fnl34}) to be very small $\mathcal{O}\left(10^{-3}\right)$.
In Fig.~\ref{fNLshape} we plot the total $f_{{\rm NL}}$ versus $N$
for squeezed ($k_{2}\ll k_{1}=k_{3}$), equilateral ($k_{1}=k_{2}=k_{3}$)
and orthogonal ($k_{1}=2k_{2}=2k_{3}$) triangles.

\begin{figure}[h!]
\centering\includegraphics[width=4in]{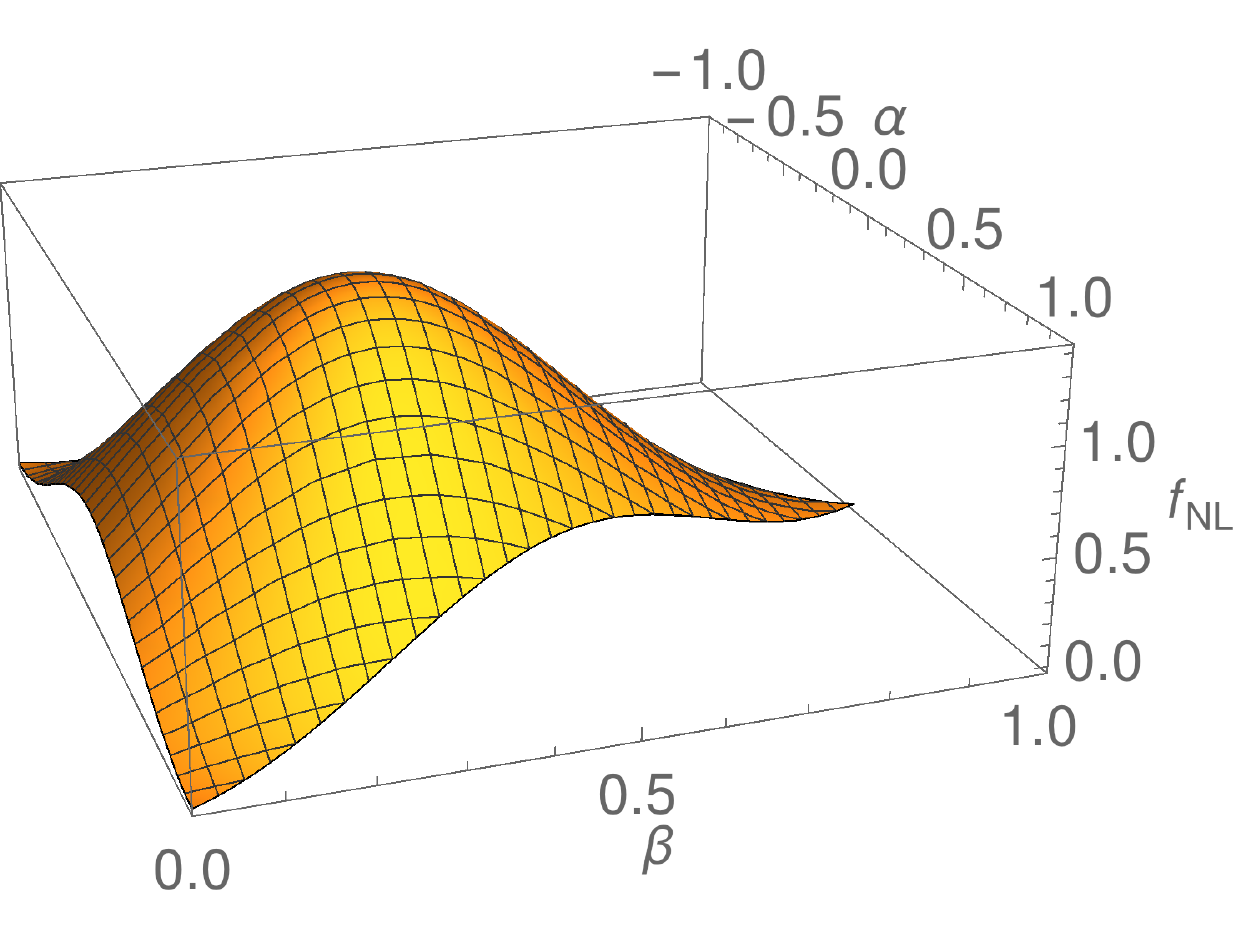} \caption{\label{fig2} Graphical representation of the non-Gaussianity shape $f_{{\rm NL}}\left(\alpha,\,\beta\right)$.
We have considered the potentials $V_{1}=V_{01}\left(\chi_{1}^{2}+b_{1}\chi_{1}^{4}\right)$
and $V_{2}=V_{20}\left(\chi_{2}^{2}+b_{2}\chi_{2}^{4}\right)$ with
$V_{01}=1,\:V_{20}=0.93,\:b_{1,2}=-0.35$ and taken the initial conditions
$\chi_{1}\left(0\right)\approx0.5763,\,\chi_{2}\left(0\right)\approx0.5766,\,\chi_{1}^{\prime}\left(0\right)=-0.000224,\,\chi_{2}^{\prime}\left(0\right)=0.00014$. }
\end{figure}

It is convenient to express the reduced bispectrum in terms of the
following independent variables \cite{Fergusson:2008ra,Fergusson:2009nv}
\begin{equation}
\alpha=\frac{k_{2}-k_{3}}{k}\quad,\quad\beta=\frac{k-k_{1}}{k}\quad\text{where}\quad k=\frac{k_{1}+k_{2}+k_{3}}{2}\,,
\end{equation}
where $0\leq\beta\leq1$ and, $-\left(1-\beta\right)\leq\alpha\leq\left(1-\beta\right)$.
In Fig.~\ref{fig2} we depict the shape of a slice through the reduced
bispectrum $f_{{\rm NL}}\left(k_{1},k_{2},k_{3}\right)$ at $N=60$
using these variables. The bispectrum shape reveals details about
the dominant interaction contributions \cite{Babich:2004gb}. In general,
the presence of a signal in the squeezed limit represents the interaction
of the long wavelength mode, which already exited the horizon, with
the short wavelength modes still being within the horizon. This can
happen in the case where more than one light scalar field drives the
period of inflation. When, instead, we observe a peak in the equilateral
limit, the dominant interaction between the fields occurs when the
modes are exiting the horizon at the same time during inflation. This
is taken to be the distinctive feature of models with a non-canonical
kinetic term or models involving higher derivative interactions
\cite{Chen:2010xka}. In the case of multiple non-canonical scalar
field inflation (which is effectively happening in the two 3-form
inflation scenario), it is possible that we would encounter a mixture
of shapes \cite{Babich:2004gb,Chen:2010xka}. Although in the example
we explored there is no significant signal in the squeezed limit.

\subsection{Summary}

\label{sum3-form} 

Let us summarize our specific results. Inflation
driven by a multifield setting, in particular by a couple of 3-form
fields, is very much still admissible within current {\it Planck} data.
This is the main assertion that this chapter indicates. Moreover, two
3-form fields with a small asymmetry (in the sense explained in this
chapter) produces better results (in terms of fitting within current
observational data) for concrete cosmological parameters, in contrast
to a symmetric configuration or to a single 3-form setting. This is
interesting if we take into consideration, the correspondence (on
dualization) between 3-form field and non-canonical (kinetic) scalar
field dynamics. In fact, a dual description of two 3-forms assists
to relate to k-inflationary models \cite{Ohashi:2011na}. We
have shown that having multiple 3-forms driven inflation brings the
inflaton mass to a lower scale, when compared with a single 3-form.
We then identified the existence of de Sitter like fixed points, where
two 3-forms inflation can mimic single 3-form inflationary scenarios,
for a suitable class of potentials. We also did a detailed numerical
study of a different type of inflationary dynamics (type II) characterized
by the dominance of a non trivial (gravity mediated) coupling, between
the two 3-form fields. The type II solution stands physically interesting
by its ability to generate substantial isocurvature perturbations
at the end of inflation. We have numerically computed the effect of
these perturbations via transfer functions. The comparison of selected
inflationary parameters against the observational data, in the case
where the 3-form fields potential have the form $\chi_{I}^{2}+b_{I}\chi_{I}^{4}$,
show that type II solutions, predicting a small variation in the speed
of sound, are in excellent agreement with the observational bounds
of running spectral indexes.

We presented a generic framework to compute primordial
non-Gaussianity in the case of multiple 3-form field inflation.
We followed the $\delta N$ formalism which is a well-known method
to study the evolution of curvature perturbations on superhorizon
scales in the case of multiple scalar fields. Because of the fact that
the 3-form fields are dual to non-canonical scalar fields, which
was shown in \cite{Mulryne:2012ax}, we developed an indirect methodology
to implement $\delta N$ formalism to 3-form fields. For a specific
case of two 3-form fields, we derived a relation between the derivatives
of $N$ with respect to unperturbed values of scalar field duals at
horizon exit $c_{s}k=aH$ and the $N$ derivatives with respect to
3-form fields. We employed a numerical finite difference approach
for this purpose. We computed the bispectrum at horizon exit for the
two 3-form field case using known expressions for 3-point
field space correlations for a general multiscalar field model. Then
using the $N$ derivatives we determined the complete superhorizon
evolution of $f_{{\rm NL}}$ for squeezed, equilateral and orthogonal
configurations until the end of inflation. Considering the
potentials $\chi_{I}^{2}+b_{I}\chi_{I}^{4}$ and specific values of model parameters that were consistent with $n_{s}\sim0.967$ and $r\sim0.0422$, we obtained the corresponding $f_{{\rm NL}}$ predictions for the
two 3-form inflationary model as $f_{{\rm NL}}^{{\rm sq}}\sim-2.6\times10^{-3},\,f_{{\rm NL}}^{{\rm equi}}\sim1.409,\,f_{{\rm NL}}^{{\rm ortho}}\sim0.495$. Therefore, the model is well within the observational bounds of {\it Planck} 2015 data and, most important to emphasize, it can be tested with the future probes \cite{Errard:2015cxa,Creminelli:2015oda,Huang:2015gca}.
\chapter{DBI Galileon inflation}
\label{DBIGc}

\begin{chapquote}{Edward Witten}
If I take the theory as we have it now, literally, I would conclude that extra dimensions really exist. They're part of nature. We don't really know how big they are yet, but we hope to explore that in various ways.
\end{chapquote}

\lhead{\bf Chapter 3. \emph{DBI Galileon inflation}} % This is for the header on each page - perhaps a shortened title

In this chapter we explore an observationally consistent inflationary scenario that involves a D-brane setting with an additional effect of induced gravity.  
In Refs.~\cite{deRham:2010eu,Goon:2010xh,Goon:2011qf}, it was observed that the motion of a D-brane in warped space generally causes an effect of induced gravity. This resultant action of D-brane with induced gravity effect comes under a class of generalized Galileon model \cite{Kobayashi:2011nu}. Therefore, this new setting is named as DBI Galileon (DBIG) model. The studies so far in literature
\cite{Mizuno:2010ag,RenauxPetel:2011dv,
Gao:2011qe,RenauxPetel:2011uk,Choudhury:2012yh,Koyama:2013wma,Renaux-Petel:2013ppa,Andre:2013afa},
are mainly focused to explore the parameter space of the single-field and multifield
DBIG model with respect to the various types of non-Gaussianities. 
Furthermore, in Ref.~\cite{Choudhury:2012yh} single field DBIG inflation is studied in the background
of SUGRA under the assumption of a Coleman-Weinberg type of
potential. In this chapter, we propose to study single-field DBIG inflation
without any particular choice of potential. More precisely, our objective is
to constrain the parameter space of the DBIG model with respect to the inflationary observables of primordial power spectrum in accordance with
latest {\it Planck} 2015 data. We mainly focus our attention in two inflationary
regimes. Namely, those with and without a constant warp factor. We
aim to identify crucial differences between these two scenarios with
respect to the corresponding inflationary predictions. In addition,
in each case, we analyze the deviation from the standard slow-roll
consistency relation $r=-8n_{t}$ due to the effect of induced gravity
on the D-brane. 

The organization of this chapter is as follows. In Sec.~\ref{BGS} we briefly describe the model and present the background equations for the DBIG inflation with non-trivial warping \cite{RenauxPetel:2011uk}. In the case of constant sound speed and warp factor, we obtain the exact background solutions. In Sec.~\ref{comp} we study the parameter space of the DBIG model by comparing its predictions in different limits with CMB data. In Sec.~\ref{varysols} we present general background solutions using two different ansatz to integrate analytically the equations of motion. A detailed computation of the approximate solutions can be found in Appendix \ref{ADBIG1}. Finally, we present our conclusions in Sec.~\ref{conclns}.

\section{DBI-Galileon inflationary model}\label{BGS}
We begin by reviewing the DBIG inflationary scenario
following Ref.~\cite{RenauxPetel:2011uk}. Such a setup
considers a D3-brane with tension $T_{3}$ evolving in a ten dimensional
geometry described by the metric, 
\begin{equation}
ds^{2}=h^{-1/2}\left(y^{K}\right)g_{\mu\nu}dx^{\mu}dx^{\nu}+h^{1/2}\left(y^{K}\right)G_{IJ}\left(y^{K}\right)dy^{I}dy^{J}\equiv H_{AB}dY^{A}dY^{B}\,,\label{10Dmetric}
\end{equation}
with coordinates $Y^{A}=\left\{ x^{\mu},y^{I}\right\} $, where $\mu=0,....3$
and $I=1,....,6$. The induced metric on the D3-brane is given by
\begin{equation}
\gamma_{\mu\nu}=H_{AB}\partial_{\mu}Y_{\left(b\right)}^{A}\partial_{\nu}Y_{\left(b\right)}^{B}\,,\label{indmetric10D}
\end{equation}
where the brane is embedded in higher dimensions by means of the functions
$Y_{\left(b\right)}^{A}\left(x^{\mu}\right)$, with the $x^{\mu}$
being the space time coordinates on the brane. In brane inflation,
the role of the inflaton is played by the radial coordinate $\left(\rho\right)$
of the brane that is moving in the extra dimensions. Since we are
only considering single-field inflation in this chapter, we choose the
brane embedding as $Y_{\left(b\right)}^{A}\left(x^{\mu}\right)=\left(x^{\mu},\varphi\left(x^{\mu}\right)\right)$.
Then, the induced metric can be written as 
\begin{equation}
\gamma_{\mu\nu}=f^{-1/2}\left(g_{\mu\nu}+f\partial_{\mu}\varphi\partial_{\nu}\varphi\right)\,,\label{indmetric5D}
\end{equation}
where $f$ and $\varphi$ are the warp factor and the scalar field
defined by 
\begin{equation}
f=\frac{h}{T_{3}}\,,\quad\varphi=\sqrt{T_{3}}\rho\,.\label{rescale}
\end{equation}
The D3-brane here is embedded in 5D geometry with the induced metric
(\ref{indmetric5D}). This introduces an additional contribution
in the action known as Galileon term \cite{Chow:2009fm}. The total action
is then given by 
\begin{equation}
S=\int d^{4}x\left[\frac{m_P^{2}}{2}\sqrt{-g}R\left[g\right]+\frac{\tilde{m}^{2}}{2}\sqrt{-\gamma}R\left[\gamma\right]+\sqrt{-g}\mathcal{L}_{brane}\right]\,,\label{actionDBIG}
\end{equation}
where $\tilde{m}$ is a parameter associated with the induced
gravity%
\footnote{$\tilde{m}$ non trivially depends on the warping $h$, see \cite{RenauxPetel:2011uk}.
In this chapter, $\tilde{m}$ is treated as a model parameter.%
} and 
\begin{equation}
\mathcal{L}_{brane}=-\frac{1}{f\left(\varphi\right)}\left(\sqrt{\mathcal{D}}-1\right)-V\left(\varphi\right)\,,\label{braneDBI}
\end{equation}
where 
\[
\mathcal{D}\equiv\textrm{det}\left(\delta_{\nu}^{\mu}+f\partial_{\mu}\varphi\partial_{\nu}\varphi\right)\,.
\]
DBIG action belongs to the particular class of generalized G-inflation {\ref{G-action}}  \cite{RenauxPetel:2011uk} with
the functions $\mathcal{F}_{s}$ and $\mathcal{G}_{s}$ that determine
the second order action for scalar perturbations are 
\begin{equation}
\begin{array}{rcl}
\mathcal{F}_{s}
(c_{\mathcal{D}},\epsilon_{\mathcal{D}},
\epsilon) &=&m_{\rm P}^{2}\left(\epsilon{\cal K}(3{\cal K}-2)+{\cal K}-1\right)+{\displaystyle \frac{\tilde{m}^{2}}{c_{\mathcal{D}}}\left[(\epsilon+\epsilon_{\mathcal{D}}){\cal K}\left(\frac{3{\cal K}}{c_{\mathcal{D}}^{2}}-2\right)+{\cal K}-c_{\mathcal{D}}^{2}\right]}\,,\\
\\
\mathcal{G}_{s}(c_{\mathcal{D}},\epsilon_{\mathcal{D}},\epsilon) &=&{\displaystyle \frac{m_{\rm P}^{2}}{c_{\mathcal{D}}^{2}}\left(\epsilon{\cal K}^{2}+3c_{\mathcal{D}}^{2}(1-{\cal K}^{2})\right)+{\displaystyle \frac{\tilde{m}^{2}}{c_{\mathcal{D}}^{3}}\left[(\epsilon+\epsilon_{\mathcal{D}}){\cal K}^{2}+3c_{\mathcal{D}}^{2}\left(1-\frac{{\cal K}^{2}}{c_{\mathcal{D}}^{4}}\right)\right]\,,}}
\end{array}\label{FsGs}
\end{equation}
where ${\cal K}\equiv\frac{m_P^{2}+c_{\mathcal{D}}^{-1}\tilde{m}^{2}}{m_P^{2}+c_{\mathcal{D}}^{-3}\tilde{m}^{2}}$. And the functions corresponds to tensor perturbations are
\begin{equation}
\mathcal{F}_{t}(c_{\mathcal{D}})\equiv m_{\rm P}^{2}+\tilde{m}^{2}c_{\mathcal{D}}\quad,\quad\mathcal{G}_{t}(c_{\mathcal{D}})\equiv m_{\rm P}^{2}+\frac{\tilde{m}^{2}}{c_{\mathcal{D}}}\,.\label{Fgt}
\end{equation}

Assuming the flat FLRW metric
%\begin{equation}
%ds^{2}=-dt{}^{2}+a^{2}(t)d\boldsymbol{x}^{2}\,.\label{FRW-metric1-1}
%\end{equation}
and allowing the warp factor $f$ to vary, the gravitational field
equations for the action in (\ref{actionDBIG}) are \cite{RenauxPetel:2011uk}
\begin{equation}\label{Friedmann}
3H^{2}m_{\rm P}^{2}+3\widehat{H}^{2}\frac{\tilde{m}^{2}}{c_{\cd}^{3}}=\frac{1}{f}\left(\frac{1}{c_{\cd}}-1\right)+V\,.
\end{equation}
\begin{equation}\label{Raychaudhuri}
-m_{\rm P}^{2}\dot{H}+\frac{\tilde{m}^{2}H^{2}}{c_{\cd}}\left[-\frac{\dot{\widehat{H}}}{H^{2}}-\frac{c_{\cd}}{h^{1/4}}\left(\frac{h^{1/4}}{c_{\cd}}\right)^{\cdot}\frac{\widehat{H}}{H^{2}}+\frac{3}{2}\left(\frac{1}{c_{\cd}^{2}}-1\right)\frac{\widehat{H}^{2}}{H^{2}}\right]=\frac{\dot{\sigma}^{2}}{2c_{\cd}}\,,
\end{equation}
where $c_{\cd}^{2}\equiv1-f\dot{\sigma}^{2}$ is the squared sound
speed%
\footnote{Note that the sound speed $c_{\cd}$ here depends not only on the
brane dynamics, (as in DBI models \cite{Chen:2005fe,Shandera:2006ax,Baumann:2006cd})
but also on the induced gravity \cite{RenauxPetel:2011uk}.}, $\widehat{H}\equiv H-\frac{\dot{f}}{4f}$ and ${\dot{\sigma}}^{2}\equiv G_{IJ}\dot{\phi}^{I}\dot{\phi}^{J}$.
The appearance of (\ref{Raychaudhuri}) can be simplified to
\begin{equation}\label{BGS-1}
\dot{H}-\lambda_{1}H^{2}+\lambda_{2}=0
\end{equation}
after introducing the functions 
\begin{eqnarray}
\lambda_{1}&\equiv & \frac{\tilde{m}^{2}}{m_{\rm P}^{2}c_{\cd}+\tilde{m}^{2}}\left[%\frac{1}{4H}\frac{d(\dot{f}/f)}{d\ln a}
\frac{\epsilon_f(\eta_f-\epsilon)}4-\frac{d\ln\left(\frac{h^{1/4}}{c_{\cd}}\right)}{d\ln a}\left(1-\frac{\epsilon_{f}}{4}\right)+\frac{3}{2}\left(\frac{1}{c_{\cd}^{2}}-1\right)
\left(1-\frac{\epsilon_{f}}{4}\right)^{2}\right]\!\!,\quad\label{lamda1}\\ 
\lambda_{2}&\equiv & \frac{1-c_{\cd}^{2}}{2f\left(m_{\rm P}^{2}c_{\cd}+\tilde{m}^{2}\right)}\,
,\label{lambda2}
\end{eqnarray}
which depend on $\tilde{m}$ and $c_{\cd}$. We also introduce the slow-roll parameters 
\begin{equation}
\epsilon\equiv-\frac{\dot{H}}{H^{2}}\,\,,\,\,\eta\equiv\frac{d\ln\epsilon}{d\ln a}\,\,,\,\,\epsilon_{\cd}\equiv\frac{d\ln c_{\cd}}{d\ln a}\,\,,\,\,\eta_{\cd}\equiv\frac{d\ln\epsilon_{\cd}}{d\ln a}\,\,,\,\,\epsilon_{f}\equiv\frac{d\ln f}{d\ln a}\,\,,\,\,\eta_{f}\equiv\frac{d\ln\epsilon_{f}}{d\ln a}\label{slwrolldef}
\end{equation}
to describe the evolution of the background geometry, the sound speed
and the warp factor. Note also that in the above we take the brane
tension $T_{3}$ to be a constant, as is usually considered.

In the following we obtain solutions to the background equations for the cases when $\lambda_{1,2}$ are constants.

\subsection{Constant sound speed and warp factor}\label{constcase}
Whenever the sound speed $\left(c_\cd\leq1\right)$ and the warp factor is constant, i.e., $\epsilon_\cd=\epsilon_f=0$, the coefficients $\lambda_{1,2}$ in (\ref{BGS-1}) are constants. Integrating (\ref{BGS-1}) in that case is straightforward. We obtain 
\begin{equation}
H^{2}=\frac{\lambda_{2}}{\lambda_{1}}+\kappa a^{2\lambda_{1}}\,,\label{Hconscase}
\end{equation}
where $\kappa\neq0$ is an arbitrary, dimensionful constant. Writing $H=\dot{a}/a$, the solution to (\ref{Hconscase}) is 
\begin{equation}
a^{2\lambda_{1}}(t)=\left(\frac{\lambda_{2}}{\lambda_{1}|\kappa|}\right)\,\exp\left[i\left(1+\sigma_{1}\right)\pi/2\right]\,{\rm sech}^{2}\left[\sqrt{\lambda_{1}\lambda_{2}}\sigma_{2}\left(t-\overline{t}\right)-i\left(1+\sigma_{1}\right)\pi/4\right]\,,\label{scalefconscase}
\end{equation}
where we introduce 
\begin{equation}
\sigma_{1}\equiv{\rm sign}(\kappa)={\rm sign}(\dot{H})\quad,\quad\sigma_{2}\equiv{\rm sign}(\dot{a})\,.\label{sigma12}
\end{equation}
The explicit time-dependence of the Hubble parameter can be obtained
from (\ref{scalefconscase}) 
\begin{equation}
H(t)=-\left(\frac{\lambda_{2}}{\lambda_{1}}\right)^{1/2}\sigma_{2}\tanh\left[\sqrt{\lambda_{1}\lambda_{2}}\sigma_{2}\left(t-\overline{t}\right)
-i\left(1+\sigma_{1}\right)\pi/4\right]\,.\label{Hsolconstcase}
\end{equation}
To study inflation we need to set $\sigma_{2}={\rm sign}\left(\dot{a}\right)=+1$,
regardless of $\sigma_{1}={\rm sign}(\dot{H})$. An increasing expansion
rate is obtained for $\sigma_{1}=+1$ ($\lambda_{2}<\lambda_{1}H^{2}$),
which corresponds to the singular behaviour of the scale factor and
the Hubble parameter at $t\to\bar{t}$ (purple line) displayed in
Fig.~\ref{fig1}. A decreasing expansion rate corresponds to $\sigma_{1}=-1$
($\lambda_{2}>\lambda_{1}H^{2}$), in which case both $a(t)$ and
$H(t)$ remain finite throughout the entire evolution (blue line).
In the context of inflation, we focus only on the decreasing expansion
rate $\sigma_{1}=-1$, for which we find a non-singular behaviour
for the scale factor $a(t)$. 
\begin{figure}[htbp]
\includegraphics[width=15cm]{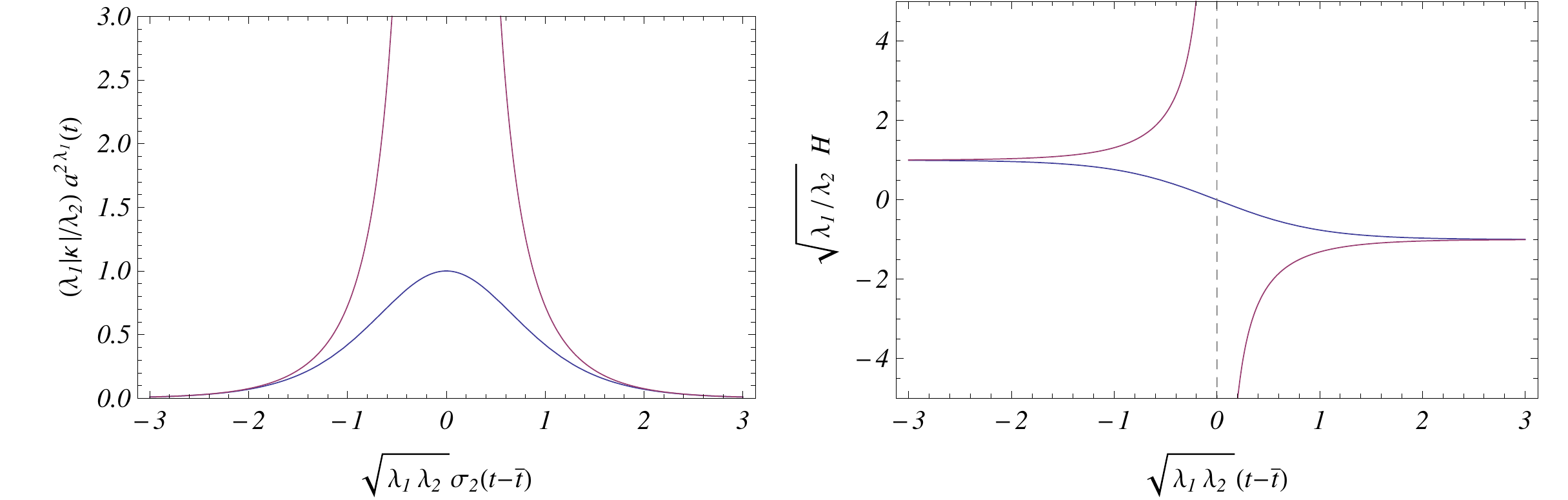}
\caption{Evolution of the scale factor according to (\ref{scalefconscase}) (left panel) and the Hubble parameter $H$, according to (\ref{Hsolconstcase})
(right panel).}
\label{fig1} 
\end{figure}

In Sec.~\ref{constcasepert} we impose the necessary conditions to obtain
an inflationary expansion in agreement with current observations.
To do so, in the next section we investigate the scalar and tensor
perturbation spectra, which depend on the slow-roll parameters $\epsilon$
and $\eta$. Using (\ref{slwrolldef}) and (\ref{Hsolconstcase})
we obtain 
\begin{eqnarray}
\epsilon(t) & = & \lambda_{1}{\rm csch}^{2}\left[\sqrt{\lambda_{1}\lambda_{2}}\sigma_{2}(t-\overline{t})-
i(1+\sigma_{1})\pi/4\right]\,,\label{epsconstcase}\\
\eta(t) & = & 2\lambda_{1}{\rm coth}^{2}\left[\sqrt{\lambda_{1}\lambda_{2}}\sigma_{2}(t-\overline{t})-i(1+\sigma_{1})\pi/4\right]\,,\label{etaconstcase}
\end{eqnarray}
from which we arrive at the relations
\begin{equation}\label{relations-eps-eta}
\eta=2\left(\epsilon+\lambda_1\right)\quad,\quad H^2=\lambda_2\left(\lambda_1+\epsilon\right)^{-1}\,,
\end{equation}
where we emphasize that the slow-roll parameter $\eta$ explicitly depends on $\lambda_{1}$. During inflation, $\eta\ll1$ implies
$\lambda_1\ll1$. Therefore, several constraints (to be discussed later on) must be imposed on the model parameters to have $\lambda_{1}\ll1$.

\section{Comparison to observations}
\label{comp}

In this section we study in detail the observational predictions of DBIG inflation and examine the status of the tensor consistency relation. We compute $n_s,\,r$ and $n_t$ by plugging (\ref{FsGs}),\,(\ref{Fgt}) in the general expressions presented in Appendix.~\ref{Intro-app}. We study the different limits of DBIG inflation and evaluate the effect of higher order corrections in
slow-roll parameters on the model predictions.

We explore the parameter space $\left(c_{\mathcal{D}}\,,\:\tilde{m},\: f\right)$ of DBIG inflation using the {\it Planck} constraints on $\left(n_{s},\: r\right)$ and the observed amplitude of the power spectrum $\mathcal{P}_{\zeta_{\ast}}\simeq2.2\times10^{-9}$ at the pivot scale $k_{\ast}=0.002\textrm{ Mpc}^{-1}$ \cite{Ade:2015lrj}. In all cases, we find that the predictions of $\left(n_{s},\: r\right)$ do not explicitly depend on the warp factor. Therefore, we first find the range of model parameters $\left(c_{\mathcal{D}},\:\tilde{m}\right)$
compatible with the observed values of $n_{s}=0.968\pm0.006$ and $r<0.1$ at the $95\%$ CL \cite{Ade:2015lrj}. After that, we calculate the tensor tilt $\left(n_{t}\right)$ for the same parameter space that was previously constrained. We expect to
find departures from the consistency relation of single-field inflation, $r=-8n_{t}$. Finally, we compare our results with the BKP+LIGO constraints on the tensor tilt $n_{t}=-0.76_{-0.52}^{+1.37}$ at the $68\%$ CL \cite{Ade:2015lrj,Huang:2015gka}.

\subsection{Constant sound speed and warp factor}
\label{constcasepert}

Let us examine the parameter space of DBIG inflation with $\epsilon_\cd=\epsilon_f=0$ in different limits.
For this we use the solutions derived in Sec.~\ref{constcase}. We focus only on the decreasing expansion
rate $\sigma_1=-1$, for which we find a non-singular behaviour for the scale factor $a(t)$.

Firstly, the number of $e$-foldings during inflation can be computed as 
\begin{equation}
N=\int_{t_{\ast}}^{t_{e}}H\, dt\,,\label{efolds}
\end{equation}
where $t_{\ast}$ is the time when cosmological scales exit the horizon and $t_{e}$ signals the end of inflation, set through the condition
$\epsilon(t_{e})=1$. According to observations, the length of the inflationary phase required to solve the flatness and horizon problems
is around $N=50$ to $N=60$. Using (\ref{Hsolconstcase}) and the condition $\epsilon=1$ to determine $t_{e}$, we integrate
(\ref{efolds}) to obtain
\begin{equation}
N=\frac{1}{\lambda_{1}}\ln\frac{{\rm cosh}\left[\sqrt{\lambda_{1}\lambda_{2}}\sigma_{2}(t_{\ast}-\overline{t})
-i(1+\sigma_{1})\pi/4\right]}{\sqrt{1+\lambda_{1}}}\,,\label{Nstar}
\end{equation}
which we can relate to the slow-roll parameters $\epsilon$ and $\eta=2\left(\epsilon+\lambda_{1}\right)$ at the time of horizon crossing 
\[
\epsilon_{\ast}=\frac{\lambda_{1}}{\left(1+\lambda_{1}\right)\exp[2\lambda_{1}]-1}\,.
\]
Using (\ref{Friedmann}) and (\ref{relations-eps-eta}), we find the scalar potential $V$ in terms of the model parameters
\begin{equation}\label{potentialpdbi}
V=\frac{3\lambda_2}{\lambda_1+\epsilon}\left[m_{\rm P}^{2}+\frac{\tilde{m}^{2}}{c_\cd^3}\right]-\frac{1}{f}\left(\frac{1}{c_{D}}-1\right)\,,
\end{equation}
which allows us to find the energy scale of inflation $V_{\ast}^{1/4}$ after evaluating at the time of horizon crossing for cosmological scales. Also, we obtain the mass squared of the inflaton
\begin{equation}
m_{\phi}^{2}=V_{,\phi\phi}=\frac{\ddot{V}}{\dot{\phi}^{2}}\label{mass2phi}\,,
\end{equation}
where 
\begin{equation}
\dot{\phi}^2=\frac{1-c_{\mathcal{D}}^{2}}{f}\label{phiddot}\,.
\end{equation}

\subsubsection{DBI limit: $\tilde{m}\rightarrow0$} \label{dbilimit}
The phenomenology of DBI inflation has been done in recent literature \cite{Weller:2011ey,Li:2013cem} assuming a particular
form of potential. We emphasize here, however, that in our study we do not assume any form of the potential. % The parameter space for this case is two dimensional $\left(c_{\mathcal{D}},f\right)$.

\begin{figure}[htbp]
\centering\includegraphics[width=7cm]{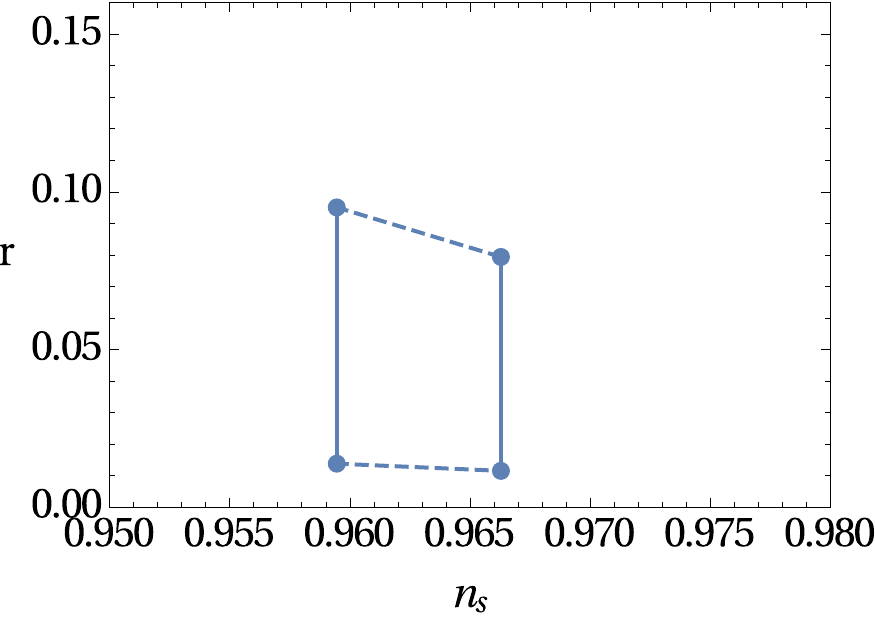}\hspace{0.25cm}\centering
\includegraphics[width=7cm]{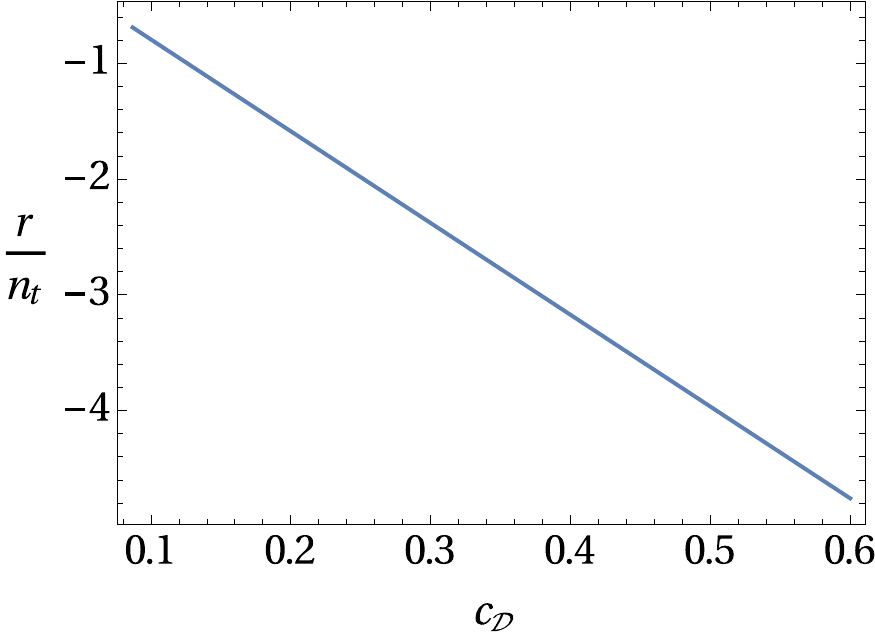}
\caption{In the left panel we depict tensor-to-scalar ratio vs. spectral index where in the plot $N$ varies from 50 to 60 (from left to right) and $c_\cd$ varies from $0.087$ to $0.6$ (from bottom to top). In the right panel we plot the ratio $r/n_t$ vs. sound speed $c_\cd$ for $N=60$.}\label{fig1-1} 
\end{figure}

In this limit $\lambda_{1}\to0$ (see (\ref{lamda1})), and we obtain the corresponding background solution from the one obtained
in Sec.~\ref{constcase} as the zeroth order in a series expansion around $\lambda_{1}=0$. Operating similarly for the number of $e$-foldings in (\ref{Nstar}) we easily obtain 
\begin{equation}
\lambda_{2}\to\frac{1-c_{\cd}^{2}}{2fm_{\rm P}^{2}c_{\cd}}\quad,\quad H^{2}\to\frac{\lambda_{2}}{\epsilon}\quad,\quad\epsilon\to\frac{1}{1+2N}\quad,\quad\eta\to2\epsilon\quad,\quad{\cal P}_{\zeta}\to\frac{H^{2}}{8\pi^{2}\epsilon c_{\cd}}\,.\label{DBIlargeN}
\end{equation}
Fixing the number of $e$-foldings and the amplitude of the perturbation
spectrum we constrain the warp factor $f$. Since we treat $c_{\cd}$
as a model parameter, we obtain its range from the prediction for
non-Gaussianity $f_{{\rm NL}}^{eq}=-\frac{35}{108}\left(\frac{1}{c_{\mathcal{D}}^{2}}-1\right)$
in DBI models \cite{Silverstein:2003hf,Alishahiha:2004eh}. Although
more accurate expressions exist in the literature \cite{Mizuno:2010ag,
RenauxPetel:2011dv,Gao:2011qe,
RenauxPetel:2011uk,Choudhury:2015yna,
Koyama:2013wma,Renaux-Petel:2013ppa},
for our purposes it suffices to consider this simple estimate. This
is appropriate since in the absence of a clear detection of non-Gaussianity
\cite{Ade:2015ava}, the use of more elaborate or complicated expressions
is, in principle, uncalled for. Therefore, in this chapter we will not
be concerned with non-Gaussian computations and will use the above
expression to constrain the sound speed $c_{\mathcal{D}}$. The analysis
of the {\it Planck} data on $r<0.1$ and $\fnl^{equi}=-4\pm43$ allows to
set a conservative bound for this $0.087\leq c_{\cd}\leq0.6$ \cite{Ade:2015lrj,Ade:2015ava}.
Note that larger values of $c_{\cd}$, albeit allowed by the bound
from non-Gaussianity, are disfavoured as they result in a tensor-to-scalar
ratio in excess of the current bound $r<0.1$. Fig.~\ref{fig1-1}
represents the viability of the DBI model. Because of the stringent
bound on $\fnl^{equi}$ the DBI inflation is not capable to induce
$r<0.01$ which is consistent with previous studies \cite{Baumann:2006cd,Peiris:2007gz}.
The range of model parameters obtained for $0.087\leq c_{\cd}\leq0.6$
can be found in Table~\ref{summary}. In Fig.~\ref{fig1-1} we depict
our results in the DBI limit. 

\subsubsection{Galileon limit: $\tilde{m}\gg m_{\rm P}$}\label{glimit-1}
Although studying this limit is not generic with
respect to the structure of DBIG, this would nevertheless be useful
to understand the role of induced gravity. Since $c_{\cd}\lesssim1$,
(\ref{BGS-1}) gives 
\begin{equation}
\lambda_{1}=\frac{3}{2}\left(\frac{1}{c_{\mathcal{D}}^{2}}-1\right)\quad,\quad\lambda_{2}\equiv\frac{1-c_{\cd}^{2}}{2c_{\cd}f\tilde{m}^{2}}\,.\label{lambdasdbi-1}
\end{equation}
The slow-roll parameters in this case which are given below 
\begin{equation}\label{swrll-G-inf}
\epsilon=\frac{3\left(1-c_\cd^2\right)}{\left(3-c_\cd^2\right)e^{\frac{3}{2}\left(\frac{1}{c_{\mathcal{D}}^{2}}-1\right)N}-2c_\cd^2}\quad,\quad
\eta=3\left(\frac{1}{c_{\mathcal{D}}^{2}}-1\right)+2\epsilon\,.
\end{equation}

\begin{figure}[htbp]
\centering\includegraphics[width=6.5cm]{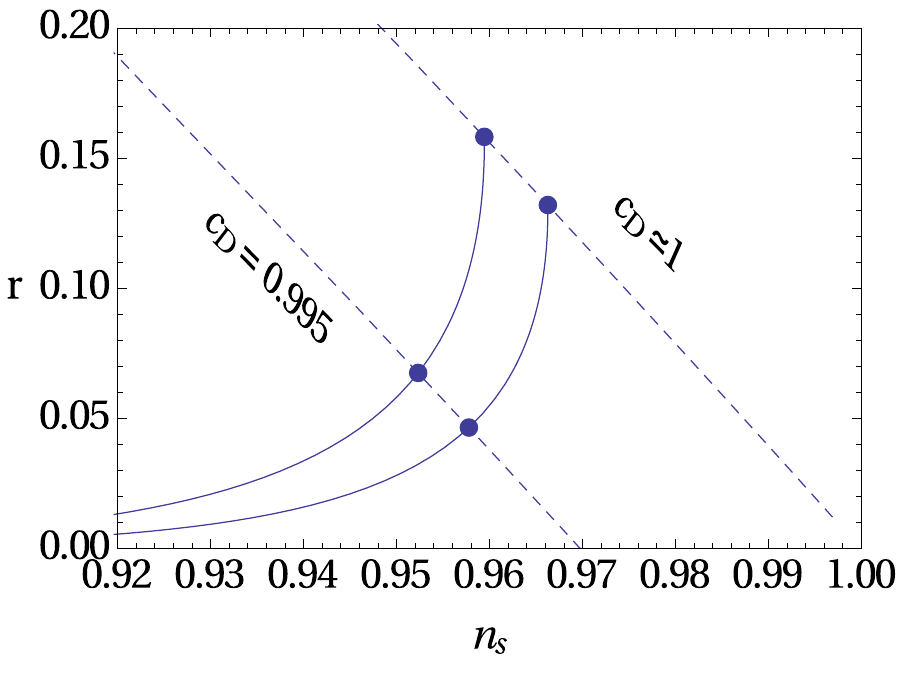}\hspace{0.5cm} \centering
\includegraphics[width=7.0cm]{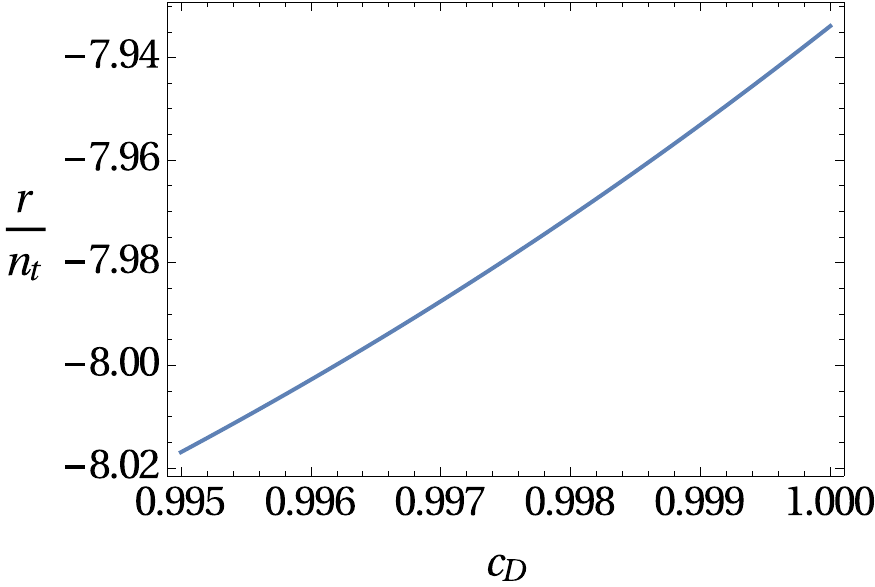}
\caption{Plots of spectral index $n_{s}$ vs. tensor-to-scalar ratio $r$ (left)
and the ratio $r/n_t$ vs. sound speed $c_\cd$ (right)
in the Galileon limit. In the left panel, we take $N$ varying
from 50 to 60 (from bottom to top). For the right panel we considered $ N=60 $.}\label{glimit} 
\end{figure}

Unlike in the DBI limit (cf.~(\ref{DBIlargeN})), in the Galileon
limit, the slow-roll parameters explicitly depend on the sound speed.
It is obvious from (\ref{swrll-G-inf}) that $c_{\mathcal{D}}\ll1$
would actually spoil the smallness of $\eta$. Therefore, in this
case we need to keep the sound speed in the narrow range $0.995\leq c_{\cd}<1$
for the results to agree with the current {\it Planck} data. Any value of
$c_{\mathcal{D}}<0.995$ would essentially spoil the prediction of
the spectral index and its value would be significantly
out of the current bounds $n_s=0.968\pm0.006$. Therefore, observationally viable inflation due to the induced gravity
term sets $c_{\cd}\lesssim1$, thus resulting in small non-Gaussianities.
This allows to discriminate between the current case and the DBI limit
previously studied. Also, the consistency of the predictions with
data becomes better as the number of $e$-foldings reduces. In particular,
for $N\sim50$ our results are perfectly consistent with current
data whereas for $N\gtrsim60$ the model is ruled out. Our results
in this case are depicted in Fig.~\ref{glimit}. The derived model
parameters for $0.995\leq c_{\cd}\leq1$ can be found in Table~\ref{summary}. 

\subsubsection{DBI-Galileon case}
\label{intercomp}

In this section we consider Einstein and Galileon gravity are on an equal footing. In this case 
\[
\lambda_{1}\equiv\frac{\tilde{m}^{2}}{m_{\rm P}^{2}c_{\cd}+\tilde{m}^{2}}\left[\frac{3}{2}\left(\frac{1}{c_{\cd}^{2}}-1\right)\right]\quad,\quad\lambda_{2}
\equiv\frac{1-c_{\cd}^{2}}{2f\left(m_{\rm P}^{2}c_{\cd}+\tilde{m}^{2}\right)}\,.
\]
The corresponding slow-roll parameters are (expressing in the units of $ m_{\rm P}=1 $)
\begin{equation}\label{eps-DBIG}
\epsilon=\frac{3\left(1-c_{\mathcal{D}}^{2}\right)\tilde{m}^{2}}{\left[2c_{\mathcal{D}}^{3}-\left(c_{\mathcal{D}}^{2}-3\right)\tilde{m}^{2}\right]e^{\frac{3\left(\frac{1}{c_{\mathcal{D}}^{2}}-1\right)\tilde{m}^{2}N}{c_{\mathcal{D}}+\tilde{m}^{2}}}-2c_{\mathcal{D}}^{2}\left(c_{\mathcal{D}}+\tilde{m}^{2}\right)}\,\,,\,\,\eta= \frac{3\left(\frac{1}{c_{\mathcal{D}}^{2}}-1\right)\tilde{m}^{2}}{\left(c_{\mathcal{D}}+\tilde{m}^{2}\right)}+2\epsilon\,.
\end{equation}

\begin{figure}[htbp]
\centering\includegraphics[width=9cm]{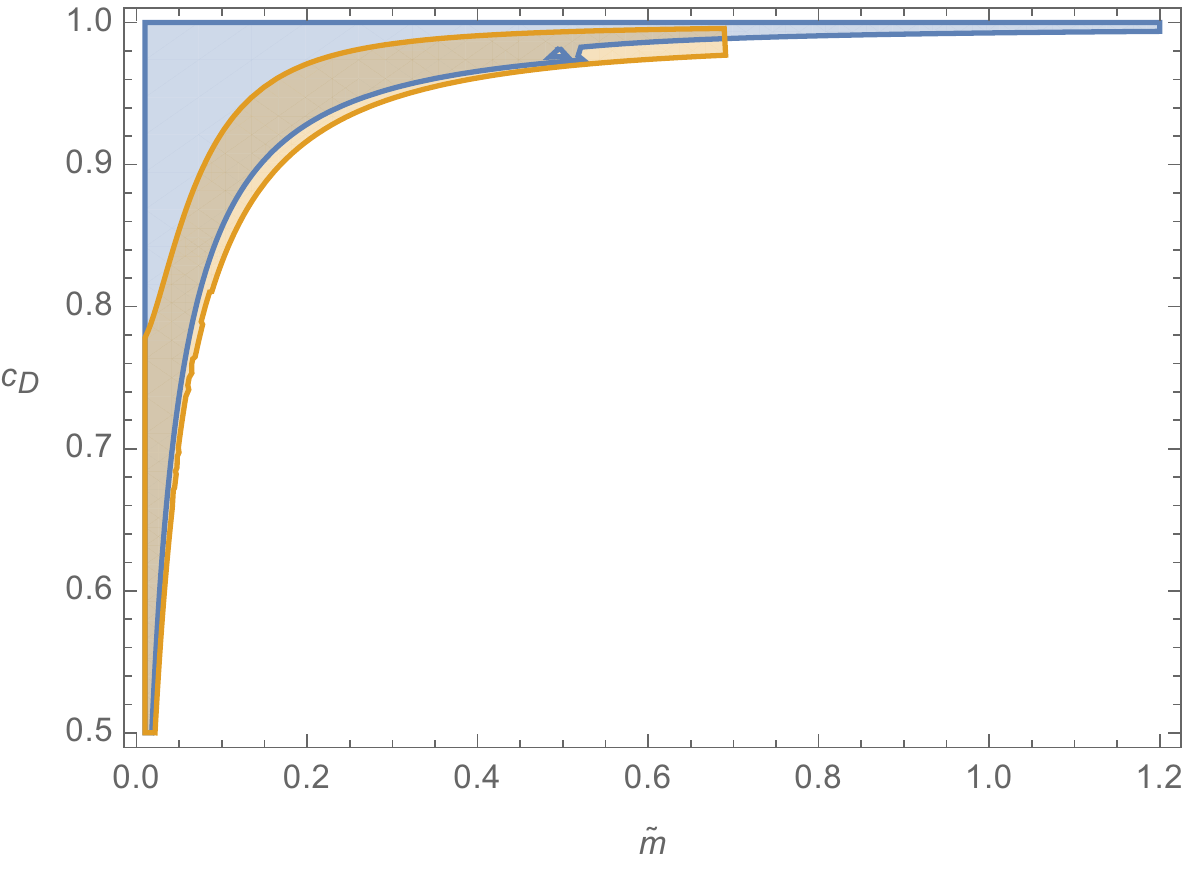}\caption{Contour plots in the plane $(\tilde m,c_\cd)$ (with $\tilde m$ in units of $m_P$). Blue and orange regions represent the space where $n_s=0.968\pm0.006$ and $0.01\leq r\leq0.1$, respectively.}\label{PspaceDBIG} 
\end{figure}

Similarly to the Galileon limit studied in Sec.~\ref{glimit-1}, the sound speed needs to be tuned to $c_{\cd}\simeq0.98-0.99$ to keep 
the slow-roll parameter $\eta$ small enough to have $n_s=0.968\pm0.006$. We find that $c_{\cd}<0.98$ would essentially spoil the prediction of scalar tilt. We also note here that if $c_{\mathcal{D}}=1$ we obtain exact scale invariance, i.e. $n_s=1$. Since the slow-roll parameter $\epsilon$ in (\ref{eps-DBIG}) depends on the parameter $\tilde{m}$, the tensor-to-scalar ratio varies for different values of the induced gravity parameter $\tilde{m}$. This allows us to identify the range of the parameters consistent with current data. In Fig.~\ref{PspaceDBIG} we study the parameter space $\left(c_{\mathcal{D}}\,,\,\tilde{m}\right)$ using the bounds on $\left(n_{s}\,,\, r\right)$. The plot shows that, in the limit $\tilde{m}\to0$, the model reduces to DBI case. Moreover, unless $\tilde m<m_P$, the effect of the induced gravity forces us to constrain the sound speed to $c_{\cd}\sim1$ in order to maintain the agreement with observations.

To constrain the model parameters $\left(c_{\mathcal{D}}\,,\,\tilde{m}\right)$
with the bounds of $\left(n_{s}\,,\, r\right)$ it is also necessary
to check if non-Gaussianities are large. Since the full study of
non-Gaussianity is beyond the scope of this chapter, we use the results
in Ref.~\cite{RenauxPetel:2011uk}, where the authors study non-Gaussianity
in the multifield DBIG inflation model. We adopt their expression for $\fnl^{equi}$
in the single-field limit, i.e. taking the adiabatic and isocurvature
mode transfer function $T_{\sigma s}\rightarrow0$. We thus constrain
our parameter space using the approximate expression \cite{RenauxPetel:2011uk}
\begin{equation}\label{fnleinter}
\fnl^{equi}=-\frac{5}{324c_{\cd}^{2}}\frac{21-404\alpha+2233\alpha^{2}-3066\alpha^{3}}{(1-5\alpha)^{2}(1-9\alpha)}\,\,,\,\,\alpha\equiv\frac{fH^{2}\tilde{m}^{2}}{c_{\mathcal{D}}^{2}}\,.
\end{equation}

\begin{figure}[htbp]
\centering\includegraphics[width=7.0cm]{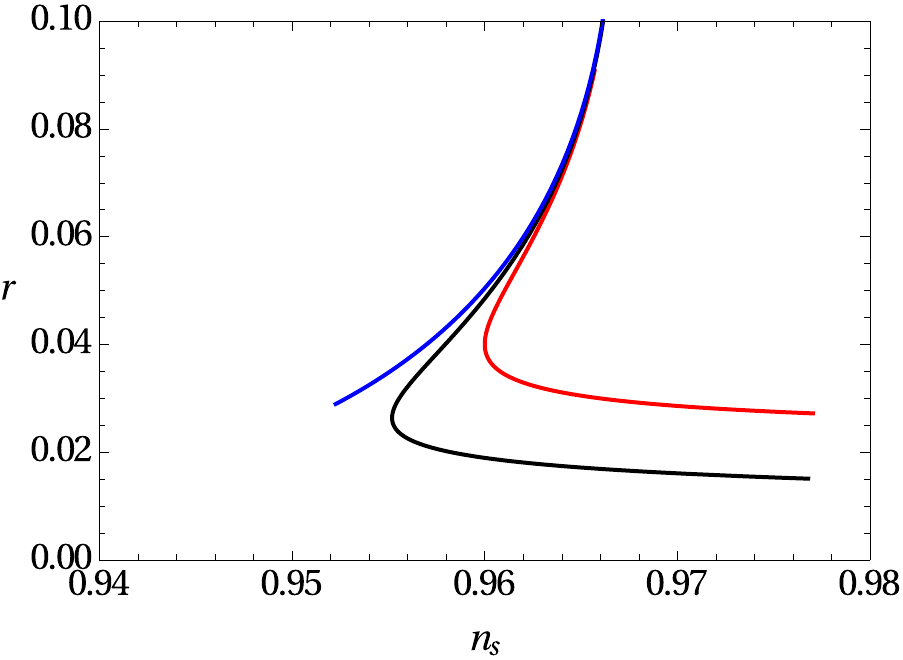}\hspace{0.5cm} \includegraphics[width=7.0cm]{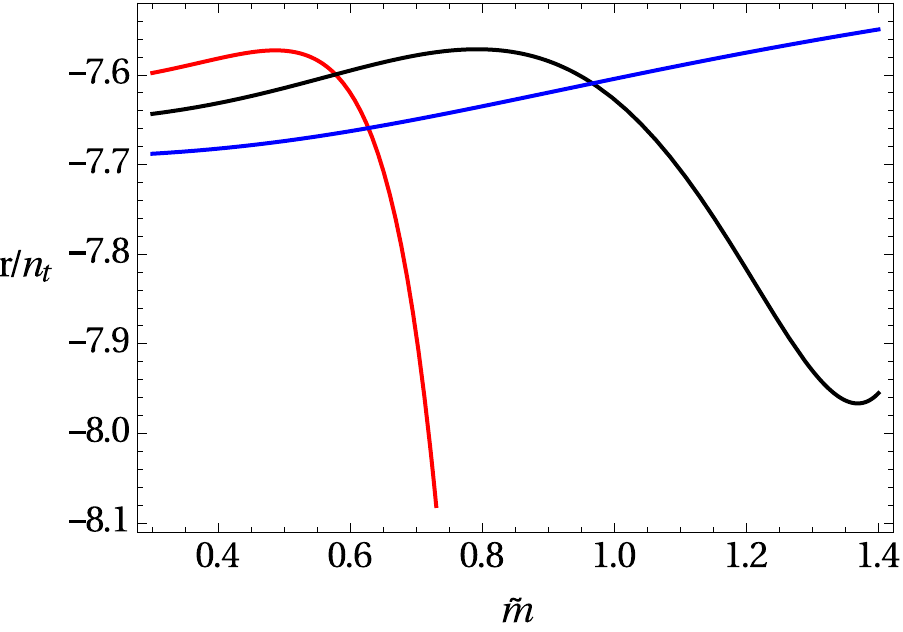}
\caption{Plots of spectral index $n_s$ vs. tensor-to-scalar ratio $r$ (left
panel) and the ratio $r/n_t$ vs. $\tilde{m}$ (with $\tilde m$ in units of $m_P$) (right panel) in the DBIG model. In the left panel we take $c_\cd=0.98$ and $0.3\leq\tilde{m}/m_P\leq0.72$ (red), $c_{\cd}=0.985$ and $0.5\leq\tilde{m}/m_P\leq1.25$ (black), $c_{\cd}=0.99$ and $0.5\leq\tilde{m}/m_P\leq1.25$ (blue). In the plotted curves $\tilde m$ increases as $r$ decreases. In the right panel, the plotted curves correspond to $c_\cd=0.98$ (red), $c_\cd=0.985$ (black) and $c_\cd=0.99$ (blue).}\label{intercase} 
\end{figure}

Setting $N=60$, in Fig.~\ref{intercase} we plot the model predictions in the plane $(n_s,r)$ (left panel) for different values of $c_\cd$ and for different ranges of $\tilde m$, as indicated. In the plotted curves, the tensor-to-scalar ratio decreases as we increase $\tilde m$. Therefore, our results show that an increase of the induced gravity lowers the tensor-to-scalar ratio. In the right panel we plot the ratio $r/n_t$ as a function of $\tilde m$. In the range of values of $c_\cd$ consistent with the observed value of the spectral index we find a slight deviation from the standard consistency relation. Nevertheless, such a deviation does not seem to be sufficiently significant to be detected with confidence.

In Fig.~\ref{fnl} we plot the mass squared of the inflaton, as obtained from (\ref{mass2phi}) evaluated at the time of horizon crossing for cosmological scales (left panel), and $\fnl^{equi}$ calculated from (\ref{fnleinter}) (right panel). From the left plot, we find that the inflaton is tachyonic, whereas for smaller values of $\tilde{m}$, we recover a potential with positive curvature, in agreement with the DBI case. In this sense, it may be worth mentioning that the authors in   Ref.~\cite{Bernardini:2013tba} have studied the possibility that the Born-Infeld tachyon be equivalent to a scalar field in an effective field theory in different warped geometries. Moreover, in Ref.~\cite{Li:2013cem} the observational constraints on tachyon and DBI inflation were studied, and the authors showed that tachyon inflation fits better with cosmological data than DBI. It is also important to notice that $n_t<0$ in all cases, which is statistically preferred by data after the {\it Planck} and BKP joint analysis \cite{Ade:2015lrj,Ade:2015tva}, Also, the joint analysis of BKP+LIGO indicates a red tensor tilt $n_{t}=-0.76_{-0.52}^{+1.37}$ at the $68\%$ CL \cite{Huang:2015gka}. In Table~\ref{summary} we report the values of the ratio $r/n_t$, which only results in a slight deviation from the standard consistency relation in most of the cases. We recall that future cosmology probes will be able to discriminate inflationary models by direct detection of primordial B-modes \cite{Creminelli:2015oda}. Finally, from the right panel of Fig.~\ref{fnl} we find that the non-Gaussianity parameter $\fnl^{equi}$ is consistent with the stringent bounds imposed by {\it Planck} data \cite{Ade:2015ava}.
\begin{figure}[htbp]
\centering\includegraphics[width=7cm]{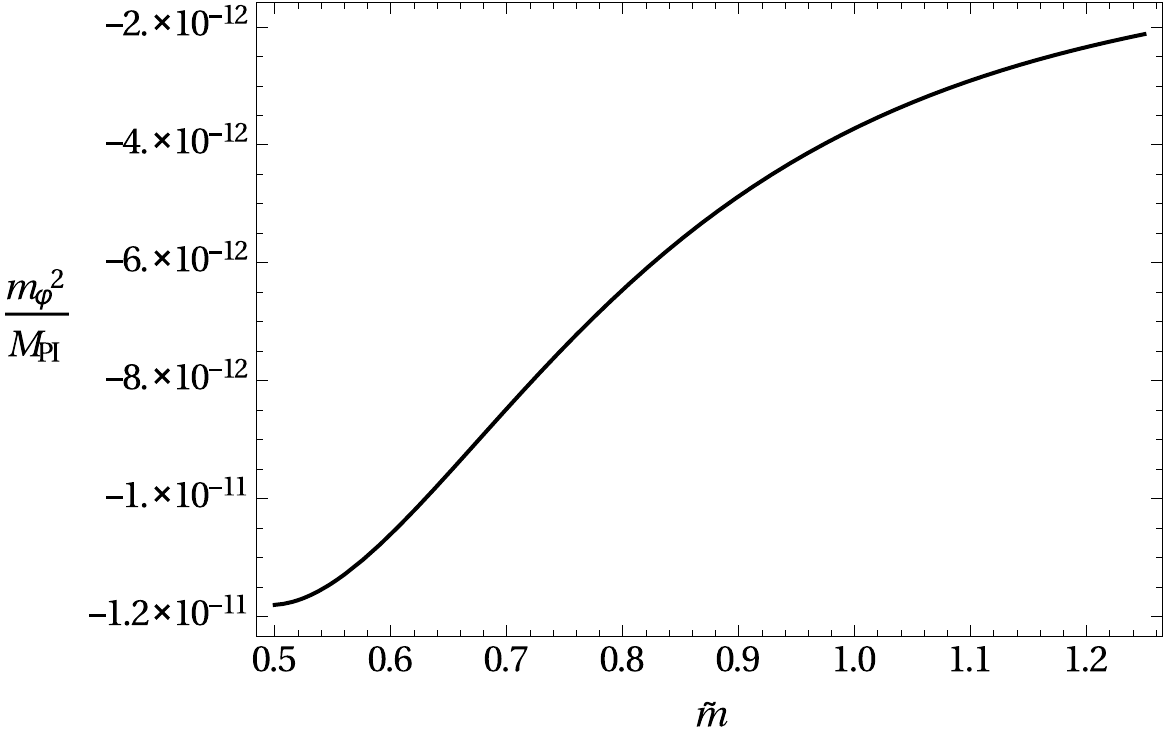}\hspace{0.5cm} \includegraphics[width=6.5cm]{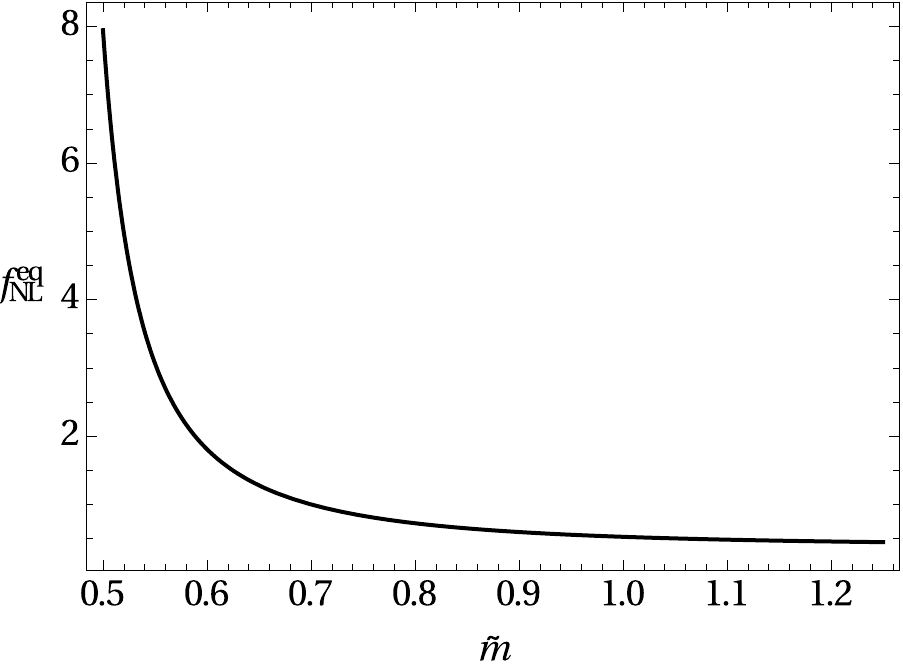}
\caption{Plots of the mass squared of the inflaton field (left panel) and the non-Gaussian parameter $\fnl^{eq}$ (right panel) as a function of $\tilde{m}$ (with $\tilde m$ in units of $m_P$). In this plot $ 0.22\leq\alpha\leq0.32 $ for $ 0.5\leq\tilde{m}\leq1.25 $. We take $c_{\cd}=0.985$ to build the plots, hence the depicted behaviour corresponds to the black line in Fig.~\ref{intercase}.}\label{fnl} 
\end{figure}

\subsection{Varying both sound speed and warp factor}\label{varycdf}
The cases considered in Sec.~\ref{constcasepert} (constant sound speed and constant warp factor) are consistent with observational data. However, it is interesting to understand the cases with varying $c_{\cd}$ and $f$. The questions we can pose in these cases are, can we get a parameter space with $r\sim\mathcal{O}\left(10^{-3}\right)$? How do the warped geometries and the scale of inflation change when $\left(c_{\cd},f\right)$ change with time? What is the nature of inflaton field is such cases? In this section, we obtain exact background solutions in two cases: a slowly varying sound speed at fixed warp factor and a slowly varying warp factor at fixed sound speed.

\subsubsection{Varying sound speed ($\epsilon_\cd\neq0,\eta_\cd=0$) and constant warp factor ($\epsilon_f=0$)}
\label{varycd}

We assume a slow variation of the sound speed, i.e. $\epsilon_{\mathcal{D}}\ll1$. Using the definition of slow-roll parameters
from (\ref{slwrolldef}), we can approximate $c_{\cd}$ in terms of $N=\ln a$ as 
\begin{equation}
c_{\mathcal{D}}=c_{d}\exp\left(\epsilon_{\mathcal{D}}N\right)\simeq c_{d}\left(1+\epsilon_{\mathcal{D}}N\right)\,,\label{cdslwvary}
\end{equation}
where $c_{d}$ is a constant whose magnitude is set some four $e$-foldings after the largest cosmological scales exit the horizon.

To integrate the background (\ref{BGS-1}) it is now convenient to rewrite it as 
\begin{equation}\label{varyspeddBG}
H^{\prime}-\lambda_{1}H+\frac{\lambda_{2}}{H}=0\,,
\end{equation}
where $\lambda_{1,2}$ are computed using the approximation in (\ref{cdslwvary}) and the prime stands for $'\equiv\frac{d}{dN}$. Integrating (\ref{varyspeddBG}) we obtain the solution $H=H(N)$. To fix the integration constant in the solution it suffices to impose that $\epsilon\equiv-\frac{H'}{H}=1$ at the end of inflation. We choose not to include here the solution $H=H(N)$ as it is a complicated expression involving imaginary error functions \cite{Abramowitz}. To constrain the model parameters we proceed as in Sec.~\ref{constcasepert}. Since in this case $\left(n_{s}\,,\, r\right)$ do not depend on warp factor $f$, we may find the range for $\left(c_{d}\,,\,\tilde{m\,},\,\epsilon_{\cd}\right)$ using the current bounds on $\left(n_{s}\,,\, r\right)$. Since we assume a slowly varying sound speed, its constraint in this case is not significantly different from the one obtained in Sec.~\ref{intercomp}. Consequently, we must tune $c_d\simeq0.98$ so that the spectral index agrees with observations. We also find that consistency with observations demands $\epsilon_{\mathcal{D}}<0$. This resembles the result of Ref.~\cite{Khoury:2008wj}, where it was shown that DBI inflation with a decreasing sound speed results in an expanding universe, in
contrast to the case of increasing sound speed. The observables in this case $\left(n_{s}\,,\, r\right)$ are not very different from
those obtained for a constant sound speed and warp factor in Sec.~\ref{intercomp}. In fact, after an extensive numerical study we find it difficult to obtain $r\sim\mathcal{O}\left(10^{-3}\right)$ in this case. Therefore, from our analysis we conclude that DBIG inflation with a varying sound speed and constant warp factor does not bring any new features.

\subsubsection{Varying warp factor $\left(\epsilon_{f}\protect\neq0,\eta_{f}=0\right)$
and constant sound speed $\left(\epsilon_{\cd}=0\right)$}\label{varywarp}
In general, the warp factor can depend on fields not stabilised during inflation. Therefore, it is feasible to expect a time-dependent warp factor while cosmological scales are exiting the horizon. For example, in Ref.~\cite{Gmeiner:2007uw}, various solutions for warped geometries were considered in the context of DBI inflation. In the following, we consider a slowly varying warp factor in the DBI-Galileon inflation model and constrain its variation using current data. Therefore, taking $\epsilon_{f}\ll1$ we approximate the warp factor as follows
\begin{equation}
f=f_{0}\exp\left(\epsilon_{f}N\right)\simeq f_{0}\left(1+\epsilon_{f}N\right)\,,\label{warpvary}
\end{equation}
where $f_{0}$ is the initial value warp factor and $\epsilon_{f}$ is constant and treated as free parameter. Similarly to the previous
case, we set the magnitude of $f_0$ four $e$-foldings after the largest cosmological scales exit the horizon.

It is important to remark that, in contrast to the previous case, where $\lambda_{1,2}=\lambda_{1,2}(N)$ and no simple analytical solution can be found for (\ref{varyspeddBG}), using $\epsilon_\cd=0$ and $\epsilon_f={\rm const.}$ gives $\lambda_1={\rm const.}$ and only $\lambda_{2}=\lambda_{2}(N)$. In turn, this allows us to find a simple solution to (\ref{BGS-1}) in terms of $N$ 
\begin{equation}
H^{2}=\frac{F_{1}}{F_{3}^{2}}\exp\left(\frac{\tilde{m}^{2}N\left(2c_{\cd}^{2}(\epsilon_{f}-3)-3\epsilon_{f}+6\right)}{2c_{\cd}^{2}(c_{\cd}m_P^{2}+\tilde{m}^{2})}\right)C_{2}
+\frac{F_{2}\left(N\right)}{f_{0}F_{3}^{2}}\label{Hsol-varyf}\,,
\end{equation}
where $C_{2}$ is an integration constant, determined by the condition $\epsilon=1$ at $N=60$, and 
\[
F_{1}=\tilde{m}^{4}\left[2c_{\mathcal{D}}^{2}\left(\epsilon_{f}-3\right)-3\epsilon_{f}+6\right]^{2}\,,
\]
\[
F_{2}\left(N\right)=2c_{\mathcal{D}}^{2}\left(c_{\mathcal{D}}^{2}-1\right)\left\{ 2c_{\mathcal{D}}^{3}m_P^{2}\epsilon_{f}+2\tilde{m}^{2}c_{\mathcal{D}}^{2}\left[N\left(\epsilon_{f}-3\right)\epsilon_{f}+3\right]-3\left(\epsilon_{f}-2\right)\left(N\epsilon_{f}-1\right)\right\} \,,
\]
\[
F_{3}=\tilde{m}^{2}\left[2c_{\mathcal{D}}^{2}\left(\epsilon_{f}-3\right)-3\epsilon_{f}+6\right]\,.
\]

In the following we find the range of parameters $\left(c_{\mathcal{D}},\tilde{m},\epsilon_{f}\right)$ using the CMB constraints on $\left(n_{s},r\right)$. Firstly, since the sound speed is constant we obtain the same constraint as in Sec.~\ref{intercomp}, namely $c_\cd\simeq0.98$ to keep $n_s$ within its observed range.
\begin{figure}[htbp]
\centering \includegraphics[width=8cm]{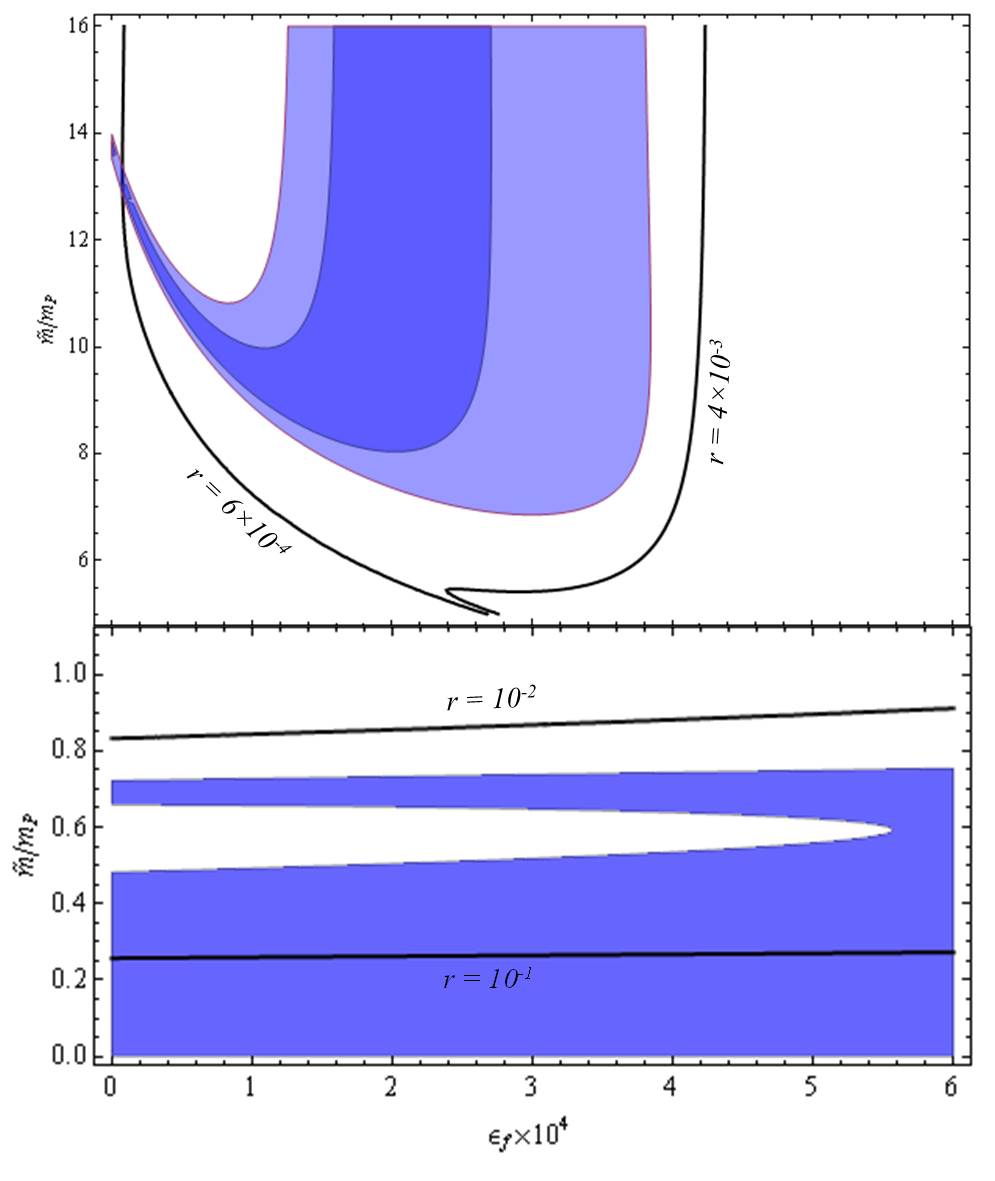}
\caption{Contour plots in the plane $\left(\tilde{m}\,,\,\epsilon_{f}\right)$. In the top panel, light and dark blue regions represent the 68\% and 95\% CL for the spectral index $n_s$, respectively. Black lines represent contours for different values of the tensor-to-scalar ratio, as indicated. In the bottom panel, the blue region depicts the 95\% CL for the spectral index $n_s$. We use $c_\cd=0.980$.}\label{varynsrntr} 
\end{figure}

In Fig.~\ref{varynsrntr} we depict the parameter space $\left(\tilde{m},\epsilon_{f}\right)$ consistent with observations of the spectral index and tensor-to-scalar ratio. Taking $c_\cd=0.98$ and enforcing \mbox{$n_s=0.968\pm0.006$}, our plot shows that it is indeed feasible to obtain a tensor-to-scalar ratio as low as \mbox{$r\simeq6\times10^{-4}$}. Nevertheless, the plot also evidences that this requires a considerable tuning between $\tilde m$ and $\epsilon_f$. We have checked that using the 2$\sigma$ interval for the spectral index does not contribute to enlarge significantly the space where $r\sim10^{-4}$. In the absence of the aforementioned tuning, expected values correspond to the range $10^{-3}\lesssim r\lesssim3\times10^{-3}$. Moreover, we have checked as well that the space where $r\sim10^{-4}$ becomes incompatible with the observed spectral index even for small deviations away from $c_\cd=0.98$. Consequently, finding $r\sim10^{-4}$ requires the combined tuning of $\tilde m,\epsilon_f$ and $c_\cd$. Nevertheless, it seems fair to say that, despite these tunings, the DBIG model of inflation represents an improvement, albeit a moderate one, with respect to the DBI model studied in Sec.~\ref{dbilimit}.

\begin{figure}[htbp]
\centering\includegraphics[width=8.0cm]{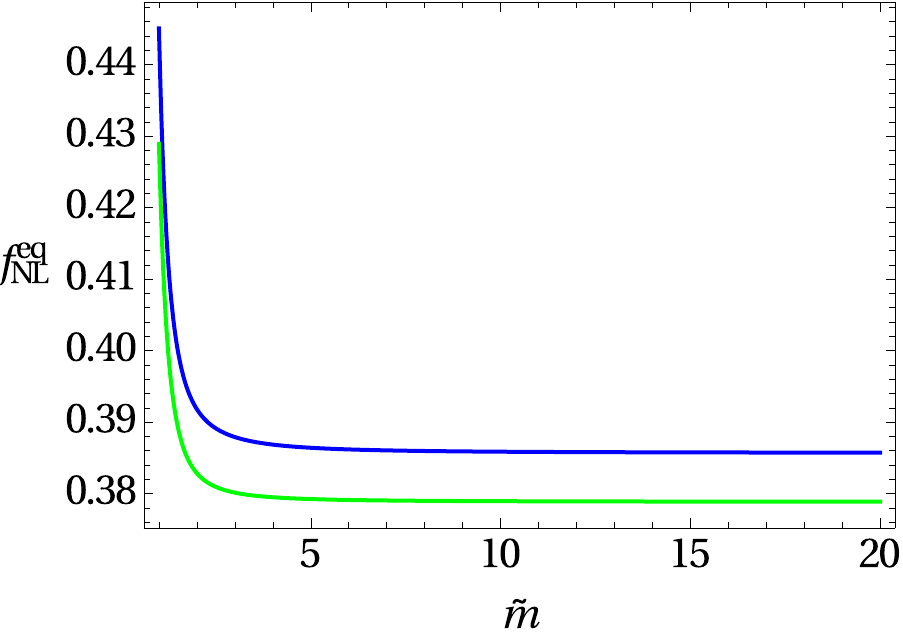}
\caption{In this plot, we depict the non-Gaussian parameter $\fnl^{equi}$ as a function of $\tilde{m}$ (with $\tilde m$ in units of $m_P$). We take $c_{\cd}=0.98$ and $ \epsilon_{f}\sim10^{-4} $ (Blue line) and $ \epsilon_{f}\sim10^{-6} $ (Green line). In this plot $ 0.326\leq\alpha\leq0.33 $ for $ 1\leq\tilde{m}\leq20 $.}\label{fnlvaryf} 
\end{figure}

In addition, we verify the equilateral non-Gaussianity by using the approximate expression for $f^{equi}_{NL}$ in (\ref{fnleinter}). Since we consider a tiny variation of the warp factor we can practically neglect its contribution to non-Gaussianity. From Fig.~\ref{fnlvaryf} we can conclude that the DBIG model with varying warp factor leads to non-Gaussianities within the current observational bounds. Consequently, we conclude that after including a varying warp factor the DBIG model of inflation could be of crucial importance with respect to B-mode detection and non-Gaussianities in future CMB experiments \cite{Creminelli:2015oda}.

\begin{figure}[htbp]
\centering\includegraphics[height=3in]{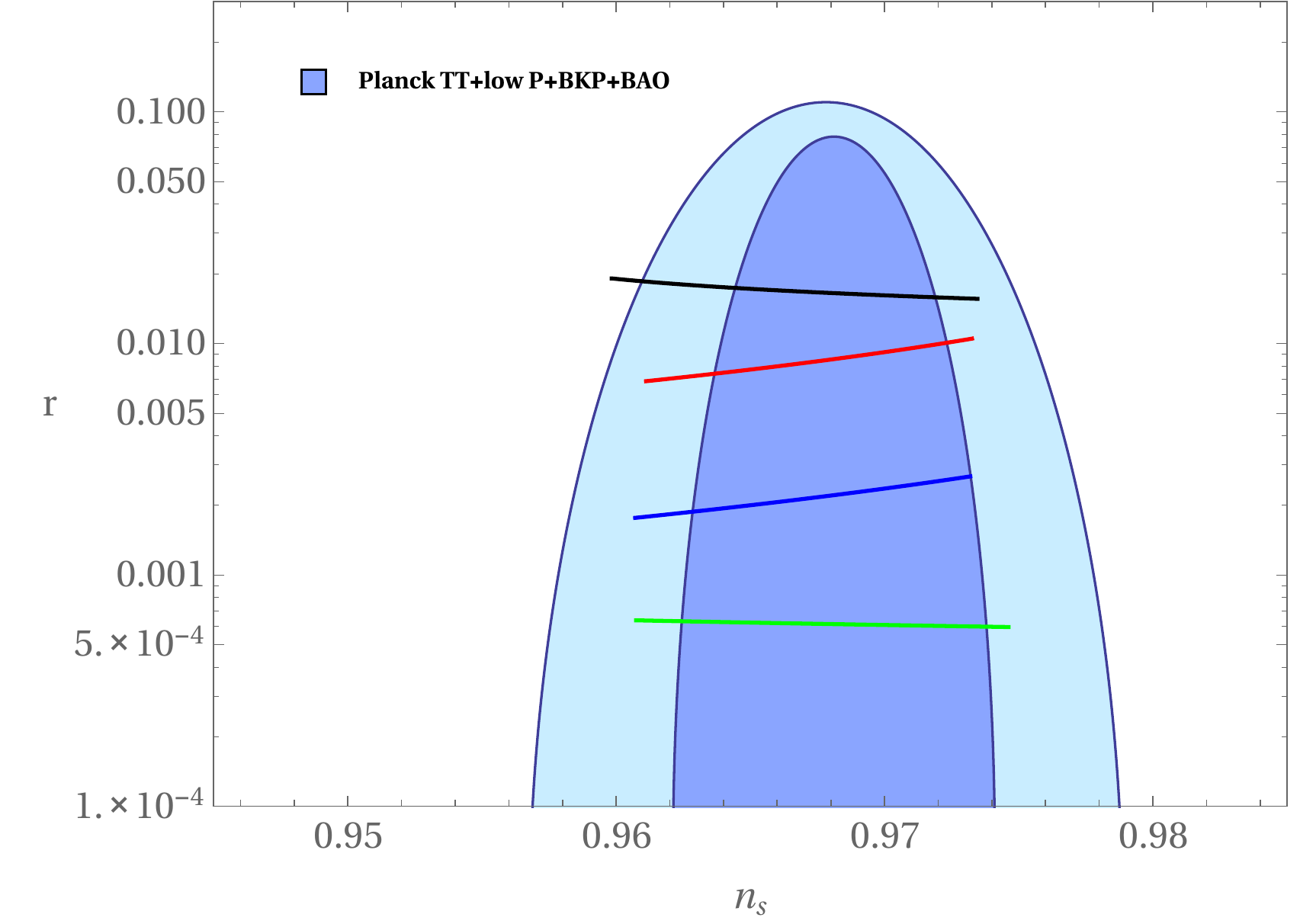}\caption{Predictions of the DBIG model for $N=60$ along with the {\it Planck} TT+lowP+BKP+BAO constraints on the space $(n_s,r)$ at the 68\% and 95\% CL. The black line represents the case with constant sound speed and warp factor ($c_\cd=0.985$, $1\leq\tilde m/m_P\leq1.25$). Different model predictions for a constant sound speed and varying warp factor are plotted in red ($c_\cd=0.985$, $\tilde m=15m_P$ and $5.1\leq10^4\epsilon_f\leq8.5$), blue ($c_\cd=0.98$, $\tilde m=15m_P$ and $1.5\leq10^4\epsilon_f\leq2.6$) and green ($c_\cd=0.98$, $\tilde m=13m_P$ and $0.07\leq10^4\epsilon_f\leq0.11$).}\label{DBIG-total} 
\end{figure}

We finish this section by depicting the predictions of DBIG inflation for different sets of values of the model parameters in Fig.~\ref{DBIG-total} and by summarizing our results in Table~\ref{summary}. We recall that the values collected in the table were obtained taking by enforcing the scalar spectral index to lie within its observed range $n_s=0.968\pm0.006$ at the 95\% CL and taking $N=60$.

\begin{center}
\begin{table}[htbp]
\begin{tabular}{|c||c|c|c|c|c|}
\hline 
{\tiny{}{}{Inflation}{}  } & {\tiny{}{}$r$  } & {\tiny{}{}$r/n_{t}$  } & {\tiny{}{}$\begin{aligned}m_{\phi}/m_{P}\end{aligned}
$  } & {\tiny{}{}$\begin{aligned}V_{\ast}^{1/4}/10^{16}\,{\rm GeV}\end{aligned}
$  } & {\tiny{}{}$f/m_{P}^{4}$ }\tabularnewline
\hline 
\hline 
{\tiny{}{}{DBI limit}{}  } & {\tiny{}{}$\left(0.01,\,0.1\right)$  } & {\tiny{}{}$\left(-4.8,-0.7\right)$  } & {\tiny{}{}$6.63\times10^{-6}$  } & {\tiny{}{}$\left(0.95,1.82\right)$  } & {\tiny{}{}$\sim10^{12}-10^{14}$}\tabularnewline
\hline 
{\tiny{}{}{Galileon limit}{}  } & {\tiny{}{}$\left(0.13,\,0.15\right)$  } & {\tiny{}{}$\left(-8.1,-7.93\right)$  } & {\tiny{}{}$m_{\phi}^{2}<0$  } & {\tiny{}{}$\left(0.64,0.70\right)$  } & {\tiny{}{}$\sim10^{9}$}\tabularnewline
\hline 
{\tiny{}{}{DBIG}{}  } & {\tiny{}{}$\left(0.01,\,0.1\right)$  } & {\tiny{}{}$\left(-7.95,-7.5\right)$  } & {\tiny{}{}$m_{\phi}^{2}<0$  } & {\tiny{}{}$\left(1.7,2.1\right)$  } & {\tiny{}{}$\sim10^{8}-10^{9}$}\tabularnewline
\hline 
{\tiny{}{}%\hline
{}{(ii)Varying $f$}{}  } & {\tiny{}{}$\left(0.0068,\,0.0095\right)$  } & {\tiny{}{}$\left(-7.95,\,-7.85\right)$ } & {\tiny{}{}$\left(2.41,\,2.9\right)\times10^{-7}$  } & {\tiny{}{}$\left(5.9,\,6.4\right)$  } & {\tiny{}{}$\left(6,\,9\right)\times10^{10}$}\tabularnewline
 & {\tiny{}{}$\left(0.0018,\,0.0027\right)$  } & {\tiny{}{}$\left(-8.01,\,-7.95\right)$ } & {\tiny{}{}$\left(3.6,\,5.2\right)\times10^{-8}$  } & {\tiny{}{}$\left(4.1,\,4.6\right)$  } & {\tiny{}{}$\left(2.2\,,\,3.4\right)\times10^{9}$}\tabularnewline
 & {\tiny{}{}$\left(0.0006,\,0.0007\right)$ } & {\tiny{}{}$\left(-7.63,\,-7.52\right)$  } & {\tiny{}{}$\left(1.52,\,1.58\right)\times10^{-8}$  } & {\tiny{}{}$\left(2.8,\,2.85\right)$  } & {\tiny{}{}$\left(0.17,\,0.18\right)\times10^{7}$}\tabularnewline
\hline 
\end{tabular}\caption{Inflationary observables in various limits of DBIG inflation.}
\label{summary} 
\end{table}
\par\end{center}

\section{On a class of background solutions}
\label{varysols}

Until now, we have explored solutions to the background (\ref{Friedmann}) and (\ref{Raychaudhuri}) in which the sound speed and warp factor are either constants or time-dependent functions with very slow variation, although not simultaneously time-dependent. This choice is motivated by the simplicity of the perturbation spectrum imprinted in the CMB, which strongly favors the simplest inflationary models. Nevertheless, it is reasonable to conjecture, and to some extent expected, that in the early stages of inflation, when the observable cosmological scales are still deep within the horizon, the background dynamics has been much different from the simple slow-roll evolution supported by CMB observations. Therefore, it is interesting to investigate what kind of inflationary dynamics does the DBI-Galileon model give rise to when the  sound speed and warp factor become time-dependent functions simultaneously. In general, however, it is not possible to integrate the equations of motion for general functions $c_\cd(t)$ and $f(t)$. Owing to this difficulty, in order to find analytical solutions of the background equations we pursue a phenomenological approach in which we consider two different ansatz for the functions $\lambda_1$ and $\lambda_2$.

%$\bullet$ In this case, however, it is not possible to integrate the equations of motion for arbitrary variations of $\lambda_{1,2}$. 

%$\lambda_1$ and $\lambda_2$ become time-dependent functions
 
%We want to make the most of the model parameters to see what background dynamics we can get by allowing a simultaneous variations of both $\lambda_1$ and  $\lambda_2$.

%$\bullet$ $\lambda$ with small variation... no, no. We will obtain solutions where $\lambda$ features a rapid variation, as may be during the earliest stages of inflation.

If we allow the sound speed $c_{\mathcal{D}}$ and warp factor $f$ to change ($\epsilon_{\cd},\epsilon_{f}\neq0$) the coefficients $\lambda_{1,2}$ become time-dependent functions. In such case, (\ref{BGS-1}) can be rewritten as 
\begin{equation}
\frac{d\ln H}{\lambda_{1}-\lambda_{2}H^{-2}}=d\ln a\,.\label{csvHintegral}
\end{equation}
In what follows, we discuss two different parameterizations for $\lambda_{1,2}$ to find approximate solutions for $a(t)$.

\subsubsection*{Parametrization 1}
\begin{figure}[htbp]
\centering\includegraphics[width=4.5cm]{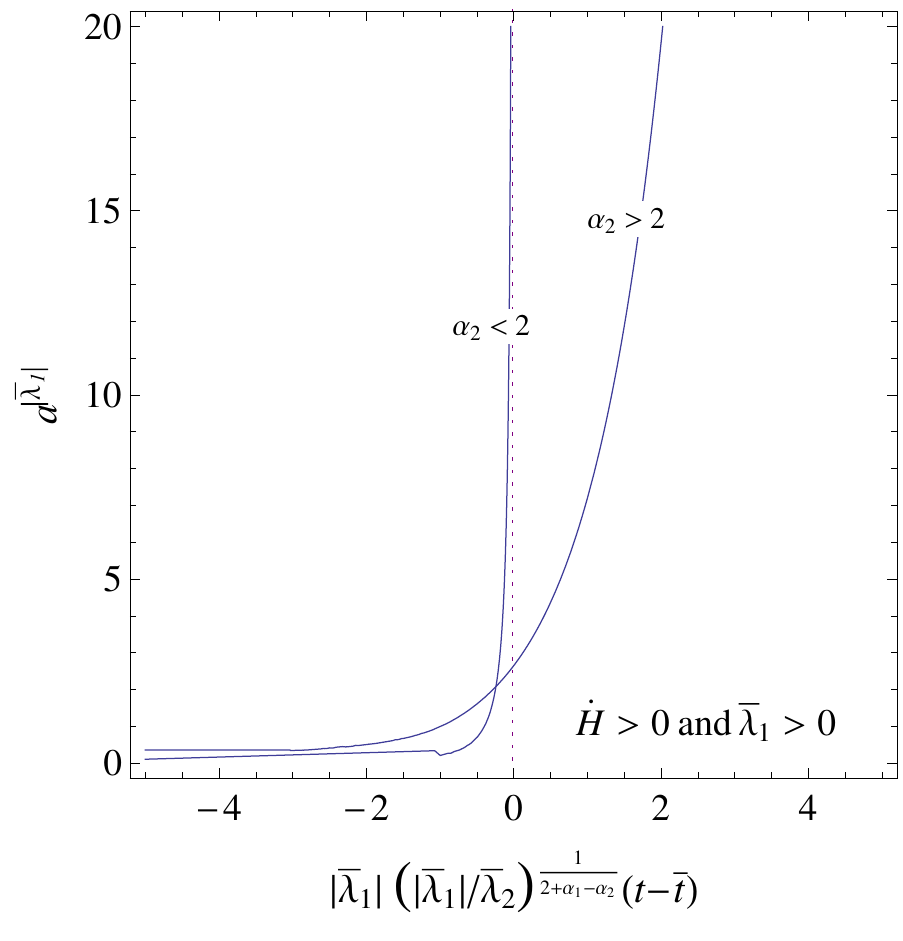}\hspace{0.25cm} \includegraphics[width=4.5cm]{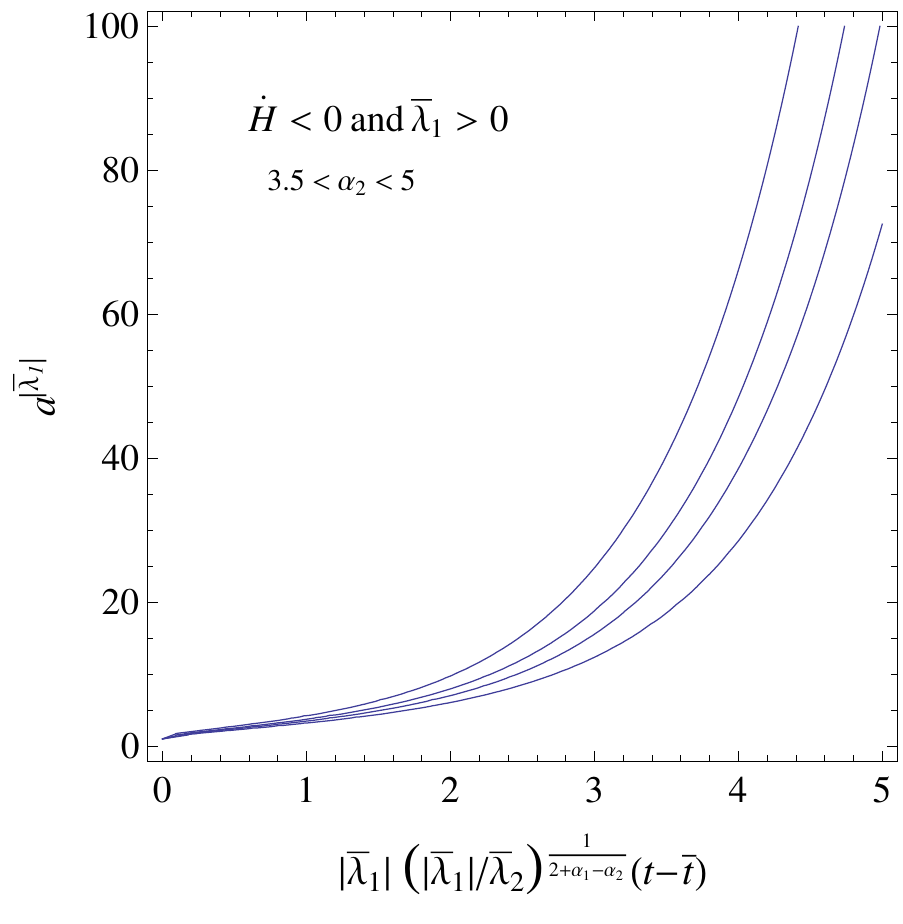}\hspace{0.25cm}\includegraphics[width=4.5cm]{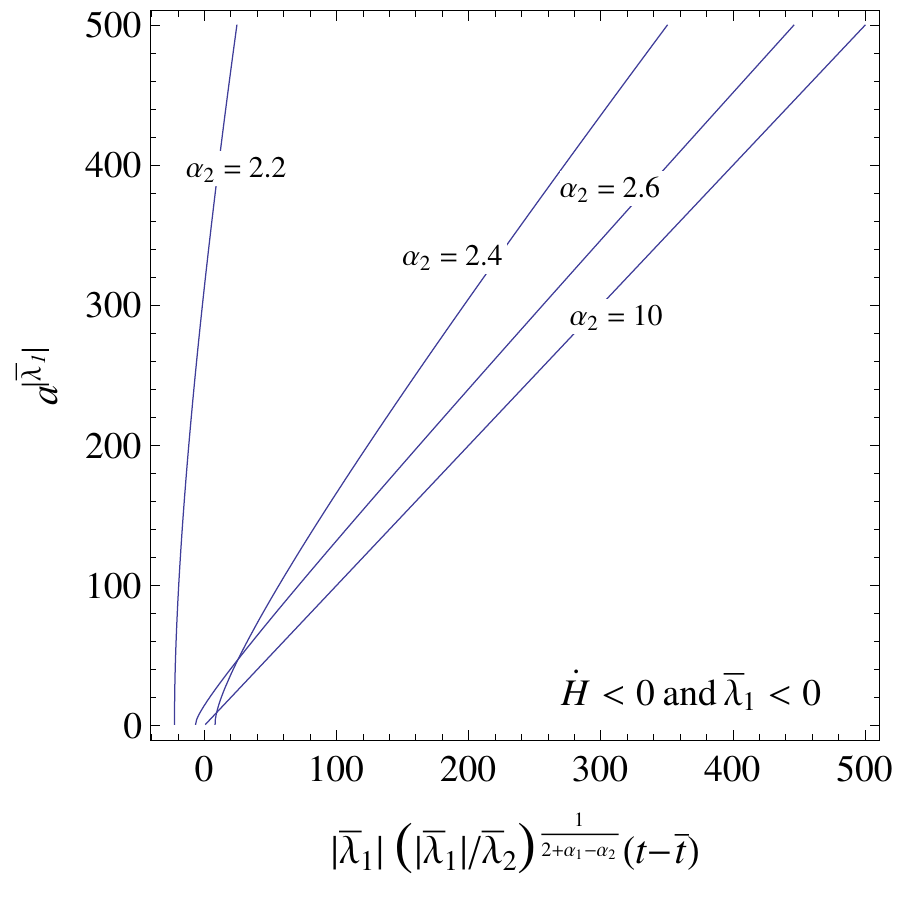}\caption{Evolution of the scale factor $a(t)$, according to (\ref{PV1h2a}),
for $\bar{\lambda}_{1},\dot{H}>0$ (left panel), for $\bar{\lambda}_{1}>0$,
$\dot{H}<0$ (central panel) and for $\bar{\lambda}_{1},\dot{H}<0$
(right panel). For simplicity we take $\kappa=1$.}\label{fig-p1} 
\end{figure}

The simplest strategy to integrate (\ref{csvHintegral}) is to
rewrite $\lambda_{1,2}$ as functions of $H$. Thus, we consider the
temporal dependence for $\lambda_{i}$ (with $i=1,2$) of the form
\begin{equation}
\lambda_{i}=\overline{\lambda}_{i}H^{\alpha_{i}}\,,\label{PV1}
\end{equation}
where $\overline{\lambda}_{i},\alpha_{i}$ are constants. Using this
ansatz, (\ref{csvHintegral}) can be integrated to give 
\begin{equation}
_{2}F_{1}\left(1,1+\beta;2+\beta;\frac{\lambda_{1}H^{2}}{\lambda_{2}}\right)H^{2}=\lambda_{2}\left(\alpha_{2}-2\right)\ln|\kappa a|\quad,\quad\beta\equiv\frac{\alpha_{1}}{\alpha_{2}-\alpha_{1}-2}\,,\label{PV1hyperG}
\end{equation}
where $_{2}F_{1}$ is the hypergeometric function and $\kappa$ is
an arbitrary constant. Note that in the limit $\alpha_{1,2}\to0$
we can use the identity $_{2}F_{1}(1,1;2;z)\, z=-\ln|1-z|$ to arrive
at (\ref{Hconscase}). Given the complexity of the above solution,
substituting $H=\dot{a}/a$ to integrate the resulting differential
equation in terms of $a(t)$ is of no practical use. Thus, it is necessary
to resort to numerical methods to integrate it. Nevertheless, if $|\beta|<1$
an approximation to the evolution equation is given by (see Appendix
\ref{ADBIG1} for details) 
\begin{equation}
\ln\left|1-\frac{\lambda_{1}H^{2}}{\lambda_{2}}\right|\simeq\ln|\kappa a|^{A}\quad{\rm with}\quad A\equiv\frac{(2-\alpha_{2})\lambda_{1}}{1+\beta}\simeq(2-\alpha_{2})\lambda_{1}\,.\label{PV1approax}
\end{equation}
For $\alpha_{2}\lesssim{\cal O}(1)$, the condition $|\beta|\ll1$
implies $|\alpha_{1}|\ll1$. Provided $H$ does not change exponentially,
which can be certainly applied to the regular solution plotted in
Fig.~\ref{fig1}, we can approximate $\lambda_{1}$ by a constant
since \mbox{$\lambda_{1}\simeq\overline{\lambda}_{1}\left(1+\alpha_{1}\ln(H/H_{\ast})+\ldots\right)$}.
This reasoning can be applied to the singular solution as well whenever
it finds itself sufficiently away from the singularity at $t=\bar{t}$.
Using (\ref{BGS-1}), we rewrite (\ref{PV1approax}) as
\begin{equation}
H^{2+\alpha_{1}-\alpha2}=\frac{\overline{\lambda}_{2}}{|\overline{\lambda}_{1}|}\,{\rm sign}(\lambda_{1})\left(1+{\rm sign}(\dot{H})|\kappa a|^{A}\right)\,,\label{PV1h2a}
\end{equation}
which can be integrated to obtain the scale factor $a(t)$ in terms
of hypergeometric functions. The implicit function (for simplicity
we present the solution for $\kappa=1$ and vanishing $\alpha_{1}$)
which defines the scale factor is given by
\begin{equation}
\begin{split}\bar{\lambda}_{1}\left(t-\bar{t}\right)\approx & -\textrm{sign}\left(\dot{H}\right)\left(a^{\left(\alpha_{2}-2\right)\bar{\lambda}_{1}}+\textrm{sign}\left(\dot{H}\right)\right)\left(-\textrm{sign}\left(\bar{\lambda}_{1}\right)\frac{\bar{\lambda}_{2}\left(a^{\left(2-\alpha_{2}\right)\bar{\lambda}_{1}}+\textrm{sign}\left(\dot{H}\right)\right)}{\left|\bar{\lambda}_{1}\right|}\right){}^{\frac{1}{\alpha_{2}-2}}\\
 & _{2}F_{1}\left(1,1;1+\frac{1}{2-\alpha_{2}};-\textrm{sign}\left(\dot{H}\right)
 a^{\left(\alpha_{2}-2\right)\bar{\lambda}_{1}}\right)\quad,\quad\alpha_{2}\neq2\,.
\end{split}
\label{GC-p1-sol}
\end{equation}
From (\ref{PV1h2a}) we easily recover the background solution
with constant sound speed and constant warp factor, (\ref{Hconscase}),
in the limit $\alpha_{1,2}\to0$. An important aspect of (\ref{PV1h2a})
is that it only requires $|\alpha_{1}|$ to be small, whereas $|\alpha_{2}|$
can be relatively large, thus allowing a significant evolution of
$\lambda_{2}$ during inflation. Note that if we consider $c_{\mathcal{D}}$
constant, for consistency with the smallness of $\alpha_{1}$, then
from (\ref{lambda2}) it follows that the evolution of $\lambda_{2}$
is to be attributed to the warp factor $f$. Below we study the behaviour
of the computed solution for different values of $\alpha_{2}$. In
view of (\ref{PV1h2a}), we may consider three cases consistent
with $H^{2}>0$: 
\begin{itemize}
\item $\bar{\lambda}_{1}>0$ and $\dot{H}>0$. This case is illustrated
in the left panel of Fig.~\ref{fig-p1}, where for $\alpha_{2}<2$
we have a singular solution when $t\rightarrow\bar{t}$ . Any other
solution with $\alpha_{2}>2$ is regular at $t=\bar{t}$. 
\item $\bar{\lambda}_{1}>0$ and $\dot{H}<0$. This regime takes place provided
$(|\kappa|a)^{A}<1$. A thorough numerical study of this scenario
shows that only for a limited range of values of $\alpha_{2}$ the
integration of (\ref{PV1h2a}) yields a well behaved physical
solution for the scale factor. In the central panel of Fig.~\ref{fig-p1}
we depict the solution for a few values of $\alpha_{2}$ in the range
$3.5<\alpha_{2}<5$ 
\item $\bar{\lambda}_{1}<0$ and $\dot{H}<0$. The constraint now is $(|\kappa|a)^{A}>1$.
This case, depicted in the right panel of Fig.~\ref{fig-p1},
possesses smooth solutions for $\alpha_{2}>2$. Moreover, for large
values of $\alpha_{2}$, the scale factor follows approximately a
power law $a\left(t\right)\sim\left(t-\bar{t}\right)^{1/\left|\bar{\lambda}_{1}\right|}$. 
\end{itemize}

\subsubsection*{Parametrization 2}
A second, simple alternative to solve (\ref{BGS-1}) with time-dependent
$\lambda_{1,2}$ is to parametrize their dependence as 
\begin{equation}
\lambda_{i}=\lambda_{i\ast}(a/a_{\ast})^{\alpha_{i}}\,,\label{pv2}
\end{equation}
where $\lambda_{i\ast},\alpha_{i}$ are constants and $a_{\ast}$ is the
scale factor when the largest cosmological scales exit the horizon.
Defining 
\begin{equation}
z\equiv\ln H\quad,\quad y\equiv\ln(a/a_{\ast})\,,\label{pv2variables}
\end{equation}
we find the exact solution to (\ref{BGS-1}) (see Appendix \ref{ADBIG1})
\begin{equation}
e^{2z}=\frac{2\lambda_{2}}{\alpha_{1}}\left(\frac{\alpha_{1}}{2\lambda_{1}}\right)^{\frac{\alpha_{2}}{\alpha_{1}}}{\rm exp}\left(\frac{2\lambda_{1}}{\alpha_{1}}\, e^{y\alpha_{1}}\right)\,\Gamma\left(\frac{\alpha_{2}}{\alpha_{1}},\frac{2\lambda_{1}}{\alpha_{1}}\, e^{y\alpha_{1}}\right)+\kappa\,{\rm exp}\left[\frac{2\lambda_{1}}{\alpha_{1}}\left(e^{y\alpha_{1}}-1\right)\right]\,,\label{pv2soln}
\end{equation}
where $\Gamma(s,x)$ is the incomplete Gamma function \cite{Abramowitz}.
In the limit $\alpha_{1,2}\to0$ we easily recover (\ref{Hconscase}),
whereas for $\alpha_{1,2}\neq0$ we can use the asymptotic formula
$\Gamma(s,x)\approx x^{s-1}e^{-x}$ when $x\gg1$. In such case, the
above equation becomes 
\begin{equation}
e^{2z}\simeq\left(\frac{\lambda_{2\ast}}{\lambda_{1\ast}}\right)e^{y(\alpha_{2}-\alpha_{1})}+\kappa\exp\left[\frac{2\lambda_{1\ast}}{\alpha_{1}}\left(e^{y\alpha_{1}}-1\right)\right]\,.\label{pv2solution}
\end{equation}
If we focus on the background evolution while cosmological scales
are exiting the horizon then $0\leq y\lesssim9$, and $y\alpha_{1}\ll1$
provided $|\alpha_{1}|\ll1$. Neglecting higher orders in $y\alpha_{1}$
we obtain 
\begin{equation}
H^{2}\simeq\left(\frac{\lambda_{2\ast}}{\lambda_{1\ast}}\right)a^{\alpha_{2}-\alpha_{1}}+\kappa a^{2\lambda_{1\ast}}\,,\label{pv2h2a}
\end{equation}
which can be integrated to obtain $a(t)$ in terms of hypergeometric
functions, and also gives (\ref{Hconscase}) in the limit $\alpha_{1,2}\to0$.
The implicit function (again, and for simplicity, we present the solution
for $\kappa=1$ and vanishing $\alpha_{1}$) that determines the scale
factor $a(t)$ is given by 
\begin{equation}
\begin{split}\left(t-\bar{t}\right)\approx- & \frac{2\lambda_{1\ast}a^{-\alpha_{2}}\sqrt{\frac{\lambda_{2\ast}a^{\alpha_{2}}}{\lambda_{1\ast}}+a^{2\lambda_{1\ast}}}\,_{2}F_{1}\left(1,\frac{\lambda_{1\ast}-\alpha_{2}}{2\lambda_{1\ast}-\alpha_{2}};\frac{-\alpha_{2}}{4\lambda_{1\ast}-2\alpha_{2}}+1;-\frac{a^{2\lambda_{1\ast}-\alpha_{2}}\lambda_{1\ast}}{\lambda_{2\ast}}\right)}{\alpha_{2}\lambda_{2\ast}}\,.\end{split}
\label{GC-p2-sol}
\end{equation}
Similarly to (\ref{PV1h2a}), $|\alpha_{2}|$ is allowed to take
on relatively large values in (\ref{pv2h2a}).

Notice that in (\ref{GC-p2-sol}) the hypergeometric function $_{2}F_{1}$ is undefined when $\left(\frac{-\alpha_{2}}{4\lambda_{1\ast}-2\alpha_{2}}+1\right)$ is a negative integer. Therefore, from a formal point of view, by taking $\alpha_{2}<0$ we avoid the regions where (\ref{GC-p2-sol}) is undefined. In addition we must also impose that $\alpha_{2}\neq2\lambda_{1\ast}$. When $\alpha_{2}<0$, we have from (\ref{pv2h2a}) that $H$ becomes singular if the scale factor $a$ goes to zero. It can be checked in (\ref{GC-p2-sol}) that $\left(t-\bar{t}\right)$ is zero whenever $a$ is zero, which amounts to have an undesirable singular solution for $H$ when $t=\bar{t}$.

In view of our results, it seems reasonable to conclude that the solutions obtained using the ansatz in (\ref{PV1}) for $\dot{H}<0$ (with either sign of $\bar{\lambda}_{1}$) provide a more appropriate qualitative evolution for $a(t)$ than those described by the ansatz in (\ref{pv2}). Therefore, our analysis \textit{demonstrates} that, within the context of DBI-Galileon inflation, it is possible to envisage an early inflationary stage during which the warp factor undergoes a significant variation. The relevance of this result is that such phase can be smoothly connected to the last phase of slow-roll while allowing a marginal variation of the warp factor and agreeing with current CMB observations. In this sense, it is very suggestive to imagine that the early phase of rapidly evolving geometric structure could be connected to the very beginning of inflation.

\section{Summary}\label{conclns}
In this chapter we studied the DBI-Galileon inflationary scenario, which constitutes a generic extension of the DBI model involving an induced gravity, and obtain the gravitational field equations allowing the sound speed $c_{\cd}$ and warp factor $f$ that the model depends on to be time-dependent. We find exact solutions to the background (\ref{Friedmann}) and (\ref{Raychaudhuri}) when $c_{\cd}$ and $f$ are constant. We obtain a singular behaviour at finite time for the scale factor and Hubble parameter when $\lambda_{2}<\lambda_{1}H^{2}$, and also a regular behaviour when $\lambda_{2}>\lambda_{1}H^{2}$ (see Fig.~\ref{fig1}). We focused on inflationary scenarios under the slow-roll approximation and constrain the model parameters using the {\it Planck} 2015 results. In addition, we constrain the warp factor in the different inflationary regimes using CMB data. Notice that the warp factor scale might be important, regarding warped string phenomenology, to understand extra dimensions and warped geometries arising from string theory.  %\cite{Underwood:2008dh,Underwood:2008zz,Kecskemeti:2006cg,Goon:2011qf,sevaral_DBI2,Kobayashi:2007hm}.
%Warped geometries could also play a crucial role in the production of cosmic
%strings in brane driven inflationary mechanisms \cite{HenryTye:2006uv,sevaral_DBI3,Slagter:2014gra}.
We found that, in general, different warped geometries give rise to distinct inflationary predictions. In the case of constant $c_{\cd}$ and $f$ (see Fig.~\ref{PspaceDBIG}), the tensor-to-scalar ratio is $r\gtrsim\mathcal{O}\left(10^{-2}\right)$. Later, we considered the DBI-Galileon model with a slowly varying warp factor and find that the tensor-to-scalar ratio can be as low as $r\simeq6\times10^{-4}$ (see Figs.~\ref{varynsrntr} and \ref{DBIG-total}). However, we find that this requires the combined tuning of $\tilde m$, $\epsilon_f$ and $c_\cd$. In any case, a varying warped geometry brings the predictions of the DBIG model closer to those of the Starobinsky model.

Another aspect of our study is the violation of the standard consistency relation of single-field inflation, $r=-8n_{t}$. Since DBIG inflation is a class of generalized G-inflation, we find deviations away from the standard consistency relation $r=-8n_{t}$. However, with the exception of the DBI limit (see Fig.~\ref{fig1-1}), the deviations found in the rest of cases under study are quite small (see Table.~\ref{summary}). This result is consistent with the status about the tensor consistency relation in Galileon models as it is described in Ref.~\cite{Unnikrishnan:2013rka}. We emphasize that a prominent detection of the B-modes, within future CMB probes devised with a greater sensitivity \cite{Errard:2015cxa,Creminelli:2015oda,Huang:2015gca},can discriminate DBIG inflation.
 
Finally, we aimed at describing an early stage of inflation taking place well before cosmological scales exit the horizon, we obtain general background solutions allowing an arbitrary time dependence for $c_{\cd}$ and $f$ by promoting the coefficients $\lambda_{1}$ and $\lambda_{2}$ in the background (\ref{BGS-1}) to time-dependent functions. To integrate the background equations analytically we pursue a phenomenological approach, making use of the ansatze in (\ref{PV1}) and (\ref{pv2}). The validity of our approximations demands that $\lambda_{1}$ remains approximately constant ($\alpha_{1}\simeq0$) for both ansatze, whereas $\lambda_{2}$ can have substantial variation since $\alpha_{2}$ is not constrained to be small (see Fig.~\ref{fig-p1}). This variation of $\lambda_{2}$, in turn, can be attributed to a variation of the geometric warp factor $f$ since $c_{\cd}$ remains approximately constant. From our numerical exploration of the approximate solution we conclude that the ansatz in (\ref{PV1}) provides a more appropriate, qualitative evolution for the scale factor. Our analysis thus provides the intriguing possibility to consider an early stage of DBI-Galileon inflation (may be even connected to its very beginning) with a significantly varying geometric structure that gives way, once the geometric structure becomes approximately stabilized, to a final phase of slow-roll in perfect agreement with current CMB observations.

\chapter{Effective models of inflation from SFT framework}
\label{SFTin}
\begin{chapquote}{Edward Witten}
Quantum mechanics brought an unexpected fuzziness into physics because of quantum uncertainty, the Heisenberg uncertainty principle. String theory does so again because a point particle is replaced by a string, which is more spread out
\end{chapquote}

\lhead{\bf Chapter 4. \emph{Effective models of inflation from SFT framework}} %

Accounting string theory as a key player in cosmological inflation,
we take an inspiration from string field theory (SFT) \cite{Ohmori:2001am,Arefeva:2001ps} and construct successful effective models
of inflation\footnote{Note that our study is different from the early attempts of considering
inflation in SFT studied with $p-$adic strings \cite{Barnaby:2006hi,Biswas:2012bp}}. Our model is based on the system of open string tachyon and closed string dilaton including the concepts of non-locality and tachyon condensation (c.f., appendix \ref{AppSFT} for a brief review). 
In this chapter, we consider a system of closed
string dilaton and open string tachyon, present in the low energy
limit and assume any higher excitations are either stabilized or not
relevant for our purposes. The open string tachyon is known to condense
rolling to its potential minimum due to brane (or brane-anti-brane pair) instability,
present in the system as it decays \cite{Sen:1999nx}. This phenomenon is called as the tachyon condensation
(TC) process. {It is important to understand that the Sen conjecture about TC i.e., the compensation
of the brane tension by the negative vacuum energy of the tachyon
in the minimum of its potential, was considered in Minkowski as the
target spacetime for strings. The TC process itself
does not require a dynamical departure from Minkowski background.
This is supported by explicit papers \cite{Moeller:2002vx,Aref'eva:2003qu}
and related studies. {Therefore, the rolling tachyon process does not have to be necessarily associated
with an inflation or other spacetime dynamics. The models \cite{Feinstein:2002aj,Leblond:2006cc} which treat the open
string tachyon as the inflaton are often effective phenomenological constructions without a
computational support in the SFT framework.} }

The novel step in this chapter is that we assume a system
of the dilaton and the open string tachyon near TC. In this regime, non-locality
enters through the tachyon potential, without introducing any dynamics
to the tachyon itself (see Appendix \ref{AppSFT} where we presented some review
of SFT, TC and non-locality). In the low energy limit of SFT, the dilaton
and the open string tachyon are coupled through the metric and the dilaton (see c.f. \cite{Koshelev:2007fi} for the so-called ``linear dilaton'' model). We will show
that this regime can only support Minkowski backgrounds as long as the brane tension of the decaying brane is compensated by the open string tachyon vacuum energy and the dilaton is stabilized.
Notwithstanding this and motivated by the fact that any higher energy modification
of this theory introduces higher order couplings between
tachyon and dilaton we claim that in general the model can support an anti-de Sitter or de
Sitter (AdS/dS) backgrounds. We continue by introducing an action that accounts
the higher order couplings of the dilaton and the open string tachyon system near the 
TC point. Although, our proposed action is not systematically derived within SFT,
it is supported by current developments beyond the linear dilaton model \cite{DiVecchia:2015oba}.
%In this paper, we stick to the notions of SFT without supersymmetry.
%Therefore, the question of superstrings is out of discussion for our
%present context.

We study the quadratic variations of our newly introduced action around
dS background which is possible in our model. We observe that dilaton perturbations acquire non-locality
from the infinite derivative terms in the tachyon potential. This
is one of the significant result of this chapter that we attach the
features of non-locality to the dilaton. Here the non-locality of
dilaton is characterized by the function $\mathcal{F}\left(\Box\right)=\overset{\infty}{\underset{n=0}{\sum}}f_{n}\Box^{n}$
where $\Box$ is the d'Alembertian. Depending on the number of roots
of of the characteristic equation $\mathcal{F}\left(z\right)=0$, following the studies of \cite{Koshelev:2007fi,Koshelev:2009ty,Koshelev:2010bf},
we can write the effective actions that are equivalent {to our proposed action} up to the quadratic
perturbations. More specifically, if $\mathcal{F}\left(\Box\right)$
has only one real root at $z_{1}$, the corresponding effective action
contains just one propagating scalar where the kinetic term contains the
parameter $\mathcal{F}^{\prime}\left(z_{1}\right)$ and any higher
derivatives can be neglected assuming the field slow-rolls on a sufficiently
flat potential. As a consequence we obtain a successful single field
inflation with controlled slow-roll dynamics through the parameter
$\mathcal{F}^{\prime}\left(z_{1}\right)$, which leads to the universal prediction of $r<0.09$ without changing $n_{s}=0.967$ for 60 $e$-foldings. If $\mathcal{F}\left(\Box\right)$
has a complex root the corresponding effective action contains two
real scalar fields, which we show to bear conformal invariance. In this
case, the two scalar fields share an opposite sign of kinetic terms.
With spontaneous breaking of conformal symmetry, we gauge fix one
of the scalar field and obtain a Starobinsky like inflation, accompanied
with a non-trivial uplifting of the inflaton potential at the minimum.

This chapter is organized as follows. In Sec.~\ref{sec:SFT1} we discuss
the low energy SFT model and show that it can only provide a Minkowski
background, upon consideration of TC due to brane decay supported
by the details presented in Appendix~\ref{AppSFT}. Then we provide
SFT heuristic motivations for a more generalized action which can
support AdS/dS backgrounds. In Sec.~\ref{sftpert}, given an action
that can support dS solution, we perform perturbations around it and
prove that dilatonic perturbations acquire non-local properties from
the tachyonic part. Then we prescribe a method to write effective
actions depending on the structure of non-locality. We study in detail
two particular effective actions which leads to interesting inflationary
scenarios. In Sec.~\ref{concdisc} we summarize and discuss open
questions which follow from our postulated SFT action and followed
by corresponding inflationary scenarios. 

%Through out the chapter, we set the metric signature $(-,+,+,+)$, small
%Greek letters are the fully covariant indexes and and the units $\hbar=1,\,c=1,\,m_{\rm P}^{2}=\frac{1}{8\pi G}$.

\section{Introducing a framework of SFT for AdS/dS backgrounds }

\label{sec:SFT1}

Before we proceed we refer to the Appendix \ref{AppSFT} for some
review of SFT and tachyon condensation (TC). In this section we start
with the well-known action of a low energy open-closd SFT coupling obtained in the framework of the linear dilaton conformal field theory (see for instance \cite{Aref'eva:2008gj}). We will show by means of a simple
computation that this regime yields only a Minkowski spacetime background as long as the open string tachyon in the minimum of its potential compensates the decaying brane tension and the dilaton field itself is stabilized.
Then we will provide generic SFT motivations to propose a generalized
action which supports AdS/dS solutions which
make it possible to construct effective models of inflation. 
 
\subsection{Low energy open-closed SFT coupling}

From closed SFT, the massless part of action containing dilaton and graviton is given
by \cite{Zwiebach:1993cs,Yang:2005rw} 
\begin{equation}
S_{c}=\int d^{4}x\sqrt{-g}\frac{m_{\rm P}^{2}}{2}e^{-2\phi}\left(R+4\pd_{\mu}\phi\pd^{\mu}\phi\right)\,.\label{action_modelc}
\end{equation}
Here $m_{\rm P}$ is the reduced Planck mass such that $8\pi G=\frac{1}{m_{\rm P}^{2}}$,
with $G$ being the Newtonian constant. The dilaton field $\phi$
is dimensionless. Notice that it is the correct sign for the dilaton
kinetic term as it appears in a closed string spectrum. Action (\ref{action_modelc})
is the zero mass level of the closed strings. We can add to consideration a $p$-form but it enters the action quadratically and we put
it to zero using its equations of motion. Direct SFT based computation
can be done to support the latter action
\cite{Yang:2005iua,Yang:2005ep}.

We however do not include neither the closed string tachyon, nor any
potential for the dilaton. Closed string tachyon, even though it is
in the spectrum of closed strings, seems to condense to a point where
the value of the field is infinite but the potential is zero (not
only its derivative) \cite{Yang:2005rx}. Additionally, it was shown
in \cite{Yang:2005rx} that this vacuum is background independent
exactly due to the fact that the field takes an infinite value. In
such a way, a closed string tachyon does not contribute to our consideration
of subsequent inflation. Regarding the dilaton potential, it was suggested
in \cite{Bergman:1994qq} that apparently no dilaton potential is
generated in the string frame. This claim finds supporting computations in the same paper
and this is known as the dilaton theorem.

Considering the open SFT sector we immediately make use of formula
(\ref{tachyonnearvac}) which is relevant to describe the open string
effects close to the end of an unstable $D$-brane decay\footnote{To avoid confusions we notice that this is in no way the so called
Vacuum SFT (VSFT) \cite{Rastelli:2001uv} but rather a linearization
of the spacetime action derived in \textit{perturbative} SFT near
the bottom of the tachyon potential. VSFT on contrary is a whole new
construction involving a different BRST operator.} due to the open string tachyon $\Tc$. If we couple (\ref{tachyonnearvac}) to dilaton field in a minimal way supported
by the linear dilaton conformal field theory
(see for instance \cite{Aref'eva:2008gj}), we obtain 
\begin{equation}
S_{o}=-\frac{{T}}{2}\int d^{4}x\sqrt{-g}e^{-\phi}[v(\Box,\Tc)+1]\,.\label{action_premodelo}
\end{equation}
The unit term represents the brane tension. This would exactly compensate
the value of the potential at the minimum in a pure open SFT in
Minkowski background where all the computation regarding the Sen's
conjecture were done in a standard SFT approach. However, since the
value of the open string tachyon field in the minimum of the potential is finite,
the minimum should be background-dependent. This means that in a curved
background the energy may not (and most likely will not) be compensated
exactly.
%Moreover, we notice that the power exponent of dilaton is
%1 here and not 2 as in the closed string part of the action, as it
%is suggested by the open SFT in a linear dilaton background \cite{Aref'eva:2008gj}.

Proceeding with a minimal gravitational coupling of (\ref{action_modelc})
and (\ref{action_premodelo}) we get 
\begin{equation}
S=S_{c}+S_{o}=\int d^{4}x\sqrt{-g}\left[\frac{m_{\rm P}^{2}}{2}\left(\Phi^{2}R+4\pd_{\mu}\Phi\pd^{\mu}\Phi\right)-\frac{{T}}{2}\Phi[v(\Box,\Tc)+1]\right].\label{action_model}
\end{equation}
Here we have redefined the dilaton field as $\Phi=e^{-\phi}$. Dilaton
gravity on its own is a well developed subject already for a long
time. See, for instance, \cite{Gasperini:2007ar} for a review. A
careful but quick analysis immediately shows that this latter action
{does not support dS background}. We can easily see that the Minkowski background is 
the only option here {that corresponds to} an exact compensation of the tension
of the initial $D$-brane by the tachyon energy at the bottom of the
potential and the dilaton is a constant. Indeed, varying with respect
to $\Phi$ and seeking for a constant dilaton solution (which cannot
be zero as the true dilaton is $\phi=-\log(\Phi)$) we get 
\begin{equation}
m_{\rm P}^{2}R=\frac{T}{2}[v(\Box,\Tc)+1]\,.\label{someR}
\end{equation}
This latter equation together with the trace of Einstein equations
gives rise to the result that the brane tension must compensate exactly
the tachyon potential value in the minimum and consequently we are
left with $R=0$. This further yields $R_{\mu\nu}=0$. 

\subsection{Action beyond the low-energy open-closed SFT coupling}

In the previous section we learned that the low-energy set-up articulated in (\ref{action_model}) is not
suitable to produce inflation, in which case we essentially require
a presence of a nearly dS background. However, including further terms in (\ref{action_model})
we may expect that other and in particular constant curvature backgrounds are possible. Such
terms may arise from a number of sources: 
\begin{itemize}
\item Once a general (not linear) conformal field theory of the dilaton is considered the above analysis would
not work. New interactions will be generated since the BRST algebra
of the primary fields will get modified. 
\item Open-closed string interactions in general contain higher vertexes
beyond the action above. These contributions generate new vertexes
involving graviton, dilaton and open string tachyon. 
\item The so called ``marginal deformation'' \cite{Yang:2005ep} excitation
in the closed strings. This operator is also of a weight zero but
in fact is non-dynamical at a low-level considerations. However, its
exclusion by equations of motion will generate additional terms to
an effective action as well. 
\end{itemize}

We propose a generalized action that includes new possible interactions of tachyon of open string and the dilaton of closed string:
\begin{equation}
S=\int d^{4}x\sqrt{-g}\left[\frac{m_{\rm P}^{2}}{2}\left(\Phi^{2}R+4\pd_{\mu}\Phi\pd^{\mu}\Phi\right)-
\frac{{T}}{2}\sum_{n=0}^{\infty}\Phi^{n+1}v_{n}\left(\Box,\Tc\right)\right]\,.\label{action_model_new}
\end{equation}
where $R$ is Ricci scalar, ${T}$ is the tension of the D-brane. {Here, the term for $v_0$ is the one appearing in (\ref{action_model}), i.e. $v_0=v(\Box,\,\Tc)+1$. The other terms $v_{n}\left(\Box,\,\Tc\right)$ for $n\geq 1$ correspond to the higher order couplings of the tachyon potential to the dilaton}
which in general depends on infinite number of d'Alembertian operators
$\left(\Box\right)$ based on the concepts of SFT (cf., Appendix~\ref{AppSFT}). 
We assume here it is possible to organize a dimensional reduction
with all moduli fields stabilized so that we are left with an action (\ref{action_model_new}) in $(1+3)$-dimensional
space time. The dimensional reduction of this kind such that an impact of the compactification is absorbed in the overall
action normalization can generically be done in a straightforward way \cite{Yang:2005rw,Yang:2005rx}. The low-energy $p$-brane action obtained from SFT is a good example here \cite{Ohmori:2001am,Arefeva:2001ps}. Also $p$-adic string theory is a model worth mentioning in this regard. It reproduces SFT properties up to and including the tree-level scattering and can be formulated in any dimension \cite{Witten:1986qs,Vladimirov:1994wi}. 

{This latter action is different from (\ref{action_model}) by new terms involving coupling of dilaton and tachyon. First we stress that we aim at establishing whether an inflation is possible in this framework keeping dilaton constant in the vacuum and as such we hunt for constant curvature solutions. This makes irrelevant to consider higher curvature terms. We will comment on this below in the next Section. Second, appearance of an explicit dilaton potential does not contradict the ``dilaton theorem'' claim as this claim was developed in pure closed string framework. Moreover, results of \cite{DiVecchia:2015oba} {indicate} that  the {open-closed SFT coupling} will waive the ``dilaton theorem'' statement. As such, the latter action is a viable attempt to account in full the open-closed strings coupling during the TC process. Explicit computation
of all such extra terms in the action within the pure SFT considerations is
beyond the scope of our present analysis.}

The dilaton is a natural candidate for the inflaton {as the present day understanding
of inflation from the point of view of collected CMB data significantly
favours models where the inflaton is coupled non-minimally to the
Ricci scalar in the action.}
%For example, neither DBI-type models \cite{Feinstein:2002aj,Leblond:2006cc} nor $p$-adic inflation model \cite{Barnaby:2006hi} have this feature.
Inflation via dilaton in (\ref{action_model_new}) can be achieved given that the string scales are higher that the brane tension which in turn is higher than the scale of inflation. In this hierarchy, inflation would start at the final stage of the brane decay.

To support this idea we have to show that action (\ref{action_model_new}) indeed may have a constant
curvature (in particular dS) background solution when dilaton field
takes a constant value and the open string tachyon condenses to its minimum. Varying (\ref{action_model_new}) with respect
to the metric $g_{\mu\nu}$, $\Tc$ and $\Phi$ we can show that the
following configuration is a solution 
\begin{equation}
\Phi=\Phi_{0}=1,~\Tc=\Tc_{0},~g_{\mu\nu}\text{ is dS with }R=R_{0}=2\frac{{T}}{m_{\rm P}^{2}}\sum_{n}v_{n,0}\,,\label{dssol}
\end{equation}
together with the following relations fulfilled 
\begin{equation}
\sum_{n}v_{n,0}^{\prime}=\sum_{n}v_{n,0}(3-n)=0\,,\label{dssolrel}
\end{equation}
where prime $^{\prime}$ is the derivative with respect to an argument
and the subscript $0$ means that the function is evaluated at $\Tc=\Tc_{0}$.
We note that $\Phi_{0}$ can be any value and is irrelevant as long
as it is finite, so we took $\Phi_{0}=1$ for simplicity. We will
pay the special account to the question how generic such configurations
(\ref{dssol}) satisfying (\ref{dssolrel}) arise in SFT in a separate
forthcoming study \cite{Koshelevetal}. We recall from (\ref{action_model})
and (\ref{someR}) that having just a single component $v_{0}\left(\Box,\,\Tc\right)=v\left(\Box,\,\Tc\right)+1$
ends up with necessity with a Minkowski spacetime. Thus, in order
to generate dS spacetime we need at least two terms with different
powers of $\Phi$ in the action.\footnote{Moreover, we notice that the generality of our construction implies
that an appearance of AdS spacetime in which the quantization of
strings is well-defined \cite{Metsaev:2001bj} also requires dilaton
potential terms like in the (SFT inspired) action (\ref{action_model_new}).} 

Hence our proposed modification of linear dilaton in (\ref{action_model_new})
supports dS solution (\ref{dssol}). We stress that our main goal is the retrieval of satisfactory inflation and subsequently
computation of inflationary observables. In the next section
we study the quadratic perturbations of the action (\ref{action_model_new}) and find the effective models of inflation.

\section{Retrieving effective models of inflation}

\label{sftpert}

The quadratic variation of our background action (\ref{action_model_new})
can be written as two parts in the following way

\begin{equation}
\delta^{(2)}S=\delta^{(2)}S_{m_{\rm P}^{2}}+\delta^{(2)}S_{int}\,.\label{d2sfteffmp2}
\end{equation}

The perturbative modes are $\varphi=\delta\Phi$, trace of the metric perturbations
$h$ (we define $\delta g_{\mu\nu}=h_{\mu\nu}$, $h=h_{\mu}^{\mu}$)
and $\tau=\delta\Tc$.
Generically, different spins do not mix in the quadratic action i.e.,
tensor modes do not mix with scalar modes. Therefore, the first part
of the quadratic action reads
\begin{equation}
\delta^{(2)}S_{m_{\rm P}^{2}}=\int d^{4}x\sqrt{-g}\frac{m_{\rm P}^{2}}{2}\left[\varphi^{2}R_{0}+4\pd\varphi^{2}-\frac{3}{32}h\left(\Box+\frac{R_{0}}{3}\right)h-\frac{3}{2}\varphi\left(\Box+\frac{R_{0}}{3}\right)h\right]\,.\label{d2sft-free}
\end{equation}
From the above action we can exclude $h$ from its equation of motion.
Due to the fact that differential operators acting on $h$ and
$\varphi$ are identical, we have $h=-8\varphi+h_{hom}$ where
$(\Box+R_{0}/3)h_{hom}=0$. Substituting this $h$ back in the quadratic
action yields 
\begin{equation}
\delta^{(2)}S_{m_{\rm P}^{2}}=\int d^{4}x\sqrt{-g}\frac{m_{\rm P}^{2}}{2}\varphi\left(2\Box+3R_{0}\right)\varphi\,.\label{d2sfteffmp2phi}
\end{equation}
The second part of the quadratic action after a Taylor expansion of
the tachyon potential $v\left(\Box,\,\Tc\right)$ around $\Tc=\Tc_{0}$
reads

\begin{equation}
\delta^{(2)}S_{int}=-\frac{T}{2}\int d^{4}x\sqrt{-g}\sum_{n}\left[(n+1)n\varphi^{2}v_{n,0}+nv'_{n,0}\varphi f(\Box)\tau+\frac{v''_{n,0}}{2}\tau e^{\gamma(\Box)}\tau\right]\,,\label{d2sft-int}
\end{equation}
where we have used (\ref{tachyonvac}). Accounting the fact that the
open string tachyon on its own is not dynamical, the function $\gamma\left(\Box\right)$
in the exponent must be an entire function but the operator $f(\Box)$
may have eigenvalues. Excluding $\tau$ by its equation of motion
is dictated by $\tau=-\frac{\sum_{n}\left(nv'_{n,o}\right)}{\sum_{n}v''_{n,0}}f(\Box)e^{-\gamma(\Box)}\varphi$.
Substituting this back into action (\ref{d2sft-int}) yields 
\begin{equation}
\delta^{(2)}S_{int}=-\frac{T}{2}\int d^{4}x\sqrt{-g}\varphi\left[\sum_{n}\left((n+1)nv_{n,0}\right)-\frac{\left(\sum_{n}nv'_{n,0}\right){}^{2}}{2\sum_{n}v''_{n,0}}f(\Box)^{2}e^{-\gamma(\Box)}\right]\varphi\,.\label{d2sfteffintphi}
\end{equation}
It is clear from the above formulae that higher curvature corrections
are not relevant for us. Indeed, suppose there is a term in the action
like 
$
\sqrt{-g}\Phi^{2}R^{2}
$,
such a term would produce contributions to $h^{2}$ and $\varphi h$
but as long as our background has constant scalar curvature and constant
dilaton field the final effect of such an additional term would be
just renormalization of constants in action (\ref{d2sfteffmp2phi}).
We see that both the spin-0 excitation of the metric and
the dilaton field are combined into one joint scalar mode. Again,
we can show by explicit computation that including other interactions,
like for instance 
$
\sqrt{-g}\Phi^{2}R^{2}w(\Box,\tau)\,,
$
will result in the same net result when all but one scalar fields
can be excluded by equations of motion which finally results in a
single (non-local) scalar excitation.\footnote{We here note that additional contributions to scalar and tensor modes
can be generated by means of adding the curvature squared corrections,
like $R_{\mu\nu}^{2}$ or $C^{2}$ where $C$ is the Weyl tensor Moreover,
following the recent studies performed in \cite{Koshelev:2016xqb,Biswas:2016egy}
one has to pay special attention in order to maintain unitarity upon
inclusion of terms which modify the Lagrangian for tensor modes beyond
the Einstein's gravity. A standard minimal structure like $C^{2}$
in the action will generate a massive spin-2 ghost (see \cite{Stelle:1976gc}
for the first comprehensive study of this question). We therefore
leave the full consideration as an open question.}

We established above
why our proposal (\ref{action_model_new}) provides a framework to
generate a dS background and we will demonstrate how it can describe inflationary effects, which require
the second variation of the action around such a background. We recall
here that the open string sector contains only the tachyon, since
higher mass fields have been integrated out, in the course of the
brane decay consideration (cf. the Appendix \ref{AppSFT}). %\subsection{dS phase and quadratic action}
Thus in the nearly dS phase when the scalar curvature does not change
considerably, we get from (\ref{d2sfteffmp2}), (\ref{d2sfteffmp2phi})
and (\ref{d2sfteffintphi}) the following action that describes the
propagation of scalar perturbations 
\begin{equation}
\delta^{(2)}S=\frac{1}{2}\int d^{4}x\sqrt{-g}\varphi\Fc(\Box)\varphi\,,\label{d2sfteffFbox}
\end{equation}
where 
\begin{equation}
\Fc(\Box)={m_{\rm P}^{2}}\left(2\Box+3R_{0}\right)-T\left[\sum_{n}\left((n+1)nv_{n,0}\right)-\frac{\left(\sum_{n}nv'_{n,0}\right){}^{2}}{2\sum_{n}v''_{n,0}}f(\Box)^{2}e^{-\gamma(\Box)}\right]\,.\label{Fboxds}
\end{equation}

To generate inflation we must have an appropriate
potential in our set-up. The linearization of (\ref{action_model_new}) and corresponding
analysis do not shed light on the form of the potential though. Rigorously
speaking, a potential would follow from SFT provided we have computational
abilities to extract one. At present, the state of the art of the
knowledge in SFT lacks established methods to do so. In the course
of this chapter we will continue by assuming potentials which do not
violate general principles of SFT construction (cf. the Appendix \ref{AppSFT}
for more discussions on this issue). This strategy can be reversed
and be used to constrain perhaps certain parameters in SFT, given
we will reach eventually the ability to do such computations directly
in the SFT framework.

Considering action (\ref{d2sfteffFbox}) for a general operator function
$\Fc(\Box)$ we cannot convey inflationary physics straightforwardly.
In general, $\Fc(\Box)$ being considered as an algebraic function
may have many roots. That is, equation 
\begin{equation}
\Fc(z)=0\label{characteristic}
\end{equation}
can have more than one solution. We name it a characteristic equation.
Because of that, the propagator for the field $\varphi$ will have
more than one pole. As such, it is equivalent to multiple degrees
of freedom. Let us therefore write a local realization of (\ref{d2sfteffFbox}).
Originally, this was done in \cite{Koshelev:2007fi} and then formalized
in \cite{Aref'eva:2007mf,Koshelev:2009ty,Koshelev:2010bf}. We use
the Weierstrass factorization \cite{Koshelev:2007fi} which prescribes
that any entire function (we recall that SFT ensures that operators
$\Fc(\Box)$ are analytic functions and in all existing computations they appear to be entire functions) can be written as 
\begin{equation}
\Fc(z)=e^{\gamma(z)}\prod_{j}\left(z-z_{j}\right){}^{m_{j}}\,,\label{weierstrass}
\end{equation}
where $z_{j}$ are roots of the characteristic equation and $m_{j}$
are their respective multiplicities. We assume hereafter that all
$m_{j}=1$ for simplicity. $\gamma(z)$ is an entire function and
as such its exponent has no roots on the whole complex plane. It was
shown in \cite{Koshelev:2007fi} that for a quadratic Lagrangian of
the type (\ref{d2sfteffFbox}), local equivalent quadratic Lagrangian
can be constructed as 
\begin{equation}
\delta^{2}S_{local}=\frac{1}{2}\int d^{4}x\sqrt{-g}\sum_{j}\Fc'\left(z_{j}\right)\varphi_{j}\left(\Box-z_{j}\right)\varphi_{j}\label{locallocal}
\end{equation}
where prime means derivative with respect to the argument $z$ with
the further evaluation at the point $z_{j}$. It is said to be equivalent,
thanks to the fact that solution for $\varphi$ which can be obtained
from equations of motion following from (\ref{d2sfteffFbox}) is connected
to solutions for $\varphi_{j}$ simply 
\begin{equation}
\varphi=\sum\varphi_{j}\,.\label{solsol}
\end{equation}

Roots $z_{j}$ become the most crucial objects in classifying our
model. Several comments are in order here: 
\begin{itemize}
\item Note that roots $z_{j}$ can be complex in general. One real root
$z_{1}$ is the simplest situation. In this case, we have just a Lagrangian
for a massive scalar. It is acceptable if $\Fc'(z_{1})>0$ in order
to evade a ghost in the spectrum.
\item More than one real root apparently seems not to be a promising scenario.
Since the function $\Fc(z)$ is analytic (and therefore continuous),
neighbouring real roots will be accompanied with $\Fc'\left(z_{j}\right)$
of opposite signs. In other words, one root is normal and the next
to it is a ghost.
\end{itemize}

\subsection{Effective model of single field inflation}

If $\mathcal{F}\left(z\right)$ has one real root, then (\ref{locallocal})
contains a single scalar degrees of freedom 

\begin{equation}
\delta^{2}S_{local}=\frac{1}{2}\int d^{4}x\sqrt{-g}\Fc'\left(z_{1}\right)\varphi\left(\Box-z_{j}\right)\varphi\,.\label{eff-single-field}
\end{equation}
The effective action which is perturbatively equivalent up to quadratic
order to (\ref{eff-single-field}) around dS background, looks
like (taking $m_{\rm P}=1$)

\begin{equation}
S_{1}=\int d^{4}x\sqrt{-g}\left[\frac{1}{2}\tilde{\Phi}^{2}R-\frac{A}{2}\pd\tilde{\Phi}^{2}-V(\tilde{\Phi})\right]\,,\label{example1}
\end{equation}
where $\tilde{\Phi}$ is an effective dilatonic field and the respective correspondence is 

\begin{equation}
\begin{split}\Fc'(z_{1}) & =6+A\\
\Fc'(z_{1})z_{1} & =3R_{0}-V''\left(\tilde{\Phi}_{0}\right)\,.
\end{split}
\label{identification}
\end{equation}
Here $ R_{0} $ is scalar curvature of the dS vacuum solution for a constant $\tilde{\Phi}$. 
%\[
%\tilde{\Phi}=\tilde{\Phi}_{0}\text{ and }R_{0}=4\frac{V(\tilde{\Phi}_{0})}{\tilde{\Phi}_{0}^{2}}=\frac{V'(\tilde{\Phi}_{0})}{\tilde{\Phi}_{0}}>0\,.
%\]
Assuming the generalized structure of from the proposed action (\ref{action_model_new}),
the potential $V(\tilde{\Phi})$ can be taken to be arbitrary. 
%The action (\ref{example1}) in the Einstein frame looks like 
%\begin{equation}
%{S}_{1E}=\int d^{4}x\sqrt{-g_{E}}\left[\frac{1}{2}R_{E}-\frac{A+6}{2\tilde{\Phi}^{2}}\left(\partial\tilde{\Phi}\right)^{2}-V_{E}\left(\tilde{\Phi}\right)\right]\,,\label{SFTonerootEin}
%\end{equation}
%where the Einstein frame potential is $V_{E}\left(\tilde{\Phi}\right)=\frac{V_{J}\left(\tilde{\Phi}\right)}{\tilde{\Phi}^{4}}$
%. Considering a generic Jordan frame potential as 
%\begin{equation}
%V_{J}\left(\tilde{\Phi}\right)=\underset{n=2}{\overset{\infty}{\sum}}c_{n}\tilde{\Phi}^{2n}\,.\label{genericpot}
%\end{equation}
%In terms of a canonically normalized field %$\tilde{\Phi}=e^{-\sqrt{\frac{1}{A+6}}\tilde{\phi}}$
%the Einstein frame potential (\ref{genericpot}) takes the form 
%\begin{equation}
%V_{E}=V_{0}\underset{n=2}{\overset{\infty}{\sum}}\left(1-e^{-(2n-2)\sqrt{\frac{1}{A+6}}\tilde{\phi}}\right)\,,\label{plateaupot}
%\end{equation}
%where we have taken $c_{2}=-c_{n-1}$. The potential in (\ref{plateaupot})
%becomes exponentially flat in the limit $\tilde{\phi}\gg1$ making
%it suitable for inflation and in principle, we can omit the higher
%order terms keeping only the leading contributions in (\ref{plateaupot}).
If we consider a potential $V_{J}\left(\tilde{\Phi}\right)=V_{0}\left(-\tilde{\Phi}^{2}+\tilde{\Phi}^{4}\right)^{2}$ which looks in the Einstein frame as 
\begin{equation}
V_{E}=\tilde{V}_{0}\left(1-e^{-\sqrt{\frac{2}
{3\left[\mathcal{F}^{\prime}\left(z_{1}\right)/6\right]}}
\tilde{\phi}}\right)^{2}\,,\label{potEmodel}
\end{equation}
where $\tilde{\phi}$ is canonically normalized field by definining $\tilde{\Phi}=e^{-\sqrt{\frac{1}{A+6}}\tilde{\phi}}$. 
The inflationary predictions corresponding to
the potential in (\ref{potEmodel}) are well known \cite{Ellis:2013nxa,Kallosh:2013yoa,Kehagias:2013mya,Carrasco:2015pla} and in particular we retrieve 
\[
n_{s}=1-\frac{2}{N}\quad,\quad r=\frac{2\mathcal{F}^{\prime}\left(z_{1}\right)}{N^{2}}\,, \label{sft-attractor}
\]
where we consider  $N=60$ number of $e$-foldings. 
We therefore conclude that provided the non-local operator $\Fc(\Box)$
contains one real root, it gives a successful inflation with a universal
prediction of $n_{s}=0.967$ and the tensor
to scalar ratio $r<0.09$. The value of $r$ can be varied to any value
by varying the non-local parameter $\mathcal{F}^{\prime}\left(z_{1}\right)$.

\subsection{Effective model of conformal inflation}

If $\mathcal{F}\left(z\right)$ has a complex root then we should
write action (\ref{locallocal}) for a scalar field and also for its
complex conjugate. So considering such a pair of complex conjugate
roots, we have 
\begin{equation}
\delta^{2}S_{local}=\frac{1}{2}\int d^{4}x\sqrt{-g}\left[\Fc'\left(z_{1}\right)\varphi_{1}\left(\Box-z_{1}\right)\varphi_{1}+\Fc'\left(\bar{z}_{1}\right)\bar{\varphi}_{1}\left(\Box-\bar{z}_{1}\right)\bar{\varphi}_{1}\right]\,,\label{pairpair}
\end{equation}
where a bar over represents the complex conjugates. To maintain the
connection with the original action (\ref{d2sfteffFbox}) we should
consider complex conjugate solutions to equations of motion, such
that $\varphi=\varphi_{1}+\bar{\varphi}_{1}$ is real. The important
feature is that the quadratic form of fields is already diagonal.
Introducing $\varphi_{1}=\chi+i\sigma$, $z_{1}=\alpha+i\beta$, $\Fc'(z_{1})=c+is$
we can rewrite action (\ref{pairpair}) in terms of real components
as 
\begin{equation}
\delta^{2}S_{local}=\int d^{4}x\sqrt{-g}\left[\chi(c\Box-c\alpha+s\beta)\chi-\sigma(c\Box-c\alpha+s\beta)\sigma-2\chi(s\Box-s\alpha-c\beta)\sigma\right]\,.\label{pairpairreal}
\end{equation}
The above action is inevitably non-diagonal and features a cross-product
of real fields $\sim\chi\sigma$. In the formulation above, note that
the two fields $\chi,\,\sigma$ share a opposite sign of kinetic term
\cite{Galli:2010qx}.  We will show that the following effective action
of two fields with conformal invariance can be perturbatively equivalent
up to quadratic order to (\ref{pairpairreal}) around dS background
\begin{equation}
\begin{split}S_{2}=\int d^{4}x\sqrt{-g} & \left[\frac{m_{\rm P}^{2}}{2}[\tilde{\alpha}\tilde{\Phi}_{1}^{2}-\tilde{\alpha}\tilde{\Phi}_{2}^{2}{-}2\tilde{\beta}\tilde{\Phi}_{1}\tilde{\Phi}_{2}]f\left(\frac{\tilde{\Phi}_{2}}{\tilde{\Phi}_{1}}\right)R\right.\\
 & +\left.\frac{A}{2}[\tilde{\alpha}\pd\tilde{\Phi}_{1}^{2}-\tilde{\alpha}\pd\tilde{\Phi}_{2}^{2}-2\tilde{\beta}\pd_{\mu}\tilde{\Phi}_{1}\pd^{\mu}\tilde{\Phi}_{2}]f\left(\frac{\tilde{\Phi}_{2}}{\tilde{\Phi}_{1}}\right)-V\left(\tilde{\Phi}_{1},\tilde{\Phi}_{2}\right)\right]\,.
\end{split}
\label{example2}
\end{equation}
where $\tilde{\Phi}_{1},\,\tilde{\Phi}_{2}$ are effective dilatonic fields. 
%and Einstein equations we can obtain a dS solution in the form 
%\[
%\begin{split}\tilde{\Phi}_{1} & =\tilde{\Phi}_{1,0},~\tilde{\Phi}_{2}=\tilde{\Phi}_{2,0}\\
%R_{0} & =4\frac{V_{0}}{m_{\rm P}^{2}I_{0}}=\frac{2\pd_{\tilde{\Phi}_{1}}V_{0}}{m_{\rm P}^{2}\pd_{\tilde{\Phi}_{1}}I_{0}}=\frac{2\pd_{\tilde{\Phi}_{2}}V_{0}}{m_{\rm P}^{2}\pd_{\tilde{\Phi}_{2}}I_{0}}>0\,,
%\end{split}
%\]

We can write the quadratic Lagrangian for the spin-0 part which contains
2 components $\tilde{\chi}{=\delta\tilde{\Phi}_{1}}$ and $\tilde{\sigma}{={\delta\tilde{\Phi_{2}}}}$
(i.e. again the spin-0 metric perturbation is excluded by equations
of motion), as 
\begin{equation}
\begin{split}\delta^{2}S_{2}=\frac{1}{2}\int d^{4}x\sqrt{-g} & \left[\tilde{\chi}\Delta_{\tilde{\chi}}\tilde{\chi}+\tilde{\sigma}\Delta_{\tilde{\sigma}}\tilde{\sigma}+\tilde{\chi}\Delta_{\tilde{\chi}\tilde{\sigma}}\tilde{\sigma}\right]\end{split}
\,,\label{s2eff}
\end{equation}
where 
\begin{eqnarray*}
\Delta_{\tilde{\chi}} & = & \frac{m_{\rm P}^{2}}{2}\left(\frac{(\pd_{\tilde{\Phi}_{1}}I_{0})^{2}}{I_{0}}(3\Box+R_{0})+\frac{\pd^{2}I_{0}}{\pd\tilde{\Phi}_{1}^{2}}R_{0}\right)-A\tilde{\alpha}f_{0}\Box-\frac{\pd^{2}V_{0}}{\pd\tilde{\Phi}_{1}^{2}}\,,\\
\Delta_{\tilde{\sigma}} & = & \frac{m_{\rm P}^{2}}{2}\left(\frac{(\pd_{\tilde{\Phi}_{2}}I_{0})^{2}}{I_{0}}(3\Box+R_{0})+\frac{\pd^{2}I_{0}}{\pd\tilde{\Phi}_{2}^{2}}R_{0}\right)+A\tilde{\alpha}f_{0}\Box-\frac{\pd^{2}V_{0}}{\pd\tilde{\Phi}_{2}^{2}}\,,\\
\Delta_{\tilde{\chi}\tilde{\sigma}} & = & \frac{m_{\rm P}^{2}}{2}\left(\frac{\pd_{\tilde{\Phi}_{1}}I_{0}\pd_{\tilde{\Phi}_{2}}I_{0}}{I_{0}}(3\Box+R_{0})+\frac{\pd^{2}I_{0}}{\pd\tilde{\Phi}_{1}\pd\tilde{\Phi}_{2}}R_{0}\right)-A\tilde{\beta}f_{0}\Box-\frac{\pd^{2}V_{0}}{\pd\tilde{\Phi}_{1}\pd\tilde{\Phi}_{2}}\,,
\end{eqnarray*}
where $R_{0}$ is the scalar curvature of dS vacuum for constant dilatonic fields $\tilde{\Phi}_{1}=\tilde{\Phi}_{1,0},\,\tilde{\Phi}_{2}=\tilde{\Phi}_{2,0}$. Here we define $I(\tilde{\Phi}_{1},\tilde{\Phi}_{2})=\left[\tilde{\alpha}\tilde{\Phi}_{1}^{2}-\tilde{\alpha}\tilde{\Phi}_{2}^{2}{-}2\tilde{\beta}\tilde{\Phi}_{1}\tilde{\Phi}_{2}\right]f\left({\tilde{\Phi}_{2}}/{\tilde{\Phi}_{1}}\right)$
and $I_{0}\equiv I(\tilde{\Phi}_{1,0},\tilde{\Phi}_{2,0})$, 
$\pd_{\tilde{\Phi}_{1}}I_{0}\equiv\pd I(\tilde{\Phi}_{1},\tilde{\Phi}_{2})/\pd\tilde{\Phi}_{1}$
are the quantities evaluated at the values of fields at dS vacuum and so on for analogous terms. 

We can make use of (\ref{pairpairreal}), which is the case of two
complex conjugate roots with the Lagrangian written in real fields.
Hence, we can try to juxtapose (\ref{pairpairreal}) and (\ref{s2eff}).
The motivation for doing this is to establish a more fundamental correspondence
for the effective model (\ref{example2}). This is, however much more
involved than in the previous Section with a single field. Essentially,
the most important is to establish $\Delta_{\tilde{\chi}}=-\Delta_{\tilde{\sigma}}$.
On this way, we can neglect the second derivatives of the potential
$V$. However, we must satisfy a number of constraints, namely, all
parameters and vacuum fields values must be real and $I_{0}$ strictly
positive. And we want to have $\tilde{\beta}\neq0$, which we will
explain why in the following. The greatly simplifying point is that
we must require such an adjustment of coefficients of $\Delta$-s
only in a single point $(\tilde{\Phi}_{1}=\tilde{\Phi}_{1,0}$, $\tilde{\Phi}_{2}=\tilde{\Phi}_{2,0})$.
On top of this we emphasize once again that we aim at retrieving a
nearly dS phase, not an exact one. These requirements are generically
satisfied altogether with the presence of a function $f\left(\frac{\tilde{\Phi}_{2}}{\tilde{\Phi}_{1}}\right)$
(apart from special situations which we discuss shortly). It is important
that being a function of the ratio of fields it cannot spoil a possible
conformal invariance.

Let us recall that our main purpose in this Section is to establish
an effective setting which can emulate (\ref{pairpairreal}). We claim
that we have such an effective model as long we can match quadratic
actions for scalar modes around a dS background. We can thus establish
a correspondence between (\ref{pairpairreal}) and (\ref{s2eff})
by means of the following: 
\begin{itemize}
\item During inflationary expansion we can assume that the scalar fields
varies slowly and the kinetic terms can be neglected. We are thus
mainly interested in whether $\Delta_{\tilde{\chi}}=-\Delta_{\tilde{\sigma}}$
for the terms proportional to $R_{0}$. To have this we should require
\begin{equation}
\frac{(\pd_{\tilde{\Phi}_{1}}I_{0})^{2}}{I_{0}}+\frac{\pd^{2}I_{0}}{\pd\tilde{\Phi}_{1}^{2}}+\frac{(\pd_{\tilde{\Phi}_{2}}I_{0})^{2}}{I_{0}}+\frac{\pd^{2}I_{0}}{\pd\tilde{\Phi}_{2}^{2}}\approx0\,.\label{condr0}
\end{equation}
\item We can check that even in the very simple case of $\tilde{\beta}=0$,
a non-constant function $f$ is required to satisfy the above relation.
A simple choice like 
\begin{equation}
f=1+f_{1}\tilde{\Phi}_{2}/\tilde{\Phi}_{1}\,,\label{condr0f}
\end{equation}
with just one free parameter $f_{1}$ is sufficient. Otherwise, for
$f=\const$ a condition $\tilde{\Phi}_{1,0}=\pm i\tilde{\Phi}_{2,0}$
arises from (\ref{condr0}). Therefore to build such an effective
model the function $f\left(\frac{\tilde{\Phi}_{2}}{\Phi_{1}}\right)$
is very useful and important. The cross-product of fields may arise
for $\tilde{\beta}=0$ but a quite involved non-polynomial function
$f$ is required. 
\item For a non-trivial $\tilde{\beta}$ the same function $f$ as above
in (\ref{condr0f}) is enough to arrange the condition (\ref{condr0}).
Moreover $\tilde{\beta}\neq0$ generates a cross-product of fields. 
\item In complete analogy we can consider the coefficients of the kinetic
terms. We have to require a non-constant function $f$. We note that
having opposite coefficients in front of d'Alembertian operators for
different fields essentially means that one of these fields is a ghost. 
\end{itemize}
Recalling expressions (\ref{locallocal}) and (\ref{pairpairreal}),
we see that the presence of a cross-product is a special feature related
to a complex root of the function $\Fc(z)$ (which defines the non-local
operator $\Fc(\Box)$). This means that the parameter $\beta$ found
in (\ref{pairpairreal}) is essentially non-zero (notice that there
is no a direct simple relation between $\tilde{\beta}$ and $\beta$).
In the limiting case of $\beta\to0$, we should see the cross-product
disappearing and this corresponds to $\tilde{\beta}\to0$ in the effective
model (\ref{example2}). Another way to recognize the effective model
(\ref{example2}) without a cross-product of fields is to consider
directly (\ref{locallocal}) with two specially tuned real roots.
This means that these roots are related as $z_{2}=-z_{1}$ and moreover
$\Fc'(z_{2})=-\Fc'(z_{1})$.

To resolve the issue of a ghost in the spectrum requires an extra
symmetry in order to gauge the ghost away. The most natural candidate
is the conformal symmetry used in the building of similar models in
\cite{Kallosh:2013xya,Kallosh:2013pby,Kallosh:2013yoa}. The conformal
invariance is restored in (\ref{example2}) if we assume $A=6$. Our
model without a cross-product resembles the conformal models studied
in \cite{Kallosh:2013hoa,Kallosh:2013daa}. We stress that the cross-product
appeared for the first time in the cosmological models and we have
here provided an imperative explanation through the non-local dilaton.

Assuming $f\left(\frac{\tilde{\Phi}_{2}}{\tilde{\Phi}_{1}}\right)\approx\mathrm{constant}$
during inflation (\ref{example2}) can be written as
\begin{equation}
\begin{split}S_{2}=\int d^{4}x\sqrt{-g} & \left[\left(\tilde{\alpha}\tilde{\Phi}_{1}^{2}-\tilde{\alpha}\tilde{\Phi}_{2}^{2}{-}2\tilde{\beta}\tilde{\Phi}_{1}\tilde{\Phi}_{2}\right)\frac{R}{12}\right.\\
 & +\left.\frac{\tilde{\alpha}}{2}\pd\tilde{\Phi}_{1}^{2}-\frac{\tilde{\alpha}}{2}\pd\tilde{\Phi}_{2}^{2}-\tilde{\beta}\pd_{\mu}\tilde{\Phi}_{1}\pd^{\mu}\tilde{\Phi}_{2}-V_{J}\left(\tilde{\Phi}_{1},\tilde{\Phi}_{2}\right)\right]\,,
\end{split}
\label{Cinf}
\end{equation}
where we have set $m_{\rm P}=1$ for simplicity and use the subscript
$J$ for the Jordan frame as before. Since the field $\tilde{\Phi}_{1}$
has a wrong sign kinetic term (assuming $\tilde{\alpha}>0$), we can
eliminate it by the choice of conformal gauge $\tilde{\Phi}_{1}=\sqrt{6}$
which spontaneously breaks the conformal invariance. To obtain a consistent
inflation within this model we consider the following potential 
\begin{equation}
V_{J}\left(\tilde{\Phi}_{1},\,\tilde{\Phi}_{2}\right)=\frac{\lambda}{4}\left(\gamma_{1}\tilde{\Phi}_{2}^{2}+\gamma_{2}\tilde{\Phi}_{1}\tilde{\Phi}_{2}+\gamma_{3}\tilde{\Phi}_{1}^{2}\right)\left(\tilde{\Phi}_{2}-\tilde{\Phi}_{1}\right)^{2}\,,\label{inflatonpotCI}
\end{equation}
where $\gamma_{1}$, $\gamma_{2}$, $\gamma_{3}$ are arbitrary constant
parameters. The potential (\ref{inflatonpotCI}) is motivated from
\cite{Kallosh:2013xya}, which we generalize here to our conformal
model with a term containing the cross-product of fields. The importance
of this generalization will be explained in what follows. Note that
if $\tilde{\beta}=\gamma_{2}=\gamma_{3}=0$ , the model reduces to
the conformal model without a cross-product of fields studied in \cite{Kallosh:2013xya}.

Rescaling the fields as $\tilde{\Phi}_{1}\to\frac{\tilde{\Phi}_{1}}{\sqrt{\tilde{\alpha}}}$
and $\tilde{\Phi}_{2}\to\frac{\tilde{\Phi}_{2}}{\sqrt{\tilde{\alpha}}}$
in action (\ref{Cinf}) and using the gauge $\tilde{\Phi}_{1}=\sqrt{6}$
we yield

\begin{equation}
\begin{split}S_{2}=\int d^{4}x\sqrt{-g} & \left[\frac{R}{2}\left(1-\frac{\tilde{\Phi}_{2}^{2}}{6}-\frac{2\tilde{\beta}}{\sqrt{6}\tilde{\alpha}}\tilde{\Phi}_{2}\right)\right.\\
 & \left.-\frac{1}{2}\partial_{\mu}\tilde{\Phi}_{2}\partial^{\mu}\tilde{\Phi}_{2}-\frac{\lambda}{4\tilde{\alpha}^{2}}\left(\gamma_{1}\tilde{\Phi}_{2}^{2}+\gamma_{2}\tilde{\Phi}_{1}\tilde{\Phi}_{2}+\gamma_{3}\tilde{\Phi}_{1}^{2}\right)\left(\tilde{\Phi}_{2}-\sqrt{6}\right)^{2}\right]\,.
\end{split}
\label{sft3}
\end{equation}

%We can rewrite the latter action as

%\begin{equation}
%\begin{split}S_{3}=\int d^{4}x\sqrt{-g} & \left[\frac{R}{2}\left[1+\frac{\tilde{\beta}^{2}}{\tilde{\alpha}^{2}}-\frac{1}{6}\left(\tilde{\Phi}_{2}+\frac{\tilde{\beta}}{\tilde{\alpha}}\sqrt{6}\right)^{2}\right]\right.\\
% & \left.-\frac{1}{2}\partial_{\mu}\tilde{\Phi}_{2}\partial^{\mu}\tilde{\Phi}_{2}-\frac{\lambda}{4\tilde{\alpha}^{2}}\left(\gamma_{1}\tilde{\Phi}_{2}^{2}+\gamma_{2}\tilde{\Phi}_{1}\tilde{\Phi}_{2}+\gamma_{3}\tilde{\Phi}_{1}^{2}\right)\left(\tilde{\Phi}_{2}-\sqrt{6}\right)^{2}\right]\,.
%\end{split}
%\label{sft4}
%\end{equation}

Performing the conformal transformation $g_{\mu\nu}\to\left[1+\frac{\tilde{\beta}^{2}}{\tilde{\alpha}^{2}}-\frac{1}{6}\left(\tilde{\Phi}_{2}
+\frac{\tilde{\beta}}{\tilde{\alpha}}\sqrt{6}\right)^{2}\right]^{-1}g_{\mu\nu}$
and shifting the field $\tilde{\Phi}_{2}\to\tilde{\Phi}_{2}+\frac{\tilde{\beta}}{\tilde{\alpha}}\sqrt{6}$,
we arrive to the Einstein frame action 
\begin{equation}
S_{2E}=\int d^{4}x\,\sqrt{-g_{E}}\left[\frac{R_{E}}{2}-\frac{\omega}{2\left(\omega-\frac{\tilde{\Phi}_{2}^{2}}{6}\right)^{2}}\partial_{\mu}\tilde{\Phi}_{2}\partial^{\mu}\tilde{\Phi}_{2}-V_{E}\left(\tilde{\Phi}_{2}\right)\right]\,,\label{sft5}
\end{equation}
where $\omega=1+\frac{\tilde{\beta}^{2}}{\tilde{\alpha}^{2}}$ and
\begin{equation}
V_{E}\left(\tilde{\Phi}_{2}\right)=\frac{9\lambda}{\tilde{\alpha}^{2}}\frac{\left[\gamma_{1}\tilde{\Phi}_{2}^{2}+\left(\gamma_{2}-2\gamma_{1}\frac{\tilde{\beta}}{\tilde{\alpha}}\right)\sqrt{6}\tilde{\Phi}_{2}+6\left(\gamma_{1}\frac{\tilde{\beta}^{2}}{\alpha^{2}}-\gamma_{2}\frac{\tilde{\beta}}{\tilde{\alpha}}+\gamma_{3}\right)\right]\left(\tilde{\Phi}_{2}-\sqrt{6}\frac{\tilde{\beta}}{\tilde{\alpha}}-\sqrt{6}\right)^{2}}{\left(6\omega-\tilde{\Phi}_{2}^{2}\right)^{2}}\,.\label{finalEpot}
\end{equation}
If $\gamma_{i}$ are chosen such that $\gamma_{2}=2\gamma_{1}\frac{\tilde{\beta}}{\tilde{\alpha}}$
and $\gamma_{1}\frac{\tilde{\beta}^{2}}{\alpha^{2}}-\gamma_{2}\frac{\tilde{\beta}}{\tilde{\alpha}}+\gamma_{3}\gtrsim0$,
we can obtain inflation with an uplifting of the potential at the
minimum.

\begin{figure}[h!]
\centering \includegraphics[height=2in]{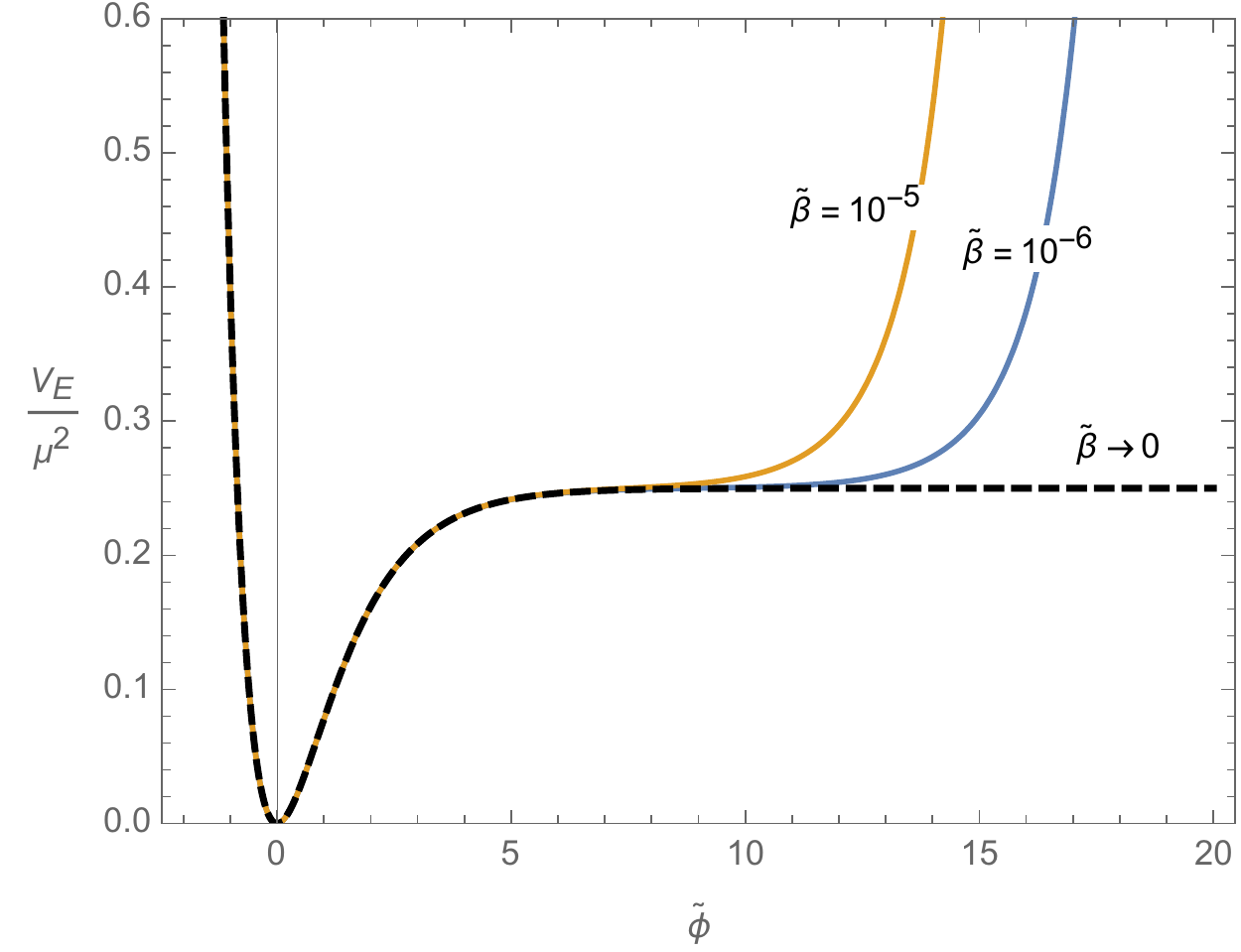}\quad{}\includegraphics[height=2in]{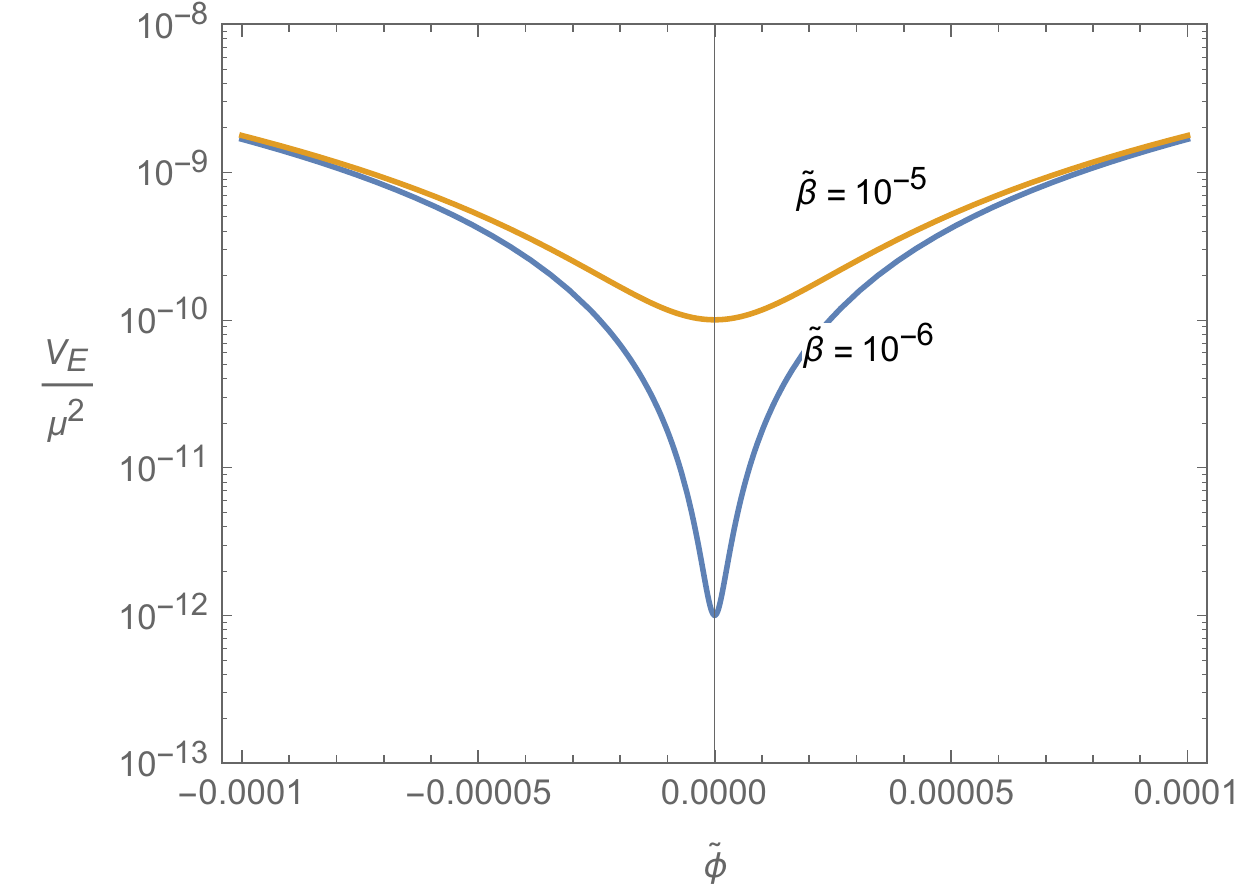}
\caption{{In the left panel we plot the potential $V_{E}\left(\tilde{\phi}\right)$
for values of $\tilde{\beta}=10^{-5},\,10^{-6}$ and $\tilde{\alpha}=1$.
In the right panel, we depict the corresponding minimum of the potential
around $\tilde{\phi}\approx0$. }}
\label{plot-pot} 
\end{figure}

For example, let us consider a simple case with $\gamma_{1}=1$ ,
$\gamma_{2}=2\frac{\tilde{\beta}}{\tilde{\alpha}}$ and $\gamma_{3}=2\frac{\tilde{\beta}^{2}}{\tilde{\alpha}^{2}}$
, for which (\ref{finalEpot}) reduces to the following form interms of canonically normalized field $\tilde{\Phi}_{2}=\sqrt{6\omega}\tanh\left(\frac{\tilde{\phi}}{\sqrt{6}}\right)$ as
%\begin{equation}
%V_{E}\left(\tilde{\Phi}_{2}\right)=\frac{9\lambda}{\tilde{\alpha}^{2}}\frac{\left[\tilde{\Phi}_{2}^{2}+6\frac{\tilde{\beta}^{2}}{\tilde{\alpha}^{2}}\right]\left(\tilde{\Phi}_{2}-\sqrt{6}\frac{\tilde{\beta}}{\tilde{\alpha}}-\sqrt{6}\right)^{2}}{\left(6\omega-\tilde{\Phi}_{2}^{2}\right)^{2}}\,,\label{finalEpot-1}
%\end{equation}
%The canonically normalized field $\tilde{\phi}$ is given by the relation
%$\tilde{\Phi}_{2}=\sqrt{6\omega}\tanh\left(\frac{\tilde{\phi}}{\sqrt{6}}\right)$.
%The potential $V_{E}$ in (\ref{finalEpot-1}) in terms of $\tilde{\phi}$
%reads 
\begin{equation}
\begin{aligned}V_{E}\left(\tilde{\phi}\right)= & \mu^{2}\left[\sinh^{2}\left(\frac{\tilde{\phi}}{\sqrt{6}}\right)+\frac{\tilde{\beta}^{2}}{
\left(\tilde{\alpha}^{2}+\tilde{\beta}^{2}\right)}\cosh^{2}\left(\frac{\tilde{\phi}}{\sqrt{6}}\right)\right] \\
& \left[\cosh\left(\frac{\tilde{\phi}}{\sqrt{6}}\right)
-\frac{1}{1+\frac{\tilde{\beta}}{\tilde{\alpha}}}\sqrt{1+\frac{\tilde{\beta}}{\tilde{\alpha}^{2}}^{2}}\sinh\left(\frac{\tilde{\phi}}{\sqrt{6}}\right)\right]^{2}\\
% & +\frac{\mu^{2}\tilde{\beta}^{2}}{\tilde{\alpha}^{2}\left(1+\frac{\tilde{\beta}^{2}}{\tilde{\alpha}^{2}}\right)}\cosh^{2}\left(\frac{\tilde{\phi}}{\sqrt{6}}\right)\left[\cosh\left(\frac{\tilde{\phi}}{\sqrt{6}}\right)-\frac{1}{1+\frac{\tilde{\beta}}{\tilde{\alpha}}}\sqrt{1+\frac{\tilde{\beta}}{\tilde{\alpha}^{2}}^{2}}\sinh\left(\frac{\tilde{\phi}}{\sqrt{6}}\right)\right]^{2}\,,
\end{aligned}
\label{final-pot}
\end{equation}
where $\mu^{2}=\frac{9\lambda\left(\tilde{\alpha}+\tilde{\beta}\right)^{2}}{\tilde{\alpha}^{2}\left(\tilde{\alpha}^{2}+\tilde{\beta}^{2}\right)}$.
In the limit $\frac{\tilde{\beta}}{\tilde{\alpha}}\ll1$ , the first
term in (\ref{final-pot}) dominates during inflation while the second
term is negligible. The potential (\ref{final-pot}) is always positive
and in particular has a non-zero value at the minimum at $\tilde{\phi}\approx0$.
In general the shape of the potential is similar to the Starobinsky-like
models in no-scale SUGRA \cite{Ellis:2013nxa}. In Fig. \ref{plot-pot}
we depict the shape of the potential for various values of $\tilde{\beta}$.
This corresponds to different values of vacuum energy $\left(\Lambda\right)$
after inflation. We can see that the smaller the value of $\tilde{\beta}$,
the greater the chance of approaching the plateau region of the Starobinsky
model, and eventually the smaller will be the value of the vacuum
energy.

Setting $\tilde{\alpha}=1$, in the limit $\tilde{\beta}\ll1$, we can approximate
the potential in (\ref{final-pot}) as
\begin{equation}
V_{E}\left(\tilde{\phi}\right)\approx\frac{\mu^{2}}{4}\left(1-e^{-\sqrt{\frac{2}{3}}\tilde{\phi}}\right)^{2}+\frac{\mu^{2}\tilde{\beta}^{2}}{4}\left(1+e^{-\sqrt{\frac{2}{3}}\tilde{\phi}}\right)^{2}\,,\label{appfinalpot}
\end{equation}
where the first term dominates when $\tilde{\phi}\gg1$ and leads
to a Starobinsky like inflation i.e., $n_{s}\sim0.967,\,r\sim0.0033$
for $N=60$ and the second term gives a non-zero vacuum energy at
the minimum of the potential\footnote{A potential of similar kind can be found in the $\alpha-$attractor
models where the inflaton potential was uplifted due to the effect
of a SUSY breaking mechanism \cite{Carrasco:2015pla}.} near $\tilde{\phi}=0$. Here $\mu\approx2\times10^{-5}$ (in Planck
units as we have set $m_{\rm P}=1$) which can be determined from the
observed amplitude of scalar perturbations $A_{s}=2.2\times10^{-9}$
at the horizon exit \cite{Ade:2015lrj}. In particular $\tilde{\beta}\sim10^{-55}$
gives a vacuum energy that reproduces the present day cosmological
constant $\Lambda\sim10^{-120}$. Therefore, we conclude that a non-locality
induced cross-product of the fields $\tilde{\Phi}_{1}$ and $\tilde{\Phi}_{2}$
in (\ref{Cinf}) naturally uplifts the inflaton potential at the minimum
and possibly explain the present day dark energy (assuming it is $\Lambda\text{CDM}$).

\section{Summary}

\label{concdisc}

In this chapter, we have investigated effective models of inflation
emerging from a framework motivated from SFT. {Our models of inflation are essentially an aftermath of TC being possible since not all the brane tension is compensated by the tachyon in a curved background. For the inflation to happen we assume that the inflation scale is below the brane tension.} In our setup, we proposed an action beyond the low-energy open-closed strings coupling in SFT containing closed string dilaton
and open open string tachyon near the tachyon condensation. We observed
that this action can contain (A)dS as background solutions. We have
studied the quadratic perturbations of this action and have shown that the
infinite derivative operators associated with tachyon induce non-locality
dilaton perturbations characterized by $\Fc(\Box)$. The cornerstone
technical question is about the roots $z_{j}$ of the characteristic
equation $\Fc(z)=0$. Moreover, the derivatives $\Fc'(z_{j})$ play
an important role. This is seen from action (\ref{locallocal}), which
describes the evolution of scalar perturbations around a dS vacuum
within a non-local context, SFT being a guide in this process. Its
importance is obvious as inflation is a dS like expansion and all
the observable quantities related to scalars can be obtained from
exploring the action for linear perturbations. A very important restriction
is that no ghosts must be in the spectrum. This selects two configurations
of roots.

First, there is a situation with one real root $z_{1}$ accompanied
with a correct sign of $\Fc'(z_{1})$. In this case there is one scalar
perturbative degree of freedom. Such a configuration can be obtained
from the effective model description (\ref{example1}). It is important
that coefficients in front of the Einstein-Hilbert term and the kinetic
term of a scalar field are independent. We therefore conclude that provided the non-local operator $\Fc(\Box)$
contains one real root, it gives a successful inflation with a universal
prediction of $n_{s}=0.967$ and tensor
to scalar ratio as in (\ref{sft-attractor}) which can be adjusted to any
value $r<0.09$ by means of the parameter
$\mathcal{F}^{\prime}\left(z_{1}\right)$.
A future more accurate detection of parameter $r$ from CMB \cite{Creminelli:2015oda} would indicate the values of $z_{1}$
and $\Fc'(z_{1})$.

Second, there was a case with two roots. They can be complex conjugate
and then we should look at (\ref{pairpairreal}) which is written
in manifestly real components. In this scenario, we inevitably get
a quadratic cross-product of fields. Moreover, one field looks like a ghost.
However, kinetic and mass terms have exactly opposite signs. This
suggests that a conformal symmetry may help exorcising the ghost.
Indeed, building an effective model (\ref{example2}) we have taken
the conformal symmetry into account and have shown that we indeed
can make use of it to remove the unwanted degrees of freedom. The
cross-product of fields naturally leads to an uplifting of the potential
in the reheating point. In principle one can get a similar two-field
model starting with two real roots which are related as $z_{1}=-z_{2}$
and $\Fc'(z_{1})=-\Fc'(z_{2})$. This latter case has no cross-product
of fields and falls into the considerations of \cite{Kallosh:2013hoa,Kallosh:2013daa}.
The novel feature here is that the conformally invariant models with
a quadratic cross-product of scalar fields appear for the first time
in a cosmological setup and can be naturally explained using the non-locality
of a dilaton.

\chapter{Non-slow-roll dynamics in $ \alpha- $attractors}
\label{CAM}
\begin{chapquote}{Peter Higgs}
I'm a fan of supersymmetry, largely because it seems to be the only route by which gravity can be brought into the scheme. If you have supersymmetry, then there are more of these particles. That would be my favourite outcome.
\end{chapquote}
\lhead{\bf Chapter 5. \emph{Non-slow-roll dynamics in $ \alpha- $attractors}} % This is for the header on each page - perhaps a shortened title

Since the first release of {\it Planck} 2013 data, two scenarios (Starobinsky
model and Higgs inflation) started to attract a lot of attention. They
have been extensively studied and realized in the context of conformal
symmetries \cite{Kallosh:2013hoa,Kallosh:2013daa}, later generalized
as $\alpha-$ and non-minimal (or) $\xi-$attractors. In addition,
these models have been embedded in SUGRA through the
use of superconformal symmetries \cite{Kallosh:2013lkr,Kallosh:2013yoa,Kallosh:2013maa,Kallosh:2013pby,Kallosh:2014rga}.
Recently, $\alpha-$ attractor models were also realized by means of the inclusion of an auxiliary vector field for the Starobinsky model \cite{Ozkan:2015iva}.
These two classes of models have also, a posteriori, been unified as
cosmological attractor models (CAM) \cite{Galante:2014ifa,Cecotti:2014ipa,Roest:2013fha}.
By varying the parameters $\left(\alpha,\xi\right)$ in CAM, on the one hand, it leads to the predictions of Starobinsky inflation 
and on the other hand it also reproduces the chaotic inflation predictions with the $m^{2}\phi^{2}$
potential. In particular, for
$\alpha=\frac{1}{9}$, we retrieve the first model of chaotic inflation
in SUGRA proposed in 1983, which is known as the Goncharov-Linde
(GL) model, and it is well consistent with the present data \cite{Linde:2014hfa,Goncharov:1983mw,
Goncharov:1985yu}.
CAM were embedded in $ \mathcal{N}=1 $ SUGRA using superconformal symmetries by introducing a 3 chiral super multiplets: a conformon $ X^{0} $, an inflaton $ X^{1}=\Phi $ and a sGoldstino $X^{3}=S$ \cite{Kallosh:2013yoa,Kallosh:2013lkr,
Kallosh:2013pby}. In this set up, single field inflation is achieved at the minimum of the superpotential by the requirement that the fields $S$ and Im\,$\Phi$ remain heavy during inflation\footnote{This mechanism has
also envisaged the multifield inflation with a curvaton, i.e, where
we can have generation of isocurvature perturbations when $S$ or
Im$\Phi$ are light and play the role of curvaton during or after
the end of inflation \cite{Kallosh:2010ug,Kallosh:2010xz,
Demozzi:2010aj}}. 
In recent studies, $ \alpha- $ attractors were realized in SUGRA\footnote{Obtaining inflation from SUGRA also brings other benefits such as, exploring SUSY breaking sector
and the presence of dark energy \cite{Abe:2014opa,Linde:2014ela,Kallosh:2015lwa,
Carrasco:2015pla,Scalisi:2015qga}.}
by only requiring a single chiral superfield  \cite{Roest:2015qya,Linde:2015uga}. A generalization
of K\"ahler potentials for viable single field models with respect to
{\it Planck} data, plus their connection to open and closed string sector has been
investigated in \cite{Roest:2013aoa}. 

In this chapter, we study non-slow-roll inflaton dynamics in the $\alpha-$attractor
model using the recently proposed approach of Gong and Sasaki (GS)
\cite{Gong:2015ypa}, which constitutes, to our knowledge, a new strategy. 
More concretely, we focus on the
non-canonical aspect of the $\alpha-$ attractor model. We start with the assumption
of GS \cite{Gong:2015ypa}, where the number of $e$-foldings $N$
which is counted backward in time is assumed to be a function of the inflaton
field $\phi$ during inflation. We retrieve the local shape of the
potential during inflation which can be steep and allowing for $60$ $e$-foldings to occur.
More precisely, we restrict our study to the region of the potential
where inflation is occurring. We emphasize that both the pre- and post-inflationary dynamics are beyond the scope of this chapter. Afterwards,
we explore the GS parametrization within our chosen inflaton dynamics showing that inflation occurs for a wider class of potentials. We further show that we can maintain the predictions of the $\alpha-$attractor model displayed in \cite{Kallosh:2013yoa}, but now herein retrieved alternatively within a non-slow-roll. Finally, we study 
the possibility of realizing this model within $\mathcal{N}=1$ SUGRA. We explore the relation between the inflaton dynamics and the corresponding K\"ahler geometry curvature. We also comment on the stability of inflaton  trajectory during inflation. 

The chapter is organized as follows: In Sec.~\ref{alphaattractor},
we revise the $\alpha-$attractor model and present arguments
supporting a non-slow-roll approach for these models. In Sec.~\ref{non-slow-roll-dynamics},
we describe GS parametrization and implement the non-slow-roll dynamics
in the context of $\alpha-$attractors. In Sec.~\ref{inflaitionary predictionsn1},
we present predictions for a specific
case of the GS parametrization. In Sec.~\ref{Largesmallattractors},
we complement the previous predictions for a wider class of non-slow-roll
dynamics and discuss on large and small field
inflation. We show that these scenarios exhibit an attractor in the $\left(n_{s},\, r\right)$
plane and discuss the (dis)similarities with standard slow-roll inflaton
dynamics. In Sec.~\ref{SUGRAembedding}, we review the SUGRA realization
of this scenario and verify the stabilization of the inflaton trajectory
during inflation.

\section{$\alpha-$attractor model}

\label{alphaattractor} In this section, we revise the essentials
of $\alpha-$attractor models which have been studied under slow-roll frameworks so far as in  \cite{Galante:2014ifa,Kallosh:2013yoa,
Kallosh:2015lwa}
and provide a baseline for our interest on these models which we will
be exploring in the rest of the manuscript from a new perspective and methodology.

The Lagrangian for $\alpha-$attractor models, in the Einstein frame, is given by%
\footnote{We assume the units $m_{\textrm{P}}=1$.%
} \cite{Kallosh:2015lwa}

\begin{equation}
\mathcal{\mathcal{L}}_{E}=\sqrt{-g}\left[\frac{R}{2}-\frac{1}{\left(1-\phi^{2}/6\alpha\right)^{2}}\frac{\left(\partial\phi\right)^{2}}{2}-f^{2}\left(\phi/\sqrt{6\alpha}\right)\right]\,,\label{alphaL}
\end{equation}
where $\alpha=1$ leads to the same prediction of the Starobinsky
model (in the Einstein frame), $\alpha=1/9$ corresponds to GL model \cite{Linde:2014hfa},
and for large $\alpha$ this model is equivalent to chaotic inflation
with quadratic potential \cite{Linde:1983gd}. In order to prevent
 negative gravity in the Jordan frame it is required to have $\vert\phi\vert<\sqrt{6\alpha}$
\cite{Kallosh:2013yoa,Kallosh:2014rga}. Furthermore, in the SUGRA
embedding of this model, the parameter $\alpha$ is shown to be related
to the curvature of K\"ahler manifold as

\begin{equation}
\mathcal{R}_{\mathcal{K}}=-\frac{2}{3\alpha}\,.\label{kalhercurvature}
\end{equation}

The Lagrangian (\ref{alphaL}) is a subclass of $ k $-inflationary
model where the kinetic term is linear%
\footnote{$K\left(\phi\right)=1$ gives the canonical kinetic term.%
} in $X$, i.e.,

\begin{equation}
P\left(X,\phi\right)=K\left(\phi\right)X-f^{2}\left(\phi/\sqrt{6\alpha}\right)\,,\label{kinflationE}
\end{equation}
where $K\left(\phi\right)=\frac{1}{\left(1-\phi^{2}/6\alpha\right)^{2}}$
and $X=-\frac{\left(\partial\phi\right)^{2}}{2}$. The speed of sound
for these class of models is $c_{s}^{2}=1$ \cite{ArmendarizPicon:1999rj}, therefore
these models are not expected to show large non-Gaussianities \cite{Chen:2006nt}.

In this theory, the Friedmann equation is

\begin{equation}
H^{2}=\frac{1}{3}\left[XK\left(\phi\right)+f^{2}\left(\frac{\phi}{\sqrt{6\alpha}}\right)\right]\,.\label{Efriedmann}
\end{equation}
The Raychaudhuri equation is

\begin{equation}
\dot{H}=-XP_{,X}\,\:\: \text{with}\:\: P_{,X}= \frac{\partial P}{\partial X},\label{RaychaudhuriE}
\end{equation}

and the equation of motion for the scalar field is given by

\begin{equation}
\frac{d}{dt}\left(K\left(\phi\right)\dot{\phi}\right)+3HK\left(\phi\right)\dot{\phi}-P_{,\phi}=0\,.\label{Eom}
\end{equation}
In the literature it is found that inflation in the $\alpha-$attractor
model has been realized in terms of a canonically normalized field $\left(\varphi\right)$ as

\begin{equation}
\frac{d\varphi}{d\phi}=\frac{1}{\left(1-\frac{\phi^{2}}{6\alpha}\right)}\Rightarrow\frac{\phi}{\sqrt{6\alpha}}=\tanh\frac{\varphi}{\sqrt{6\alpha}}\,.\label{canonical field}
\end{equation}
In this case, flat potentials are natural and subsequent slow-roll
dynamics of $\varphi$ lead to viable inflationary scenario with respect
to the observational data. The predictions of $\left(n_{s},\, r\right)$
for these models are shown to be solely determined by the order and
residue of the Laurent series expansion leading pole in the kinetic
term \cite{Galante:2014ifa}. The slow-roll inflationary predictions
of $\alpha-$attractor models are

\begin{equation}
n_{s}=1-\frac{2}{N}\quad r=\frac{12\alpha}{N^{2}}\,.\label{sweetspot-1}
\end{equation}
In terms of this canonically normalized field $\left(\varphi\right)$ the
equation of motion (\ref{Eom}) becomes

\begin{equation}
\ddot{\varphi}+3H\dot{\varphi}+V_{,\varphi}=0\,.\label{canonicalEOM}
\end{equation}
Therefore, under slow-roll assumption this reduces to

\begin{equation}
3H\dot{\varphi}\simeq V_{,\varphi}\,.\label{slowroll}
\end{equation}

Our purpose is to obtain viable inflationary
predictions, by means of extending $ \alpha- $ attractors towards non-slow-roll dynamics. Therefore, in the present work, we restrict
ourselves to the range $\phi^{2}<6\alpha$. We will emphasize similarities
and of course the differences with the (canonically normalized
field) slow-roll inflation case. In the following section we unveil the context of non-slow-roll towards $\alpha-$attractors.

\section{Non-slow-roll dynamics}

\label{non-slow-roll-dynamics} 
The recent work by Gong \& Sasaki (GS)
\cite{Gong:2015ypa} points out a cautionary remark on applying slow-roll approximation
in the context of k-inflation. The argument, presented there, lies
in the fact that the second derivative term in the equation of motion
(\ref{Eom}) may not be negligible in general. In this regard,
the authors introduce a new parameter
\begin{equation}
p=\frac{\dot{P}_{,X}}{HP_{,X}}\,,
\end{equation}
which could bring significant differences in the local non-Gaussianity.
They have illustrated the role of this new parameter and observationally
viable inflationary scenarios in the context of some non-trivial examples. 

Let us implement the aforementioned procedure here in the context
of $\alpha-$attractors as $\phi$ is a non-canonical scalar field
given by (\ref{kinflationE}). This new approach enable us to
study the $\alpha-$attractors in the context of non-slow-roll by assuming
that the inflaton field during inflation behaves as%
\footnote{We start with a similar parametrization as the one used in Sec.3.2 of \cite{Gong:2015ypa}.%
}

\begin{equation}
\phi=n\exp\left(\beta N\right)\,,\label{sasakiparametrization}
\end{equation}
where $N=\ln a\left(t\right)$ is the number of efoldings counted
backward in time from the end of inflation and $n$ is treated as
a free parameter that specifies the value of the field at $N\rightarrow0$.
We assign (\ref{sasakiparametrization}) as GS parametrization
for subsequent reference. This parametrization is particularly useful in the cases of non-canonical scalar field models, whereas in Refs.~\cite{Martin:2012pe,Motohashi:2014ppa} a different parametrization was applied to the case of canonical scalar field inflation. We declare here that our study of inflation in $ \alpha-$ attractor model is based on the dynamics for the inflaton assumed in (\ref{sasakiparametrization}) parametrized by $ \left(n,\,\beta\right) $. Therefore, we label our approach for the $ \alpha- $attractor framework as non-slow-roll, following the same terminology used in Ref.~\cite{Gong:2015ypa}. Being more precise, in this chapter we do not impose any slow-roll approximation in particular. We note at this point that non-slow-roll does not mean a non-smallness of conventional parameters $ \epsilon,\,\eta $ (see Ref.~\cite{Gong:2015ypa} for more details). Moreover, and we stress that this is a most important point in our
study, we completely relax the choice of the inflaton potential and rather concentrate on the inflaton dynamics that can give rise to viable observational predictions.

Substituting $\phi$ from (\ref{sasakiparametrization})
in the Raychaudhuri equation we obtain

\begin{equation}
H^{\prime}=\frac{\alpha^{2}H\left(N\right)}{2}\phi^{2}K\left(\phi\right)\,,\label{rayintegrate}
\end{equation}
where the prime  $ ^{\prime} $ denotes differentiation with respect to $N$. Integrating
(\ref{rayintegrate}), we get

\begin{equation}
H\left(N\right)=\lambda e^{-\frac{9\beta\alpha^{2}}{\phi^{2}-6\alpha}}\,,\label{HEsol}
\end{equation}
where $\lambda$ is the integration constant. At this point, we should mention that our calculations are similar to the Hamilton-Jacobi like formalism found in \cite{Muslimov:1990be,Salopek:1990jq,Motohashi:2014ppa}.

Inserting the aforementioned
solution in (\ref{Efriedmann}), we can express the
local shape of the potential during inflation as

\begin{equation}
f^{2}\left(\frac{\phi}{\sqrt{6\alpha}}\right)=\lambda \exp\left({-\frac{18\beta\alpha^{2}}{\phi^{2}-6\alpha}}\right)\left[3-\frac{\beta^{2}\phi^{2}}{2\left(1-\frac{\phi^{2}}{6\alpha}\right)^{2}}\right]\,.\label{potential}
\end{equation}

It should therefore be noted that the suitable choice of potentials considered in the case of slow-roll $ \alpha- $attractors are quite different, namely, power law type $ V\sim\phi^{2n} $ in terms of original scalar field (or) T-models, i.e.,$  \,V\sim\tanh^{2n}{\frac{\varphi}{\sqrt{6\alpha}}} $ in terms of canonically normalized field \cite{Galante:2014ifa,Kallosh:2013yoa,
Kallosh:2015lwa}. In Ref.~\cite{Kallosh:2014rga} the power law potentials were generalized to the following form of power series

\begin{equation}
f^{2}\left(\frac{\phi}{\sqrt{6\alpha}}\right)= \sum_{n} c_{n}\phi^{n}\,,
\label{seriespot}
\end{equation}
where $ c_{n} $ are non-zero constants and it was argued to be $ c_{0}\ll1 $. In this class of potentials the inflaton slow-rolls towards the potential minimum\footnote{It has been studied in the Ref.~\cite{Cespedes:2015jga} that the slow-roll inflation in T-models can be interrupted abruptly in some cases of matter couplings to inflaton field.} which is located at $ \phi=0 $.

In the subsequent sections, with the assumed GS parametrization, we will show that non-slow-roll inflation occurs to be near the pole of the kinetic term i.e., $\vert\phi\vert\to\sqrt{6\alpha} $. Therefore, we can observe from (\ref{potential}) that the local shape of the potential in the non-slow-roll approach is different from the power-law (or) T-models and also the power series form given in (\ref{seriespot}). In this regard, our study about the non-slow-roll approach widens the scope for different shapes of inflationary potentials in $ \alpha- $ attractors.  

Subsequently, for the conventional parameters general definitions\footnote{The sign difference in the definition of parameters $ \epsilon,\,\eta $ is due
to $N$ which is counted backward in time from the end of inflation (see (\ref{epsilonnsl})). % 
}

\begin{equation}
\epsilon=\frac{H^{\prime}}{H}\quad,\quad\eta=-\frac{\epsilon^{\prime}}{\epsilon}\,,
\label{epsilonnsl}
\end{equation}
substituting the Hubble parameter from (\ref{HEsol}) and demanding the end of inflation $\epsilon=1$ at $N=0$
we get

\begin{equation}
\alpha=\frac{n^{2}}{3\sqrt{2}\beta n+6}\,.\label{alphafix}
\end{equation}
Consequently, constraining the parameter space $\left(n,\,\beta\right)$
automatically gives the values of $\alpha$. In the next sections we show that the $ \beta $ parameter determines the value of scalar spectral index $ n_{s} $, whereas as the parameter $ n $, which indicates the value of inflaton field at the end of inflation, regulates the tensor to scalar ratio $ r $. From (\ref{sasakiparametrization}),
(\ref{potential}) and (\ref{alphafix}), we can say that the local
shape of the potential, the inflaton dynamics and the parameter $\alpha$
are interconnected. In other words, identifying $\alpha$ as the curvature
of K\"ahler geometry given by (\ref{kalhercurvature}), we can establish
a web of relations, 

\vspace{1cm}
\hspace{1cm}
\begin{picture}(0,0)%
   
    \put(65,0){\color[rgb]{0,0,0}\makebox(0,0)[lb]{K\"ahler Geometry}}%
    \put(165,0){\makebox(0,0)[lb]{{\Huge{$\leftrightharpoons$}}}}%
    \put(205,0){\color[rgb]{0,0,0}\makebox(0,0)[lb]{Inflaton Dynamics}}%
    
    \put(115,-15){\color[rgb]{0,0,0}\rotatebox{-45}{\makebox(0,0)[lb]
    {\Huge{$\leftrightharpoons$}}}}%
    \put(225,-5){\color[rgb]{0,0,0}\rotatebox{-135}{\makebox(0,0)[lb]
    {\Huge{$\leftrightharpoons$}}}}%
     \put(125,-45){\color[rgb]{0,0,0}\makebox(0,0)[lb]{Local shape of the potential}}%

\end{picture}%

\vspace{2cm}

From the above schematic diagram we can decipher that the class of potentials which are obtained by allowing different values for
$\left(n,\,\beta\right)$ is related to the family of K\"ahler geometries,
which determine the dynamics of inflaton during inflation. In the next section, we derive the scalar and tensor power spectrum for this model. 

\section{Inflationary predictions for $n=1$}

\label{inflaitionary predictionsn1}

In this section, we study the inflationary predictions of the model taking $n=1$. We constrain the parameter $\beta$ to obtain the predictions of $\left(n_{s}\,,\,r\right)$  within current observational range. 

\begin{figure}[t]
\centering\includegraphics[height=2.5in]{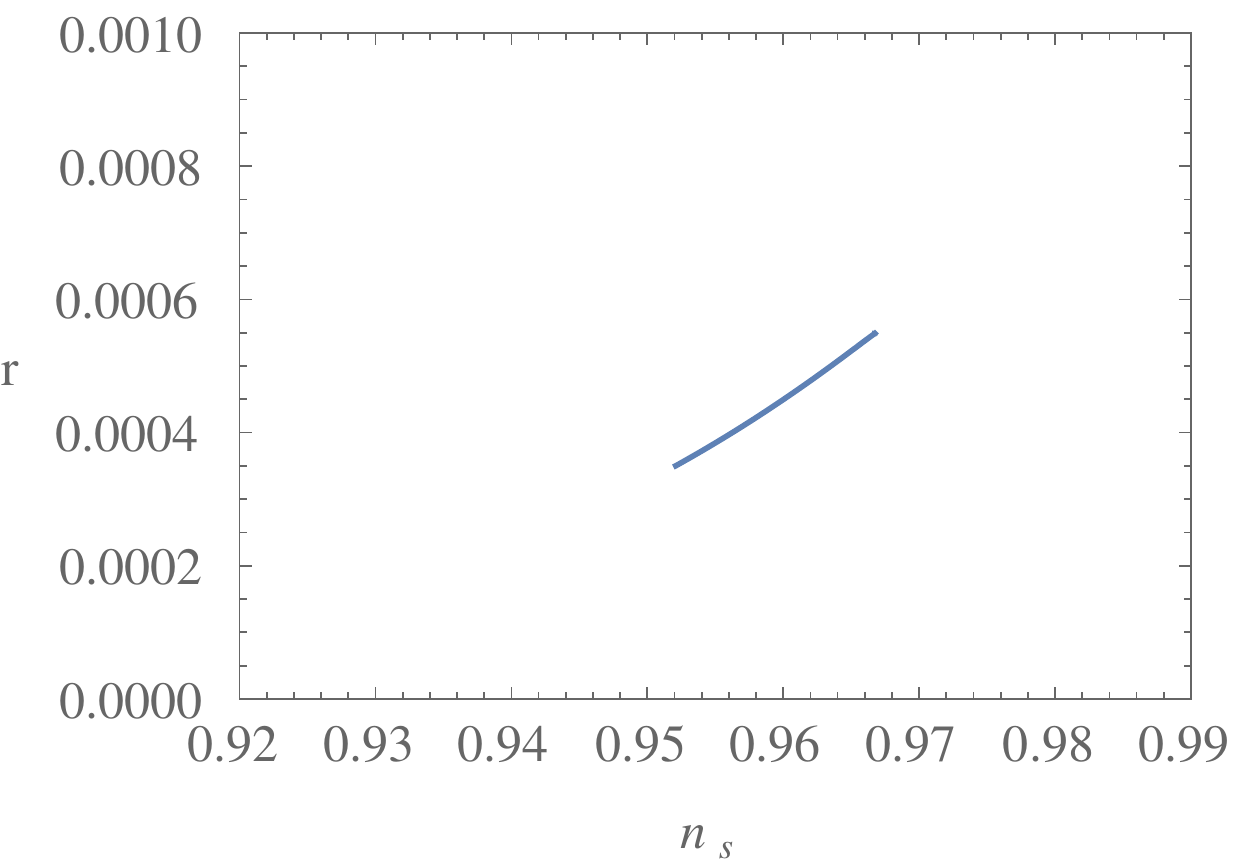}\quad{}\caption{Parametric plot of spectral index $\left(n_{s}\right)$ verses tensor
scalar ratio $\left(r\right)$. We have considered 60 number of efoldings with $n=1\,,\,-0.03<\beta<-0.001$ (or equivalently $0.166\lesssim\alpha\lesssim0.17$).}
\label{ns-r-nt-t} 
\end{figure}

Imposing the spectral index $n_{s}=0.968\pm0.006$, we obtain the constraint
$\vert\beta\vert\sim\mathcal{O}\left(10^{-3}\right)$ (or equivalently,
from (\ref{alphafix}), $\alpha\sim\mathcal{O}\left(10^{-1}\right)$).
However, we verify that the inflaton dynamics for the case $\beta>0$
violates the requirement that $\phi^{2}<6\alpha$. Therefore, we only
consider the case with $\beta<0$ as a viable inflationary paradigm
complying with $\phi^{2}<6\alpha$ during inflation. In this case,
we find that inflation occurs while approaching asymptotically the
kinetic term pole at $\vert\phi\vert\rightarrow\sqrt{6\alpha}$. The
predictions of $\left(n_{s},\, r\right)$ are depicted in the Fig.~\ref{ns-r-nt-t}. 

\begin{figure}[t]
\centering\includegraphics[height=2in]{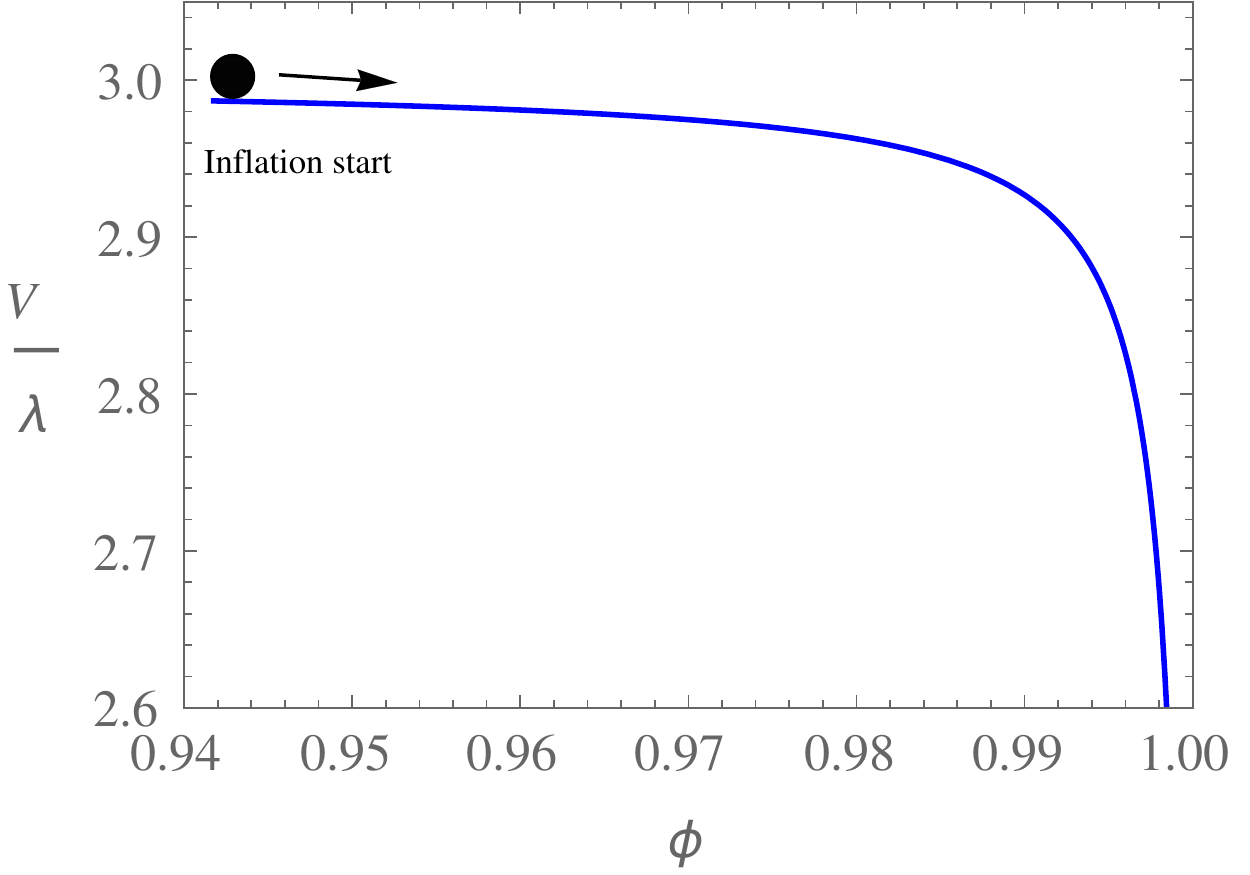}\quad{}\includegraphics[height=2in]{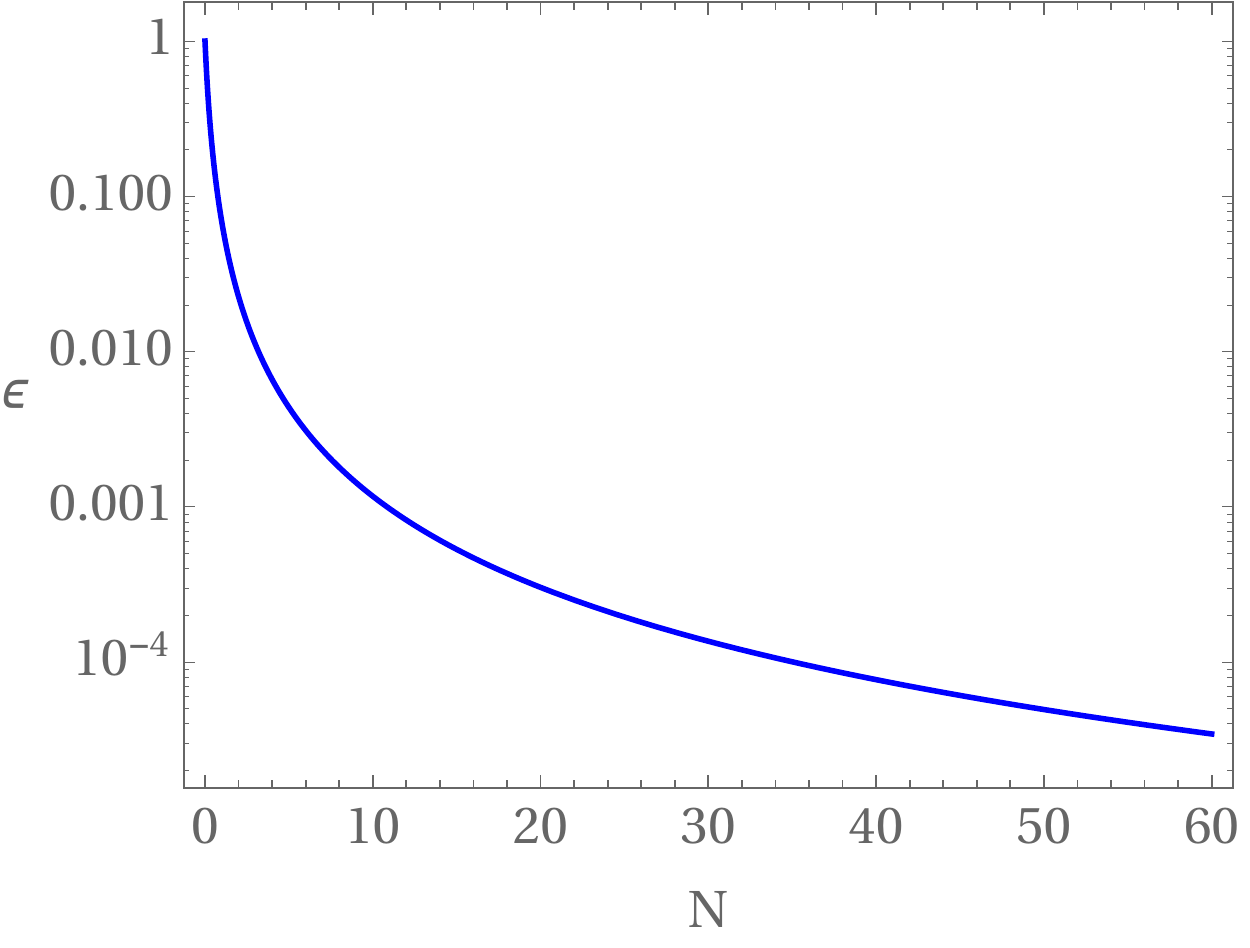}\quad{}\caption{The left panel is the graphical presentation of the local shape of the potential verses scalar field during inflation. The right panel
depicts the parameter $\epsilon\;$verses$\;N$. We have taken $\beta=-0.001$ (or equivalently $\alpha=0.167$) for both plots.}

\label{pot-slow} 
\end{figure}

The left panel of Fig.~\ref{pot-slow} depict
the shape of the potential during which inflation is happening in the non-slow-roll context. 
In the right panel of Fig.~\ref{pot-slow}, we plot the parameter $\epsilon\;$verses$\;N$ 
for a particular value of $\alpha$ corresponds to $ n=1 $. 

In addition, we compute the energy scale of inflation and mass of
the inflaton $\left(m_{\phi}^{2}\right)$ by computing the $V_{*}^{1/4}$
and the $\partial_{\phi}^{2}V_{*}$ where $V_{*}$ is the the potential
evaluated at horizon exit. In this context, the shape of the potential
during inflation is given by (\ref{potential}), consequently
we obtain,

\begin{equation}
f_{*}^{1/2}\sim1.2\times10^{17}\:\textrm{GeV}\quad,\quad m_{\phi}^{2}<0\,.\label{energymass}
\end{equation}

Therefore, since the energy scale of inflation appears to be greater than
GUT scale but still below Planck scale, this naturally justify the embedding
of this model in SUGRA. Since the mass squared of the inflaton is
negative, inflation is driven by a tachyonic field.

\section{Non-slow-roll $\alpha-$attractor}

\label{Largesmallattractors}In Sec.~\ref{inflaitionary predictionsn1},
we have studied non-slow-roll inflation with GS parametrization and
$n=1$, in this case we obtained $r\sim\mathcal{O}\left(10^{-4}\right)$.
The objective, at this point, is to assess inflationary scenarios
with any value of $r<0.09$, by allowing $n\neq1$ in (\ref{alphafix}). 

\subsection{Conditions for small field and large field inflation}

In this section, we study the parameter space of the model allowing the inflaton to do large and small
field excursions during inflation. We address the possibility of large and small field inflation in the context of non-slow-roll
dynamics in $\alpha-$attractors.

\begin{figure}[t]
\centering\includegraphics[height=2.4in]{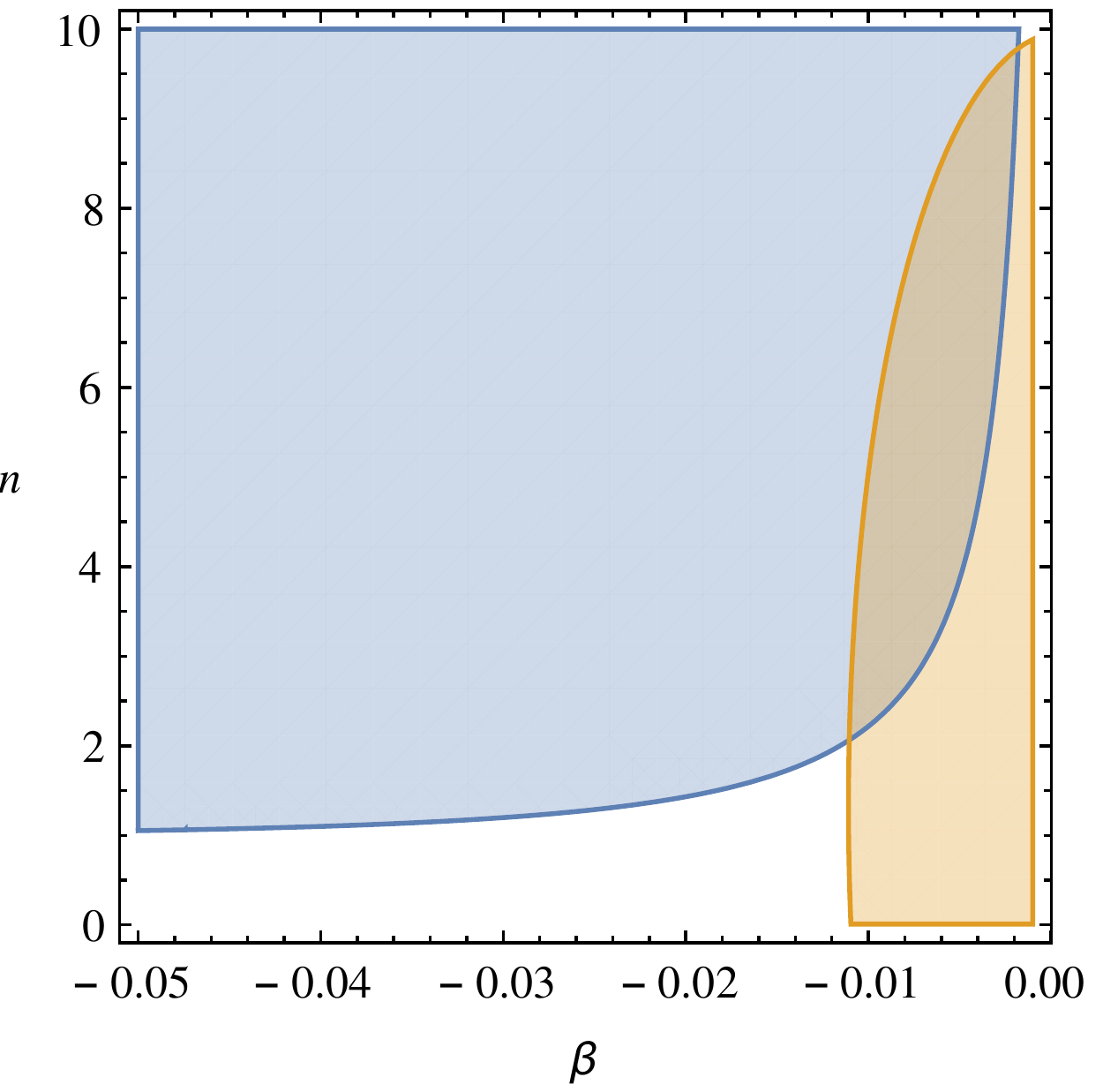}\quad{}\includegraphics[height=2.4in]{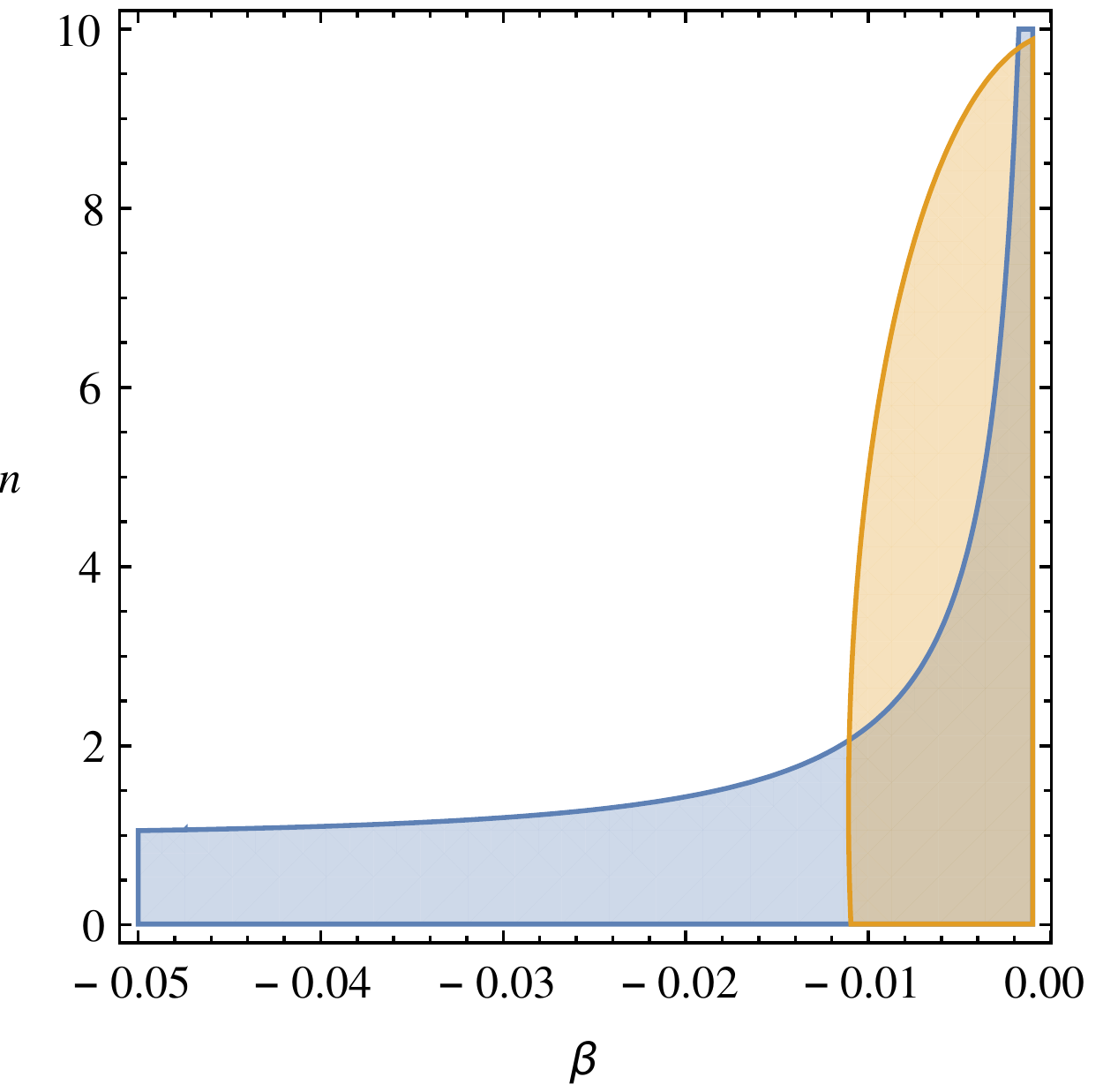}\quad{}\caption{In both plots orange shaded region corresponds to the constraint $ 0.962<n_{s}<0.974 $. The blue shaded region in the left panel is for large field $ \Delta\phi>1 $ whereas in the right panel is for small field $ \Delta\phi<1 $.
We have considered $N=60$.}
\label{nalphalargesmall} 
\end{figure}

Using the parametrization from (\ref{sasakiparametrization})
the field excursion during the period of inflation is given by

\begin{equation}
\Delta\phi=n\left(1-\exp\left(60\beta\right)\right)\,.\label{fieldexcursion}
\end{equation}
The above relation allows us to identify the parameter space of $\left(n,\,\beta\right)$
to explicit the region of large field $\left(\Delta\phi>1\right)$ and
small field $\left(\Delta\phi<1\right)$ inflation (see Fig.~\ref{nalphalargesmall}).
We further constrain the parameter space, by imposing $0.962<n_{s}<0.974$ which is the $95\%$ CL region
given by {\it Planck} 2015. This constraint on spectral index confine $-0.001<\beta<-0.01$,
and precisely $\beta\sim-0.002$ corresponds to the central value of $n_{s}\sim0.967$.

\begin{figure}[t]
\centering\includegraphics[height=2in]{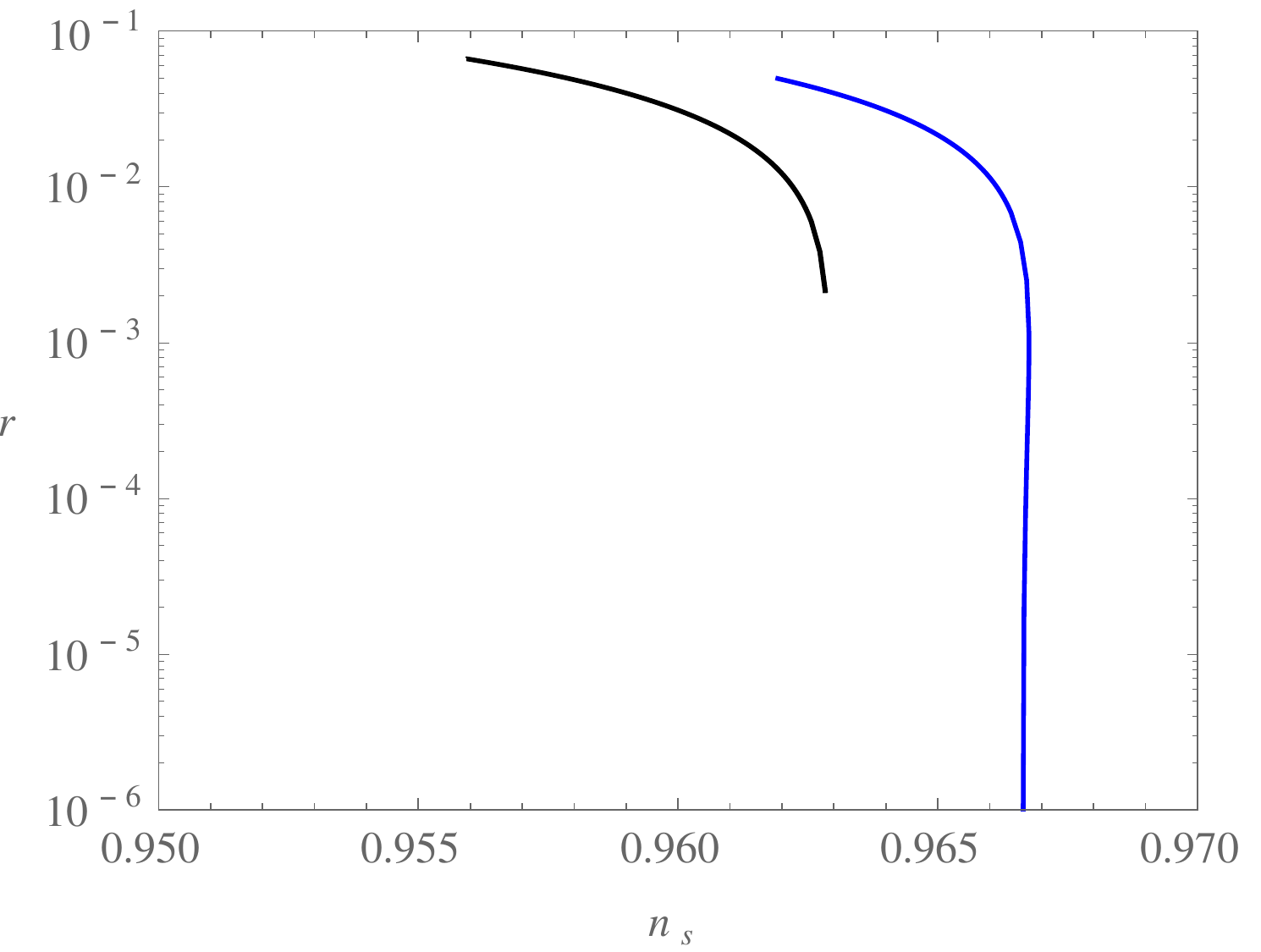}\quad{}\includegraphics[height=2in]{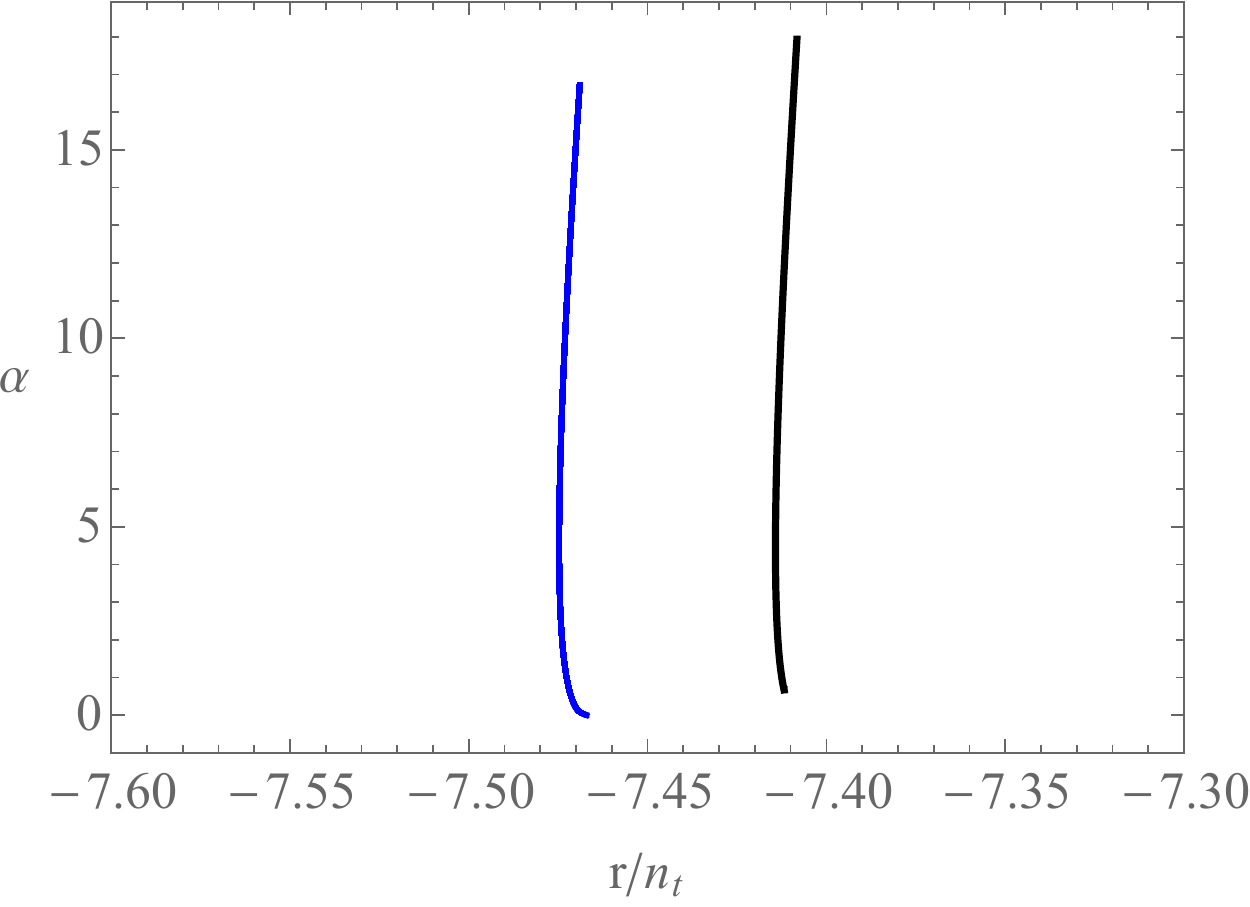}\quad{}\caption{Parametric plots of spectral index $\left(n_{s}\right)$ verses tensor
scalar ratio $\left(r\right)$ (left panel),  $\alpha$
verses the ratio of tensor scalar ratio and tensor tilt (right panel). In these plots the blue line
denote predictions for small field inflation for which we take $\beta\sim-0.002$
and $0<n<10$. In this case $r\rightarrow0$ as $n\rightarrow0$ (equivalently
$\alpha\rightarrow0$). The black line denote predictions for large
field inflation for which $\beta\sim-0.01$ and $2<n<10$. In this
case $r\gtrsim\mathcal{O}\left(10^{-3}\right)$. We have considered
$N=60$. }
\label{large-small-nsrntr} 
\end{figure}

The relation between tensor to scalar ratio and field excursion
during the period of inflation is defined by the Lyth bound \cite{Lyth:1996im} which is 

\begin{equation}
\Delta\phi>\sqrt{\frac{r}{8}}\left(\frac{N}{60}\right)\,.\label{lyth bound}
\end{equation}
%where $N_{e}=60$ which is the number of $e$-foldings before the end of inflation. 
We can see from the above relation that $r>0.002$
implies $\Delta\phi>\textrm{M}_{\textrm{Pl}}$, i.e, large field inflation. However, this bound gets modified for the $ k $-inflationary models \cite{Baumann:2006cd}. In this case, the generalization of (\ref{lyth bound}) is given by

\begin{equation}
\Delta\phi>\int_{0}^{N_{e}}\sqrt{\frac{r}{8}\:\frac{1}{c_{s}\:P_{,X}}}dN\,.\label{newlyth}
\end{equation}
where the sound speed $ c_{s}=1 $ in the case of $ \alpha- $ attractor model. In (\ref{newlyth}) the term $ P_{,X}=\left(1-\frac{\phi^{2}}{6\alpha}\right)^{-2} $ affects Lyth bound depending on the value of the parameter $ \alpha $. From (\ref{alphafix}) we know that the $ \alpha $ parameter is directly related to the inflaton dynamics. In Fig.~\ref{nalphalargesmall}, we depict the parameter space for large and small field inflation overlapped on the  region where $ 0.962<n_{s}<0.974 $. Here, we explicitly characterize the possibility of super planckian excursion of the field $ \phi $ attributing to the field value at the end of inflation $ n\gtrsim2 $ and the parameter $ \beta\sim-0.01 $ (see left panel of Fig.~\ref{nalphalargesmall}). The field $ \phi $ is sub planckian for  $ 0<n<\mathcal{O}\left(10\right) $ and the parameter $ \beta\sim-0.002 $ (see right panel of Fig.~\ref{nalphalargesmall}). We present the corresponding predictions in Fig.~\ref{large-small-nsrntr}, where we found that the large field inflation in the non-slow-roll context can give rise to the tensor to scalar ratio $0.003\lesssim r<0.09$ and the spectral index $ 0.955\lesssim n_{s}\lesssim 0.964 $. Whereas in the case of small field we obtain $0\lesssim r<0.09$ and the spectral index $ 0.96\lesssim n_{s}\lesssim 0.967 $.

\begin{figure}[t]
\centering\includegraphics[height=2.2in]{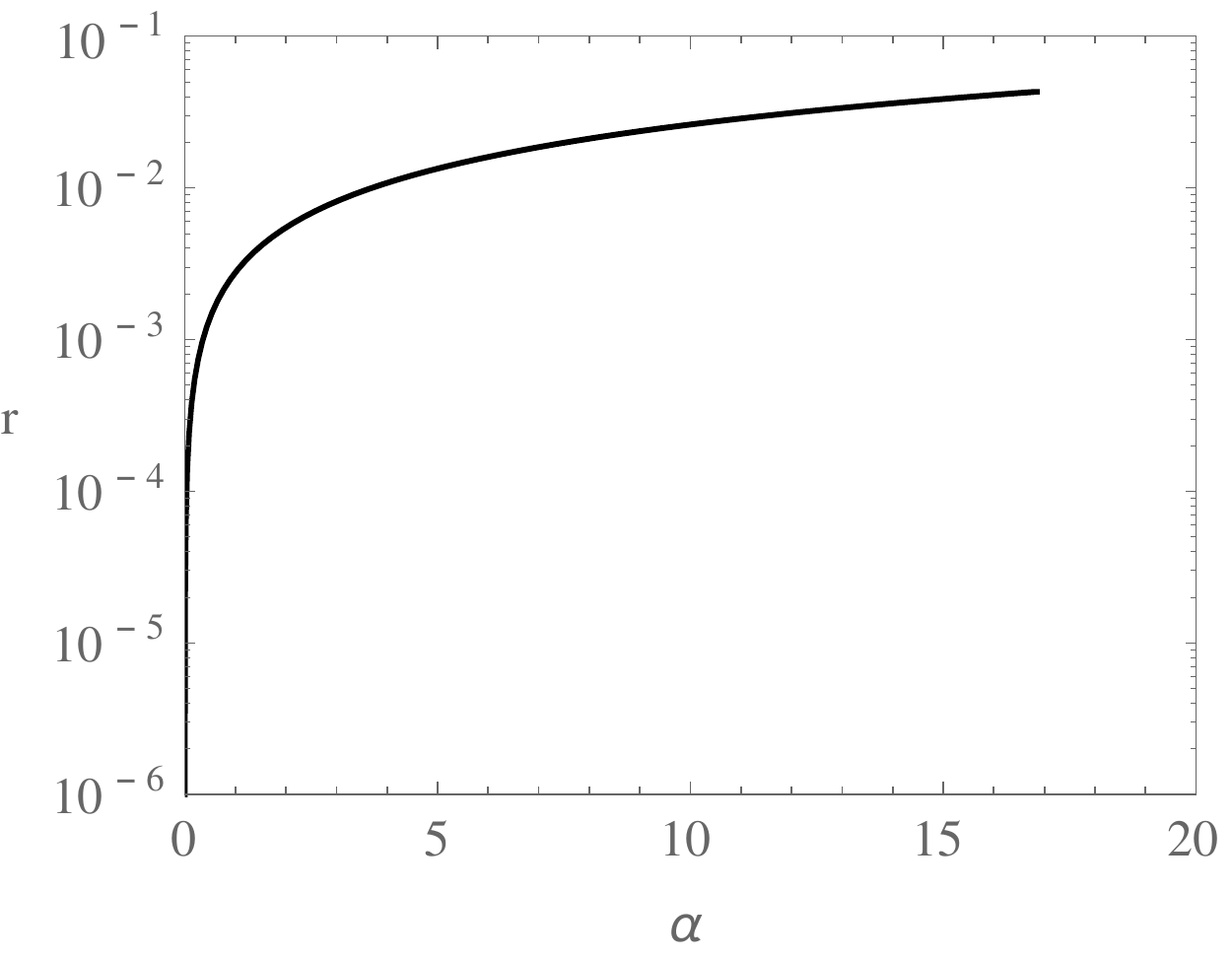}\quad{}\caption{Plot of tensor scalar ratio $\left(r\right)$ verses $\alpha$. Here we
have taken $\beta\sim-0.002$ and $0<n<10$. This plot is for $N=60$.}
\label{rattractor} 
\end{figure}

The parametrization used in (\ref{sasakiparametrization}) leads
to an attractor starting at $r\sim5.5\times10^{-4}$ which is the
prediction for $n=1$. We find that $r\rightarrow0$ as $n\rightarrow0$
(or equivalently $\alpha\rightarrow0$). We depict this behavior in
Fig.~\ref{rattractor}. This attractor behaviour resembles with the
recently studied E-models \cite{Carrasco:2015rva}. The most interesting feature of our study is that, even with non-slow-roll dynamics of the inflaton, $\alpha-$attractors still appear to be the most promising models in the $\left(n_{s},\, r\right)$ plane. Including the higher order corrections in (\ref{nusfull}) and (\ref{nut}) we have undetectably small deviation from the standard consistency relation $r=-8n_{t}$ as presented in the right panel of Fig.~\ref{large-small-nsrntr}. However, the validity of the standard consistency relation remains
an open question and not even expected to be tested in any future CMB observations \cite{Errard:2015cxa}.

\section{Embedding in $\mathcal{N}=1$ SUGRA}

\label{SUGRAembedding}In this section, we revise the embedding of
$\alpha-$attractor within $\mathcal{N}=1$ SUGRA \cite{Kallosh:2013yoa}
and verify the stability of inflaton trajectory \cite{Kallosh:2010xz,Kallosh:2010ug}
in the context of non-slow-roll dynamics. 

The $\alpha-$attractor model can be embedded in SUGRA using 3 chiral
multiplets: a conformon $X^{0}$, an inflaton $X^{1}=\Phi=\frac{\phi+i\sigma}{\sqrt{2}}$
and a sGoldstino $X^{2}=S$. In order to extract a Poincar\'e SUGRA
conformon is gauge fixed as $X^{0}=\overline{X^{0}}=\sqrt{3}$. We
write the K\"ahler and superpotential in the similar way as studied
in Refs.~\cite{Kallosh:2013yoa,Kallosh:2015lwa},

\begin{equation}
\mathcal{K}=-3\alpha\log\left(1-Z\overline{Z}-\frac{S\overline{S}}{3\alpha}+\frac{g}{3\alpha^{2}}\frac{\left(S\overline{S}\right)^{2}}{\left(1-Z\overline{Z}\right)}-\frac{\gamma}{3\alpha^{2}}\frac{S\overline{S}\left(Z-\overline{Z}\right)^{2}}{\left(1-Z\overline{Z}\right)^{2}}\right)\,,\label{kalher}
\end{equation}

\begin{equation}
W=Sf\left(Z\right)\left(1-Z^{2}\right)^{\left(3\alpha-1\right)/2}\,,\label{superpotential}
\end{equation}
where $Z=\frac{X^{1}}{X^{0}}=\frac{\Phi}{\sqrt{6\alpha}}$ and $f\left(Z\right)$
is an arbitrary function and the square of which serves as the inflaton
potential along $S=$Im$\Phi=0$. In the K\"ahler potential in (\ref{kalher})
we added an extra term $\frac{S\overline{S}\left(Z-\overline{Z}\right)^{2}}{\left(1-Z\overline{Z}\right)^{2}}$
in order to stabilize the inflaton trajectory in the direction of
$\textrm{Im}\Phi$ for any value of $\alpha$.
Although in some cases it is not required to add this extra term \cite{Kallosh:2015lwa,Carrasco:2015rva}.
In our case, we only focus our attention to the form of K\"ahler potential
given by (\ref{kalher}). 

The mass squares of $S$ and Im$\Phi$ for a given K\"ahler potential
are given by \cite{Kallosh:2010xz},

\begin{equation}
\begin{aligned}m_{\sigma}^{2} & =2\left(1-\mathcal{K}_{\Phi\overline{\Phi}S\overline{S}}\right)f^{2}+\left(\partial_{\Phi}f\right)^{2}-f\partial_{\Phi}^{2}f\\
m_{s}^{2} & =-\mathcal{K}_{S\overline{S}S\overline{S}}f^{2}+\left(\partial_{\Phi}f\right)^{2}\,,
\end{aligned}
\,,\label{supermasses}
\end{equation}
where all the terms in (\ref{supermasses}) are to be evaluated
along the inflaton trajectory $S=\textrm{Im\ensuremath{\Phi}=0. }$And
here $\mathcal{K}_{a\overline{b}c\overline{d}}=\partial_{a}\partial_{\overline{b}}\partial_{c}\partial_{\overline{d}}\mathcal{K}$.
For the stability of the inflaton trajectory it is required to have
$m_{\sigma}^{2}\,,\, m_{s}^{2}\gg H^{2}$ during inflation, in order
to ensure the absence of isocurvature perturbations and therefore
to have inflation solely driven by a single field \cite{Kallosh:2010xz}. 

For the K\"ahler potential given by (\ref{kalher}) we obtain

\begin{equation}
\mathcal{K}_{\Phi\overline{\Phi}S\overline{S}}=-\frac{36\alpha^{2}\left(6\left(\alpha-2\gamma\right)+\phi^{2}\right)}{\left(\phi^{2}-6\alpha\right)^{3}}\,,\qquad\mathcal{K}_{S\overline{S}S\overline{S}}=\frac{24\alpha(1-6g)}{\left(\phi^{2}-6\alpha\right)^{2}}\,.\label{kderivatives}
\end{equation}

\begin{figure}[t]
\centering\includegraphics[height=2.2in]{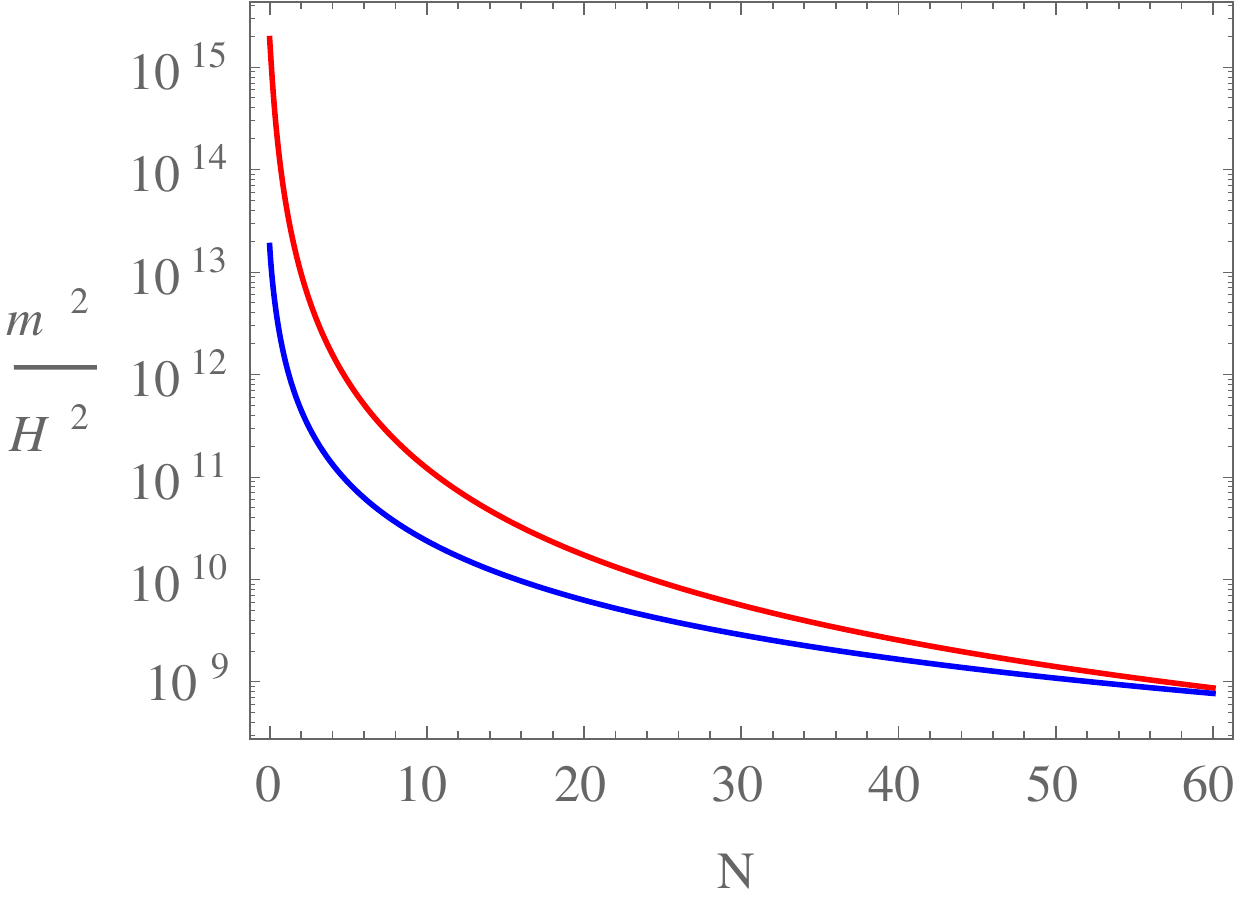}\quad{}\caption{In this figure we depict the ratio of the square of masses to the
square of Hubble parameter $H^{2}$. The red line indicates for Im$\Phi$
and the blue line is for $S$. We have taken $n=1$, $\alpha=0.167$,
$g=0.5$ and $\gamma=0.2$. }

\label{stability} 
\end{figure}

Evaluating the masses $m_{\sigma}^{2}$ and $m_{s}^{2}$ for the local
shape of inflaton potential given by (\ref{potential}) for $n=1$,
we obtain $m_{s}^{2},m_{\sigma}^{2}\gg H^{2}$ for $g\,,\,\gamma\geq0.2$
and for $\alpha\sim0.17$. For example, in Fig.~\ref{stability},
we depict the ratio of inflaton mass square to Hubble parameter square
during inflation for a chosen values of $\left(g\,,\,\gamma\right)$. 

We can similarly verify the stability of the inflaton trajectory for
$n\neq1$ by appropriate choice of free parameters $\left(g\,,\,\gamma\right)$.

\section{Summary}

In this chapter we have considered the $\alpha-$attractor models from a new perspective, more precisely, employing the framework of non-slow-roll approach in the way it was recently proposed
by Gong and Sasaki \cite{Gong:2015ypa}. We found that the $\alpha-$attractor models are quite compatible
in the $\left(n_{s},\, r\right)$ plane of {\it Planck} 2015 within non-slow-roll inflaton dynamics. We showed that such a particular inflationary scenario predicts an attractor at $n_{s}\approx0.967$ and $r\approx5.5\times10^{-4}$. We further found that the model can in principle predict any $r<0.09$. In addition,  we have extracted relation (\ref{alphafix}) between the $ \alpha- $ parameter, to the curvature of K\"ahler geometry, and to the inflaton dynamics. In other words, in our model, the curvature of the K\"ahler geometry defines the  local shape of the inflaton potential during inflation. This constitutes an interesting phenomenon which might be useful to understand the pre-inflationary physics. Furthermore, we also studied the possibility of large and small field inflation in the non-slow-roll context and contrasted them in terms of the predictions of the tensor to scalar ratio. 

%In this chapter we have not considered any particular form for the inflaton potential, since the assumption of $N=N\left(\phi\right)$ during inflation provides all the necessary ingredients for studying inflation. It would be interesting to look at the reheating phase in this model, given an adequate assumption about the shape of the potential in the post inflationnary epoch. In the view of the recent literature about the reheating process regarding Starobinsky inflation \cite{Terada:2014uia}, it would also  be interesting to consider the role of the superfields and their various possible decay modes, as well as of the inflaton. There are other possible directions to be explored, in the context of non-slow-roll dynamics in $\alpha-$attractors with respect to SUGRA, such as studying the various mechanisms for SUSY breaking and the origin of dark energy.  
\chapter{Conformal GUT inflation}
\label{CGUTin}
\begin{chapquote}{Murray Gell-Mann}
The history of the Universe is co-determined by the basic mathematical law of beauty and an unimaginably long sequence of accidents
\end{chapquote}

\lhead{\bf Chapter 6. \emph{Conformal GUT inflation}} %

Since the inflationary scale
is in general expected to be $\sim10^{16}\,\text{GeV}$, it is natural to
consider the inflaton to be a scalar field associated with grand unified
theory (GUT) groups, such as $\text{SU}(5)$ and $\text{SO}(10)$.
Shafi-Vilenkin (SV) model \cite{Shafi:1983bd} is one of the first
realistic model of inflation which was based on $\text{SU}(5)$ GUT
\cite{Georgi:1974sy}. In this framework, inflation is a result of the
spontaneous breaking of $\text{SU}(5)\to\textrm{SU}(3)_{c}\times\textrm{SU}(2)_{L}\times\textrm{U}(1)_{Y}$
by a GUT field ($\textbf{24-\text{plet}}$ adjoint Higgs) and a inflaton,
which is a SU(5) singlet that rolls down to a vacuum expectation value
(VEV). The success of the SV model is that it can lead to a successful
baryogenesis after inflation and predicts proton life time above
the current lower bound \cite{Shafi:2006cs,Rehman:2008qs}. In this model,
the scalar field potential is of a Coleman-Weinberg (CW) form,
according to which primordial gravitational waves are constrained
by $0.02\le r\le0.1$ \cite{Okada:2014lxa}. Although the SV model is
well within the current bounds of {\it Planck} 2015, several extensions of this
model were studied to get smaller values of tensor to scalar
ratio. In \cite{Cerioni:2009kn,Panotopoulos:2014hwa,Barenboim:2013wra},
CW inflation was studied in the context of induced gravity, non-minimal
coupling and brane-world scenario, where the tensor to scalar ratio
was obtained to be $r\sim\mathcal{O}\left(10^{-2}\right)-\mathcal{O}\left(10^{-3}\right)$.
After all these modifications necessarily introduce an additional
parameter into the theory that is responsible for the flatness of the potential. 

Moreover, extensions of the SV model within particle physics offer
rich physics beyond the SM. Therefore, the SV model is embedded
in a higher gauge group as $\text{SO}\left(10\right)$, which can
be broken to SM via an intermediate group $\text{G}_{422}=\text{SU}(4)_{c}\times\text{SU}\left(2\right)_{L}\times\text{SU}\left(2\right)_{R}$
\cite{Lazarides:1984pq,Lazarides:1991wu}. Obtaining successful inflation
in $\text{SO}\left(10\right)$, is more realistic with additional benefits
to explain physics beyond SM, such as neutrino physics, matter anti-matter
asymmetry through non-thermal leptogenesis, monopoles and dark matter
(DM) \cite{Rehman:2008qs}. For example, Ref.~\cite{Boucenna:2014uma}
considered a complex singlet scalar being coupled to RHNs followed by implementing type I seesaw mechanism. This
approach unified inflation with Majorana DM together with the scheme
of generating neutrino masses. In \cite{Okada:2013vxa} an additional
$\text{U}(1)_{B-L}$ symmetry was considered in the SM i.e., $\textrm{SU}(3)_{c}\times\textrm{SU}(2)_{L}\times\textrm{U}(1)_{Y}\times\text{U}\left(1\right)_{B-L}$,
where\footnote{Here $B$, $L$ stands for Baryon number and Lepton number respectively.} $B-L$ symmetry can be spontaneously broken when the scalar
field takes the VEV. In this setup, we can explain baryon asymmetry
of the Universe through non-thermal leptogenesis \cite{Lazarides:1991wu,Asaka:2002zu,Senoguz:2003hc,Senoguz:2005bc}.
Recently, CW inflation was studied in an extension with $\text{SO}(10)$
and $\text{E}_{6}$ groups, pointing out the possibilities of observing primordial
monopoles \cite{Senoguz:2015lba}. 

The main goal of this chapter is to generalize the SV model in order to
achieve $r\sim\mathcal{O}\left(10^{-3}\right)$ without introducing
any additional parameters for inflaton potential
flatness\footnote{Our construction is different from the models with non-minimally coupled
scalars where a flat potential comes from requiring $\xi\gg1$ \cite{Broy:2016rfg}. }. Instead, we consider an additional conformal invariance (or local scale
invariance) in our GUT model. It was long ago shown by Wetterich \cite{Wetterich:1987fm}
that scale symmetries play a crucial role in the construction of realistic
cosmological models based on particle physics. Moreover, scale symmetries
successfully explain the hierarchy of different scales such as the Planck
and Higgs mass \cite{Hooft:2014daa,Quiros:2014hua,Scholz:2012ev,Bars:2013yba}.
Therefore, it is natural to consider scale invariance in constructing an
inflationary scenario, through which we can obtain dynamical generation of the
Planck mass, inflationary scale and particle physics scales beyond SM. In this regard, we
introduce two complex singlet fields $\left(\bar{X},\,\Phi\right)$
of $\text{SU}(5)$ or $\text{SO}(10)$ and couple them to Ricci scalar
and adjoint Higgs field $\left(\Sigma\right)$ such that the total
action would be conformally invariant. We promote inflation as a result
of spontaneous breaking of conformal and GUT symmetries. The former
occurs due to gauge fixing of one singlet field to a constant for all spacetime and the
latter occurs due to $\Sigma$ field takes its GUT VEV. Here the inflaton
is identified with the real part of the second singlet ($\phi=\sqrt{2}\mathfrak{Re}\left[\Phi\right]$),
whereas the imaginary part is the corresponding Nambu-Goldstone boson,
is assumed to pick up a mass due to the presence of small explicit
soft lepton number violation terms in the scalar potential \cite{Boucenna:2014uma}.
Here, we assume $\Phi$ carries two units of lepton number and coupled to the right handed neutrinos (RHNs) in such a way that the coupling is
highly suppressed during inflation\footnote{This will be explained in detail in the due course of this chapter.}.
Near the end of inflation, the inflaton is supposed to reach its VEV and
also the global lepton number is violated. Thereafter, we study the
dominant decay of inflaton into heavy RHNs producing non-thermal
leptogenesis. We compute the corresponding reheating temperature
and also discuss the issue of producing observed baryon asymmetry.
We provide an observationally viable inflationary
scenario, predicting proton life time, neutrino masses and producing
non-thermal leptogenesis from heavy RHNs. 

The chapter is briefly organized as follows. In Sec.~\ref{SCmodel-sec},
we describe toy models with conformal and scale invariance. We identify
the interesting aspects of spontaneous symmetry breaking leading to
viable inflationary scenario. In Sec.~\ref{SU5CW-sec}, we briefly
present the SV model and the computation of proton
life time. In Sec.~\ref{twofieldmodel-sec} we propose our generalization
of SV model by introducing an additional conformal symmetry. We report
the inflationary predictions of the model together with estimates
of proton life time. In Sec.~\ref{seesawSec} we later explore the
nature of inflaton couplings to the SM Higgs, singlet RHNs
through type I seesaw mechanism. We constrain the Yukawa couplings
of the inflaton field compatible with the generation of light neutrino
masses. In Sec.~\ref{ReheatSec} we implement non-thermal leptogenesis
and compute the reheating temperatures corresponding to the dominant
decay of inflaton to heavy RHNs. We additionally comment on
the necessary requirements for the production of observed baryon asymmetry
through CP violation decays of RHNs. In Sec.~\ref{conclusions}
we summarize our results pointing future steps. 

\section{Conformal vs Scale invariance}

\label{SCmodel-sec}

Models with global and local scale invariance (Weyl invariance (or)
conformal invariance) are often very useful to address the issue
of hierarchies in both particle physics and
cosmology \cite{Englert:1976ep,Deser:1970hs,Hooft:2014daa,Shaposhnikov:2008xi,Scholz:2012ev,Quiros:2014hua}.
Models with these symmetries contains no mass input mass parameters.
The spontaneous breaking of those symmetries induced by the VEV's
of the scalar fields present in the theory, generates a hierarchy of
mass scales e.g., Planck mass, GUT scale and neutrino masses\footnote{For example, single scalar field models with the spontaneously broken scale invariance due to
the 1-loop corrections to the tree level potential were studied in
\cite{Rinaldi:2015uvu,Rinaldi:2015yoa,Csaki:2014bua}. In \cite{Ferreira:2016vsc}
two field model with the spontaneously broken scale invariance was studied to
generate hierarchy of mass scales and the dynamical generation of the
Planck mass from the VEV's of the scalar fields. Recently in \cite{Kannike:2016wuy},
some constraints were derived on these models from Big Bang Nucleosynthesis
(BBN). }. Moreover, it is a generic feature that scale or conformal symmetry breaking
induce a flat direction in the scalar field potential \cite{Wetterich:1987fm}
which makes these models even more interesting in the context of inflation.
Another motivation to consider scale invariance for inflationary model
building comes from CMB power spectra which is found to be nearly
scale invariant \cite{Ade:2015lrj}.

In this section, we discuss firstly a toy model (with two fields) that
is (global) scale invariant and present the generic form of (scale invariant)
potentials and their properties. We review the presence of massless
Goldstone boson that appears as a result of spontaneous breaking of global scale invariance. In the following, we discuss the two field conformally invariant
model, in which case the presence of a massless Goldstone boson can be removed
by appropriate gauge fixing. The resultant Spontaneous Breaking
of Conformal Symmetry (SBCS) turns to be very useful to obtain a Starobinsky
like inflation\footnote{Toy models of conformal inflation were studied in \cite{Kallosh:2013lkr,Kallosh:2013daa}
and were embedded in $\mathcal{N}=1$ SUGRA. Furthermore, in a recent study
conformal models were shown to be motivated in the context of string
field theory \cite{Koshelev:2016vhi}.}. We will later explore the role of SBCS in a more realistic
inflationary setting based on GUTs. 

\subsection{Scale invariance }

\label{sec-SI}

Here we discuss a toy model with two scalar fields (in view of Refs.
\cite{Wetterich:1987fk,Wetterich:1987fm,Ghilencea:2015mza,Ferreira:2016vsc})
and point out interesting features that we later utilize
in our construction.

A generic two field global scale invariant action can be written as

\begin{equation}
S_{global}=\int d^{4}x\,\sqrt{-g}\left[\frac{\alpha}{12}\phi^{2}R+\frac{\beta}{12}\chi^{2}R-
\frac{1}{2}\partial^{\mu}\phi\partial_{\mu}\phi-\frac{1}{2}\partial^{\mu}\chi\partial_{\mu}\chi-\phi^{4}f\left(\rho\right)\right]\,,\label{SIaction}
\end{equation}
where $\alpha,\,\beta$ are constants and $\rho=\frac{\phi}{\chi}$,
the generic function $f\left(\frac{\phi}{\chi}\right)$ here can be
treated as quartic self coupling of the field $\phi$ \cite{Wetterich:1987fm,Ghilencea:2015mza}.
The action (\ref{SIaction}) is scale invariant, i.e., invariant under
global scale transformations $g_{\mu\nu}\to e^{-2\lambda}g_{\mu\nu}\,,\,\phi\to e^{\lambda}\phi\,,\,\chi\to e^{\lambda}\chi$
for any constant $\lambda$ (dilatation symmetry). 

Since the potential $V\left(\phi,\,\chi\right)=\phi^{4}f\left(\rho\right)$
is homogeneous, it must satisfy the following constraint \cite{Ghilencea:2015mza,Ferreira:2016vsc}

\begin{equation}
\phi\frac{\partial V}{\partial\phi}+\chi\frac{\partial V}{\partial\chi}=4V\,.\label{homo-cd}
\end{equation}
The extremum conditions for $V$, i.e., $\partial_{\phi}V=\partial_{\chi}V=0$
can also be written as $f\left(\rho\right)=f^{\prime}\left(\rho\right)=0$.
One of the conditions fix the ratio of VEV's of fields, while the
other gives a relation between couplings (if $\langle\phi\rangle\neq0$
and $\langle\chi\rangle\neq0$). The most important and crucial point
here is that if $\langle\phi\rangle\propto\langle\chi\rangle$ there
exists a flat direction for the field $\phi$ (see \cite{Wetterich:1987fm}
for detailed analysis). This will be more clearer in the due course
of this chapter, when we show this property turns out to be maintained
and more useful in the context of local scale invariant model.

Lets consider a scale invariant potential of the form

\begin{equation}
V_{1}=\frac{\lambda_{\phi}}{4}\phi^{4}+\frac{\lambda_{m}}{2}\phi^{2}\chi^{2}+\frac{\lambda_{\chi}}{4}\chi^{4}\,,\label{pot-coupling}
\end{equation}
where the couplings can in general depend on the ratio of two fields
i.e., $\phi/\chi$. If for example, we assume the couplings are independent
of the ratio of two fields and consider the spontaneous breaking of
scale symmetry i.e., the case with $\langle\phi\rangle\neq0,\,\langle\chi\rangle\neq0$,
thus, as a result of minimizing the potential, we arrive at \cite{Ghilencea:2015mza}

\begin{equation}
\frac{\langle\phi\rangle}{\langle\chi\rangle}=-\frac{\lambda_{m}}{\lambda_{\phi}}\quad,\quad V=\frac{\lambda_{\chi}}{4}\left(\chi^{2}+\frac{\lambda_{m}}{\lambda_{\chi}}\phi^{2}\right)^{2}\,,\label{pot-SI1}
\end{equation}
with $\lambda_{m}^{2}=\lambda_{\phi}\lambda_{\chi}$ and $\lambda_{m}<0$.

In (\ref{pot-SI1}) we can re-define the coupling as

\begin{equation}
\bar{\lambda}_{\chi}=\lambda_{\chi}\left(1+\frac{\lambda_{m}}{\lambda_{\chi}}\frac{\phi^{2}}{\chi^{2}}\right)^{2}\,,\label{coupling-SI}
\end{equation}
then the potential (\ref{pot-SI1}) looks like a simple quartic potential

\begin{equation}
V_{1}=\frac{\bar{\lambda}_{\chi}}{4}\chi^{4}\,.\label{pot-SI2}
\end{equation}
We can also alternatively have the potential of the form 

\begin{equation}
V_{2}=\frac{\tilde{\lambda}_{\phi}}{4}\phi^{4}\,,\quad\tilde{\lambda}_{\phi}=\lambda_{\phi}\left(1-\frac{\phi^{2}}{\chi^{2}}\right)^{2}\,,\label{pot-real}
\end{equation}
which also satisfies the constraint (\ref{homo-cd}) and is slightly
different from (\ref{pot-coupling}). We will later see that the form
of potential in (\ref{pot-real}) gives viable inflationary scenario. From (\ref{pot-SI1}) -(\ref{pot-real}) we can crucially
learn that how to define couplings as a function of ratio of
two fields in a scale invariant model. Of course, we only considered
here a simple toy model. However, we note that such field dependent
couplings can be expected to arise in string theory and were applied
in the context of early Universe \cite{Kao:1990tp}. 

The spontaneous breaking of scale symmetry occurs when one of the fields develops a VEV (let us take the field $\chi$).
This leads to an emergence of a corresponding massless Goldstone
boson (dilaton) defined by $\tilde{\chi}=\sqrt{6}M\ln\left(\frac{\chi}{\sqrt{6}M}\right)$
with an arbitrary mass scale $M\propto m_{\rm P}$
\cite{Wetterich:1987fm}. By performing a Weyl rescaling of the metric
$g_{\mu\nu}\to\tilde{g}_{\mu\nu}=\left(\frac{\chi}{\sqrt{6}M}\right)^{2}g_{\mu\nu}$
and $\phi\to\tilde{\phi}=\frac{M}{\sqrt{6}\chi}\phi$ we indeed observe
that the field $\tilde{\chi}$ is massless since the
potential becomes independent of the field $\tilde{\chi}$

\begin{equation}
V\left(\phi,\,\chi\right)=\phi^{4}f\left(\frac{\phi}{\chi}\right)=\tilde{\phi}^{4}f\left(\frac{\tilde{\phi}}{M}\right)\,.\label{pot-SIB}
\end{equation}
Although interesting cosmology and particle physics can be developed
based on the scale invariant models, we need to constrain the
implications of the massless dilaton present in the system \cite{Bars:2013yba}.
It was shown that the dilaton can be gauged away if we consider a
model with local scale symmetry \cite{Bars:2012mt}. 

\subsection{Conformal invariance }

\label{CS}

A general action that is invariant under local scale transformations
$g_{\mu\nu}\to\Omega^{-2}\left(x\right)g_{\mu\nu}\,,\,\phi\to\Omega(x)\phi\,,\,\chi\to\Omega(x)\chi$
can be written as 

\begin{equation}
S_{local}=\int d^{4}x\,\sqrt{-g}\left[\frac{\left(\chi^{2}-\phi^{2}\right)}{12}R+\frac{1}{2}\partial^{\mu}\chi\partial_{\mu}\chi-\frac{1}{2}\partial^{\mu}\phi\partial_{\mu}\phi-\phi^{4}f\left(\frac{\phi}{\chi}\right)\right]\,,\label{toy}
\end{equation}
where the potential in the above action should also satisfy the condition (\ref{homo-cd}).

From the above action we can define an effective Planck mass $m_{eff}^{2}=\frac{\chi^{2}-\phi^{2}}{6}$
which evolves with time. In these theories, we would recover the standard
Planck scale $m_{\rm P}$ when the fields reach their VEV. Note that the
field $\chi$ contains a wrong sign for kinetic term but it is not a
problem as we can gauge fix the field at $\chi=\text{constant}=\sqrt{6}M$
for all spacetime where $M\sim\mathcal{O}\left(m_{\rm P}\right)$. This
particular gauge choice is called $c-$gauge\footnote{It was first realized in the SUGRA models \cite{Bars:2012mt} and shown to be useful to gain geodesic completeness of the theory.} which spontaneously breaks
the conformal symmetry. It was argued that the theories in this
gauge are of interest especially in cosmological models based on particle
physics \cite{Bars:2013yba}. In the inflationary models based on GUTs it
natural that the field $\phi$ takes a non-zero VEV, i.e., $\langle\phi\rangle\neq0$
in which case it is useful to assume $6M^{2}-\langle\phi\rangle^{2}=6m_{\rm P}^{2}$
in order to generate Planck mass. Moreover, it is also necessary to
keep the evolution of the field $\phi\lesssim\sqrt{6}M$ in order
to avoid an anti-gravity regime.

Considering $f\left(\frac{\phi}{\chi}\right)=\lambda\left(1-\frac{\phi^{2}}{\chi^{2}}\right)^{2}$
in (\ref{toy}), SBCS via gauge fixing $\chi=\sqrt{6}m_{\rm P}$ leads
to the Einstein frame action in terms of a canonically normalized field
$\phi=\sqrt{6}m_{\rm P}\,\tanh\left(\frac{\varphi}{\sqrt{6}m_{\rm P}}\right)$
and it is written as 

\begin{equation}
S_{local}=\int d^{4}x\,\sqrt{-g}\left[\frac{m_{\rm P}^{2}}{2}R-\frac{1}{2}\partial^{\mu}\varphi\partial_{\mu}\varphi-\lambda m_{\rm P}^{4}\tanh^{4}\left(\frac{\varphi}{\sqrt{6}m_{\rm P}}\right)\right]\,.\label{toy-1}
\end{equation}
We can see that the above action leads to a Starobinsky like inflation
as the potential acquires a plateau when $\varphi\gg m_{\rm P}$ (i.e., $\phi\to\sqrt{6}m_{\rm P}$).
In this case the inflaton rolls down to zero VEV by the end of inflation and consequently, because of the gauge fixing $\chi=\sqrt{6}m_{\rm P}$, Einstein gravity is recovered.

In the next sections, we will study realistic GUT inflationary models where inflaton
rolls down to non-zero VEV and sources interesting implications in particle physics sector. 

\section{Coleman-Weinberg GUT inflation}

\label{SU5CW-sec}

In this section, we briefly review the Shafi-Vilenkin model \cite{Shafi:1983bd,Linde:2005ht}.
It is one of the first realistic model of inflation which was based
on SU(5) GUT. In this framework a new scalar
field $\phi$, a SU(5) singlet was considered and it weakly interacts
with the GUT symmetry breaking field (adjoint) $\Sigma$ and fundamental
Higgs field $H_{5}$. The tree level scalar potential is given by

\begin{equation}
\begin{aligned}V\left(\phi,\,\Sigma,\,H_{5}\right)= & \frac{1}{4}a\left(\textrm{Tr}\Sigma^{2}\right)^{2}+\frac{1}{2}b\textrm{Tr}\Sigma^{4}-\alpha\left(H_{5}^{\dagger}H_{5}\right)\textrm{Tr}\Sigma^{2}+\frac{\beta}{4}\left(H_{5}^{\dagger}H_{5}\right)^{2}\\
 & +\gamma H_{5}^{\dagger}\Sigma^{2}H_{5}+\frac{\lambda_{1}}{4}\phi^{4}-\frac{\lambda_{2}}{2}\phi^{2}\textrm{Tr}\Sigma^{2}+\frac{\lambda_{3}}{2}\phi^{2}H_{5}^{\dagger}H_{5}\,.
\end{aligned}
\label{TreeSU5}
\end{equation}
where the coefficients $a,\,b,\,\alpha$ and $\beta$ are taken to
be of the order of\footnote{The field $\Sigma$ interacts with vector boson $X$ with a coupling
constant $g$.} $g^{2}$, therefore the radiative corrections in $\left(\Sigma,\,H_{5}\right)$
sector can be neglected. The coefficient $\gamma$ takes a relatively
smaller value and $0<\lambda_{i}\ll g^{2}$ and $\lambda_{1}\lesssim\textrm{max}\left(\lambda_{2}^{2},\,\lambda_{3}^{2}\right)$.

{The GUT field $\Sigma$ which is a $5\times5$ matrix
can diagonalized as }

{
\begin{equation}
\begin{aligned}\Sigma_{i}^{j} & =\delta_{i}^{j}\sigma_{i}\,\\
\sum_{i=1}^{5}\sigma_{i} & =0\,.
\end{aligned}
\label{Sigma-diag}
\end{equation}
where $i,\,j=1,...,5$.

Various symmetry breaking patterns of $\text{SU}(5)$ were studied in
\cite{Magg:1979pf}, among which the one with $\textrm{SU}(5)$ symmetry
is broken to $\textrm{SU}(3)_{c}\times\textrm{SU}(2)_{L}\times\textrm{U}(1)_{Y}$
corresponds to }

{
\begin{equation}
\langle\Sigma\rangle=\sqrt{\frac{1}{15}}\sigma.\textrm{diag}\left(1,\,1,\,1,-\frac{3}{2},\,-\frac{3}{2}\right)\,,\label{GUTfieldVEV}
\end{equation}
where $\sigma$ is scalar field that emerges from spontaneous breaking
of $\text{SU}(5)$. Substituting it in (\ref{TreeSU5}) the equations
of motion for the $\sigma$ field reads as}

\begin{equation}
\Box\sigma+\frac{\lambda_{c}}{4}\sigma^{3}-\frac{\lambda_{2}}{2}\sigma\phi^{2}=0\,,\label{sigmeq}
\end{equation}
where $\lambda_{c}=a+\frac{7}{15}b$. Taking $\lambda_{2}\ll\lambda_{c}$, the $\sigma$ field quickly evolves to its local minimum 
of the potential given by
\begin{equation}
\sigma^{2}=\frac{2\lambda_{2}}{\lambda_{c}}\phi^{2}\,,\label{Vevsigma}
\end{equation}
%much faster than inflaton field $\phi$ to its VEV}

Adding the radiative corrections due to the couplings $-\frac{\lambda_{2}}{2}\phi^{2}\textrm{Tr}\Sigma^{2}$
and $\frac{\lambda_{3}}{2}\phi^{2}H_{5}^{\dagger}H_{5}$, 
the effective potential of $\phi$ gets to the CW form given by \cite{Linde:2005ht,Shafi:1983bd}

\begin{equation}
V_{eff}\left(\phi\right)=A\phi^{4}\left[\ln\left(\frac{\phi}{\mu}\right)+C\right]+V_{0}\,,\label{GUTCW}
\end{equation}
where 
\begin{equation}
A=\frac{\lambda_{2}^{2}}{16\pi^{2}}\left(1+\frac{25}{16}\frac{g^{4}}{\lambda_{c}^{2}}+\frac{14}{9}\frac{b^{2}}{\lambda_{c}}\right)\,.\label{AGUT}
\end{equation}
The $\left(\phi\,,\,\sigma\right)$ sector of effective potential
is given by

\begin{equation}
V_{eff}=\frac{\lambda_{c}}{16}\sigma^{4}-\frac{\lambda_{2}}{4}\sigma^{2}\phi^{2}+A\phi^{4}\left[\ln\left(\frac{\phi}{\mu}\right)+C\right]+V_{0}\,.\label{effphiad}
\end{equation}
and $\mu=\langle\phi\rangle$ denotes the VEV of $\phi$ at the minimum.
$V_{0}=\frac{A\mu^{4}}{4}$ is the vacuum energy density i.e., $V\left(\phi=0\right)$.
$C$ is a constant which we can chose such that $V\left(\phi=\mu\right)=0$.
Therefore, the effective potential (\ref{effphiad}) can be written as 

\begin{equation}
V_{eff}=A\phi^{4}\left[\ln\left(\frac{\phi}{\mu}\right)-\frac{1}{4}\right]+\frac{A\mu^{4}}{4}\,.\label{CWpot}
\end{equation}
{Following (\ref{Vevsigma}) the GUT field $\sigma$
reaches its global minimum only when the inflaton field reach its VEV
by the end of inflation.} The inflationary predictions of
this model were reported in detail in \cite{Shafi:2006cs,Rehman:2008qs}. This
model was shown to be in good agreement with spectral index $n_{s}=0.96-0.967$
and the tensor to scalar ratio $0.02\le r\le0.1$, which is well consistent
with the {\it Planck} 2015 data \cite{Ade:2015lrj,Okada:2014lxa}. 

From the VEV of the singlet field $\phi$ we can compute the masses
of superheavy gauge bosons as 

\begin{equation}
M_{X}=\sqrt{\frac{5\lambda_{2}g^{2}}{3\lambda_{c}A^{1/2}}}V_{0}^{1/4}\,.\label{Xboson}
\end{equation}
Taking $A\sim\frac{\lambda_{2}^{2}}{16\pi^{2}}$ the mass of gauge
bosons are approximately 2-4 times larger than the scale
of vacuum energy $\left(V_{0}^{1/4}\right)$. The key prediction of
GUT models is proton decay $\left(p\to\pi^{0}+e^{+}\right)$ mediated
by $X,\,Y$ gauge bosons. The life time of proton can be computed
using 

\begin{equation}
\tau_{p}=\frac{M_{X}^{4}}{\alpha_{G}^{2}m_{pr}^{5}}\,,\label{prlifetime}
\end{equation}
where $m_{pr}$ is proton mass and $\alpha_{G}\sim1/40$ is the GUT
coupling constant. The current lower bound on proton life time is given
by $\tau_{p}>1.6\times10^{34}$ years indicates $M_{X}\sim4\times10^{15}\,\text{GeV}$
\cite{Nishino:2009aa,Miura:2016krn}. 

\section{GUT inflation with conformal symmetry }

\label{twofieldmodel-sec}

As discussed in Sec.~\ref{SCmodel-sec}, conformal symmetry is useful
to generate flat potentials and the hierarchy of mass scales. Therefore,
embedding conformal symmetry in GUT inflation is more realistic and helpful to generate
simultaneously a Planck scale $m_{\rm P}$ along with the mass scale of X Bosons $M_{X}\sim10^{15}\,\text{GeV}$ that sources proton decay.
In this section, we extend the previously discussed CW inflation by
means of introducing conformal symmetry in SU(5) GUT theory. We then
obtain an interesting model of inflation by implementing spontaneous
breaking of conformal symmetry together with GUT symmetry\footnote{We note that conformal symmetry was considered in GUT inflation \cite{Esposito:1992xf,CervantesCota:1994zf,Buccella:1992rk}
but in those models inflaton was fundamental Higgs field of SU(5) whereas in our case inflaton is GUT singlet weakly coupled to fundamental Higgs.}. We start with two complex singlet fields\footnote{Complex singlet is required to implement type I mechanism which we
later explain in Sec.~\ref{seesawSec}. } of $\text{SU}(5)$ $\left(\Phi,\,\bar{X}\right)$ where the real
part of $\Phi$ ($\phi=\sqrt{2}\mathfrak{Re}\left[\Phi\right]$) is
identified as inflaton. Gauge fixing the field $\bar{X}$ causes SBCS as discussed in Sec.~\ref{SCmodel-sec}. It is worth to note that the same framework we study here based on $SU(5)$ GUT 
can be easily realized in the $SO(10)$ GUT. Therefore, the two complex singlets of $SU(5)$ considered here are also singlets of $SO(10)$ \cite{Lazarides:1991wu,Rehman:2008qs}.

The conformally invariant action with complex SU(5) singlet fields
$\left(\Phi,\,\bar{X}\right)$ can be written as

\begin{equation}
\begin{split}S_{G}= & \int d^{4}x\,\sqrt{-g}\Bigg[\left(\vert\bar{X}\vert^{2}-\vert\Phi\vert^{2}-\textrm{Tr}\Sigma^{2}\right)\frac{R}{12}-\frac{1}{2}\left(\partial\Phi\right)^{\dagger}\left(\partial\Phi\right)+\frac{1}{2}\left(\partial\bar{X}\right)^{\dagger}\left(\partial\bar{X}\right)\\
 & -\frac{1}{2}\text{Tr}\left[\left(D^{\mu}\Sigma\right)^{\dagger}\left(D_{\mu}\Sigma\right)\right]-\frac{1}{4}\text{Tr}\left(\boldsymbol{F}_{\mu\nu}\boldsymbol{F}^{\mu\nu}\right)-V\left(\Phi,\,\bar{X},\,\Sigma\right)\Bigg]\,,
\end{split}
\label{CFTSU(5)}
\end{equation}
where $D_{\mu}\Sigma=\partial_{\mu}\Sigma-ig\left[\boldsymbol{A}_{\mu},\,\Sigma\right]$, $\boldsymbol{A}_{\mu}$ are the 24 massless Yangmills fields
with Field strength defined by $\boldsymbol{F}_{\mu\nu}\equiv\boldsymbol{\nabla}_{[\mu}\boldsymbol{A}_{\nu]}-ig\left[\boldsymbol{A}_{\mu},\,\boldsymbol{A}_{\nu}\right]$.
Here we assume the Higgs field $H_{5}$ is not very relevant during
inflation. We consider that the singlet field $\Phi$ is weakly coupled
to the adjoint field $\Sigma$ through the following tree level potential

\begin{equation}
V\left(\Phi,\,\bar{X},\,\Sigma\right)=\frac{1}{4}a\left(\textrm{Tr}\Sigma^{2}\right)^{2}+\frac{1}{2}b\textrm{Tr}\Sigma^{4}-\frac{\lambda_{2}}{2}\Phi^{2}\textrm{Tr}\Sigma^{2}f\left(\frac{\Phi}{\bar{X}}\right)+\frac{\lambda_{1}}{4}\Phi^{4}f^{2}\left(\frac{\Phi}{\bar{X}}\right)\,,\label{potCFTSU5}
\end{equation}
where the coefficients $a\sim b\sim g^{2}$(gauge couplings $g^{2}\sim0.3$).
Following the discussion in section \ref{SCmodel-sec} we assume the
coupling constants are field dependent, i.e., in (\ref{potCFTSU5})
the coupling constants can be read as $\tilde{\lambda}_{2}=\lambda_{2}f\left(\frac{\Phi}{\bar{X}}\right),\,\tilde{\lambda}_{1}=\lambda_{1}f^{2}\left(\frac{\Phi}{\bar{X}}\right)$
which depend on the ratio of fields $\left(\Phi,\,\bar{X}\right)$.
We consider 
\begin{equation}
f\left(\frac{\Phi}{\bar{X}}\right)=\left(1-\frac{\vert\Phi\vert^{2}}{\vert\bar{X}\vert^{2}}\right)\,.\label{fPchi}
\end{equation}
With the tree level potential in (\ref{potCFTSU5}) the action (\ref{CFTSU(5)})
is conformally invariant under the following transformations

\begin{equation}
g_{\mu\nu}\to\Omega\left(x\right)^{2}g_{\mu\nu}\quad,\quad\bar{X}\to\Omega^{-1}\left(x\right)\bar{X}\quad,\quad\Phi\to\Omega^{-1}\left(x\right)\Phi\quad,\quad\Sigma\to\Omega^{-1}\left(x\right)\Sigma\,.\label{ConformtrSU5}
\end{equation}
{The SBCS occurs
with gauge fixing $\bar{X}=\bar{X}^{*}=\sqrt{3}M$, where $M\sim\mathcal{O}\left(m_{\rm P}\right)$.
We assume inflation to happen in a direction $\text{I}m\Phi=0$.
Therefore, for the inflaton trajectory to be stable we require the
mass of $\text{I}m\Phi$ to be}\footnote{{Where $H_{inf}$ is the Hubble parameter during inflation.}}{{}
$m_{\text{Im}\Phi}^{2}\gg H_{inf}^{2}$. To arrange this, we can add
a new term to the potential (\ref{potCFTSU5}) as }

{
\begin{equation}
V_{S}=V\left(\Phi,\,\bar{X},\,\Sigma\right)+\frac{\lambda_{im}}{4}\left(\Phi-\Phi^{\dagger}\right)^{2}\left(\Phi+\Phi^{\dagger}\right)^{2}\,,\label{stability-Pot}
\end{equation}
such that the mass of the $\text{Im}\Phi$ in the inflationary direction
$\text{Im}\Phi=0$ is $m_{\text{Im}\Phi}^{2}=\frac{\partial^{2}V_{S}}{\partial\text{Im}\Phi^{2}}=\lambda_{im}\left(\Phi+\Phi^{*}\right)^{2}$.
Therefore, If $\lambda_{im}\gg\lambda_{1,2}$ we can have $m_{\text{Im}\Phi}^{2}\Big\vert_{\text{Im}\Phi=0}\gg H_{inf}^{2}$
during inflation. In this way, we can successfully obtain the stability
of the inflaton trajectory during inflation \cite{Kallosh:2010xz}. Similarly to the SV model, here also we consider $\textrm{SU}(5)\to\textrm{SU}(3)_{c}\times\textrm{SU}(2)_{L}\times\textrm{U}(1)_{Y}$
by}

{
\begin{equation}
\langle\Sigma\rangle=\sqrt{\frac{1}{15}}\sigma.\textrm{diag}\left(1,\,1,\,1,-\frac{3}{2},\,-\frac{3}{2}\right)\,,\label{GUTfieldVEV-1}
\end{equation}
Likewise to the SV model, we assume $\lambda_{1}\ll\lambda_{2}\ll a,\,b$
and due to the coupling $-\frac{\lambda_{2}}{2}\phi^{2}\textrm{Tr}\Sigma^{2}f\left(\frac{\phi}{\sqrt{6}M}\right)$,
the GUT field $\sigma$ reaches to its local field dependent minimum
given by}\footnote{{The similar scenario happens in the context of Hybrid
inflationary scenario discussed in \cite{Buchmuller:2014dda}.}}

\begin{equation}
\sigma^{2}=\frac{2}{\lambda_{c}}\lambda_{2}\phi^{2}f\left(\frac{\phi}{\sqrt{3}M}\right)\,.\label{sigma2field}
\end{equation}
{Note that the above local minimum of the GUT field remains the same even though there is non-minimal coupling with the Ricci scalar. We can easily understand this by conformally transforming the action (\ref{CFTSU(5)}) into the Einstein frame.}

After SU(5) symmetry breaking, the X gauge Bosons become superheavy
whereas the field $\sigma$ continues to follow the behavior of the
field $\phi$. The tree level potential for $\left(\phi,\,\sigma\right)$
sector is given by

\begin{equation}
V=\left[\frac{\lambda_{c}}{16}\sigma^{4}
-\frac{\lambda_{2}}{4}\sigma^{2}\phi^{2}f\left(\frac{\phi}{\sqrt{3}M}\right)
+\frac{\lambda_{1}}{4}\phi^{4}f^{2}\left(\frac{\phi}{\sqrt{3}M}\right)\right]\,.\label{treelevel3fieldpot}
\end{equation}
Substituting (\ref{sigma2field}) in (\ref{CFTSU(5)}) and
rescaling the field $\phi\to\sqrt{1+\frac{\lambda_{2}}{\lambda_{c}}}\phi$
we obtain

\begin{equation}
\begin{aligned}S_{G}=\int d^{4}x\,\sqrt{-g}\Bigg\{ & \left(6M^{2}-\phi^{2}\right)\frac{R}{12}
-\frac{1}{2}\left(\partial\phi\right)^{2}\\
 & -\left[\frac{\lambda_{c}}{16}\sigma^{4}-\frac{\bar{\lambda}_{2}}{4}\sigma^{2}\phi^{2} f\left(\frac{\phi}{\sqrt{3}M}\right)+\frac{\bar{\lambda}_{1}}{4}\phi^{4}f^{2}\left(\frac{\phi}{\sqrt{3}M}\right)\right]\Bigg\}\,,
\end{aligned}
\label{actionphi}
\end{equation}
where $\bar{\lambda}_{1,2}=\lambda_{1,2}\sqrt{\frac{1}{1+\frac{\lambda_{2}}{\lambda_{c}}}}$.

%and we consider the coefficient of Ricci scalar should lead to the effective Planck mass $\left(\bar{m}_{p}\right)$

%\begin{equation}
%\bar{X}^{2}-\Phi^{2}=6\bar{m}_{p}^{2}\,.\label{dynPlanck}
%\end{equation}
%From the action (\ref{actionphi}) it is natural to expect the CW 1-loop corrections to the tree level potential due to the coupling of the field $\Phi$ to the GUT field $\sigma$. Since 1-loop Quantum corrections any way spoil the conformal invariance 

Since $\lambda_{1}\ll\lambda_{2}$, the effective potential for the inflaton field $\phi$ due to the radiative corrections become

\begin{equation}
V_{eff}\left(\phi\right)=V+\delta V+m_{\sigma}^{4}\ln\left(\frac{m_{\sigma}^{2}}{\mu^{2}}\right)+V_{0}\,,\label{vef}
\end{equation}
where $\delta V$ is the counter term, $\mu$ is the VEV of the field $\phi$ and $V_{0}$ is a constant.
Using (\ref{sigma2field}), choosing an appropriate $\delta V=\frac{\delta\bar{\lambda}_{2}}{4}\sigma^{2}\phi^{2}f^{2}\left(\frac{\phi}{\sqrt{6}M}\right)$,
a normalization constant such that $V_{eff}\left(\phi=\mu\right)=0$
and the vacuum energy density such that $V\left(\phi=0\right)=V_{0}=\frac{A\mu^{4}}{4}$,
we obtain

{
\begin{equation}
V_{eff}\left(\phi\right)=A\phi^{4}f^{2}\left(\frac{\phi}{\sqrt{3}M}\right)\ln\left(\left(\frac{6\phi^{2}M^{2}f\left(\frac{\phi}{\sqrt{3}M}\right)}
{\mu^{2}m_{\rm P}^{2}}\right)-\frac{1}{4}\right)+\frac{A\mu^{4}}{4}\,,\label{vef-1}
\end{equation}
where $A\sim\frac{\bar{\lambda}_{2}^2}{16\pi^{2}}$.}

We note here that the CW potential we considered is the standard one
obtained from 1-loop correction in Minkowski spacetime. In the de
Sitter background 1-loop corrections are in principle different and
their significance was discussed in literature \cite{Boyanovsky:2005px,Boyanovsky:2009xh,Destri:2009wn}.
In a recent Ref.~\cite{Jain:2015jpa}, it was argued that during slow-roll
inflation we can neglect the contribution of 1-loop corrections in
the gravity sector. In addition, the contributions from higher loops
can also be neglected by the consideration of slow-rolling scalar
field \cite{Kirsten:1993jn,Markkanen:2012rh}. 
%Refs.~\cite{Kirsten:1993jn,Markkanen:2012rh} provide quantum corrections calculated for the cases of non-minimally coupled scalar fields.

{In order to get Planck mass $m_{\rm P}$ dynamically generated by the end of inflation, we should take the corresponding
VEV of the inflaton field as}
\begin{equation}
\langle\phi\rangle=\mu=\sqrt{6M^{2}-6m_{\rm P}^{2}}\,.\label{phiVev}
\end{equation}

Taking the function $f\left(\frac{\phi}{\sqrt{6}M}\right)$ from (\ref{fPchi})
and by doing a conformal transformation of the action (\ref{actionphi})
into Einstein frame, we obtain {(expressing in the
units of $m_{\rm P}=1$)}

{
\begin{equation}
S_{G}^{E}=\int d^{4}x\,\sqrt{-g_{E}}\Bigg[\frac{1}{2}R_{E}-\frac{1}{2M^{2}\left(1-\frac{\phi^{2}}{6M^{2}}\right)^{2}}\partial^{\mu}\phi\partial_{\mu}\phi-\frac{V_{eff}\left(\phi\right)}{36M^{4}f^{2}\left(\frac{\phi}{\sqrt{3}M}\right)}\Bigg]\,.\label{GUT starobinsky}
\end{equation}
Under the conformal transformation the mass scales in the Einstein
frame must be redefined as $\mu^{2}\to\mu^{2}\left(6M^{2}-\phi^{2}\right)^{-1}$.
This is very much an equivalent procedure to the 1-loop analysis of Higgs
inflation. See Refs.~\cite{Bezrukov:2009db,George:2015nza,Fumagalli:2016lls,Pallis:2014cda}
for a detailed discussion on the equivalence between Jordan and Einstein
frames, which exactly matches if we redefine the mass scales accordingly
by conformal factor. Subsequently, substituting (\ref{vef-1}) in
(\ref{GUT starobinsky})}

\begin{figure}[h]
\centering\includegraphics[height=2.5in]{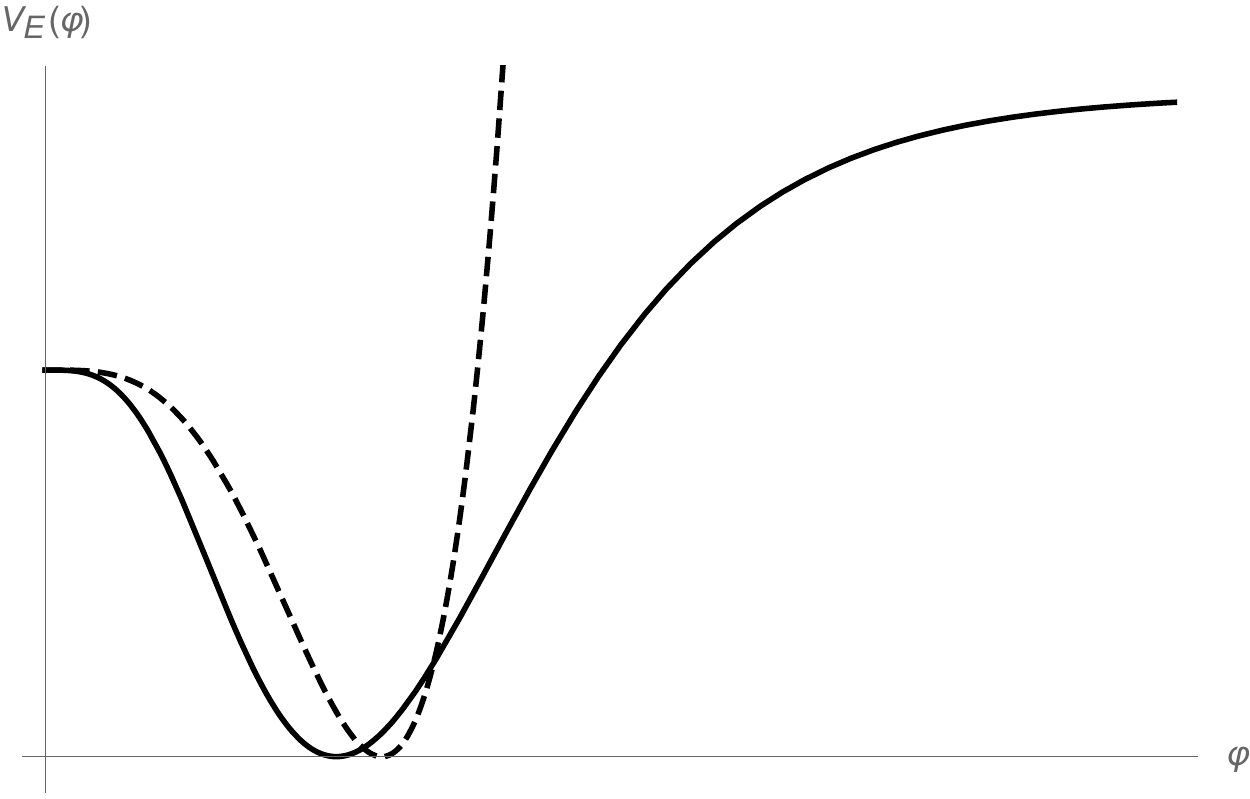}\caption{The dashed line denotes the CW potential in SV model. The full line
indicates the shape of the potential obtained in (\ref{varphipot}) which
comes from the insertion of conformal symmetry in SU(5). When $\varphi\gg\mu$
the above VEV branch of the potential approaches the plateau of Starobinsky
model. }

\label{CWflat} 
\end{figure}

{
\begin{equation}
S_{G}^{E}=\int d^{4}x\,\sqrt{-g}\left\{ \frac{1}{2}R_{E}-\frac{1}{2M^{2}\left(1-\frac{\phi^{2}}{6M^{2}}\right)^{2}}\partial^{\mu}\phi\partial_{\mu}\phi
-A\phi^{4}\left[\ln\left(\frac{\phi^{2}}{\mu^{2}}\right)-\frac{1}{4}\right]-\frac{A\mu^{4}}{4}\right\}\,.\label{conformalTaction}
\end{equation}
The kinetic term of (\ref{conformalTaction}) is similar the no-scale
models \cite{Ellis:2013nxa}.} Canonically normalizing the scalar
field as $\phi=\sqrt{6}M\tanh\left(\frac{\varphi}{\sqrt{6}}\right)$
yields the Einstein frame potential 

\begin{equation}
V_{E}\left(\varphi\right)=A\tanh^{4}\left(\frac{\varphi}{\sqrt{6}}\right)\left(\log\left(\frac{\sqrt{6}M\tanh\left(\frac{\varphi}{\sqrt{6}}\right)}{\mu}\right)-\frac{1}{4}\right)+\frac{A\mu^{4}}{4}\,.\label{varphipot}
\end{equation}
The corresponding VEV of the canonically normalized field is $\langle\varphi\rangle=\sqrt{6}\arctan\left(\frac{\mu}{\sqrt{6}M}\right)$.
The potential in (\ref{varphipot}) is a flattened version of CW potential
(\ref{CWpot}). Due to SBCS the shape of the potential above VEV $\phi>\mu$
significantly gets flattened. In Fig.~\ref{CWflat} we compare the
CW potential of the SV model with the modified form (\ref{varphipot})
we obtained in our case. The shape of the potential reaches a plateau
like in Starobinsky model when $\varphi\gg\mu$ i.e., $\phi\to\sqrt{6}M$.
{Inflation always starts near the plateau and continues
to evolve as $\phi\lesssim\sqrt{6}M$, therefore $f\left(\frac{\phi}{\sqrt{3}M}\right)$
defined in (\ref{fPchi}) is always positive and consequently that avoids an anti-gravity regime.} {Note that
the flat potential (\ref{varphipot}) is significantly different
from the one of CW inflation studied with positive non-minimal coupling
in \cite{Panotopoulos:2014hwa}.} {In the next
subsection we show that the inflationary observables for the potential
(\ref{varphipot}) exactly match that of Starobinsky and Higgs
inflation. }

%\begin{figure}[h]
%\centering\includegraphics[height=2.5in]{Figures/CGUT/ns-r.pdf}
%\caption{In this plot we depict the inflationary predictions for the potential in (\ref{varphipot}) for the values of $M\gtrsim m_{\rm P}$. The inflationary predictions are almost independent of the inflaton VEV $\langle\phi\rangle=\mu$.}
%\label{ns-r} 
%\end{figure}

\subsection{Inflationary predictions and proton lifetime}

We assume the standard FLRW background. Let us define the general definitions
of slow-roll parameters as 

\begin{equation}
\epsilon=\frac{H^{\prime}}{H}\quad,\quad\eta=
-\frac{\epsilon^{\prime}}{\epsilon}\quad,\quad\delta_{1}=
-\frac{\eta^{\prime}}{\eta}\quad,\quad\delta_{2}=-\frac{\delta_{1}^{\prime}}{\delta_{1}}\,,\label{slwrolls}
\end{equation}
where $H$ is the Hubble parameter and the prime $^{\prime}$ denotes
derivative with respect to e-folding number $N=\ln a\left(t\right)$
before the end of inflation.

The scalar power spectrum is given by

\begin{equation}
{\cal P}_{\mathcal{R}}=\frac{\gamma_{s}H^{2}}{8\pi^{2}\epsilon}\Biggr|_{k=aH}\quad,\quad\gamma_{s}\equiv2^{2\nu_{s}-3}\frac{\Gamma\left(\nu_{s}\right){}^{2}}{\Gamma(3/2)^{2}}\left(1-\epsilon\right)^{2}\,.\label{pwrspectrum}
\end{equation}
%The scalar power spectrum amplitude at pivot scale $k=0.002\,Mpc^{-1}$ is measured to be $P_{\mathcal{R}_{*}}=2.2\times10^{-9}$ \cite{Ade:2015lrj}.

The scalar spectral index up to the first orders in slow-roll parameters
is given by 

\begin{equation}
n_{s}-1=3-2\nu_{s}\,,\label{ns}
\end{equation}
 where $\nu_{s}=-\epsilon-\eta/2$. 

The running the spectral index can be expressed
as \cite{vandeBruck:2016rfv}

\begin{equation}
\alpha_{s}\equiv\frac{dn_{s}}{d\ln k}\Biggr|_{k=aH}\simeq-2\epsilon\eta-\delta_{1}\delta_{2}\,.
\label{runnings}
\end{equation}
The ratio of tensor to scalar power spectrum is 

\begin{equation}
r=16\epsilon\Big\vert_{k=aH}\,.\label{r-}
\end{equation}
{The potential (\ref{varphipot}) when $\varphi\gg\mu$ 
can be approximated as }

{
\begin{equation}
\begin{aligned}V_{E}\left(\varphi\right) & \simeq A\left(1-e^{-\sqrt{2/3}\varphi}\right)^{4}\ln\left(\frac{\sqrt{6}M\left(1-e^{-\sqrt{2/3}\varphi}\right)}{\mu}\right)\\
 & \approx A\left(1-e^{-\sqrt{2/3}\varphi}\right)^{4}\ln\left(\frac{\sqrt{6}M}{\mu}\right)\,.
\end{aligned}
\label{vaphiplotapp}
\end{equation}
The equation of motion of the canonically normalized field is }

{
\[
\ddot{\varphi}+3H\dot{\varphi}+V_{E,\varphi}=0\,,
\]
 which during the slow-roll regime reduces to }

{
\begin{equation}
\frac{\partial\varphi}{\partial N}\approx\frac{V_{E,\varphi}}{V_{E}}=4\sqrt{\frac{2}{3}}e^{-\sqrt{\frac{2}{3}}\varphi}\,,\label{slwapp}
\end{equation}
where we use the fact that $H_{inf}\approx\frac{V_{E}\left(\varphi\right)}{3}$.
Integrating (\ref{slwapp}) and expressing the slow-roll parameter
$\epsilon\left(N\right),\,\eta\left(N\right)$ when $N\gg1$ we get }

{
\begin{equation}
\begin{aligned}\epsilon=\frac{\partial\ln H}{\partial N} & \approx\frac{1}{2}\left(\frac{V_{E,\varphi}}{V_{E}}\right)^{2}\approx\frac{3}{4N^{2}}\,,\quad\eta=-\frac{\partial\epsilon}{\partial N}\approx\frac{2}{N}\,.\end{aligned}
\label{epsilon}
\end{equation}
Using (\ref{epsilon}) we can write the predictions for the scalar
tilt (\ref{ns}) and tensor to scalar ratio (\ref{r-}) as }

{
\begin{equation}
n_{s}\approx1-\frac{2}{N}\,,\quad r=\frac{12}{N^{2}}\,,\label{ns-r-1}
\end{equation}
which exactly match with the predictions of Starobinsky and Higgs
inflation \cite{Starobinsky:1980te,Bezrukov:2007ep}. We emphasize
that the predictions of our model in (\ref{ns-r-1}) are independent
of the VEV of the inflaton field $\langle\phi\rangle=\mu$. In Table.
\ref{tab1} we support this result by numerically solving equation
of motion of the field $\varphi$ and the Friedmann equations. }

\begin{figure}[h]
\centering\includegraphics[height=2.2in]{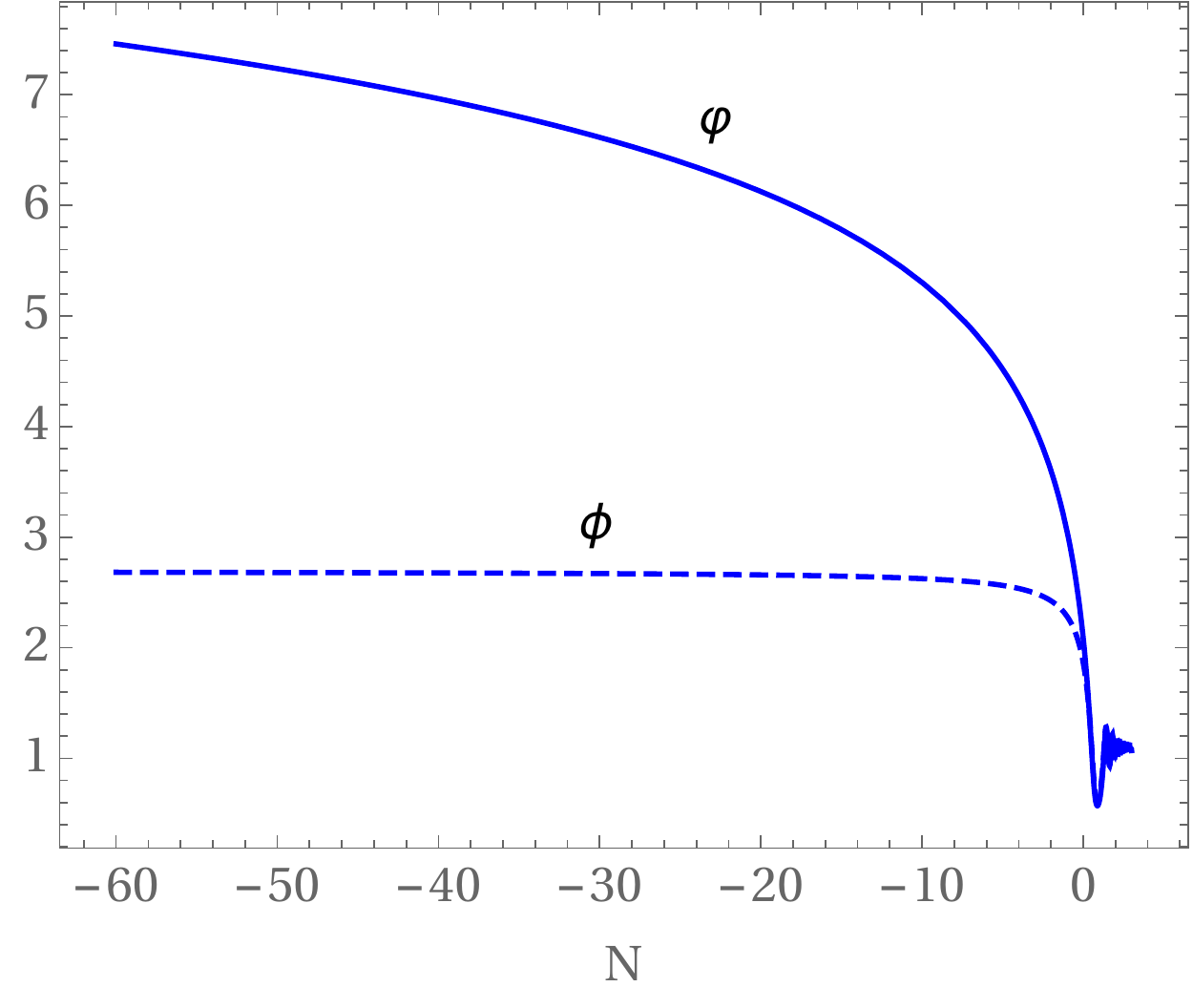}\quad{}\includegraphics[height=2.2in]{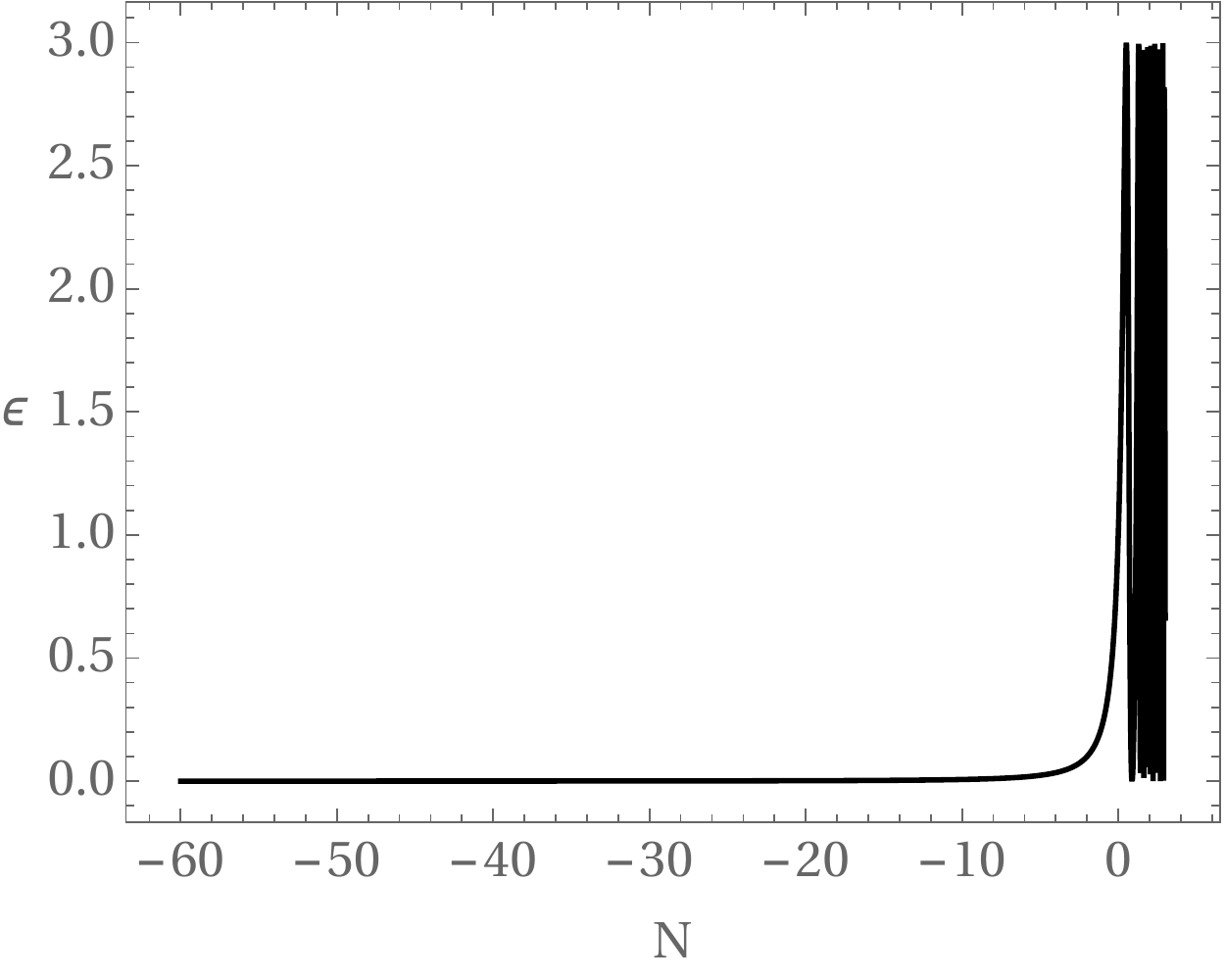}
\caption{In the left panel we depict the evolution of scalar field during inflation
verses the e-folding number. The solid blue line indicates the evolution
of canonically normalized field $\varphi$, whereas the dotted blue
line is for the original field $\phi$. In the right panel we plot
the corresponding slow-roll parameter $\epsilon$ verses $N$. Inflation
ends when $\epsilon=1$. For both plots we have taken $\mu=1.12m_{\rm P}$. }
\label{fieldevolution} 
\end{figure}

In Table. \ref{tab1} we present the inflationary predictions of the
model together with the corresponding $X$ bosons mass and proton
life time using (\ref{Xboson}) and (\ref{prlifetime}). In Table.
\ref{tab1} we present results for the case when the inflaton field
rolls from above VEV (AV) i.e., when $\phi>\mu$. The predictions
of below VEV (BV) branch i.e., when $\phi<\mu$ are not very interesting
as those are nearly same in the original CW inflation without any
conformal symmetry \cite{Rehman:2008qs}. This is evident from Fig.~\ref{CWflat} 
where we can see only the AV branch of the potential
significantly different in our case, whereas the BV branch is nearly
same as in the SV model. Therefore, our interest in this chapter
is restricted to AV branch. For this case, from Table. \ref{tab1}
we can see that the inflationary predictions of the model are extremely
stable with respect to the choice of VEV and any value of $M$. In
%Fig.~\ref{ns-r} we depict the generic predictions of $\left(n_{s},\,r\right)$ of our model which are almost independent of the choice of $M$ or $\mu$. 
In Fig.~\ref{fieldevolution} we depict the evolution of field $\phi$
(also for the canonically normalized field $\varphi$) and slow-roll parameter
$\epsilon$ for particular parameter values. 

\begin{center}
\begin{table}[!h]
\centering{\scriptsize{}{}}%
\begin{tabular}{|c|c|c|c|c|c|c|c|c|c|c|}
\hline 
{\scriptsize{}{}\enskip{}$M$\enskip{}}%
 & %
{\scriptsize{}{}$\enskip A\enskip$}%
 & %
{\scriptsize{}{}$H_{inf}$}%
 & %
{\scriptsize{}{}\enskip{}$N$\enskip{}}%
 & %
{\scriptsize{}{}\enskip{}$\varphi_{0}$\enskip{}}%
 & %
{\scriptsize{}{}\enskip{}$\varphi_{e}$\enskip{}}%
 & %
{\scriptsize{}{}$\enskip n_{s}\enskip$}%
 & %
{\scriptsize{}{}$\enskip r\enskip$}%
 & %
{\scriptsize{}{}\enskip{}$-\alpha_{s}$\enskip{}}%
 & %
{\scriptsize{}{}$\enskip M_{X}\enskip$}%
 & %
{\scriptsize{}{}$\enskip\tau_{p}\enskip$}%
\tabularnewline
{\scriptsize{}{}$\left(m_{\rm P}\right)$}%
 & %
{\scriptsize{}{}$\left(10^{-12}\right)$}%
 & %
{\scriptsize{}{}$\left(10^{13}\,\text{Gev}\right)$}%
 & %
 & %
{\scriptsize{}{}$\left(m_{\rm P}\right)$}%
 & %
{\scriptsize{}{}$\left(m_{\rm P}\right)$}%
 & %
 & %
 & %
{\scriptsize{}{}$\left(10^{-4}\right)$}%
 & %
{\scriptsize{}{}$\left(\sim10^{16}\,\text{Gev}\right)$}%
 & %
{\scriptsize{}{}$\left(\text{years}\right)$ }%
\tabularnewline
\hline 
{\scriptsize{}{}$1.1$}%
 & %
{\scriptsize{}{}$4.79$}%
 & %
{\scriptsize{}{}$1.74$}%
 & %
{\scriptsize{}{}$50$}%
 & %
{\scriptsize{}{}$7.24$}%
 & %
{\scriptsize{}{}$2.10$}%
 & %
{\scriptsize{}{}$0.960$}%
 & %
{\scriptsize{}{}$0.0048$}%
 & %
{\scriptsize{}{}$8.07$}%
 & %
{\scriptsize{}{}$0.57$}%
 & %
{\scriptsize{}{}$5.0\times10^{34}$}%
\tabularnewline
 & %
{\scriptsize{}{}$3.95$}%
 & %
{\scriptsize{}{}$1.59$}%
 & %
{\scriptsize{}{}$55$}%
 & %
{\scriptsize{}{}$7.35$}%
 & %
{\scriptsize{}{}$2.10$}%
 & %
{\scriptsize{}{}$0.963$}%
 & %
{\scriptsize{}{}$0.0039$}%
 & %
{\scriptsize{}{}$6.67$}%
 & %
{\scriptsize{}{}$0.54$}%
 & %
{\scriptsize{}{}$4.2\times10^{34}$}%
\tabularnewline
 & %
{\scriptsize{}{}$3.32$}%
 & %
{\scriptsize{}{}$1.46$}%
 & %
{\scriptsize{}{}$60$}%
 & %
{\scriptsize{}{}$7.46$}%
 & %
{\scriptsize{}{}$2.10$}%
 & %
{\scriptsize{}{}$0.966$}%
 & %
{\scriptsize{}{}$0.0033$}%
 & %
{\scriptsize{}{}$5.61$}%
 & %
{\scriptsize{}{}$0.52$}%
 & %
{\scriptsize{}{}$3.6\times10^{34}$}%
\tabularnewline
\hline 
{\scriptsize{}{}$1.5$}%
 & %
{\scriptsize{}{}$6.87$}%
 & %
{\scriptsize{}{}$1.71$}%
 & %
{\scriptsize{}{}$50$}%
 & %
{\scriptsize{}{}$7.95$}%
 & %
{\scriptsize{}{}$3.093$}%
 & %
{\scriptsize{}{}$0.960$}%
 & %
{\scriptsize{}{}$0.0046$}%
 & %
{\scriptsize{}{}$7.88$}%
 & %
{\scriptsize{}{}$1.53$}%
 & %
{\scriptsize{}{}$2.6\times10^{36}$}%
\tabularnewline
 & %
{\scriptsize{}{}$5.69$}%
 & %
{\scriptsize{}{}$1.56$}%
 & %
{\scriptsize{}{}$55$}%
 & %
{\scriptsize{}{}$8.07$}%
 & %
{\scriptsize{}{}$3.093$}%
 & %
{\scriptsize{}{}$0.964$}%
 & %
{\scriptsize{}{}$0.0038$}%
 & %
{\scriptsize{}{}$6.52$}%
 & %
{\scriptsize{}{}$1.46$}%
 & %
{\scriptsize{}{}$2.1\times10^{36}$}%
\tabularnewline
 & %
{\scriptsize{}{}$4.79$}%
 & %
{\scriptsize{}{}$1.43$}%
 & %
{\scriptsize{}{}$60$}%
 & %
{\scriptsize{}{}$8.17$}%
 & %
{\scriptsize{}{}$3.093$}%
 & %
{\scriptsize{}{}$0.967$}%
 & %
{\scriptsize{}{}$0.0032$}%
 & %
{\scriptsize{}{}$5.48$}%
 & %
{\scriptsize{}{}$1.39$}%
 & %
{\scriptsize{}{}$1.8\times10^{36}$}%
\tabularnewline
\hline 
{\scriptsize{}{}$2$}%
 & %
{\scriptsize{}{}$7.59$}%
 & %
{\scriptsize{}{}$1.70$}%
 & %
{\scriptsize{}{}$50$}%
 & %
{\scriptsize{}{}$8.63$}%
 & %
{\scriptsize{}{}$3.897$}%
 & %
{\scriptsize{}{}$0.960$}%
 & %
{\scriptsize{}{}$0.0045$}%
 & %
{\scriptsize{}{}$7.79$}%
 & %
{\scriptsize{}{}$2.47$}%
 & %
{\scriptsize{}{}$1.6\times10^{37}$}%
\tabularnewline
 & %
{\scriptsize{}{}$6.29$}%
 & %
{\scriptsize{}{}$1.55$}%
 & %
{\scriptsize{}{}$55$}%
 & %
{\scriptsize{}{}$8.75$}%
 & %
{\scriptsize{}{}$3.897$}%
 & %
{\scriptsize{}{}$0.964$}%
 & %
{\scriptsize{}{}$0.0037$}%
 & %
{\scriptsize{}{}$6.45$}%
 & %
{\scriptsize{}{}$2.30$}%
 & %
{\scriptsize{}{}$1.3\times10^{37}$}%
\tabularnewline
 & %
{\scriptsize{}{}$5.29$}%
 & %
{\scriptsize{}{}$1.52$}%
 & %
{\scriptsize{}{}$60$}%
 & %
{\scriptsize{}{}$8.85$}%
 & %
{\scriptsize{}{}$3.897$}%
 & %
{\scriptsize{}{}$0.967$}%
 & %
{\scriptsize{}{}$0.0032$}%
 & %
{\scriptsize{}{}$5.42$}%
 & %
{\scriptsize{}{}$2.21$}%
 & %
{\scriptsize{}{}$1.1\times10^{37}$}%
\tabularnewline
\hline 
{\scriptsize{}{}$3$}%
 & %
{\scriptsize{}{}$7.92$}%
 & %
{\scriptsize{}{}$1.68$}%
 & %
{\scriptsize{}{}$50$}%
 & %
{\scriptsize{}{}$9.61$}%
 & %
{\scriptsize{}{}$5.956$}%
 & %
{\scriptsize{}{}$0.960$}%
 & %
{\scriptsize{}{}$0.0044$}%
 & %
{\scriptsize{}{}$7.73$}%
 & %
{\scriptsize{}{}$3.99$}%
 & %
{\scriptsize{}{}$1.2\times10^{38}$}%
\tabularnewline
{\scriptsize{}{}$ $}%
 & %
{\scriptsize{}{}$6.57$}%
 & %
{\scriptsize{}{}$1.53$}%
 & %
{\scriptsize{}{}$55$}%
 & %
{\scriptsize{}{}$9.72$}%
 & %
{\scriptsize{}{}$5.956$}%
 & %
{\scriptsize{}{}$0.964$}%
 & %
{\scriptsize{}{}$0.0037$}%
 & %
{\scriptsize{}{}$6.40$}%
 & %
{\scriptsize{}{}$3.81$}%
 & %
{\scriptsize{}{}$1\times10^{38}$}%
\tabularnewline
{\scriptsize{}{}$ $}%
 & %
{\scriptsize{}{}$5.54$}%
 & %
{\scriptsize{}{}$1.41$}%
 & %
{\scriptsize{}{}$60$}%
 & %
{\scriptsize{}{}$9.82$}%
 & %
{\scriptsize{}{}$5.956$}%
 & %
{\scriptsize{}{}$0.967$}%
 & %
{\scriptsize{}{}$0.0031$}%
 & %
{\scriptsize{}{}$5.39$}%
 & %
{\scriptsize{}{}$3.65$}%
 & %
{\scriptsize{}{}$8.5\times10^{37}$}%
\tabularnewline
\hline 
{\scriptsize{}{}$5$}%
 & %
{\scriptsize{}{}$8.07$}%
 & %
{\scriptsize{}{}$1.68$}%
 & %
{\scriptsize{}{}$50$}%
 & %
{\scriptsize{}{}$12.5$}%
 & %
{\scriptsize{}{}$7.95$}%
 & %
{\scriptsize{}{}$0.960$}%
 & %
{\scriptsize{}{}$0.0044$}%
 & %
{\scriptsize{}{}$7.69$}%
 & %
{\scriptsize{}{}$6.95$}%
 & %
{\scriptsize{}{}$7.8\times10^{38}$}%
\tabularnewline
{\scriptsize{}{}$ $}%
 & %
{\scriptsize{}{}$6.70$}%
 & %
{\scriptsize{}{}$1.53$}%
 & %
{\scriptsize{}{}$55$}%
 & %
{\scriptsize{}{}$12.7$}%
 & %
{\scriptsize{}{}$7.95$}%
 & %
{\scriptsize{}{}$0.964$}%
 & %
{\scriptsize{}{}$0.0037$}%
 & %
{\scriptsize{}{}$6.37$}%
 & %
{\scriptsize{}{}$6.63$}%
 & %
{\scriptsize{}{}$9.2\times10^{38}$}%
\tabularnewline
{\scriptsize{}{}$ $}%
 & %
{\scriptsize{}{}$5.65$}%
 & %
{\scriptsize{}{}$1.41$}%
 & %
{\scriptsize{}{}$60$}%
 & %
{\scriptsize{}{}$12.8$}%
 & %
{\scriptsize{}{}$7.95$}%
 & %
{\scriptsize{}{}$0.967$}%
 & %
{\scriptsize{}{}$0.0031$}%
 & %
{\scriptsize{}{}$5.35$}%
 & %
{\scriptsize{}{}$6.35$}%
 & %
{\scriptsize{}{}$1.3\times10^{40}$}%
\tabularnewline
\hline 
{\scriptsize{}{}$10$}%
 & %
{\scriptsize{}{}$8.13$}%
 & %
{\scriptsize{}{}$1.68$}%
 & %
{\scriptsize{}{}$50$}%
 & %
{\scriptsize{}{}$12.5$}%
 & %
{\scriptsize{}{}$7.95$}%
 & %
{\scriptsize{}{}$0.960$}%
 & %
{\scriptsize{}{}$0.0044$}%
 & %
{\scriptsize{}{}$7.68$}%
 & %
{\scriptsize{}{}$14.1$}%
 & %
{\scriptsize{}{}$1.9\times10^{40}$}%
\tabularnewline
\hline 
{\scriptsize{}{}$ $}%
 & %
{\scriptsize{}{}$6.75$}%
 & %
{\scriptsize{}{}$1.53$}%
 & %
{\scriptsize{}{}$55$}%
 & %
{\scriptsize{}{}$12.7$}%
 & %
{\scriptsize{}{}$7.95$}%
 & %
{\scriptsize{}{}$0.964$}%
 & %
{\scriptsize{}{}$0.0037$}%
 & %
{\scriptsize{}{}$6.35$}%
 & %
{\scriptsize{}{}$13.5$}%
 & %
{\scriptsize{}{}$1.6\times10^{40}$}%
\tabularnewline
\hline 
{\scriptsize{}{}$ $}%
 & %
{\scriptsize{}{}$5.69$}%
 & %
{\scriptsize{}{}$1.41$}%
 & %
{\scriptsize{}{}$60$}%
 & %
{\scriptsize{}{}$12.8$}%
 & %
{\scriptsize{}{}$7.95$}%
 & %
{\scriptsize{}{}$0.967$}%
 & %
{\scriptsize{}{}$0.0031$}%
 & %
{\scriptsize{}{}$5.33$}%
 & %
{\scriptsize{}{}$12.9$}%
 & %
{\scriptsize{}{}$1.3\times10^{40}$}%
\tabularnewline
\hline 
\end{tabular}\caption{{Inflationary predictions of the AV branch solutions for different
parameter values. }}

\label{tab1} 
\end{table}
\par\end{center}

\section{Type I seesaw mechanism and neutrino masses}

\label{seesawSec}

In this section, we further extend our model through type I seesaw mechanism with global lepton number symmetry, whose spontaneous breaking leads to the generation of neutrino masses. In this framework, we suppose the singlet field $\Phi$ carries two units of lepton number
and is coupled to the three generation of singlet right handed Majorana neutrinos (RHNs), from \cite{Boucenna:2014uma}

\begin{equation}
V_{N}=V\left(\Phi,\,\bar{X},\,\Sigma\right)+Y_{D}^{ij}\bar{l_{L}^{j}}i\tau_{2}H^{\star}\nu_{R}^{i}+\frac{1}{2}Y_{N}^{i}\Phi f\left(\frac{\Phi}{\bar{X}}\right)  \overline{\nu_{R}^{ic}}\nu_{R}^{i}+h.c,\label{neutrino-1}
\end{equation}
where $l$ is the lepton doublet, $\tau_{2}$ is the second Pauli
matrix. Here $Y_{D}$ is the Yukawa coupling matrix of the SM Higgs
coupling to the left handed neutrinos and $Y_{N}$ is the coupling
matrix of the singlet field to the three generations of Majorana right
handed neutrinos $\left(\nu_{R}^{i}\right)$. In principle, we can
also weakly couple the inflaton with the SM Higgs boson as

\begin{equation}
V_{h}=V_{N}+\lambda_{h}f\left(\frac{\Phi}{\bar{X}}\right)\Phi^{\dagger}\Phi H^{\dagger}H\,.\label{Reheapot}
\end{equation}
We note that even with the new potential in (\ref{Reheapot}), conformal
symmetry in (\ref{CFTSU(5)}) can be preserved by the following additional transformations\footnote{The kinetic terms and couplings of SM Higgs and RHNs to the Ricci scalar are irrelevant here and can neglected in comparison with the inflaton dynamics.}

\begin{equation}
l_{L}^{i}\to\Omega^{3/2}l_{L}^{i}\,,\quad\nu_{R}^{i}\to\Omega^{3/2}\nu_{R}^{i}\,,\quad H\to\Omega H\,.\label{newcon}
\end{equation}
Applying SBCS via $\bar{X}=\bar{X}^*=\sqrt{3}M$ and computing 1-loop corrections due to the additional couplings to neutrinos
(\ref{neutrino-1}) and SM Higgs, the effective potential of the field $\phi$ becomes

\begin{equation}
V_{f}^{eff}=\frac{36A_{f}M^{4}}{m_{\rm P}^{4}}f^{2}\left(\frac{\phi}{\sqrt{3}M}\right)\phi^{4}\ln\left(\frac{\phi^{2}f\left(\frac{\phi}{\sqrt{3}M}\right)}{\mu_{f}^{2}}
-\frac{1}{4}\right)+\frac{A_{f}\mu_{f}^{4}}{4}\,,\label{v1eff}
\end{equation}
where $A_{f}=\frac{\beta_{f}}{32\pi^{2}}$ and
\begin{equation}
\beta_{f}=20\bar{\lambda}_{2}^{2}+2\lambda_{h}^{2}+2\bar{\lambda}_{2}\sum_{i}\left(Y_{N}^{i}\right)^{2}-\sum_{i}\left(Y_{N}^{i}\right)^{4}\,.\label{betaf}
\end{equation}
In (\ref{betaf}) we assume the coupling constant $Y_{N}^{i}$ to
be at least $\mathcal{O}\left(10\right)$ smaller than $\bar{\lambda}_{2}$
and $\lambda_{h}\ll Y_{N}^{i}$, such that $\beta_{f}\sim20\bar{\lambda}_{2}^{2}$
and $\mu_{f}\sim\mu$. Therefore during inflation the coupling of a singlet field to the adjoint scalar $\Sigma$ dominates, consequently
the inflationary predictions in Table. \ref{tab1} are unaffected by
this additional couplings to Higgs and singlet neutrinos. However,  since we impose $\lambda_{h}\ll Y_{N}^{i}$, the inflaton field dominantly decays to RHNs rather than to SM Higgs. 

Let us consider that the lepton number violation happens at a scale
when $\langle\phi\rangle=\mu$. Computing the mass matrix of singlet
and doublet neutrinos in the basis of $\nu_{L},\,\nu_{R}$ using the
Einstein frame potential of (\ref{neutrino-1}), we have

\begin{equation}
\mathcal{M}_{\nu}=\left[\begin{matrix}0 & Y_{D}v_{2}\\
Y_{D}^{T}v_{2} & \frac{m_{\rm P}^{2}}{M^{2}}\frac{\langle\phi\rangle Y_{N}}{\sqrt{2}}
\end{matrix}\right]\,,\label{massmatrix}
\end{equation}
where $v_{2}=246\,\text{GeV}$ is the Electroweak vacuum. The light neutrino
mass can be obtained from perturbative diagonalization of (\ref{massmatrix})
as

\begin{equation}
m_{\nu_{L}}\simeq\sqrt{2}Y_{D}Y_{N}^{-1}Y_{D}^{T}\frac{v_{2}^{2}}{\mu}\frac{M^{2}}{m_{\rm P}^{2}}\,.\label{lightneutrino}
\end{equation}
The mass of heavy RHNs is given by 
\begin{equation}
m_{\nu_{R}}=\frac{Y_{N}\langle\phi\rangle}{\sqrt{2}}\frac{m_{\rm P}^{2}}{M^{2}}\,.\label{RHmass}
\end{equation}
The essence of seesaw mechanism is the generation of neutrino masses
resulting light left handed neutrinos and heavy right handed neutrinos.
Both here are related to the VEV of the inflaton field. 

The current {\it Planck} data indicates the sum of light neutrino masses
constrained as $\sum m_{\nu_{i}}<0.23\,eV$ \cite{Ade:2015xua}. Therefore
considering the light neutrino mass to be $m_{\nu_{L}}\sim\mathcal{O}(0.1)\,\text{eV}$,
(\ref{lightneutrino}) gives a relation

\begin{equation}
Y_{N}\simeq6\sqrt{2}Y_{D}^{2}\frac{10^{14}\,GeV}{\mu}\frac{M^{2}}{m_{\rm P}^{2}}\,.\label{YNYD}
\end{equation}
Taking $Y_{D}\sim\mathcal{O}\left(10^{-1}\right)$ and from Table.
\ref{tab1} imposing $\mu\sim1.2m_{\rm P}-24.37\,m_{\rm P}$, we get $2.5\times10^{-6}\lesssim Y_{N}^{i}\lesssim1.0\times10^{-5}$.
This supports our previous assumptions after (\ref{betaf}) that the
couplings to the RHNs have negligible effect for inflation.
Our generalization of the SV model successfully fits into explaining the origin of neutrino masses. We can also take $Y_{D}<\mathcal{O}\left(10^{-1}\right)$
which results in smaller values for $Y_{N}<\mathcal{O}\left(10^{-6}\right)$.
Taking $Y_{N}\sim10^{-6}$, the heavy RHN mass will
be around $m_{\nu_{R}}\sim4\times10^{12}\,\text{GeV}.$ For $Y_{N}<\mathcal{O}\left(10^{-6}\right)$
we can lower the masses of RHNs. In the next section
we aim to study reheating in our inflationary scenario, taking into
account the constraints we have derived so far. 

\section{Reheating and non-thermal leptogenesis}

\label{ReheatSec}

We consider reheating through a dominant decay of the inflaton into heavy
RHNs which requires $m_{\varphi}\gtrsim 2m_{\nu_{R}}$.
The mass of the canonically normalized field $\varphi$ at the minimum
of the potential is given by the second derivative of the potential
(\ref{varphipot}) 
\[
m_{\varphi}=\sqrt{V_{\varphi,\,\varphi}^{E}}\Big\vert_{\varphi=\langle\varphi\rangle}=2\times10^{-6}\mu,
\]
where we have taken a value for $A\sim5\times10^{-12}$ from Table~\ref{tab1}.
%In Fig.~\ref{massphi} we plot the range inflaton mass for different
%values of $M\sim1.1m_{\rm P}-10m_{\rm P}$.

%\begin{figure}[H]
%\centering\includegraphics[height=2.5in]{inflatonmass}\caption{In this plot we depict inflaton mass $m_{\varphi}$ for $M=1.1-10m_{\rm P}$.
%We have taken $\bar{A}\sim5\times10^{-12}$. }

%\label{massphi} 
%\end{figure}

\begin{figure}[h]
\centering\includegraphics[height=2.5in]{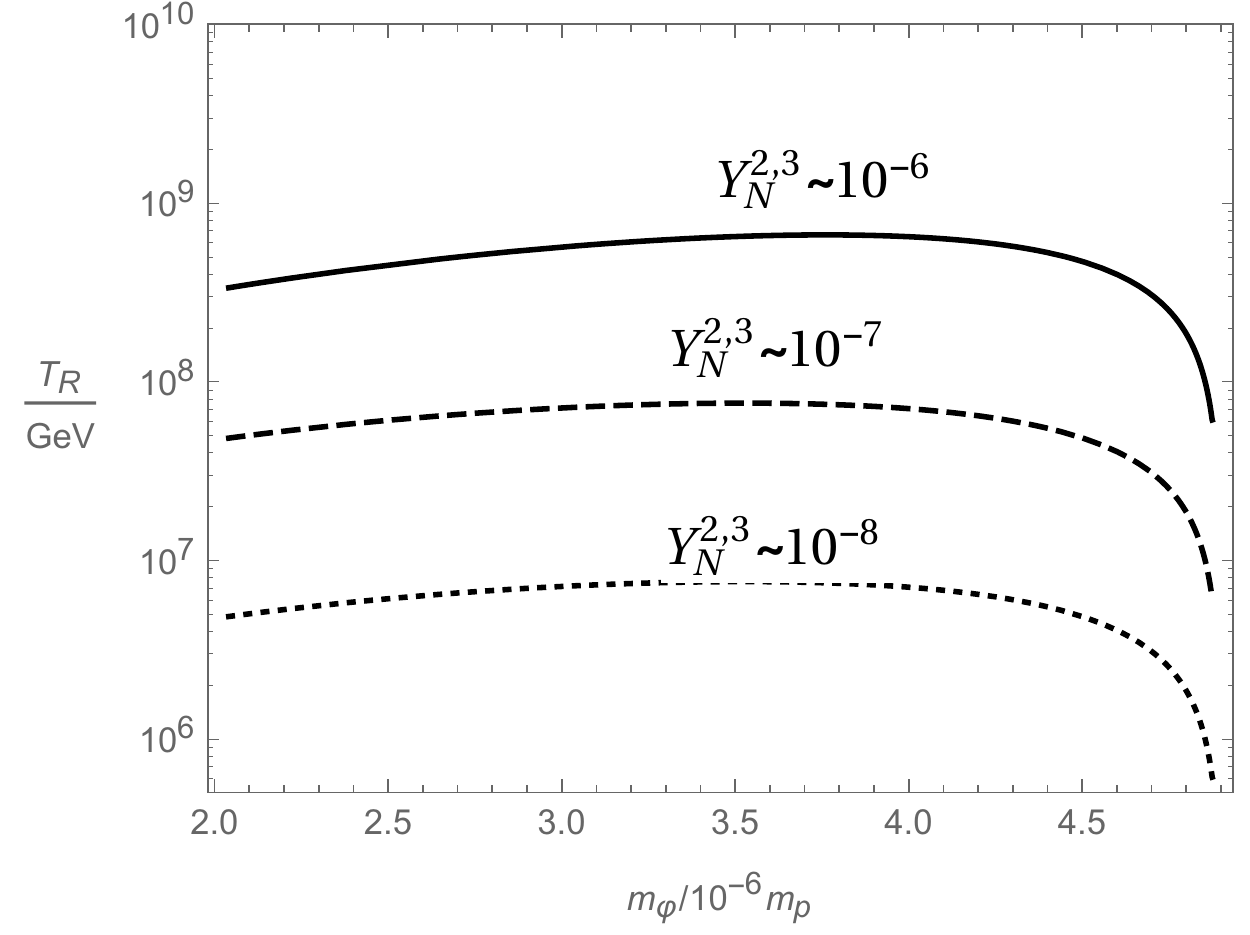}\caption{In this plot we depict the reheating temperatures $T_{R}$ Vs. $m_{\varphi}$
for the values of couplings $Y_{N}^{2,3}\sim10^{-8}-10^{-6}$.}

\label{TRm} 
\end{figure}

We implement the scheme of non-thermal leptogenisis proposed in \cite{Fukugita:1986hr,Lazarides:1991wu}
which can give rise to baryogenesis through CP violating decays
of RH Majorana neutrinos. In this section we closely follow
in \cite{Asaka:2002zu,Senoguz:2003hc,Senoguz:2005bc}. We consider:
\begin{itemize}
\item Hierarchical masses for RHNs $m_{\nu_{R}^{1}}\ll m_{\nu_{R}^{2}}\sim m_{\nu_{R}^{3}}$.
To arrange this we require the coupling constants to be $Y_{N_{1}}\ll Y_{N_{2}}\sim Y_{N_{3}}$.
{We assume that the inflaton decays equally into the two
heavy RHNs $\nu_{R}^{2,3}$ and the corresponding
reheating temperature can be computed using }\cite{Asaka:2002zu,Okada:2013vxa}
\end{itemize}
\begin{equation}
T_{R}=\left(\frac{90}{\pi^{2}g_{*}}\right)^{1/4}\sqrt{\Gamma_{\varphi}\left(\varphi\to\nu_{R}^{i}\nu_{R}^{i}\right)m_{\rm P}}\,,\label{TRexp}
\end{equation}
where $g_{*}=105.6$ is the number of relativistic degrees of freedom
and the decay rate is given by 

\[
\Gamma_{\varphi}\left(\varphi\to\nu_{R}^{i}\nu_{R}^{i}\right)\simeq\frac{m_{\varphi}}{4\pi}\sum_{i=1}^{3}c_{i}^{2}\left(\frac{m_{\nu_{R}^{i}}}{m_{\rm P}}\right)^{2}\left(1-\frac{4m_{\nu_{R}^{i}}^{2}}{m_{\varphi}^{2}}\right)^{3/2}\,.
\]
{The masses of heavy RHNs are $m_{\nu_{R}^{2,3}}\sim\frac{Y_{N}^{2,3}}{\sqrt{2}}$,
which for $Y_{N}^{2,3}\sim10^{-8}-10^{-6}$ we have $m_{\nu_{R}^{2,3}}\sim10^{10}-10^{12}$
GeV. In Fig.~\ref{TRm} we plot the possible reheating temperatures
of our case taking $c_{1}\approx0$ and $c_{2}=c_{3}=1$.}

\begin{itemize}
\item The decays of RH Majorana neutrinos $\nu_{R}^{i}$ break the lepton number
conservation and leads to CP violation. There are two decay channels
\end{itemize}
\begin{equation}
\Gamma_{i}:\nu_{R}^{i}\to H+l_{i}\,,\quad\bar{\Gamma}_{i}:\nu_{R}^{i}\to H^{\dagger}+\bar{l}_{i}\,,\label{channels-nuR}
\end{equation}
where $H$ and $l$ denote the Higgs field and the lepton doublets
of the SM. The lepton asymmetry generated by the CP violation decays
of $\nu_{R}^{i}$ is measured by the following quantity 

\begin{equation}
\epsilon_{i}\equiv\frac{\Gamma_{i}-\bar{\Gamma}_{i}}{\Gamma_{i}+\bar{\Gamma}_{i}}\lll1\,.\label{epsi}
\end{equation}
{CP asymmetry $\epsilon_{i}$ can be computed for
the dominant decays of $\nu_{R}^{2,3}$ using \cite{Flanz:1994yx,Buchmuller:1997yu,Hamaguchi:2002vc,Senoguz:2003hc}}

{
\begin{equation}
\epsilon_{i}=-\frac{1}{8\pi}\frac{1}{\left(Y_{D}Y_{D}^{\dagger}\right)_{11}}\sum_{i=2,3}\text{Im}\left[\left\{ \left(Y_{D}Y_{D}^{\dagger}\right)_{1i}\right\} ^{2}\right]\left[f\left(\frac{m_{\nu_{R}^{i}}^{2}}{m_{\nu_{R}^{1}}^{2}}\right)+g\left(\frac{m_{\nu_{R}^{i}}^{2}}{m_{\nu_{R}^{1}}^{2}}\right)\right]\,,\label{epsilon_i-exp}
\end{equation}
where
\[
f\left(y\right)=\sqrt{y}\left[-1+\left(y+1\right)\ln\left(1+\frac{1}{y}\right)\right]\,,\quad g\left(y\right)=\frac{\sqrt{y}}{y-1}\,.
\]
Here we only aim to constrain the range of values for $\epsilon_{i}$
leaving for future the explicit computation of constraining Yukawa matrix $Y_{D}^{ij}$ \cite{Asaka:2002zu}. }

The lepton asymmetry is given by

\begin{equation}
\frac{n_{L}}{s}=\sum_{i=1}^{3}\epsilon_{i}\text{Br}_{i}\frac{3T_{R}}{2m_{\varphi}}\,,\label{nLs}
\end{equation}
where $n_{L}$ is the difference between number of leptons and anti-leptons
and $s$ indicates the entropy density, $\text{Br}_{i}$ denotes the
branching ratio 
\begin{itemize}
\item The production of RH Majorana neutrinos happens non-thermally and sufficiently
late so that the produced lepton asymmetry sources the baryon asymmetry
at a later stage. This essentially requires $m_{\nu_{R}^{1}}\gtrsim T_{R}$
so that the later decay of lightest RH Majorana neutrino $\nu_{R}^{1}$
does not wash away the produced lepton asymmetry by the heavy ones.
We assume there is an accidental $B-L$ conservation\footnote{$B,\,L$ refers to baryon number and lepton number respectively. }
such that sphaleron process is active which brings a part of the above
lepton asymmetry into the baryon asymmetry (see Ref.~\cite{Khlebnikov:1988sr,Harvey:1981yk,Harvey:1990qw}
for details). {As the reheating temperature in our
case is $T_{R}\sim10^{6}-10^{9}\,\text{GeV}$ (see Fig.~\ref{TRm}) we take
$Y_{N}^{1}\sim10^{-10}-10^{-9}$ such that $m_{\nu_{R}^{1}}\sim10^{8}-10^{9}\,\text{GeV}$
. Therefore, with values $m_{\nu_{R}^{2,3}}\sim10^{10}-10^{12}\,\text{GeV}$
, $m_{\nu_{R}^{1}}\sim10^{8}-10^{9}\,\text{GeV}$ and $T_{R}\sim10^{6}-10^{9}\,\text{GeV}$
we have met the conditions for successful leptogenesis, $m_{\nu_{R}^{2}}\sim m_{\nu_{R}^{3}}\gg m_{\nu_{R}^{1}}$
and $m_{\nu_{R}^{1}}\gtrsim T_{R}$.}
\end{itemize}
Baryon asymmetry is proportional to the lepton asymmetry as 

\begin{equation}
\begin{aligned}\frac{n_{B}}{s}\simeq & \frac{28}{79}\frac{n_{L}}{s}\\
\simeq & \frac{42}{79}\sum_{i=1}^{3}\epsilon_{i}\text{Br}_{i}\frac{T_{R}}{m_{\varphi}}\,.
\end{aligned}
\label{nBsnls}
\end{equation}

The baryon asymmetry which is measured by the ratio of the difference
between the number of baryons minus the anti-baryons $n_{B}$ to the
entropy density in the present Universe, is constrained \cite{Ade:2015xua} in the following form

\begin{equation}
\frac{n_{B}}{s}=\left(6.05\pm0.06\right)\times10^{-10}\,.\label{nBs}
\end{equation}
Considering branching ratios $\text{Br}_{1}=0$ and $\text{Br}_{2}=\text{Br}_{3}=\frac{1}{2}$
with $\epsilon_{1}\ll\epsilon_{2}\sim\epsilon_{3}$ we have 

\begin{equation}
\frac{n_{B}}{s}\approx\epsilon_{2}\frac{T_{R}}{m_{\varphi}}\,.\label{nbsapp}
\end{equation}
{From Fig.~\ref{TRm} we can read that $\frac{T_{R}}{m_{\varphi}}\sim10^{-7}-10^{-4}$
, which indicates the CP violation in the decay of RH Majorana neutrinos
$\left(\epsilon_{i}\right)$ must be in the range $6\times10^{-6}\lesssim\epsilon_{2,3}\lesssim6\times10^{-3}$
to have the observed baryon asymmetry.} 

\section{Summary}

\label{conclusions}

Coleman-Weinberg inflation \cite{Shafi:1983bd} has been a successful
and realistic model based on GUT and is consistent with current {\it Planck}
data with $r\gtrsim0.02$ \cite{Okada:2014lxa}. In this chapter, we
have further generalized the framework of CW inflation with an additional
conformal symmetry. Spontaneous breaking of conformal symmetry is useful to 
create a hierarchy of mass scales, therefore it is natural to realize this symmetry in GUT models. 
In this respect, two complex singlet fields of
$\text{SU}(5)$ or $SO(10)$ are considered and are coupled to the GUT fields in
a suitable manner. We have showed that this setup, upon spontaneous
breaking of GUT and conformal symmetry, leads to an interesting inflationary
scenario driven by the real part of the singlet field. In our model,
the above VEV branch of CW potential gets flattened to a Starobinsky
plateau allowing for $n_{s}\sim0.96-0.967$ and $r\sim0.003-0.005$
for $50-60$ number of $e$-foldings. We found that these predictions
are independent of the VEV of the inflaton field. However, values
of inflaton VEV affect the masses of the superheavy gauge bosons
that mediate the proton decay. We calculated the corresponding estimates
for proton life time above the current lower bound from Super-K
data $\tau_{p}\left(p\to\pi^{0}+e^{+}\right)>1.6\times10^{34}$. In
the next step, we introduced a coupling between the complex singlet field
with the generation of three singlet RHNs, where the singlet
field is assumed to carry two units of lepton number. We implemented
type I seesaw mechanism where spontaneous symmetry breaking of global
lepton number results in generating neutrino masses. We put an upper
bound to the inflaton couplings to RHNs assuming inflation is dominated by inflaton 
couplings to GUT field. For the non-thermal
leptogenesis to happen, we have considered dominant decay of inflaton
into some of the RHNs and obtained the corresponding reheating
temperatures as $10^{6}\text{ GeV}\lesssim T_{R}<10^{9}$ GeV. In
summary, our new development of CW inflation can be tested within
future CMB data \cite{Creminelli:2015oda}. 

%In this chapter, we mainly restricted the non-SUSY construction of CW with conformal symmetry. It would be interesting to consider this model in SUGRA with superconformal symmetries which we defer for future investigations.  
\chapter{Conclusions and outlook}
\label{concouttese}

\begin{chapquote}{Sir Arthur Conan Doyle, \textit{Sherlock Holmes}}
% when you have eliminated the impossible, whatever remains, however improbable,must be the truth
Never theorize before you have data. Invariably, you end up twisting facts to suit theories, instead of theories to suit facts
\end{chapquote}

\lhead{\bf Chapter 7. \emph{Conclusions and outlook}} %

\section*{Conclusions}

In this thesis, we have studied inflationary scenarios in string theory, SUGRA and particle physics.
We have covered aspects of inflationary models following a top-down or bottom-up motivations as we described in the introduction. It is important to understand the physics of inflation from the point of view of UV completeness as well as from the point of view of physics beyond SM. Both of these motivations are naturally appealing on their own. In the scope of the latest CMB data from {\it Planck} 2015 and the upcoming ground based and space based CMB probes, it is a greater necessity than before that we not only construct interesting models of inflation but test them observationally. Moreover, we need to concentrate on developing theoretical frameworks of inflation towards generality/naturality rather than simplicity. In this respect, this thesis uncovers models beyond the conventionality, towards realistic features within the fundamental theories. This perspective is made concrete within our model by model brief appraisal in what follows. 

3-form fields are viable alternative to conventional scalar fields. 3-forms were known to have different dynamics than scalar fields, which allow inflation as an attractor phenomenon \cite{Koivisto:2009fb}. In chapter \ref{In-3-forms}, bearing the fact that multifields are more natural in string theory settings, we have studied the multiple 3-form inflation. We have explored possible dynamics of two 3-form fields with suitable choice of potentials. We have put the model for test with $(n_{s},\,r)$, running of $n_{s}$ and detailed study of non-Gaussianities. Moreover, we must notice that even though {\it Planck} 2015 data favours single field inflation, there is a wide scope of interesting dynamics and parameter space for multifield inflation \cite{Elliston:2013afa}. Therefore, in the future it would be interesting to consider a wider study of inflationary scenarios with 3-form fields with more complicated choice of potentials than we have studied herein.

Our study of DBI Galileon model in chapter \ref{DBIGc}, is also beyond the conventional DBI model. We studied DBIG model as it is indeed a natural framework in string theory, where the motion of the D-brane in the bulk space imparts effects of induced gravity \cite{deRham:2010eu}. We significantly scanned the parameter space of the model with respect to $n_{s}$ and $r$. We also contemplated our study of parameter space with available results on ``equilateral'' shape of non-Gaussianities. Overall, we have shown that the model is observationally improved over DBI inflation. However, we must note that it is important to understand the role of the geometry of the bulk space, in which the motion of the D-brane induces the inflationary expansion. We mostly relied on numerical analysis but theoretical studies regarding warped geometries and inflaton potential remain to be done. Moreover, a detailed study of non-Gaussianities, especially of orthogonal shape \cite{RenauxPetel:2011dv}, is very much required for this model, to allow it to be rigorously tested in the future observations.

Identifying SFT as a crucial part of string theory to be UV complete, in the framework of SFT, we have proposed in chapter \ref{SFTin}, a class of effective models based on the phenomenon of TC and non-locality. This study introduces a very new framework of inflation driven by closed string dilaton in SFT, where we consider the TC to happen above the inflationary energy scale, which is very different from popular models of inflation driven by tachyon \cite{Feinstein:2002aj}. Within our considerations, we obtained single field inflation with the potential (\ref{flatpot}), where the parameter $B$ is obtained witin our SFT setup. The interesting part of this study is that we have demonstrated how conformal symmetry can emerge in our SFT framework, given that conformal inflationary models are one of the best fit with current data. A further study of non-Gaussianity in this setup is essential to distinguish these models from the other competitive scenarios in the literature. The models in this chapter are, though viable with respect to observational data, quite speculative. More theoretical progress has to be done within SFT, to strengthen this framework.

In the SUGRA framework, we have explored the quite well known $\alpha-$attractor model in chapter \ref{CAM}. Our study, that follows a non-slow-roll approach, highlighted the importance of its inflaton's non-canonical kinetic term  and we also found a new class of potentials. With our efforts, the connection between K\"ahler manifold and the inflaton dynamics is more clear. However, as far as inflationary predictions are concerned our approach is indistinguishable from the original model with the respective potentials. Perhaps, only a study of the subsequent reheating might help in this regard, to falsify our approach.

Realizing conformal symmetry as the fundamental symmetry of nature \cite{Hooft:2014daa}, in the GUT model studied in chapter \ref{CGUTin}, we generalized CW inflation \cite{Shafi:1983bd}. We have shown that the inflationary predictions $\left(n_{s},\,r\right)$ in this model are the same as with the Starobinsky and Higgs inflation. Moreover, we obtain several predictions in particle physics context such as proton life time, neutrino masses and leptogenesis. Therefore, this model can be tested outside of the CMB observations. It would be interesting to extend this model in SUGRA with superconformal symmetries to attain UV completion which we defer to future studies. 

In summary, the thesis presents many facets of inflationary cosmology in fundamental theories and all of them fit successfully with {\it Planck} 2015 data. Moreover, we developed several theoretical aspects which opens new routes for interesting research in future. In the next section, we present an outlook focusing on recent trends in inflationary cosmology and point out to an outlook regarding the successful models we have studied so far. 

\section*{Outlook}

Although the inflationary paradigm is successful and observationally consistent with CMB data so far, the physical origin of inflation is still uncertain. Even after {\it Planck} 2015 data, by means of which simple inflationary models were ruled out, there are still many models being viable \cite{Martin:2013tda}. Moreover, studying inflation is so far the only way to probe observationally the physics of very high energy scales (e.g., GUT scale). Hence, it is quite natural to hope and look in inflationary cosmology for the signatures of string theory, SUGRA and beyond SM. Being more precise, inflationary cosmology is playing a vital role in the broad area of high energy physics, strengthening our efforts to ultimately build a UV complete theory as well as its connection to the current understanding of Universe.

We summarize here some lines of future research which could be useful to further distinguish not only the models we have studied in thesis but also among the other still viable models in literature, by means of upcoming CMB probes \cite{Martin:2013tda}:

\begin{itemize}	
\item Reheating/preheating after inflation and connecting to SM is crucial, especially in string/SUGRA based models. Although it is hard to observationally probe the reheating epoch, it might be crucial for the completeness of a model \cite{Cook:2015vqa,Martin:2016oyk}. In this aspect, we must study the dynamics of inflaton field at the reheating epoch and consider possible decays of inflaton field into SM or beyond SM degrees of freedom \cite{Kofman:1997yn}. 
\item {\it Planck} 2015 data has strongly indicated the presence of certain anomalies in the CMB such as power suppression at low multipoles,  hemispherical asymmetry and the regions of hot and cold spots \cite{Ade:2015hxq}, which has no consistent theoretical explanation so far. These could indicate new physics and perhaps it is time to intensify our theoretical efforts to explain these anomalies \cite{Sanchez:2016lfw,Gruppuso:2015xqa}. 
\item A more careful understanding is required for the case of single field inflation addressing the issue of the so called $\eta$-problem \cite{StGC-cqg}. If inflation is believed to be originated in the UV complete theories such as string theory or SUGRA, ubiquitous presence of heavy fields might leave non-trivial imprints in the primordial bispectrum and power spectrum \cite{Arkani-Hamed:2015bza,Chen:2014cwa,Chen:2015lza}. 
\item Studying tensor scalar cross correlations have recently gained much of interest and indeed they are a powerful tool to classify several models. Although observation of cross correlation spectra is yet not viable, it is worth to invest on these studies \cite{Bordin:2016ruc}. 
\item It is worthy to consider inflationary scenarios when addressing the origins of dark matter, leptogenesis, baryogenesis and neutrino masses. This would enable testing inflationary models outside CMB i.e., at collider and astrophysical observations \cite{Valle:2015mma}. 
\item To expand the scope of testability for inflationary models, it is useful to build unified models of inflation and dark energy such as Higgs-dilaton model \cite{Bezrukov:2012hx}. 
\end{itemize}

%----------------------------------------------------------------------------------------
%	THESIS CONTENT - APPENDICES
%----------------------------------------------------------------------------------------

\addtocontents{toc}{\vspace{2em}} % Add a gap in the Contents, for aesthetics

\appendix % Cue to tell LaTeX that the following 'chapters' are Appendices

% Include the appendices of the thesis as separate files from the Appendices folder
% Uncomment the lines as you write the Appendices

\chapter{Inflationary observables} % Main appendix title
%\label{AppendixB} % For referencing this appendix elsewhere, use \ref{AppendixA}
%\label{ADBIG}
\label{Intro-app} 
\lhead{Appendix A. Introduction} % This is for the header on each page - perhaps a shortened ti

%\section*{Appendices}
%
%\vspace*{10mm}
%%%%%%%%%%%%%%%%%%%%%%%%%%%%%%%%%%%%%%%%%%%%%%%%%%%%%%%%%%

This appendix provides complementary information concerning chapter  \ref{Intro}.

%%%%%%%%%%%%%%%%%%%%%%%%%%%%%%%%%%%%%%%%%%%%%%%%%%%%%%%%%%%%%%%%%%%%
%%%%%%%%%%%%%%%%%%%%%%%%%%%%%%%%%%%%%%%%%%%%%%%%%%%%%%%%%%%%%%%%%

\section{General definitions}

\label{curvedef}

Consider FLRW spacetime

\begin{equation}
ds^{2}=-dt{}^{2}+a^{2}(t)d\mathbf{x}^{2},\label{FLRW}
\end{equation}
where $a(t)$ is the scale factor with $t$ being the cosmic time, $\mathbf{x}$ is three dimensional space vector. 

The curvature perturbation $\zeta\left(t,\,\mathbf{x}\right)$ in the comoving gauge is defined by the perturbation of the spatial part of the metric 

\begin{equation}
\delta g_{ij}=a^{2}\left(h_{ij}+2\zeta\left(t,\,\mathbf{x}\right)\delta_{ij}\right) \,, \label{curvpert}
\end{equation}
where $h_{ij}$ denotes the tensor fluctuation in the 3+1 ADM (Arnowitt-Deser-Misner) decomposition of the metric \cite{Arnowitt:1962hi,Baumann:2009ds}. 

The second and higher order (quantum) correlations function of $\zeta$ relates to the properties of temperature anisotropies $\frac{\Delta T}{T}$ at a point in the (CMB) sky $\mathbf{x}$. A two-point function correlates the density or temperature fluctuations at two points in space, measured by the power spectrum  $P_{\zeta}(k)$ whose distribution is in general Gaussian 

\begin{equation}
\langle\zeta(\mathbf{x})\zeta(\mathbf{x}^{\prime})\rangle\vert_{t=t_{f}}\equiv\int \frac{d^{3}k}{(2\pi)^{3}}\mathcal{P}_{\zeta}(k)e^{i\mathbf{k}.(\mathbf{x}-\mathbf{x}^{\prime})}\,,
\label{two-point}
\end{equation}
In the Fourier space (\ref{two-point}) can be written as 
\begin{equation}
\langle\zeta\left(\mathbf{k_{1}}\right)\zeta\left(\mathbf{k_{2}}\right)\rangle\vert_{t=t_{f}} =\left(2\pi\right)^{5}\delta^{3}\left(\mathbf{k_{1}}+\mathbf{k_{2}}\right)
\frac{1}{2k_{1}^{3}}{P}_{\zeta}\left(k_{1}\right)\,.
\end{equation}
where $t_{f}$ is the time corresponds to superhorizon scales, the power spectrum $\mathcal{P}_{\zeta}(k)=\frac{k^{3}}{2\pi^{2}}\vert \zeta_{k}\left(k,\,t\right)\vert^{2}$ and $\zeta_{k}\left(k,\,t\right)$ is the mode function of curvature perturbation 
$\zeta_k=\int d^{3}x\,\zeta\left(\mathbf{x},\,t\right)\,e^{i\mathbf{k}.\mathbf{x}}$, which is computed usually from the second order perturbation of inflationary action (see Sec. \ref{G-infapp}).

In general, scalar and the tensor power spectrum scales in the power-law form as \cite{Baumann:2009ds,Ade:2015ava}
\begin{equation}
\mathcal{P}_{\zeta}(k)=\mathcal{P}_{\zeta}(k_*) \Big(\frac{k}{k_*} \Big)^{n_s-1}~\,,\quad \mathcal{P}_{t}(k)=\mathcal{P}_{t}(k_*) \Big(\frac{k}{k_*} \Big)^{n_t}.
\end{equation}
where $k_*=0.002\,\text{Mpc}^{-1}$ is the pivot scale. From the Planck 2015 data \cite{Ade:2015xua}, the pivot scale power spectrum is measured as $P_{\zeta}(k_*)\approx 2.2\times 10^{-9}$. Here $n_{s}$ and $n_{t}$ are named scalar and tensor spectral tilt (or spectral index) respectively. The ratio of tensor to scalar power spectra $r=\frac{\mathcal{P}_{t}}{\mathcal{P}_{\zeta}}$ is another important inflationary observable. 

The 3-point, 4-point correlation functions of curvature perturbation are useful to further characterize the nature of inflaton and these are defined, respectively, by 
\begin{eqnarray}
\langle\zeta\left(\mathbf{k_{1}}\right)\zeta\left(\mathbf{k_{2}}\right)\zeta\left(\mathbf{k_{3}}\right)\rangle & = & \left(2\pi\right)^{3}\delta\left(\mathbf{k_{1}}+\mathbf{k_{2}}+\mathbf{k_{3}}\right)\mathcal{B}_{\zeta}\left(k_{1},k_{2},k_{3}\right)\,,\\
\langle\zeta\left(\mathbf{k_{1}}\right)\zeta\left(\mathbf{k_{2}}\right)\zeta\left(\mathbf{k_{3}}\right)\zeta\left(\mathbf{k_{4}}\right)\rangle & = & \left(2\pi\right)^{3}\delta\left(\mathbf{k_{1}}+\mathbf{k_{2}}+\mathbf{k_{3}}+\mathbf{k_{4}}\right)
\mathcal{T}_{\zeta}\left(k_{1},k_{2},k_{3},k_{4}\right)\,,
\end{eqnarray}
where $B_{\zeta}\left(k_{1},k_{2},k_{3}\right)$, $\mathcal{T}_{\zeta}\left(k_{1},k_{2},k_{3},k_{4}\right)$ are called the bispectrum and the trispectrum\footnote{The discussion on trispectrum is beyond the scope of this thesis since the current constraints are far less stringent \cite{Ade:2015ava}.} respectively. 
Often the bispectrum is normalized to form the reduced
bispectrum $\fnl\left(k_{1},k_{2},k_{3}\right)$ as
\begin{equation}
%\begin{aligned}\end{aligned}
B_{\zeta}\left(k_{1},k_{2},k_{3}\right)=\frac{6}{5}f_{{\rm NL}}(k_{1},k_{2},k_{3})\biggl[P_{\zeta}\left(k_{1}\right)P_{\zeta}\left(k_{2}\right)+P_{\zeta}\left(k_{2}\right)P_{\zeta}\left(k_{3}\right)+P_{\zeta}\left(k_{3}\right)P_{\zeta}\left(k_{1}\right)\biggr]\,.%{aligned}
\label{bispectrum}
\end{equation}
And $f_{{\rm NL}}(k_{1},k_{2},k_{3})$ signifies the shape of the bispectrum, it is also called the non-linear or non-Gaussianity parameter. 

The 3-point correlation function usually computed in $in-in$ formalism which is the standard method of quantum field theory. For this we require the interaction Hamiltonian $(\mathcal{H}_{int})$ which can be derived from the 3rd order perturbation of the inflationary action, 

\begin{equation}
\langle\zeta\left(t,\,\mathbf{k_{1}}\right)\zeta\left(\mathbf{t,\,k_{2}}\right)\zeta\left(t,\,\mathbf{k_{3}}\right)\rangle =
-i\int^{t}_{t_{0}} dt^{\prime}\,\langle\left[\zeta\left(t,\,\mathbf{k_{1}}\right)\zeta\left(\mathbf{t,\,k_{2}}\right)\zeta\left(t,\,\mathbf{k_{3}}\right),\,
\mathcal{H}_{int}(t^{\prime})\right] \,.
\end{equation}
The computation of 3-point function is well known for various models, can be read from \cite{Maldacena:2002vr,Chen:2005fe,Gao:2008dt,DeFelice:2013ar}.

\section{Single field consistency relations}

\label{slacs}

The following two consistency relations can ultimately test (standard) scalar field inflation\footnote{These relations are also valid for Starobinsky and Higgs inflation as their action can be written in the form of the standard scalar action, with the assistance of a conformal transformation \cite{Mukhanov:1990me,Bezrukov:2007ep}.}. 

\begin{itemize}
\item Tensor consistency relation that corresponds to the relation between tensor to scalar ratio and the tensor tilt as $r=-8n_{t}$ \cite{Baumann:2009ds}.  
\item Maldacena consistency relation \cite{Maldacena:2002vr,Creminelli:2004yq} that relates the bispectrum in the squeezed limit to the scalar spectral index as $\langle\zeta_{\mathbf{k_{1}}}\zeta_{\mathbf{k_{2}}}\zeta_{\mathbf{k_{3}}}\rangle= (2\pi)^{2}\delta^{3}\left(\sum \mathbf{k}_{i}\right)(1-n_{s})\mathcal{P}_{\zeta_{k_{1}}}\mathcal{P}_{\zeta_{k_{2}}}$. In other words, the "local" shape of non-Gaussianity is proportional to the scalar tilt as $\fnl^{local}=\frac{5}{12}\left(1-n_{s}\right)$ and it was argued that this relation holds not only for standard single but for any general single scalar field  inflation \cite{Creminelli:2004yq}. 
\end{itemize}

%Consequently, equations of motion in the FRLW background read as
%\begin{equation}
%\ddot{\varphi}+3H\dot{\varphi}+V^{\prime}(\varphi)=0\,,\label{inflEQ}
%\end{equation}
%where the energy density and pressure of the scalar field $\rho_{\varphi}\,,P_{\varphi}$ relates to the Hubble parameter $(H=\frac{\dot{a}}{a})$, as
%\begin{align}
%H^2=&\frac{m_{\rm P}^{2}}{3}\rho_{\varphi}=\frac{m_{\rm P}^{2}}{3}\left[\frac{1}{2}\dot{\varphi}^2+V\right]\,,\\
%\dot{H}=&-\frac{1}{2}\dot{\varphi}^2\,.\label{InflHubble}
%\end{align}

%\subsection{Slow-roll approximation}
%\label{slacs}

%%%%%%%%%%%%%%%%%%%%%%%%%%%% G-inflation 

%In this appendix we portray the generalized approach for calculating the scalar and tensor power spectrum as described
%in the context of generalized G-inflation or most general scalar tensor theory in (\ref{G-action}) \cite{Kobayashi:2011nu,Escamilla-Rivera:2015ova}. We present the calculations of spectral index $\left(n_{s}\right)$,
%tensor-to-scalar ratio $\left(r\right)$ and tensor tilt $\left(n_{t}\right)$ up to the slow-roll approximation.

\section{Power spectra in generalized G-inflation}

\label{G-infapp}

The most general scalar-tensor theory in 4D with second order field equations\footnote{That is free from Ostrogradski instabilities \cite{Woodard:2015zca}.} \cite{Horndeski:1974wa,Kobayashi:2011nu} is given by the Lagrangian 
\begin{equation}
\mathcal{S}_G=\int d^{4}x\sqrt{-g}\biggl[\frac{m_{\text P}^{2}}{2}\, R+P(\phi,X)-G_{3}(\phi,X)\,\Box\phi+\mathcal{L}_{4}+\mathcal{L}_{5}\biggr]\,,\label{G-action}
\end{equation}
where 
\begin{align}
\mathcal{L}_{4} = & G_{4}(\phi,X)\, R+G_{4,X}\,[(\Box\phi)^{2}-(\nabla_{\mu}\nabla_{\nu}\phi)\,(\nabla^{\mu}\nabla^{\nu}\phi)]\,,\\
\mathcal{L}_{5} = & G_{5}(\phi,X)\, G_{\mu\nu}\,(\nabla^{\mu}\nabla^{\nu}\phi)-\frac{1}{6}G_{5,X}[(\Box\phi)^{3}-3(\Box\phi)\,(\nabla_{\mu}\nabla_{\nu}\phi)
\,(\nabla^{\mu}\nabla^{\nu}\phi) \nonumber \\ 
&+2(\nabla^{\mu}\nabla_{\alpha}\phi)\,(\nabla^{\alpha}\nabla_{\beta}\phi)
\,(\nabla^{\beta}\nabla_{\mu}\phi)]\,.
\label{GST-action}
\end{align}
Here $P$ and $G_{i}$'s ($i=3,4,5$) are functions in terms of $\phi$
and $X=-\partial^{\mu}\phi\partial_{\mu}\phi/2$ with the partial
derivatives $G_{i,X}\equiv\partial G_{i}/\partial X$, and $G_{\mu\nu}=R_{\mu\nu}-g_{\mu\nu}R/2$. 

\section*{Scalar power spectrum}

The second order action of (\ref{G-action}) for scalar perturbations is given by \cite{Kobayashi:2011nu,Escamilla-Rivera:2015ova},
\begin{equation}
S_{s}^{(2)}=\int\, dt\, d^{3}x\left(a^{3}\mathcal{G}_{s}\right)\left[\dot{\zeta}-\frac{\mathcal{F}_{s}/\mathcal{G}_{s}}{a^{2}}\left(\nabla\zeta\right){}^{2}\right]\,,\label{S2pert}
\end{equation}
where $\mathcal{F}_{s}$, $\mathcal{G}_{s}$ are arbitrary functions of
time\footnote{Whose definitions can be found in \cite{Kobayashi:2011nu}.} and \mbox{$c_{s}\equiv(\mathcal{F}_{s}/\mathcal{G}_{s})^{1/2}$}
is the sound speed for scalar perturbations. In addition to the slow-roll conditions in (\ref{SLRconditions}), we introduce the following new parameters which has to be sufficiently small during inflation \cite{Kobayashi:2011nu}. 
\begin{equation}
f_{s}\equiv\frac{d\ln\mathcal{F}_{s}}{d\ln a}\quad,\quad f_{s}^{(2)}\equiv\frac{d\ln f_{s}}{d\ln a}\quad,\quad g_{s}\equiv\frac{d\ln\mathcal{G}_{s}}{d\ln a}\quad,\quad g_{s}^{(2)}\equiv\frac{d\ln g_{s}}{d\ln a}\,.\label{ssroll}
\end{equation}
Moreover, using the definition of $c_{s}$ we have 
\begin{equation}
\epsilon_{s}\equiv\frac{d\ln c_{s}}{d\ln a}=\frac{1}{2}\,\left(f_{s}-g_{s}\right)\quad,\quad\eta_{s}\equiv\frac{d\ln\epsilon_{s}}{d\ln a}=\frac{1}{2\epsilon_{s}}\,\left(f_{s}f_{s}^{(2)}-g_{s}g_{s}^{(2)}\right)\quad.\label{sroolscs}
\end{equation}

To quantify the amplitude and tilt of the spectrum we introduce the
variables $dy_{s}\equiv\frac{c_{s}}{a}\, dt$, $z_{s}\equiv\sqrt{2}a(\mathcal{F}_{s}\mathcal{G}_{s})^{1/4}$
and $u\equiv z_{s}\zeta$, using which the action (\ref{S2pert})
can be canonically normalized 
\begin{equation}
S_{s}^{(2)}=\frac{1}{2}\int dy_{s}\, d^{3}x\left[(u')^{2}-(\nabla u)^{2}+\frac{z_{s}''}{z_{s}}u^{2}\right]\,.\label{SS2}
\end{equation}

Imposing the Bunch-Davies vacuum initial condition in the subhorizon limit
$c_{s}k\gg aH$, the solution for perturbation mode $u$ is given by
\begin{equation}
u_{k}=\frac{\sqrt{\pi}}{2}\sqrt{-y_{s}}\, H_{\nu_{s}}(-ky_{s})\quad,\quad\nu_{s}^{2}-\frac{1}{4}\equiv y_{s}^{2}\frac{z_{s}''}{z_{s}}\,.\label{B2}
\end{equation}
Using now $\zeta_{k}=u_{k}/z_{s}$, we obtain the the scalar power spectrum as

\begin{equation}
{\cal P}_{\zeta}=\frac{k^{3}}{2\pi^{2}}\vert\zeta_{k}\vert^{2}=\frac{\gamma_{s}}{2}\frac{\mathcal{G}_{s_{\ast}}^{1/2}}{\mathcal{F}_{s_{\ast}}^{3/2}}\frac{H_{\ast}^{2}}{4\pi^{2}}\quad,\quad\gamma_{s}\equiv2^{2\nu_{s}-3}\frac{\Gamma\left(\nu_{s}\right){}^{2}}{\Gamma(3/2)^{2}}\left(1-\epsilon_{\ast}+\frac{g_{s_{\ast}}}{2}-\frac{f_{s_{\ast}}}{2}\right)^{2}\,,\label{pwrspectrum}
\end{equation}
where ``$\ast$'' labels the time of sound horizon crossing when $ky_{s}=-1$.  

To compute the spectral index of the scalar perturbations 
\begin{equation}
n_{s}-1\equiv3-2\nu_{s}\,,
\end{equation}
first we need to compute $\nu_{s}$. Using the definition of $z_{s}$
we find 
\begin{equation}
\begin{aligned}
\frac{z_{s}''}{z_{s}}= & \left(\frac{Ha}{c_{s}}\right)^{2}\Big[\left(1+\frac{f_{s}+g_{s}}{4}\right)^{2}+\left(1-\epsilon-\frac{f_{s}}{2}+\frac{g_{s}}{2}\right)\left(1+\frac{f_{s}+g_{s}}{4}\right)\\
&-\frac{f_{s}f_{s}^{(2)}-g_{s}g_{s}^{(2)}}{4}\Big]\,.
\end{aligned}
\label{B3}
\end{equation}
The next step is to integrate $dy_{s}=(c_{s}/a)\, dt$. Assuming small
and constant $\eta$ and $\eta_{s}$ to neglect second order terms
and integrating by parts we obtain 
\begin{equation}
y_{s}=-\frac{c_{s}}{(1-\epsilon-\epsilon_{s})\, aH}\left(1+\frac{\epsilon\eta+\epsilon_{s}\eta_{s}}{(\epsilon+\epsilon_{s}-1)^{2}}\right)\,.\label{B4}
\end{equation}
%which is valid for somewhat large values of $\epsilon$ and $\epsilon_{s}$ provided $\eta,\eta_{s}$ are sufficiently small. Using now (\ref{B2})-(\ref{B4})
%Neglecting second order terms in $\eta,\eta_{s}$ we arrive at

%\begin{equation}
%\frac{z_{s}^{\prime\prime}}{z_{s}}=\left(\frac{aH}{c_{s}}\right)^{2}\left[\left(1+\frac{f_{s}+g_{s}}{4}\right)\left(1-\epsilon-\frac{f_{s}}{2}+\frac{g_{s}}{2}\right)+\left(1+\frac{f_{s}+g_{s}}{2}\right)^{2}-\frac{\left(f_{s}f_{s}^{(2)}-g_{s}g_{s}^{(2)}\right)}{4}\right]\label{zspp}
%\end{equation}

If the slow-roll
parameters are sufficiently small, in the linear approximation 
\cite{Khoury:2008wj,Ribeiro:2012ar}, the scalar spectral
index can be written as 
\begin{equation}
n_{s}-1\simeq\frac{4\epsilon_{\ast}+3f_{s_{\ast}}-g_{s_{\ast}}}{-2+2\epsilon_{\ast}+f_{s_{\ast}}-g_{s_{\ast}}}\,\,.\label{nsL}
\end{equation}

Using (\ref{nsL}) we can compute the running index $n'_{s}\equiv\frac{dn_{s}}{d\ln k}$.
Since we assumed $\eta,\eta_{s}$ approximately constant and small,
the use of (\ref{B4}) allows us to write 
\begin{equation}
n'_{s}=-\frac{y_{s}aH}{c_{s}}\,\frac{dn_{s}}{d\ln a}\simeq\frac{1}{(1-\epsilon-\epsilon_{s})}\left(1+\frac{\epsilon\eta+\epsilon_{s}\eta_{s}}{(\epsilon+\epsilon_{s}-1)^{2}}\right)\,\frac{dn_{s}}{d\ln a}\simeq\frac{1}{(1-\epsilon-\epsilon_{s})}\,\frac{dn_{s}}{d\ln a}\,.\label{B7}
\end{equation}
Provided $\eta$ and $\eta_{s}$ are small and approximately constant,
we can expand (\ref{nsL}) to first order in $\eta,\eta_{s}$. Using (\ref{sroolscs}) and (\ref{B7}) the running index becomes
\[
n_{s}'\simeq\frac{2\epsilon_{\ast}f_{s_{\ast}}
(4-f_{s_{\ast}}+g_{s_{\ast}})-2g_{s_{\ast}}g_{s_{\ast}}^{(2)}(1+\epsilon_{\ast})}{(2-2\epsilon_{\ast}-f_{s_{\ast}}+g_{s_{\ast}})^{2}}\,.
\]

For the action with $P(X,\,\phi)=K(\phi)X-V(\phi)$ and $\mathcal{L}_{4}=\mathcal{L}_{5}=0$, $\nu_{s}$ can be computed up to the third order in the parameters $ \epsilon,\,\eta $, by using the definition of $z_{s}$ and (\ref{B4}), we obtain\footnote{The expression (\ref{nusfull}) we use it in chapter \ref{non-slow-roll-dynamics}.} \cite{Kumar:2015mfa}
\begin{equation}
\begin{aligned}
\nu_{s}= & \left(\frac{3}{2}+\epsilon+\epsilon^{2}
+\epsilon^{3}\right)+\left(\frac{1}{2}+2\epsilon+\frac{29\epsilon^{2}}{6}+\frac{82\epsilon^{3}}{9}\right)\eta+\left(-\frac{1}{6}+\frac{23\epsilon}{18}+\frac{1069\epsilon^{2}}{108}+\frac{5807\epsilon^{3}}{162}\right)\eta^{2}\\
 & +\left(\frac{1}{18}+\frac{23\epsilon}{54}+\frac{707\epsilon^{2}}{108}+\frac{19633\epsilon^{3}}{486}\right)\eta^{3}+\mathcal{O}\left(\epsilon^{4}\,,\,\eta^{4}\right)\,.
\end{aligned}
\label{nusfull}
\end{equation}

\section*{Tensor power spectrum}
Similarly to the case of scalar perturbations, the second order action
for tensor perturbations can be written as \cite{Kobayashi:2011nu}
\begin{equation}
S_{t}^{(2)}=\frac{1}{8}\int\, dt\, d^{3}x\left(a^{3}\mathcal{G}_{t}\right)\left[\dot{h}_{ij}^{2}-\frac{\mathcal{F}_{t}/\mathcal{G}_{t}}{a^{2}}(\nabla h_{ij})^{2}\right]\,,\label{st2}
\end{equation}
where $\mathcal{F}_{t}$ and $\mathcal{G}_{t}$ are functions of time
and $c_{t}\equiv(\mathcal{F}_{t}/\mathcal{G}_{t})^{1/2}$ is the sound
speed for tensor perturbations. 

Similarly to (\ref{ssroll}), we consider now the additional
slow-roll parameters 
\begin{equation}
f_{t}\equiv\frac{d\ln\mathcal{F}_{t}}{d\ln a}\quad,\quad f_{t}^{(2)}\equiv\frac{d\ln f_{t}}{d\ln a}\quad,\quad g_{t}\equiv\frac{d\ln\mathcal{G}_{t}}{d\ln a}\quad,\quad g_{t}^{(2)}\equiv\frac{d\ln g_{t}}{d\ln a}\,,\label{trolls}
\end{equation}
and using the definition of $c_{t}$ we also have 
\begin{equation}
\epsilon_{t}\equiv\frac{d\ln c_{t}}{d\ln a}=\frac{1}{2}\,\left(f_{t}-g_{t}\right)\quad,\quad\eta_{t}\equiv\frac{d\ln\epsilon_{t}}{d\ln a}=\frac{1}{2\epsilon_{t}}\,\left(f_{t}f_{t}^{(2)}-g_{t}g_{t}^{(2)}\right)\,.\label{trollsct}
\end{equation}

Similarly to the case of the scalar spectrum, we introduce the variables
$dy_{t}\equiv\frac{c_{t}}{a}\, dt$, $z_{t}\equiv\frac{a}{2}(\mathcal{F}_{t}\mathcal{G}_{t})^{1/4}$
and $u_{ij}\equiv z_{t}h_{ij}$ so that the action in (\ref{st2})
can be canonically normalized 
\begin{equation}
S_{t}^{(2)}=\frac{1}{2}\int dy_{t}\, d^{3}x\left[(u_{ij}')^{2}-(\nabla u_{ij})^{2}+\frac{z_{t}''}{z_{t}}u_{ij}^{2}\right]\,.\label{B9}
\end{equation}
Imposing the Bunch-Davies vacuum initial condition as in (\ref{B2})
we find 
\begin{equation}
u_{ij}=\frac{\sqrt{\pi}}{2}\sqrt{-y_{t}}\, H_{\nu_{t}}^{(1)}(-ky_{t})\, e_{ij}\,\quad,\quad\nu_{t}^{2}-\frac{1}{4}\equiv y_{t}^{2}\frac{z_{t}''}{z_{t}}\,.\label{B10}
\end{equation}
where $e_{ij}$ is the polarization tensor and 

\begin{equation}
\begin{aligned}
\frac{z_{t}^{\prime\prime}}{z_{t}}=&\left(\frac{aH}{c_{s}}\right)^{2}\Big[\left(\frac{f_{t}+g_{t}}{4}+1\right)\left(-\frac{f_{t}}{2}+\frac{g_{t}}{2}-\epsilon+1\right)+\left(\frac{f_{t}+g_{t}}{2}+1\right)^{2} \\
&-\frac{\left(f_{t}f_{t}^{(2)}-g_{t}g_{t}^{(2)}\right)}{4}\Big]\,.
\end{aligned}
\label{zttz}
\end{equation}

Using that $h_{ij}=u_{ij}/z_{t}$
and taking into account the two polarization states, we arrive at
the tensor power spectrum given by 

\begin{equation}
{\cal P}_{t}=8\gamma_{t}\frac{\mathcal{G}_{t_{\ast}}^{1/2}}{\mathcal{F}_{t_{\ast}}^{3/2}}\frac{H_{\ast}^{2}}{4\pi^{2}}\quad,\quad\gamma_{t}\equiv2^{2\nu_{t}-3}\frac{\Gamma\left(\nu_{t}\right){}^{2}}{\Gamma(3/2)^{2}}\left(1-\epsilon_{\ast}+\frac{g_{t_{\ast}}}{2}-\frac{f_{t_{\ast}}}{2}\right)^{2}\,.\label{ptspectrum}
\end{equation}

The spectral index of the tensor spectrum is 
\begin{equation}
n_{t}\equiv3-2\nu_{t}\,,\label{nt}
\end{equation}
Similar to the scalar tilt, if slow-roll parameters are sufficiently small we can write the tensor tilt as \cite{Khoury:2008wj,Ribeiro:2012ar}
\begin{equation}
n_{t}\simeq\frac{4\epsilon_{\ast}+3f_{t_{\ast}}-g_{t_{\ast}}}{-2+2\epsilon_{\ast}+f_{t_{\ast}}-g_{t_{\ast}}}\,\,,\label{ntL}
\end{equation}
where the subindex ``$\ast$'' indicates the time of sound horizon
crossing, determined by the condition $ky_{t}=-1$.

In the case of $P(X,\,\phi)=K(\phi)X-V(\phi)$ and $\mathcal{L}_{4}=\mathcal{L}_{5}=0$, calculating $\nu_{t}$ up to the third order in the parameters $ \epsilon,\,\eta $,
by using the definition of $z_{t}$, we obtain \cite{Kumar:2015mfa}
\begin{equation}
\begin{aligned}\nu_{t}= & \left(\frac{3}{2}+\epsilon+\epsilon^{2}+\epsilon^{3}\right)+\left(\frac{4\epsilon}{3}+\frac{37\epsilon^{2}}{9}+\frac{226\epsilon^{3}}{27}\right)\eta+\left(\epsilon+\frac{227\epsilon^{2}}{27}+\frac{875\epsilon^{3}}{27}\right)\eta^{2}+\\
 & \left(\frac{28\epsilon^{2}}{9}+\frac{6491\epsilon^{3}}{243}\right)\eta^{3}+\mathcal{O}\left(\epsilon^{4}\,,\,\eta^{4}\right)\,.
\end{aligned}
\label{nut}
\end{equation}

Finally, the tensor to scalar ratio in generalized G-inflation is 
\begin{equation}
r\equiv\frac{{\cal P}_{t_{\ast}}}{{\cal P}_{\zeta_{\ast}}}=16\frac{\gamma_{t}}{\gamma_{s}}\left(\frac{\mathcal{G}_{t_{\ast}}}{\mathcal{G}_{s_{\ast}}}\right)^{1/2}\left(\frac{\mathcal{F}_{s_{\ast}}}{\mathcal{F}_{t_{\ast}}}\right)^{3/2}\,.\label{r}
\end{equation}
From (\ref{ntL}) and (\ref{r}) we observe that the standard single-field inflationary consistency relation, $r=-8n_{t}$, is in general violated. In Ref.~\cite{Unnikrishnan:2013rka} it has been shown that, in the case of power law G-inflation, we can have either $r>-8n_{t}$ or $r\leq-8n_{t}$ depending on the model parameters. However, the requirement of subluminal propagation speed of the scalar perturbations restricts $r\leq-\frac{32}{3}n_{t}$.

% Appendix A
%\chapter{Inflation driven by 3-form fields} % Main appendix title
%\label{AppendixA} % For referencing this appendix elsewhere, use \ref{AppendixA}
\chapter{Stability of type I fixed points}
\label{AI3-forms}

\lhead{Appendix B: Inflation driven by 3-form fields}% Inflation driven by 3-form fields} % This is for the header on each page - perhaps a shortened title

This appendix provides a complementary feature for chapter  \ref{In-3-forms}. %it includes a review of the gravitational collapse of a standard scalar field. 
%%

%\vspace*{3mm}

%\section{Stability of type I fixed points}

%\vspace*{3mm}

Let us now discuss the stability of these fixed points for specific
choice of potentials. The eigenvalues of ${\cal M}_{ij}$ corresponding
to the fixed point $\left(\chi_{1c},\, w_{1c}\right)$ are $\zeta_{1}=-3$,
$\zeta_{2}=0$. Since the second eigenvalue is zero, we cannot decide
on the stability of this fixed point. The eigenvector for the null
eigenvalue is given by 
\begin{equation}
v_{0}=\left(\begin{array}{c}
\sqrt{2/3}\\
1
\end{array}\right).\label{eigenv-1}
\end{equation}
Let us consider the nonlinear order perturbation in the expansion
\begin{equation}
\delta r'=\mu^{\left(n\right)}\delta r^{n}\label{pertur-eq.}
\end{equation}
where $\delta r=\sqrt{2/3}\delta \chi_{1}+\delta w_{1}$ is the perturbation
along the direction of the eigen vector (\ref{eigenv-1}). The general
solution of (\ref{pertur-eq.}) at order $n$ is 
\begin{equation}
\dfrac{\delta r^{\left(-n+1\right)}}{\left(-n+1\right)}=\mu^{\left(n\right)}N+\dfrac{\delta r_{0}^{\left(-n+1\right)}}{\left(-n+1\right)}\quad\textrm{with}\quad\delta r_{0}=\delta r\,\left(N=0\right).\label{pertur-eq-sol}
\end{equation}
For $n>1$, an initial negative perturbation $\delta r_{0}<0$ will
decay if $\mu^{\left(n\right)}$ is positive, with $n$ even, or $\mu^{\left(n\right)}$
is negative and $n$ odd. If the initial perturbation is positive,
then it will decay for $\mu^{\left(n\right)}$ is negative, for all
$n>1$. If we require that $\mu^{\left(1\right)}=1$ in (\ref{pertur-eq.}),
we must have $\delta \chi_{1}=\sqrt{3/2}\delta r/2$ and $\delta w_{1}=\delta r/2$.
The procedure consists in evaluating 
\begin{equation}
\delta r'=\sqrt{2/3}\delta \chi_{1}'+\delta w_{1}'\label{pertur-eq-exp}
\end{equation}
and collecting the second order terms of the expansion of (\ref{Dyn-x1})
and (\ref{Dyn-w1}) when the dynamical system is perturbed around
the fixed point 
\begin{eqnarray*}
\chi_{1} & = & \sqrt{\frac{2}{3}}\cos\theta+\delta \chi_{1}\,,\\
w_{1} & = & \cos\theta+\delta w_{1}.
\end{eqnarray*}
As the constraint~(\ref{eps -constr-1}) imposes that the dynamical
system, near the fixed point, can only be subjected to a small negative
perturbation, thus, we will consider an initial negative perturbation
$\delta r_{0}<0$. Otherwise, a positive perturbation, that would
slightly increase the value of the two fields above the fixed point,
would imply that the Friedmann constraint (\ref{2Quad-Fried}) would
blow up to infinity.

As it is seen from (\ref{Dyn-x1}) and (\ref{Dyn-w1}), the
presence of the functions $\lambda_{1}$ and $\lambda_{2}$, which,
in turn, depend on the potentials and their derivatives, does not
allow to study in general the stability of the type I solutions. Therefore,
we illustrate this study for some simple and suitable choice of potentials.

\section{Identical quadratic potentials}

\label{Quadrid}Let us consider the simple case when the two fields
are under the influence of identical quadratic potentials, i.e., $V\left(\chi_{1}\right)=\chi_{1}^{2}$
and $V\left(\chi_{2}\right)=\chi_{2}^{2}$. In this situation, (\ref{Dyn-w1})
and (\ref{Dyn-w2}) exhibit type I solutions for any $0<\theta<\pi/2$.
The fixed points for these solutions are constrained by (\ref{Type I omega}).
Collecting the second order term in (\ref{pertur-eq-exp}) we have
\begin{equation}
\mu^{\left(2\right)}=-\frac{9}{4}\biggl(3\cos\theta+\cos3\theta\biggr)\,,\label{c-quad-quad1}
\end{equation}
which is always negative for $0<\theta\leq\pi/4$. This means that
all fixed points with $0<\theta<\pi/4$ are unstable. If $\theta$
gets larger than $\pi/4$ then the fixed point coordinates $\chi_{2c}>\chi_{1c}$
and we can also collect the second order terms in $\delta r'=\sqrt{2/3}\delta \chi_{2}'+\delta w_{2}'$
for a negative perturbation $\chi_{2}=\sqrt{\frac{2}{3}}\sin\theta+\delta \chi_{2}$.
The coefficient yields 
\begin{equation}
\mu^{\left(2\right)}=-\frac{9}{4}\biggl(3\sin\theta-\sin3\theta\biggr)\,,\label{c-quad-quad2}
\end{equation}
which is always negative for $\pi/4\leq\theta<\pi/2$. When the angle
$\theta$ is close to $\pi/2$, then the 3-form field $\chi_{1}$ approaches
zero and (\ref{c-quad-quad1}) produces positive values for $\mu^{\left(2\right)}$.
This means that in the asymmetric situation where $\chi_{1}\approx0$
and $\chi_{2}\approx\sqrt{2/3}$, the solution $\chi_{1}\left(N\right)$
converges to zero, however $\chi_{2}\left(N\right)$ will be unstable.
In fact, from (\ref{c-quad-quad2}), the second field will eventually
diverge from $\sqrt{2/3}$, when subjected to a small negative perturbation.
Furthermore, the decrease in the value of $\chi_{2}$ implies that the
variable $w_{2}$ will start to fall faster, as (\ref{Dyn-x2})
suggests. The decrease of $\chi_{2}$ will proceed until it reaches zero.
At this point, we can show that both fields will start to oscillate
around zero with a damping factor. The discussion for the situation
where $\theta$ is near zero, is the same, in the sense that the roles
of $\chi_{1}$ and $\chi_{2}$, in the previous discussion, are interchanged.
In Fig.~\ref{fig1} (left panel), the behavior of the two fields,
at the end of the inflationary period, when the angle $\theta$ is
close to $\pi/2$, are shown. Therein, we see that the two fields
are going to a damped oscillatory regime, after the divergence of
$x_{2}$ from its fixed point. The herein analytical description is
numerically confirmed.

\section{Quadratic and quartic potentials}

\label{Quadr-quartic} When the two fields are subjected to the potentials
$V\left(\chi_{1}\right)=\chi_{1}^{2}$ and $V\left(\chi_{2}\right)=\chi_{2}^{4}$,
the evolution is generally of the type II. However, (\ref{Dyn-w1})
and (\ref{Dyn-w2}) exhibit type I solutions when the condition (\ref{cond-V1-V2})
holds, which in this case becomes 
\begin{equation}
\left(\frac{1}{\frac{3}{4}\left(\cot\theta\right){}^{2}\csc\theta+\sin\theta}\right)^{2}+
\left(\frac{6\cos\theta}{6-\cos2\theta+\cos4\theta}\right)^{2}=1\,.
\end{equation}
This last condition is satisfied for $\theta\rightarrow\pi/3$ , $\theta\rightarrow\pi/2$
and at $\theta\rightarrow0$. Collecting the second order term in
(\ref{pertur-eq-exp}) we have 
\begin{equation}
\mu^{\left(2\right)}=-\frac{3}{5}\,,
\end{equation}
which is negative. This means that the fixed point with $\theta=\pi/3$
is unstable. At $\theta=0$,i.e, the scenario with the quadratic term
dominance, we must go to third order since, $\mu^{\left(2\right)}=0$.
In that case, collecting the third order terms we have $\mu^{\left(3\right)}=0.28$,
which means that the fixed point is unstable. At $\theta=\pi/2$,
scenario with the quartic term dominance, $\mu^{\left(2\right)}=-7.5$,
which means that the fixed point is unstable.
\chapter{Analytical approximations} % Main appendix title
%\label{AppendixB} % For referencing this appendix elsewhere, use \ref{AppendixA}
\label{ADBIG1}

\lhead{Appendix C. DBI Galileon inflation} % This is for the header on each page - perhaps a shortened ti

%\section*{Appendices}
%
%\vspace*{10mm}
%%%%%%%%%%%%%%%%%%%%%%%%%%%%%%%%%%%%%%%%%%%%%%%%%%%%%%%%%%

This appendix constitutes a complement for chapter  \ref{DBIGc}.

%%%%%%%%%%%%%%%%%%%%%%%%%%%%%%%%%%%%%%%%%%%%%%%%%%%%%%%%%%%%%%%%%%%%
%%%%%%%%%%%%%%%%%%%%%%%%%%%%%%%%%%%%%%%%%%%%%%%%%%%%%%%%%%%%%%%%%

%\section{Analytical approximations}

%\label{ADBIG1} 

\section*{Parametrization 1}

Using the definition of the hypergeometric function \cite{Abramowitz}
we have 
\begin{equation}
_{2}F_{1}\left(1,1+\beta;2+\beta;z\right)=\frac{\Gamma(2+\beta)}{\Gamma(1+\beta)}\sum_{n=0}^{\infty}\frac{\Gamma(1+n)\Gamma(1+\beta+n)}{\Gamma(2+\beta+b)}\frac{z^{n}}{n!}=(1+\beta)\sum_{n=0}^{\infty}\frac{z^{n}}{1+\beta+n}\,.\label{A1}
\end{equation}
For $\beta<1$ we can approximate 
\begin{equation}
\frac{1}{1+\beta+n}=\left(\frac{1}{1+n}\right)\frac{1}{1+\frac{\beta}{1+n}}\simeq\frac{1}{1+n}\left(1-\frac{\beta}{1+n}\right)=\frac{n+1-\beta}{(n+1)^{2}}\,,\label{A2}
\end{equation}
and substituting in (\ref{A1}) we arrive at 
\begin{equation}
_{2}F_{1}\left(1,1+\beta;2+\beta;z\right)\simeq(1+\beta)\sum_{n=0}^{\infty}\frac{n+1-\beta}{(n+1)^{2}}\, z^{n}=-\frac{(1+\beta)}{z}\left(\ln|1-z|+\beta\,{\rm Li}_{2}(z)\right)\,,\label{A3}
\end{equation}
where ${\rm Li}_{n}(z)=\sum_{k=1}^{\infty}k^{-n}z^{k}$ is the polylogarithm
function \cite{Abramowitz}. Despite its being an excellent approximation
for $\beta<1$, substituting the above into (\ref{PV1hyperG})
leads to a differential equation still too complicated (to solve for
$a(t)$) due to the polylogarithmic function ${\rm Li}_{2}(z)$. Our
aim, therefore, is to find a simple analytical solution reproducing
the qualitative behaviour of the scale factor. The simplest manner
to achieve this is to neglect the term in the polylogarithm function
in (\ref{A3}). This simplification can be justified after approximating
\begin{equation}
\sum_{n=0}^{\infty}\frac{z^{n}}{1+\beta+n}\simeq\sum_{n=0}^{\infty}\frac{z^{n}}{1+n}=-\frac{\ln|1-z|}{z}\label{A5}
\end{equation}
in (\ref{A1}), which holds provided $\beta\ll1$. In that case,
after substituting $z\to\lambda_{1}H^{2}/\lambda_{2}$, the resulting
background equation (\ref{PV1approax}) has the advantage of
being relatively simple.

\section*{Parametrization 2}

Using the variables $z$ and $y$ defined in (\ref{pv2variables}),
our (\ref{BGS-1}) becomes 
\begin{equation}
\lambda_{1\ast}e^{y\alpha_{1}}e^{2z}-\lambda_{2\ast}e^{y\alpha_{2}}-e^{2z}z'(y)=0\,.\label{A6}
\end{equation}
After multiplying by $\mu(y)=\exp\left[-(2\lambda_{1\ast}/\alpha_{1})e^{y\alpha_{1}}\right]$,
(\ref{A6}) becomes an exact differential equation 
\begin{equation}
df=P(y,z)\, dy+Q(y,z)\, dz=0\,,\label{A7}
\end{equation}
where 
\begin{equation}
P(y,z)=\mu(y)\left[\lambda_{1\ast}e^{y\alpha_{1}}e^{2z}-\lambda_{2\ast}e^{y\alpha_{2}}\right]\quad\textrm{and}\quad Q(y,z)=-\mu(y)\, e^{2z}\,.\label{A8}
\end{equation}
Integral curves are of the form: $f(y,z)=\kappa$, where $\kappa$
is a constant. Integrating $f$ with respect to $y$ in the first
place we have 
\begin{equation}
f(y,z)=\int P(y,z)\, dy+g(z)\,,\label{A9}
\end{equation}
where $g(z)$ is to be computed by demanding $\partial_{z}f(y,z)=Q(y,z)$.
After integrating and solving for $g(z)$ we find that the integral
curves $f(y,z)=\kappa$ are determined by (\ref{pv2soln}).
% Appendix C
%\chapter{Effective models of inflation from an SFT inspired framework} % Main appendix title
%\label{AppendixC} % For referencing this appendix elsewhere, use \ref{AppendixC}
\chapter{A review of SFT and Tachyon condensation}
\label{AppSFT}

\lhead{Appendix D. Effective models of inflation from an SFT inspired framework} % This is for the header on each page - perhaps a shortened title

This appendix assists for chapter  \ref{SFTin}. %it includes a review of the gravitational collapse of a standard scalar field. 
%%

%\vspace*{3mm}

%\section{A review of SFT and Tachyon condensation}

%\vspace*{3mm}

In generic words SFT is an off-shell description of interacting strings
\cite{Witten:1985cc,Witten:1986qs,Zwiebach:1993cs,Arefeva:2001ps,Berkovits:1998bt,Berkovits:2004xh}.
It describes a string by means of a string field $\Psi$. This object
is a shorthand for encoding all the string excitations in one instance.
The corresponding action for open string field\footnote{An action for a closed SFT can be written only in a non-polynomial
form, even for the bosonic strings \cite{Saadi:1989tb,Sonoda:1989sj}.} can be written as 
\begin{equation}
S=\frac{1}{g_{o}^{2}}\left(\frac{1}{2}\int\Psi\star Q\Psi+\frac{1}{3}\int\Psi\star\Psi\star\Psi\right)\,,\label{osft}
\end{equation}
where $\star$ and $\int$ are Witten product and integral for string
fields respectively. $Q$ is the BRST charge. The first term clearly
corresponds to the motion of free strings while the second term represents
the interaction. The second term is the three-string vertex responsible
for the non-perturbative physics. $g_{o}$ is the open string coupling
constant, it is dimensionless.

It has been understood \cite{Sen:1998sm,Sen:1999xm,Sen:1999nx,Berkovits:2000hf,Aref'eva:2000mb}
that the tachyon of open strings is responsible for the decay of unstable
$D$-branes or $D$-brane-anti-$D$-brane pairs. The corresponding
process is the condensation of the tachyon (TC) to a non-perturbative
minimum. Upon the TC the unstable brane (or pair) decays. It is the
cornerstone of Sen's conjecture regarding TC that the depth of the
tachyon potential minimum is exactly the tension of an unstable brane
to which the string is attached to. The decay of a brane represents
a configuration in which open strings must not exist, because the
brane, to which they were attached, has decayed \cite{Rastelli:2000hv,Rastelli:2001jb}.
This being said, let us assume Sen's conjecture, which prescribes
the disappearance of open string excitations. The latter phenomenon
of open strings extinction can be formalized as follows in the field-theoretical
language. Given a field $\varphi$ the following quadratic Lagrangians
are non-dynamical 
\begin{equation}
L=-m^{2}\varphi^{2}\text{ or }L=\varphi e^{\gamma(\Box)}\varphi\,.\label{nondyn}
\end{equation}
The left Lagrangian is clearly a mass term without any dynamics. In
the right Lagrangian, $\Box$ is the space-time d'Alembertian and
$\gamma$ is an entire function. Although it may look like $\Box$
produces dynamics as it is a differential operator, as long as we
require that the function in the exponent is an entire function, the
whole exponent has no eigenvalues as an operator. This means that
the inverse of such an exponent gives no poles in the propagator and
effectively we have no dynamics at all.

We further notice that the right Lagrangian in (\ref{nondyn}) is
an essentially non-local Lagrangian. It is obviously non-dynamical
on the quadratic level and as long as the field $\varphi$ is alone.
However, novel and unusual effects can be generated upon coupling
to other fields or in the non-linear physics \cite{Sen:2002nu,Moeller:2002vx,Aref'eva:2003qu,Barnaby:2006hi,Koshelev:2007fi}.

The essence of SFT is that as long as a string interaction is involved
then the non-locality of the above type emerges. Technically, we can
understand this as follows. Strings are extended objects by construction.
When a field-theoretic model describes strings, this property of an
extended object is encoded in the non-locality of interactions. SFT
straightforwardly creates vertex terms of the form 
\begin{equation}
\sim\left(e^{\alpha'\Box}\varphi_{1}\right)\left(e^{\alpha'\Box}\varphi_{2}\right)\left(e^{\alpha'\Box}\varphi_{3}\right)\label{sftvertex}
\end{equation}
Here $\alpha'$ is the string length squared (which may be different
from the inverse of the Planck mass squared). We aim to convey in
the course of this paper\footnote{Computing any process in SFT leads to much more complicated results
than presented above. TC is not an exclusion. Schematically, to describe
the TC we should first compute an effective action in which all massive
modes of a string with positive mass square are integrated out. Upon
this computation a non-local interaction of several tachyons arise.
The non-local operators are not just identical exponents but rather
algebraic combinations of them. This effective action is enough to
test Sen's conjecture for both, depth of the potential and absence
of dynamics at the bottom of the potential. It is worth noting that
actual computations in SFT are indeed difficult and technical performed
by means of a level truncation scheme (i.e. including only fields
up to a given mass $m$ and the next iteration includes fields up
to mass $m+1$, etc. \cite{Kostelecky:1988ta}). This scheme was proven
to be convergent \cite{Kostelecky:1988ta}.} that non-locality indeed proves crucial in constructing (SFT inspired)
cosmological models.

It is sufficient for the purposes of the present paper only to note
that upon lengthy computations \cite{Arefeva:2001ps}, the quadratic
Lagrangian of the open string tachyon $\Tc$ near the vacuum is non-dynamical
of the form 
\begin{equation}
L_{\Tc}=-\frac{{T}}{2}v(\Box,\Tc)\,.\label{tachyonnearvac}
\end{equation}
For zero momenta, i.e. when $\Box=0$ the resulting $v(0,\Tc)$ is
exactly the tachyon potential. The dependence on $\Box$ is analytic
and being linearized near the vacuum value of field $\Tc=\Tc_0+\tau$ it produces
\begin{equation}
L_{\tau}=-\frac{{T}}{2}\frac{v''(\Tc=\Tc_{0})}{2}\tau e^{\gamma(\Box)}\tau\,,\label{tachyonvac}
\end{equation}
with some entire function $\gamma(\Box)$. The coupling ${T}$ is
nothing but the tension of the unstable $D$-brane given as 
\begin{equation}
{T}=\frac{1}{2\pi^{2}g_{o}^{2}(\alpha')^{\frac{p+1}{2}}}\,,\label{tachyoncoupling}
\end{equation}
where $\alpha'$ is the string length squared, $g_o$ is the open string coupling constant and $p$ comes from
the dimensionality of the $Dp$-brane. Thus, as expected for a 3-brane,
$T$ has a dimension $[\mathrm{length}]^{-4}$ and the tachyon field
$\tau$ is dimensionless.

\addtocontents{toc}{\vspace{2em}} % Add a gap in the Contents, for aesthetics

\backmatter

%----------------------------------------------------------------------------------------
%	BIBLIOGRAPHY
%----------------------------------------------------------------------------------------

%----------------------------------------------------------------------------------------
%	BIBLIOGRAPHY
%----------------------------------------------------------------------------------------
\clearpage
\newpage
%\backmatter
\parskip 0mm
%\bibliographystyle{Classes/jmb}
%\bibliographystyle{ama}

%\lhead{References}
%\renewcommand{\bibname}{References} % changes default name Bibliography to References

%\newpage
%\phantomsection \label{bibliography}
%\addcontentsline{toc}{chapter}{References} %adds References to contents page
\bibliographystyle{utphys}
%\bibliography{RefThesis}
\providecommand{\href}[2]{#2}\begingroup\raggedright\endgroup

%\label{Bibliography}

%\lhead{\emph{Bibliography}} % Change the page header to say %"Bibliography"

%\bibliographystyle{unsrtnat} % Use the "unsrtnat" BibTeX style %for formatting the Bibliography

%\bibliography{Bibliography} % The references (bibliography) %information are stored in the file named "Bibliography.bib"

\end{document}